\documentclass[11pt,a4paper]{article}
\pdfoutput=1
\usepackage{jheppub}
\usepackage[T1]{fontenc}
\usepackage{amsfonts}
\usepackage{amsmath}
\usepackage{mathtools}
\usepackage{color}
\usepackage[normalem]{ulem}
\usepackage{verbatim}
\usepackage{subcaption}
\usepackage{multirow}

\newcommand{\plus}[1]{\Bigl[\frac{\theta(#1)}{#1}\Bigr]_+}
\newcommand{\pluss}[1]{\Bigl[\frac{\theta(#1)\ln #1}{#1}\Bigr]_+}
\newcommand{\asCF}[1]{\frac{\alpha_s(#1)C_F}{4\pi}}
\newcommand{\Oas}[1]{\mathcal{O}(\alpha_s^{#1})}
\newcommand{\dd}[2]{\frac{\mathrm{d}^{#1}#2}{(2\pi)^{#1}}}
\newcommand{\as}{\alpha_s}
\newcommand{\Gcusp}{\Gamma^{\mathrm{cusp}}}

\newcommand{\Herwig}{\textsc{Herwig}}
\newcommand{\Matchbox}{\textsc{Matchbox}}

\newcommand{\gsim}{\raisebox{-0.13cm}{~\shortstack{$>$ \\[-0.07cm]
			$\sim$}}~}
\newcommand{\lsim}{\raisebox{-0.13cm}{~\shortstack{$<$ \\[-0.07cm]
			$\sim$}}~}

\title{On the Cutoff Dependence of the Quark Mass Parameter in Angular Ordered Parton Showers}

\author[a,b]{Andr\'e H. Hoang}
\author[a]{Simon Pl\"atzer}
\author[a]{Daniel Samitz}

\affiliation[a]{University of Vienna, Faculty of Physics, \\Boltzmanngasse 5, A-1090 Wien, Austria}
\affiliation[b]{Erwin Schr\"odinger International Institute for Mathematical Physics,\\
	University of Vienna, Boltzmanngasse 9, A-1090 Wien, Austria}

\preprint{UWTHPH-2018-20 MCnet-18-17}

\abstract{
We show that the presence of an infrared cutoff $Q_0$ in the parton shower (PS) evolution for massive quarks 
implies that the generator quark mass corresponds to a $Q_0$-dependent 
short-distance mass scheme and is therefore not the pole mass.  
Our analysis considers an angular ordered parton shower 
based on the coherent branching formalism for quasi-collinear stable heavy quarks and splitting functions at next-to-leading logarithmic (NLL) order, and 
it is based on the analysis of the peak of hemisphere jet mass distributions. We show that NLL shower 
evolution is sufficient to describe the peak jet mass at full next-to-leading order (NLO). 
We determine the relation of this short-distance mass to the pole mass at NLO.
We also show that the shower cut $Q_0$ affects soft radiation in a universal way for 
massless and quasi-collinear massive quark production. The basis of our analysis is 
(i) an analytic solution of the PS evolution based on the coherent branching formalism, 
(ii) an implementation of the infrared cut $Q_0$ of the angular ordered shower into factorized analytic 
calculations in the framework of Soft-Collinear-Effective-Theory (SCET) and 
(iii) the dependence of the peak of the jet mass distribution on the shower cut. 
Numerical comparisons to simulations with the \Herwig\ 7 event generator confirm our findings. 
Our analysis provides an important 
step towards a full understanding concerning the interpretation of top quark mass measurements based on 
direct reconstruction.
	}

\begin{document}

\maketitle

\section{Introduction}
\label{sec:intro}

\subsection{Prelude and review}
\label{sec:prelude}

A precise determination of the top quark mass $m_t$ represents one of the most important measurements in the context of studies of the Standard Model (SM) as well as of new physics, in particular in the context of electroweak symmetry breaking. The most precise top mass measurements are obtained from template and matrix element fits which are based on the idea of accessing $m_t$ by directly reconstructing the kinematic properties of a top quark ``particle''. These types of measurements naturally yield a very high sensitivity to the top quark mass because they involve endpoints, thresholds or resonant structures in kinematic distributions which substantially reduces the impact of uncertainties that affect poperties such as their normalization. 
The most recent reconstruction measurements are $m_t^{\rm MC} = 172.44(49)$ GeV (CMS)~\cite{Khachatryan:2015hba}, $m_t^{\rm MC} = 172.84(70)$ GeV (ATLAS)~\cite{Aaboud:2016igd} and  $m_t^{\rm MC} = 174.34(64)$ GeV (Tevatron)~\cite{Tevatron:2014cka}.

The characteristic property of these measurements, however, is that the observables employed for the reconstruction analyses are too complicated to be calculated in a systematically improvable way and, in addition, involve sizeable perturbative and non-perturbative corrections due to soft gluon emission which, in the vicinity of kinematic endpoints or thresholds, are not power-suppressed. The theoretical computations used for these measurements are therefore based on multi-purpose Monte Carlo (MC) event generators since they can produce predictions for essentially any conceivable observable. As a consequence, in these direct mass measurements the top mass parameter $m_t^{\rm MC}$ of the MC generator employed in the analyses is determined. The experimental collaborations provide estimates of the theoretical uncertainty in the extracted value of $m_t^{\rm MC}$ concerning the quality of the modelling of non-perturbative effects, e.g.\ by using different tunes or MC generators, or concerning theoretical uncertainties, e.g.\ by variations of theory parameters. The improvement of the theoretical basis of MC event generators and of methods to estimate their uncertainties is an ongoing effort~\cite{Bellm:2016rhh,Bendavid:2018nar,Dasgupta:2018nvj}. 

However, the intrinsic, i.e.\ quantum field theoretic meaning of $m_t^{\rm MC}$ has up to now not been rigorously specified. Since this matter goes beyond the task of properly estimating or reducing MC modelling uncertainties and is also tied to the constructive elements incorporated to the MC's perturbative and non-perturbative components, it is much harder to quantify. Issues one has to consider do not only involve the truncations of perturbative QCD expansions, but also MC specific implementations such as the cut on the PS evolution or even modifications that are formally subleading but play numerically important roles in reaching better agreement with data or are part of the implementation of the hadronization model. It should also be remembered that the level of theoretical rigor of MC event generators depends on the observable. 
Since the theoretical description of thresholds and endpoints in general involves the resummation of QCD radiation to all orders, the perturbative aspect of how to interpret $m_t^{\rm MC}$ thus significantly depends on the implementation of the parton showers that are used in the MC generators and to the extent that NLO fixed-order QCD corrections have been systematically implemented for the observables that are relevant for the reconstruction analyses. Apart from that, the interface between the perturbative components and the hadronization models, which involves the structure of the infrared cut of the shower evolution,  $Q_0\sim 1$~GeV, or the treatment of the top quarks finite width, $\Gamma_t\sim 1.4$~GeV, and other finite lifetime effects can play essential roles. Finally, it should also be mentioned that $m_t^{\rm MC}$ may also be affected by non-perturbative MC modelling effects as a consequence of the tuning process partly compensating for approximations and model-like features implemented into the MC perturbative components.

So, although $m_t^{\rm MC}$ is by construction closely related to the concept of a kinematic top quark mass, the identification to a particular kinematic mass scheme is far from obvious - also because there are several options for kinematic masses including schemes such the pole mass $m_t^{\rm pole}$ or short-distance threshold masses as they are employed for the top pair threshold cross section at a future Linear Collider~\cite{Baer:2013cma,Asner:2013hla,Vos:2016til} or in the context of massive quark initiated jets~\cite{Jain:2008gb,Butenschoen:2016lpz}. As shown in Ref.~\cite{Hoang:2000yr}, these kinematic mass schemes can differ by more than $1$~GeV. Given that the reconstruction analyses have reached uncertainties at the level of $0.5$~GeV it appears evident that systematic and quantitative examinations on the field theoretic meaning of the MC top mass $m_t^{\rm MC}$ are compulsory. This scrutiny may involve examinations of different MC generators, as well as the respective interplay of their perturbative and non-perturbative components. 

So far, only a limited number of theoretical considerations dedicated to this issue exist in the literature. In Ref.~\cite{Hoang:2008xm}, based on the analogy of the MC components to the QCD factorization for boosted top quark initiated jet masses in the peak region derived in the factorization framework of Soft-Collinear-Effective Theory (SCET) and boosted Heavy-Quark-Effective-Theory (bHQET)~\cite{Fleming:2007qr,Fleming:2007xt}, it was conjectured that the
relation between $m_t^{\rm MC}$ and the pole mass is given by
$m_t^{\rm pole} - m_t^{\rm MC} = R_{\rm sc} (\alpha_s/\pi)$, where the scale $R_{\rm sc}$ should be closely related to the shower cut $Q_0$. The conjecture was based on general considerations how an infrared cut affects perturbative calculations but did not provide a precise quantitative relation. It was, however, argued that the uncertainty in the relation is unlikely to exceed the level of $1$~GeV. A similar conclusion was drawn in Ref.~\cite{Hoang:2014oea} where it was argued that $m_t^{\rm MC}$, due to the effects of the hadronization models, may have properties analogous to the mass of a top (heavy-light) meson. Based on the concepts of heavy quark symmetry~\cite{Neubert:1993mb,Manohar:2000dt} the relation $m_t^{\rm MC} = m_t^{\rm MSR}(R) + \Delta_{t,{\rm MC}}(R)$ was conjectured, where $m_t^{\rm MSR}$ is the MSR mass~\cite{Hoang:2008yj,Hoang:2017suc}, the term $\Delta_{t,{\rm MC}}$ contains perturbative as well as non-perturbative corrections and $R=1$~GeV represents a factorization scale separating perturbative and non-perturbative effects. From a comparison of $B$ meson and bottom quark masses, and using heavy quark symmetry, it was concluded that  $\Delta_{t,{\rm MC}}$ could in principle be at the level of $1$~GeV. We also refer to Ref.~\cite{Nason:2017cxd} for a related discussion.

In Ref.~\cite{Butenschoen:2016lpz} the concrete numerical relation $m_t^{\rm MC}=m_t^{\rm MSR}(1\,\mbox{GeV})+(0.18\pm 0.22)$~GeV was obtained from fitting NNLL (next-to-next-to-leading logarithmic) and ${\cal O}(\alpha_s)$ matched factorized hadron level predictions for the 2-jettiness distribution in the peak region for boosted top production in $e^+e^-$ annihilation~\cite{Fleming:2007qr,Fleming:2007xt} to corresponding pseudo-data samples obtained by PYTHIA~8.2~\cite{Sjostrand:2014zea} with the default Monash tune~\cite{Skands:2014pea} correctly accounting for the dominant top quark width effects in the factorized calculation. Here the quoted error is the theoretical uncertainty of the factorized NNLL$+{\cal O}(\alpha_s)$ prediction and also includes an estimate for the intrinsic uncertainty of the PYTHIA~8.2 calculation. 
Using the pole mass scheme in the factorized NNLL$+{\cal O}(\alpha_s)$ prediction, the corresponding analysis yielded  $m_t^{\rm MC}=m_t^{\rm pole}+(0.57\pm 0.28)$~GeV.
While this analysis provided a concrete numerical result, it can only be generalized to LHC measurements if one makes the additional assumption that the MC top mass has a universal meaning covering in particular also the LHC environment and the substantially more complicated observables included in the direct mass measurements, for which currently no first principle calculations exist. In addition, systematic uncertainties in the modelling of non-perturbative effects at hadron colliders, such as multi parton interactions, or the description of the pile-up effects are much harder to control. An analogous analysis for the LHC environment was subsequently carried out in Ref.~\cite{Hoang:2017kmk} using factorized NLL soft-drop groomed~\cite{Dasgupta:2013ihk,Larkoski:2014wba} hadron level jet mass distributions showing results that are compatible with, but less precise than those of~\cite{Butenschoen:2016lpz}. We also refer to Ref.~\cite{Kieseler:2015jzh} for a related analysis.

Aside from the previously mentioned examinations, recently, a number of complementary studies were conducted focusing on various sources of uncertainties in the perturbative description of top production and decay and the non-perturbative modelling of final states involved in top mass measurements. While these studies mainly aimed at examining the potential size of uncertainties in top mass determinations from reconstruction as well as from alternative methods (see Refs.~\cite{Kim:2017rve,Vos:2016tof,Adomeit:2014yna} and references therein), some of their findings may also be relevant for addressing the question how $m_t^{\rm MC}$ obtained from reconstruction should be interpreted field theoretically.   

In Ref.~\cite{Corcella:2017rpt} the sensitivity of $m_t^{\rm MC}$ determinations from exclusive hadronic variables such as
the B-meson energy $E_B$~\cite{Agashe:2016bok}, the B-lepton invariant mass $m_{B\ell}$~\cite{Biswas:2010sa} or the transverse mass variables $m_{T_2}$~\cite{Lester:1999tx,Matchev:2009ad,CMS:2012eya,Sirunyan:2017idq} to variations of the parameters of the MC hadronization models in PYTHIA~8 and \Herwig\ 6 was studied. They found that for top mass determinations based on these distributions to be competitive with direct reconstruction methods these hadronization parameters would have to be constrained significantly more precisely than what is possible from usual multi-purpose tuning. In addition, they made the observation that the top mass dependent endpoints of these distributions are, compared to the overall shape of the distributions, largely insensitive to variations of the hadronization parameters, indicating that these kinematic endpoints only depend on global and inclusive properties of the final state dynamics. 

In Ref.~\cite{Heinrich:2017bqp} top mass determinations from distributions such as the $b$-jet and lepton invariant mass $m_{b_j\ell}$ and the variable $m_{T_2}$~\cite{Lester:1999tx} were analyzed within fixed-order perturbation theory comparing the full NLO QCD result for $W^+W^-b\bar b$ production with different approximations in the narrow width approximation (NWA) concerning NLO QCD corrections in the production and the decay of the top quarks as well as using the parton shower from SHERPA~\cite{Gleisberg:2008ta} after top production. Using pseudo-data fits they found that the extracted top mass can depend significantly (at the level of $1$~GeV or even more) on the approximation used, indicating that incomplete descriptions of finite-lifetime effects can lead to systematic deviations in the value of the extracted top mass of order $\Gamma_t$.  

In Ref.~\cite{Ravasio:2018lzi} the NLO-PS matched  POWHEG~\cite{Frixione:2007vw,Alioli:2010xd} top production generators $hvq$~\cite{Frixione:2007nw}, $t\bar tdec$~\cite{Campbell:2014kua} and $b\bar b4\ell$~\cite{Jezo:2016ujg} interfaced to PYTHIA~8.2~\cite{Sjostrand:2014zea} and \Herwig\ 7.1~\cite{Bellm:2015jjp,Bellm:2017bvx} were studied comparatively examining the peak position of the particle level $b$-jet and $W$ invariant mass $m_{b_jW}$, the peak of the $b$-jet energy $E_{b_j}$~\cite{Agashe:2016bok} and moments of various lepton observables~\cite{Frixione:2014ala} in view of an extraction of the top quark mass. They found that the $m_{b_jW}$ peak is largely insensitive to variations of the generators and the shower MC as well as to input quantities such as the strong coupling and the PDFs or the b-jet definition, and concluded that changes in the top mass due to these variations do not exceed $200$~MeV in the absence of experimental resolution effects. They also indicated that the good agreement between the three POWHEG generators may imply that $m_{b_jW}$ is not sensitive to additional finite lifetime effects. Once the smearing due to experimental resolution effects is accounted for, however, they found an increased sensitivity to the differences in the parton showers of PYTHIA~8.2 and \Herwig\ 7.1 that correspond to variations in the extracted top mass at the level of $1$~GeV or more. For $E_{b_j}$ the dependence of the extracted top mass on the shower type and on the b-jet definition is in general at the level of $1$~GeV. For the leptonic observables variations of this size arise from PDF uncertainties and from changing the shower type. 

\subsection{About this work}
\label{sec:introthiswork}

The aim of this work is to initiate dedicated individual examinations of the different components of MC event generators 
with the aim to gain insights concerning the field theoretic meaning and potential limitations of the MC top mass parameter $m_t^{\rm MC}$ {\it from first principles}. In this paper we start with an examination of the parton shower evolution with respect to the dependence on the infrared shower cut $Q_0$. 

Apart from the perturbative hard interaction matrix elements that encode the basic hard process that can be described by MC generators, the parton shower describes the parton branching for energies below the hard interaction scale and represents the perturbative component of MC generators responsible for the low energy dynamics in MC predictions. While common analytic calculations in perturbative QCD are carried out in the limit of a vanishing infrared regulator, event generators based on parton showers rely on the existence of an infrared cut in order to prevent infinite parton multiplicities and to ensure that the parton shower description does not leave the realm of perturbation theory.

From the field theoretic point of view, $Q_0$ represents a factorization scale that separates the perturbative components of MC event generators and their hadronization models. While it is generally accepted that a finite value for $Q_0$ restricts the amount of real radiation and multiplicity generated by the shower evolution, it is not per se obvious to which extend it may also affect the meaning of QCD parameters such as the MC quark mass parameters.  Due to the unitarization property of the shower evolution which is responsible for the coherent summation of real as well as infrared virtual radiative corrections for scales above $Q_0$, it is also plausible that the MC top mass parameter $m_t^{\rm MC}$ should acquire a dependence on the value of $Q_0$ unless one makes the additional assumption the $Q_0$ effects are negligible. In this work we examine this dependence and find that is is not negligible. We emphasize that in the discussions of this paper we ignore all issues related to (the shower cut dependence of) hadronization because the primary aim is to concentrate on the perturbative aspects of the relation between $m_t^{\rm MC}$ and field theoretic mass schemes. We are aware that the properties of the hadronization modelling in MC event generators may have a significant impact on the interpretation of $m_t^{\rm MC}$, but we believe that examining perturbative and non-perturbative MC components separately in this respect is essential to gain full conceptual insight. 

Because the top quark has color charge its mass is - following the principles of heavy quark symmetry - linearly affected by the momenta of {\it ultra-collinear} gluons~\cite{Fleming:2007qr,Fleming:2007xt}, which are the gluons that are soft in the top quark rest frame. The role of these ultra-collinear gluons turns out to be essential for our conceptual considerations concerning the shower cut dependence of the top quark mass. Compared to the radiation pattern of massless quarks the additional effects coming from the ultra-collinear gluons is for example responsible for the dead cone effect~\cite{Dokshitzer:1991fc,Dokshitzer:1991fd} which is generally considered as coming from the top mass regulating the emergence of collinear singularities in the quasi-collinear limit. The radiation in the dead-cone region, however, is still non-zero and to the extent that it is unresolved becomes part of the energy (i.e.\ mass) of the measured top quark state. It is this quantum mechanical feature that goes beyond the classic picture of an unambiguous top quark ``particle'' whose total energy could be determined in the direct mass measurements. Since the parton showers in all state-of-the-art MC generators account for the dead cone effect~\cite{Maltoni:2016ays}, it appears obvious that the meaning of $m_t^{\rm MC}$ should naturally have a linear dependence on the shower cut $Q_0$ restricting the ultra-collinear radiation -- unless there is a mechanism that leads to a power suppressed effect of order $Q_0^2$ or higher which we may then safely neglect for the case of the top quark.
Therefore, to examine the intrinsic field theoretical meaning of the MC top quark mass parameter $m_t^{\rm MC}$ it is essential to start with a careful examination of the production of the top quarks and the ultra-collinear gluons. From this point of view, studies of the top decay and the treatment of the observable final states are important to quantify to which extent the ultra-collinear gluons are unresolved and how they enter a particular observable. 

In this work we aim to focus primarily on the production aspect, and we are therefore studying an observable that is maximally insensitive on details of the final state dynamics and its theoretical modeling. This observable is the peak (i.e.\ resonance) position of hemisphere jet masses in $e^+e^-$ annihilation, explained in more detail in Sec.~\ref{sec:observable}. The basic outcome of our considerations concerning the field theoretic meaning of $m_t^{\rm MC}$, however, should be general and shall be systematically extended to other types of observables and to the LHC environment in subsequent work. As a further simplification we consider the narrow width approximation (NWA), i.e.\ the case of quasi-stable top quarks which allows to rigorously factorize top production and decay, the case of boosted (i.e.\ large-$p_T$) top quarks and the coherent branching formalism which is related to angular ordered showers, see Refs.~\cite{Marchesini:1983bm,Marchesini:1987cf,Catani:1990rr} for massless and Refs.~\cite{Gieseke:2003rz} for massive quarks, and also Refs.~\cite{Krauss:2003cr,Rodrigo:2003ws}. Since the limit of stable and quasi-collinear heavy quarks is the theoretical basis of all parton shower formulations for top (and other heavy) quarks, it is natural to investigate the physics in this limit first to avoid that the conclusions are affected by the additional approximations that need to be made in the attempt to account for the effects of slow and unstable top quarks. Our current focus on angular ordered showers is, on the other hand, of purely practical nature: Our considerations here require explicit analytic solutions of the shower evolution, and angular ordered showers based on the coherent branching formalism can be more easily tackled by well known analytic methods~\cite{Catani:1992ua} applicable to global event shapes. So our current results are directly relevant for the \Herwig\ MC generator which employs an angular ordered PS. Generalizations to other MC generators shall be treated elsewhere.  

In this context our paper is structured around the following three questions:
\begin{itemize}
\item[(A)] 
Can state-of-the-art partons showers in principle describe the single top resonance mass and related thresholds with NLO precision?
\item[(B)] 
What is the impact of the shower cut $Q_0$ on the resonance value of the jet masses?
\item[(C)]
Does the shower cut imply that the MC top quark mass parameter $m_t^{\rm MC}$ is a low-scale threshold short-distance mass, 
and how can this be proven from first principles at the field theoretic level?
\end{itemize}
Question~A is relevant because, only if parton showers can describe the threshold or resonance mass with NLO precision, the question of which mass scheme is employed can be addressed systematically in a meaningful way. In the course of our examination we show that this is indeed the case as long as NLL order logarithmic terms are resummed, and we also show that the additional NLO corrections implemented by NLO matched parton showers do not further increase the precision. Question~B concerns the dependence of the resonance value of the jet mass on $Q_0$. We show that the jet mass at the resonance peak depends linearly on $Q_0$ which means that for the field theoretic meaning of $m_t^{\rm MC}$ the finite shower cut is essential and cannot be neglected. Finally, question~C addresses to which extent the linear dependence on $Q_0$ must be interpreted as a $Q_0$-dependence of the MC top quark mass. As we will show,
only a part of the linear $Q_0$-dependence of the peak jet mass is related to ultra-collinear radiation and thus to the top quark mass. Overall, the shower cut also restricts the radiation of large angle soft gluons unrelated to the top quark and the ultra-collinear radiation. Only the latter is related to the top quark mass, and its dominant linear $Q_0$-dependence caused by the shower cut can be shown to automatically imply a mass redefinition which differs from the pole mass by a term proportional to $\alpha_s(Q_0)\,Q_0$. This result implies that $m_t^{\rm MC}$ is equivalent to the top quark pole mass, in the limit $Q_0=0$ which is however practically inaccessible for parton showers. 
In the formal limit $m_t\to 0$ the effects of the ultra-collinear radiation and its $Q_0$-dependence vanish and only the cutoff dependence on the soft radiation remains. This cutoff dependence represents the factorization interface between perturbative soft radiation and hadronization effects governed by the MC hadronization models. Since in this context the understanding of the shower cut dependence of the soft radiation is a prerequisite to the examination of the ultra-collinear radiation, we also analyze carefully the case of massless quark production in parallel to our discussions on the top quark. 

\subsection{Outline and reader instruction}
\label{sec:contentthiswork}

The outline of our paper is as follows: In 
Sec.~\ref{sec:observable} we set up our theoretical framework by explaining the hemisphere mass observable  
$\tau$ and reviewing the corresponding NLL and hadron level factorized QCD predictions in the resonance region
for massless as well as massive quark production. We also provide details on the hadronization model shape function
which is important for the numerical analyses carried out in the subsequent sections. In this section we also prove,
using the factorized predictions, that NLL resummation of logarithms is sufficient to achieve NLO precision for the
position of the peak in the $\tau$ distribution.

In Sec.~\ref{sec:coherentbranching} we review the coherent branching formalism, 
provide the analytic evolution equation for the jet mass distribution for massless and massive quark production
at NLL order and give some details on the practical implementation of the angular order parton shower based on the
coherent branching formalism in the \Herwig\ 7 event generator.

In Sec.~\ref{sec:CBnocut} we show -- in the absence of any infrared cutoff and in the context of strict perturbative 
computations -- that the NLL predictions for the hemisphere mass $\tau$ distribution in the resonance region 
obtained from the coherent branching formalism are fully equivalent to the NLL factorized 
QCD predictions for massless quark production as well as for massive quark production in the pole mass scheme.  
This result proves, that 
in the context of strict perturbative computations for massive quarks in the limit $Q_0=0$ the MC generator mass 
is equivalent to the pole mass. This conclusion, however, does not apply for MC event generators because their
parton shower algorithm strictly requires a finite shower cut $Q_0$ in order to terminate and to avoid infinite multiplicities. 

The impact of the shower cut $Q_0$ is then analyzed in detail in Sec.~\ref{sec:unreleased}, which represents
the core of this work. Here we analyze the power counting of the relevant modes entering the 
hemisphere mass in the resonance region in the massless and massive quark case and we focus on a coherent view
of the factorized QCD and the coherent branching approach. We calculate analytically the NLO corrections caused 
by the shower cut $Q_0$ in comparison with the results without any cut in the coherent branching formalism
and the factorized QCD approach focusing on the dominant effects linear in $Q_0$. We show that the results
obtained for the linear $Q_0$ contributions in coherent branching and factorized QCD are compatible, and we use
the direct connection of the factorized QCD computation to field theory to unambiguously distinguish shower cut
effects related to soft hadronization corrections and the quark mass parameter. By coherently examining massless and massive
quark production we prove that using a finite shower cut $Q_0$ in the coherent branching formalism -- and thus also in angular ordered parton showers -- automatically implies that one employs
a short-distance mass scheme different from the pole mass, called the \emph{coherent branching (CB) mass, $m^{\rm CB}(Q_0)$}. We explicitly calculate the relation of the coherent branching mass to the pole mass
at NLO, i.e.\ ${\cal O}(\alpha_s)$.

In Sec.~\ref{sec:intermediatesummary} the conceptual results obtained in the previous sections are summarized
coherently to set up the numerical examinations we carry out in Sec.~\ref{sec:herwigcompare}.
In Sec.~\ref{sec:herwigcompare} we compare the results obtained in Sec.~\ref{sec:unreleased} with analytic methods and conceptual 
considerations with numerical results running simulations for the hemisphere mass variable $\tau$ 
in \Herwig\ 7 using different values of the shower cut $Q_0$. Focusing mostly on the peak position of $\tau$ we show
that the simulations are in full agreement with our conceptual results. We also show explicitly that 
NLO corrections added in the context of NLO matched parton showers have extremely small effects in the resonance
location and do not modify any of the previous results, confirming that NLL accurate parton showers are already
NLO accurate as far as the resonance region is concerned. Furthermore, we also demonstrate that the results we have obtained
in the context the hemisphere mass variable $\tau$ are also compatible with numerical simulations for 
the more exclusive kinematic variables $m_{b_j\ell}$ and $m_{b_jW}$ supporting the view that our results 
are universal.

Finally, Sec.~\ref{sec:conclusions} contains our conclusions and an outlook for some of the remaining questions
that should be addressed in the future. There we also provide a brief numerical analysis how $m^{\rm CB}(Q_0)$ is related to other mass renormalization schemes. The paper also contains four appendices containing some supplemental 
material relevant for our work. In App.~\ref{app:thrustformulae} we collect all parton level results for
NLL+${\cal O}(\alpha_s)$ factorized QCD predictions of the $\tau$ distribution in the resonance region for 
massless and massive quark production. In App.~\ref{app:thrustformulaewithcut} we provide details on the 
computations of the effects of the shower cut $Q_0$ in the context of the factorized QCD predictions and
in App.~\ref{app:integrals} we collect results for loop integrals in the presence of the shower cut $Q_0$. 
Finally, in App.~\ref{app:Settings} we give information on the \Herwig\ 7 settings we have employed for our 
simulation studies.

To the reader mainly interested in the phenomenological implications of our discussions in the context of
the Monte-Carlo top mass problem: We recommend to go through our paper by starting with Sec.~\ref{sec:intro} and 
Sec.~\ref{sec:observable} for all basic information concerning our examinations and in particular for important elementary 
knowledge concerning the hemisphere mass variable $\tau$ in the resonance region and its theoretical description. 
One may then jump directly to Sec.~\ref{sec:intermediatesummary}, where all of our conceptual results are 
summarized and continue with our simulations studies in Sec.~\ref{sec:herwigcompare} and the conclusions 
in Sec.~\ref{sec:conclusions}.

\section{The observable: squared hemisphere mass sum}
\label{sec:observable}

The observable we consider in this work is the sum of the squared hemisphere masses defined with respect to the thrust axis in $e^+e^-$-collisions normalized to the square of the c.m.\ energy $Q$,
\begin{equation}
\label{eq:taudef}
\tau\, \equiv \, \frac{M_1^2+M_2^2}{Q^2}.
\end{equation}
In the lower endpoint region the $\tau$ distribution has a resonance peak which is dominated by back-to-back 2-jet configurations which arise from LO quark-antiquark production, and
it is the location of the resonance, $\tau_{\rm peak}$, which we focus on mostly in our study. 
For massless quarks this resonance region is located close to $\tau=0$ and represents the threshold region for dijet production. Non-perturbative effects shift the observable peak towards positive $\tau$ values by an mount of ${\cal O}(\Lambda/Q)$, where $\Lambda$ is a scale of around $1$~GeV.
For massive quark production the resonance region and the peak are located close to $\tau= 2m_Q^2/Q^2$, and for the case of the top quark for $Q\gg m_t$ is dominated by boosted back-to-back top quark initiated jets. As for the case of massless quark production non-perturbative effects shift the observable peak towards positive $\tau$ values by an mount of ${\cal O}(\Lambda/Q)$.
The scale of $\Lambda\approx 1$~GeV is generated from non-perturbative effects, but its value is numerically larger than 
$\Lambda_{\rm QCD}$ because it accounts for the cumulative hadronization effect from both hemispheres~\cite{Abbate:2010xh}. 
In the peak region, $\tau$ is closely related to the classic thrust variable~\cite{Farhi:1977sg} in the case of massless quark production~\cite{Catani:1992ua}, and to 2-Jettiness~\cite{Stewart:2010tn} for massive quarks~\cite{Fleming:2007qr}. To be concrete, concerning the structure of large logarithms and of terms singular in the $\tau\to\tau_{\rm min}$ limit, which dominate the shape and position of the peak, the hemisphere mass variable $\tau$, thrust and 2-jettiness are equivalent for large $Q$. We therefore frequently refer to $\tau$ simply as "thrust" in this paper.

For our examinations for top quarks we also consider the rescaled thrust variable
\begin{align}
\label{eq:Mtaudef}
M_\tau \, \equiv \, \frac{Q^2\tau}{2m_Q}\,.
\end{align}
The variable $M_\tau$ is peaked close to $M_\tau = m_Q$ and allows for a more transparent interpretation
of the shower cut $Q_0$-dependence from the point of view of the top quark mass than $\tau$. 
Note that the scheme dependence of the quark mass parameter $m_Q$ appearing in the definition~(\ref{eq:Mtaudef}) represents an effect that is ${\cal O}(\as^2)$-suppressed in the context of our examinations and therefore irrelevant at the order we are working. 

An essential aspect of the examinations in this work is that for boosted top quarks events related to top decay products being radiated outside the parent top quark's hemisphere are $(m_t/Q)^2$ power suppressed~\cite{Fleming:2007qr}. So, because thrust depends on the sum of momenta in each hemisphere, effects of the top quark decay in the thrust distribution are power suppressed as well, and the situation of a finite top quark width is smoothly connected to the NWA and the stable top quark limit. This is compatible with the factorized treatment of top production and decay used in contemporary parton showers and also allows us to carry out analytic QCD calculations for stable top quarks which are essential for the chain of arguments we use. In this way thrust is an ideal observable for the examinations made in this work since it allows to study the mass of the top quark accounting in particular for the contribution of the unresolved ultra-collinear gluon cloud around it.    

However, in thrust the effects of large angle soft radiation are maximized, and the impact of the shower cut $Q_0$ on the meaning of the top quark mass parameter interferes with that $Q_0$ has on large angle soft radiation. Since the latter is not related to the top quark mass, but represents the interface to hadronization effects~\cite{Catani:1989ne,Catani:1992ua}, it is important that both effects are disentangled unambiguously. As we will show, for thrust in the peak region this can be carried out in a straightforward way owing to soft-collinear factorization~\cite{Korchemsky:1999kt,Berger:2003iw}. Since the structure of large angle soft radiation is equivalent for the production of massless quarks and boosted massive quarks~\cite{Fleming:2007qr,Fleming:2007xt}, we discuss the case of massless quark production before we examine boosted top quarks.

Since our discussion requires the analytic comparison of 
the thrust distribution determined from the parton shower evolution based on the coherent branching formalism at NLL order (where we follow the approach of~\cite{Catani:1992ua,Davison:2008vx}) and of corresponding resummed QCD calculations based on soft-collinear factorization, we briefly review the latter in the following two subsections for massless and massive quark production. 

\subsection{Factorized QCD cross section: massless quarks}
\label{sec:factorizationtheoremmassless}

Resummed calculations for the thrust distribution in the peak region require the summation of terms that are logarithmically enhanced and singular in the limit $\tau\to \tau^{\rm min}=0$, where the partonic threshold is located. In the context of conventional perturbative QCD, factorized calculations for massless quarks have been carried out in Ref.~\cite{Berger:2003iw} at NLL order. In the context of SCET the  corresponding results have been obtained at NLL$+{\cal O}(\alpha_s)$ in Ref.~\cite{Schwartz:2007ib} and were extended to N$^3$LL$+{\cal O}(\alpha_s^3)$ in Ref.~\cite{Becher:2008cf,Abbate:2010xh}. 
Using the notations from Ref.~\cite{Abbate:2010xh} the observable hadron level thrust distribution in the peak region can be written in the form
\begin{align}
\label{eq:thrustmassless1}
\frac{\mathrm{d}\sigma}{\mathrm{d}\tau}(\tau,Q)=
\int\limits_0^{Q\tau}\!\mathrm{d}\ell\; 
\frac{\mathrm{d}\hat\sigma_s}{\mathrm{d}\tau}\Big(\tau-\frac{\ell}{Q},Q\Big)\,\,S_{\rm mod}(\ell) 
\end{align}
where $\mathrm{d}\hat\sigma_s/\mathrm{d}\tau$ contains the factorized resummed singular partonic QCD corrections (containing $\delta$-function terms of the form $\alpha_s^n\delta(\tau)$ and plus-distributions of the form $\alpha_s^n [\ln^k(\tau)/\tau]_+$) and $S_{\rm mod}(\ell)$ is the soft model shape function that describes the non-perturbative effects. It has support for positive values of $\ell$, exhibits a peaked behavior for $\ell$ values around $1$ GeV and is strongly falling for larger values. We further assume that it vanishs at zero momentum, $S_{\rm mod}(0)=0$.\footnote{The typical scale of the non-perturbative function $S^{\rm mod}$ is about twice the typical hadronization scale $\Lambda_{\rm QCD}\lsim 0.5$~GeV as it accounts for non-perturbative from both hemispheres~\cite{Abbate:2010xh}. The property $S_{\rm mod}(0)=0$ is assumed for all shape functions treated in the literature and physically motivated from the hadronization gap.} Due to the smearing caused by the non-perturbative function the visible peak of the thrust distribution is shifted to positive values by an amount of order $(1~\mbox{GeV})/Q$. The dominant perturbative corrections to the factorized cross section in Eq.~(\ref{eq:thrustmassless1}) are coming from so-called non-singular contributions containing terms of the form $\alpha_s^n\ln^k(\tau)$. For our considerations in the resonance region these corrections are power-suppressed by a {\it additional factor} of order $(1~\mbox{GeV})/Q$, i.e.\ they cause a shift in the peak position by an amount $(1~\mbox{GeV})^2/Q^2$ which we can safely neglect. 

The resummed factorized singular partonic QCD cross section has the form
\begin{align}
\label{eq:thrustmassless2}
 \frac{1}{\sigma_0}\frac{\mathrm{d}\hat\sigma_s}{\mathrm{d}\tau}(\tau,Q)=Q\,H_{Q}&(Q,\mu_H)  \int\limits_0^{Q^2\tau}\!\!\mathrm{d}s
  \,\int\limits_0^{s}\mathrm{d}s^\prime\,
 U_J(s^\prime,\mu_H,\mu_J)\,J^{(\tau)}(s-s^\prime,\mu_J)\, 
 \\ \nonumber & \times
 \int\limits_0^{Q\tau-s/Q}\hspace{-3mm}\mathrm{d}k\,\,\, U_S(k,\mu_H,\mu_S)\,
 S^{(\tau)}\Big(Q\tau-\frac{s}{Q}-k, \mu_S\Big)
\end{align}
where $\sigma_0$ is the total partonic $e^+e^-$ tree-level cross section. The term $H_Q$ is the hard function describing effects at the production scale $Q$, $J^{(\tau)}$ is the jet function describing the distribution of the squared invariant mass $s$ due to collinear radiation coming from {\it both} jets and $S^{(\tau)}$ is the soft function containing the effects of large angle soft radiation. They depend on the renormalizations scales $\mu_H\sim Q$, $\mu_J\sim Q\sqrt{\tau}$ and $\mu_S\sim Q\tau$, which are chosen such that no large logs appear in hard, jet and soft functions respectively. Large logarithmic contributions are resummed in the different $U$ factors which are evolved from the corresponding renormalization scale $\mu_H$, $\mu_J$ or $\mu_S$ to a common renormalization scale.
Since it most closely resembles the analytic form of the resummation formulae obtained in the coherent branching formalism, we have set in Eq.~(\ref{eq:thrustmassless2}) the common renormalization scale equal to the hard scale $\mu_H$, so that there is no evolution factor $U_H$ for the hard function. 
So, $U_J$ sums logarithms between the jet scale $\mu_J$ and the hard scale $\mu_H$, and $U_S$ sum logarithms between the soft scale $\mu_S$ and the hard scale. 
For our discussions we need the  expressions for the $U$ factors at NLL and the hard, soft and jet function at ${\cal O}(\alpha_s)$. These formulae (and also the renormalization group equations for the $U$ factors) are provided for convenience in App.~\ref{app:masslessthrust}.

Expanded to first order in the strong coupling and setting $\mu_H=\mu_J=\mu_S=\mu$ in Eq.~(\ref{eq:thrustmassless2}) we obtain the well-known ${\cal O}(\alpha_s)$ singular fixed-order thrust distribution
\begin{align}
\label{eq:thrustmasslessFO}
\frac{1}{\sigma_0}\frac{\mathrm{d}\hat\sigma_s}{\mathrm{d}\tau}(\tau,Q)=\delta(\tau) +\frac{\as C_F}{4\pi}\biggl\{-8\biggl[\frac{\theta(\tau)\ln\tau}{\tau}\biggr]_+-6\biggl[\frac{\theta(\tau)}{\tau}\biggr]_++\Bigl(\frac{2\pi^2}{3}-2\Bigr)\delta(\tau)\biggr\}+\Oas{2}
\,.
\end{align}

Transforming the partonic massless quark thrust distribution of Eq.~(\ref{eq:thrustmassless2}) to Laplace space with the convention
\begin{align}
\label{eq:sigmaLaplacemassless}
\tilde \sigma(\nu,Q) \, = \, \int\limits_{0} ^\infty {\rm d}\tau\,e^{-\nu\tau}\,
\frac{1}{\sigma_0}\frac{\mathrm{d}\hat\sigma_s}{\mathrm{d}\tau}(\tau,Q)
\end{align}
the NLL thrust distribution can be written in the condensed form
\begin{align}
\label{eq:Laplacemassless}
\tilde\sigma(\nu,Q)\,=\,&\,\mathrm{exp}\biggl[K(\Gamma_J,\mu_{H,\nu},\mu_{J,\nu})+K(\Gamma_S,\mu_{H,\nu},\mu_{S,\nu})
\notag\\
& \quad +\,\frac{1}{2}\Bigl(\omega(\gamma_J,\mu_{H,\nu},\mu_{J,\nu})+\omega(\gamma_S,\mu_{H,\nu},\mu_{S,\nu})\Bigr)\biggr]\,,
\end{align}
where the evolution functions $K$ and $\omega$ have the form
\begin{align}
\label{eq:Komega}
&K(\Gamma,\mu,\mu_0)=2\int_{\as(\mu_0)}^{\as(\mu)}\frac{\mathrm{d}\as}{\beta[\as]}\;\Gamma[\as]\int_{\as(\mu_0)}^{\as}\frac{\mathrm{d}\as^{\prime}}{\beta[\as^{\prime}]} \,,\\
&\omega(\Gamma,\mu,\mu_0)=2\int_{\as(\mu_0)}^{\as(\mu)}\frac{\mathrm{d}\as}{\beta[\as]}\;\Gamma[\as] \,.
\end{align}
The QCD beta function is defined as
\begin{align}
\label{eq:QCDbetafct}
\mu\frac{\mathrm{d}\as(\mu)}{\mathrm{d}\mu}=\beta[\as(\mu)] = -2\alpha_s(\mu)\sum_{n=0}\beta_n\Bigl(\frac{\as(\mu)}{4\pi}\Bigr)^{n+1}
\end{align}
with $\beta_0 = 11-\frac{2}{3}n_f$ and $\beta_1=102-\frac{38}{3}n_f$  , and the cusp and non-cusp anomalous dimensions read
\begin{align}
\label{eq:cusp1}
\Gamma_J[\as] &\,=\,-2\Gamma_S[\as]=4\,\Gcusp[\as]\,,\nonumber\\
\gamma_J[\as]&\, = \, 12 C_F \Bigl(\frac{\as}{4\pi}\Bigr)\,,\nonumber\\
\gamma_S[\as]&\, = \, 0
\end{align}
with
\begin{align}
\label{eq:cusp0}
\Gcusp[\as] &\, = \, \Gcusp_0 \Bigl(\frac{\as}{4\pi}\Bigr)+\Gcusp_1 \Bigl(\frac{\as}{4\pi}\Bigr)^2\,,\nonumber\\
\Gcusp_0&=4C_F \,, \nonumber \\
\Gcusp_1&=C_F\Bigl[C_A\Bigl(\frac{268}{9}-\frac{4\pi^2}{3}\Bigr)-\frac{80}{9}T_Fn_f\Bigr] \,,
\end{align}
The scales $\mu_{H,\nu}$, $\mu_{J,\nu}$ and $\mu_{S,\nu}$ are given by 
\begin{align}
\label{eq:salesmassless}
\mu_{H,\nu} = \, Q\,,\qquad
\mu_{J,\nu}=\, Q(\nu\, \mathrm{e}^{\gamma_E})^{-1/2} \,,\qquad
\mu_{S,\nu}=\, Q(\nu\, \mathrm{e}^{\gamma_E})^{-1} \,.
\end{align}
These scales are fixed to the expressions shown and arise from the combination of the renormalization scale dependent NLL $U$ evolution factors {\it and} the Laplace transformed ${\cal O}(\alpha_s)$ corrections in the hard, jet and soft functions shown in Eqs.~(\ref{eq:hardfunction}), (\ref{eq:scetjetfunction}) and (\ref{eq:scetsoftfunction}) that are logarithmic or plus-distributions. Dropping a $\pi^2$ term arising in the Laplace transform of the $(\ln\tau/\tau)_+$ distributions, in this combination the dependence on the renormalization scales $\mu_H$, $\mu_J$ and $\mu_S$ cancels and the result shown in Eq.~(\ref{eq:Laplacemassless}) with the physical scales given in Eqs.~(\ref{eq:salesmassless}) emerges. Since the structure of these ${\cal O}(\alpha_s)$ corrections is already unambiguously known from the NLL renormalization properties, {\it we consider them part of the NLL logarithmic contributions}.
(We refer to Ref.~\cite{Almeida:2014uva} for an extensive discussion on this issue.) 
Using in Eq.~(\ref{eq:Laplacemassless}) the renormalization scales $\mu_{i}$ instead of the scales $\mu_{i,\nu}$ ($i=H,J,S$) one recovers the renormalization scale dependent results coming from the $U$ evolution factors alone.

As we show in Sec.~\ref{sec:CBnocutmassless} all terms displayed in Eq.~(\ref{eq:Laplacemassless}) are also precisely obtained by the coherent branching formalism at NLL order.

\subsection{Factorized QCD cross section: massive quarks}
\label{sec:factorizationtheoremmassive}

In the case of boosted massive quark production the thrust distribution has been determined at NNLL$+{\cal O}(\alpha_s)$ in Refs.~\cite{Fleming:2007qr,Fleming:2007xt,Butenschoen:2016lpz}. Adopting the {\it pole mass scheme}, the $\tau$ distribution as defined in Eq.~(\ref{eq:taudef}) has its partonic threshold at
\begin{equation}
\label{eq:taumin}
\tau_{\rm min}^{\rm pole}\, = \, \frac{2 (m^{\rm pole})^2}{Q^2}\,.
\end{equation}
The observable thrust distribution in the resonance region for $\tau\gsim\tau_{\rm min}^{\rm pole}$,  can be written in a form analogous to the case of massless quarks and has the form
\begin{align}
\label{eq:thrustmassive1}
\frac{\mathrm{d}\sigma}{\mathrm{d}\tau}(\tau,Q,m^{\rm pole})=
\int\limits_0^{Q\tau}\!\mathrm{d}\ell\,\,
\frac{\mathrm{d}\hat\sigma_s}{\mathrm{d}\tau}\Big(\tau-\frac{\ell}{Q},Q,m^{\rm pole}\Big)\,\,S_{\rm mod}(\ell)\,,
\end{align}
where $\mathrm{d}\hat\sigma_s/\mathrm{d}\tau$ is the resummed singular massive quark partonic QCD cross section, which contains terms of the form $\alpha_s^n\delta(\tau-\tau_{\rm min}^{\rm pole})$ and $\alpha_s^n [\ln^k(\tau-\tau_{\rm min}^{\rm pole})/(\tau-\tau_{\rm min}^{\rm pole})]_+$). 
The non-singular corrections to the factorized cross section in Eq.~(\ref{eq:thrustmassive1}) are coming from terms of the form $\alpha_s^n\ln^k(\tau-\tau_{\rm min}^{\rm pole})$.
In the resonance region these corrections are power-suppressed by a {\it additional factor} of order $(1~\mbox{GeV})/Q$ or $(1~\mbox{GeV})/m$ and can, in analogy to the case of massless quark production, be safely neglected for top quark production. In an arbitrary mass scheme $m$ with $\delta m =m^{\rm pole} - m$ we can write the observable thrust distribution in the form 
\begin{align}
\frac{\mathrm{d}\sigma}{\mathrm{d}\tau}(\tau,Q,m,\delta m)=
\int\limits_0^{Q\tau}\!\mathrm{d}\ell\,\,
\frac{\mathrm{d}\hat\sigma_s}{\mathrm{d}\tau}\Big(\tau-\frac{\ell}{Q},Q,m,\delta m\Big)\,\,S_{\rm mod}(\ell)\,,
\end{align}
where the additional argument $\delta m$ indicates the dependence on the mass scheme changing contributions in the perturbation series for the partonic cross section.

For the rescaled thrust variable defined in Eq.~(\ref{eq:Mtaudef}) the relation analogous to Eq.~(\ref{eq:thrustmassive1}) reads
\begin{align}
\label{eq:Mtau1}
\frac{\mathrm{d}\sigma}{\mathrm{d}M_\tau}(M_\tau,Q,m^{\rm pole}) &\, =\,
\int\limits_0^{2 m^{\rm pole} M_\tau/Q}\hspace{-6mm}\mathrm{d}\ell\,\,\,
\frac{\mathrm{d}\hat\sigma_s}{\mathrm{d}M_\tau}\Big(M_{\tau}-\frac{Q\ell}{2m^{\rm pole}},Q,m^{\rm pole}\Big)\,\,S_{\rm mod}(\ell)\,,
\end{align}
where
\begin{align}
\label{eq:Mtau1_1}
\frac{\mathrm{d}\hat\sigma_s}{\mathrm{d}M_\tau}(M_{\tau},Q,m^{\rm pole})
 &\, \equiv \, \frac{2m^{\rm pole}}{Q^2}\, \,
 \frac{\mathrm{d}\hat\sigma_s}{\mathrm{d}\tau}\Big(\frac{2m^{\rm pole}M_\tau}{Q^2},Q,m^{\rm pole}\Big)\,.
\end{align}
The generalization of Eqs.~\eqref{eq:Mtau1} and~\eqref{eq:Mtau1_1} to an arbitrary mass scheme is straightforward.

The singular partonic cross section in the resonance region can be written in the factorized form
\begin{align}
\label{eq:thrustmassive2}
\frac{\mathrm{d}\hat\sigma_s}{\mathrm{d}\tau}(\tau,Q,& m^{\rm pole})\,=\,\sigma_0\, Q\,H_{Q}(Q,\mu_H) U_H(Q,\mu_H,\mu_m) 
H_m(Q,m^{\rm pole},\mu_m) U_m\Big(\frac{Q}{m^{\rm pole}},\mu_m,\mu_H\Big)\nonumber
\\ \nonumber & \times \int\limits_0^{Q^2(\tau-\tau_{\rm pole}^{\rm min})}\hspace{-6mm}\mathrm{d}s\,\,\,\,
\,\int\limits_0^{s/m}\hspace{-1mm}\mathrm{d}\hat s^\prime\,\,\,
U_{J_B}(\hat s^\prime,\mu_H,\mu_B)\,J_B^{(\tau)}\Bigl(\frac{s}{m^{\rm pole}}-\hat s^\prime,m^{\rm pole},\delta m=0,\mu_B\Bigr)\, 
\\  & \times
\int\limits_0^{Q(\tau-\tau_{\rm pole}^{\rm min})-s/Q}\hspace{-9mm}\mathrm{d}k\hspace{3mm} U_S(k,\mu_H,\mu_S)\,
S^{(\tau)}\Big(Q(\tau-\tau_{\rm min}^{\rm pole})-\frac{s}{Q}-k, \mu_S\Big)\,,
\end{align}
where $\sigma_0$ is again the total partonic $e^+e^-$ tree-level cross section. The hard function $H_Q$, the soft function $S^{(\tau)}$ and the soft evolution factor $U_S$, as well as the soft model function $S_{\rm mod}$ in Eq.~(\ref{eq:thrustmassive2}) are identical to the case of massless quarks~\cite{Fleming:2007qr,Fleming:2007xt} 
(see App.~\ref{app:masslessthrust} for their respective expressions at NLL and ${\cal O}(\as)$). 
Their effects are universal for massless and boosted massive quarks, because large angle soft radiation cannot distinguish between the color flow associated to massless and boosted massive quarks. The relation of the soft function renormalization scale to $\tau$ is, however, modified to the form $\mu_S\sim Q(\tau-\tau_{\rm min})$ because the quark mass shifts the $\tau$ threshold from zero to $\tau_{\rm min}$.
For the other components of the factorization formula the quark mass represents an additional intermediate scale which leads to modifications. The term $J_B^{(\tau)}(\hat{s})$ is the bHQET jet function~\cite{Fleming:2007qr,Fleming:2007xt} which describes the linearized distribution of the invariant mass of {\it both} jets {\it with respect to the partonic threshold}, 
\begin{align}
\label{shatdef}
\hat s=\frac{s-(2m^{\rm pole})^2}{m^{\rm pole}}\,,
\end{align}
due to ultra-collinear gluon radiation in the region where $\hat s$ is much smaller than the mass, $\hat s\ll m$. It depends on the renormalization scale  $\mu_B\sim Q \mu_S/m \sim Q^2(\tau-\tau_{\rm min})/m$, and its expression at ${\cal O}(\as)$ in an arbitrary mass scheme $m$, $J_B^{(\tau)}(\hat{s},m,\delta m,\mu_B)$, with $\delta m=m^{\rm pole}-m\neq 0$ is shown in Eq.~(\ref{eq:JBoneloop}). At NLL+$\mathcal{O}(\alpha_s)$ the bHQET jet function completely controls the quark mass scheme dependence of the singular partonic cross section. So at this order the singular partonic cross section in an arbitrary mass scheme, $\frac{\mathrm{d}\hat\sigma_s}{\mathrm{d}\tau}(\tau,Q, m,\delta m)$, is obtained from Eq.~\eqref{eq:thrustmassive2}, by employing the bHQET jet function $J_B^{(\tau)}(\hat{s},m,\delta m,\mu_B)$ and setting $m^{\rm pole}\to m$ everywhere else. This is because $J_B^{(\tau)}$ has mass sensitivity already at tree level through the dependence on $\tau_{\rm min}$, see Eq.~(\ref{eq:taumin}).
Physically the ultra-collinear radiation is, owing to heavy quark symmetry, related to the soft radiation governing the mass of heavy-light mesons. The mass mode factor $H_m$ contains fluctuations at the scale of the quark mass $\mu_m\sim m$ coming from the massive quark field fluctuations that are off-shell in the resonance region and integrated out. Its expression at ${\cal O}(\as)$ is shown in Eq.~(\ref{eq:Hmoneloop}) and a detailed discussion on its definition and properties can be found in Ref.~\cite{Fleming:2007xt}.
The factor $U_{J_B}$ sums logarithms between the ultra-collinear jet scale $\mu_B$ and the hard scale $\mu_H$, $U_S$ sums logarithms between the soft scale $\mu_S$ and the hard scale, and $U_m$ sum logarithms between the quark mass scale $\mu_m$ and the hard scale. Their formulae are for convenience also provided in App.~\ref{app:massivethrust}.

From a physical point of view it appears more appropriate to evolve the factors $U_{J_B}$, $U_S$ and $U_m$ to the quark mass scale $\mu_m$ (at which point the factor $U_m$ could be dropped) rather than the hard scale. This is because the logarithms resummed in $U_{J_B}$ and $U_m$ physically arise from scales below the quark mass. The form we have adopted here is equivalent due to renormalization group consistency conditions~\cite{Fleming:2007xt} and matches better to the form of the log resummations obtained from the coherent branching formalism as discussed in Sec.~\ref{sec:CBnocutmassive}. 
For our examinations we need the  expressions for the $U$ factors at NLL and the hard, mass matching, soft and the bHQET jet functions at ${\cal O}(\alpha_s)$. 
Expanding to first order in the strong coupling and setting $\mu_H=\mu_B=\mu_S=\mu$ we obtain the  ${\cal O}(\alpha_s)$ singular fixed-order massive quark thrust distribution in the pole mass scheme ($L_m=\ln \frac{(m^{\rm pole})^2}{Q^2}$):
\begin{align}
\label{eq:thrustmassiveFO}
\frac{1}{\sigma_0}\frac{\mathrm{d}\hat\sigma_s}{\mathrm{d}\tau}(\tau,Q,m^{\rm pole})=&\delta(\tau-\tau_{\rm min}^{\rm pole})+\frac{\as C_F}{4\pi}\biggl\{-8(1+L_m)\biggl[\frac{\theta\bigl(\tau-\tau_{\rm min}^{\rm pole}\bigr)}{\tau-\tau_{\rm min}^{\rm pole}}\biggr]_+\notag \\
&+(4L_m^2+2L_m+2\pi^2)\delta\bigl(\tau-\tau_{\rm min}^{\rm pole}\bigr)\biggr\}+\Oas{2}
\,.
\end{align}
In Eq.~(\ref{eq:thrustmassiveFO}), changing to another mass scheme $m$ leads to the additional term $ \,\delta^\prime(\tau-\tau_{\rm min}^{\rm pole})\,4 m\,\delta m/Q^2$ on the RHS, and this term has to be counted as a NLL contributions as well.

We note that
in Eq.~(\ref{eq:thrustmassiveFO}) the dead cone effect~\cite{Dokshitzer:1991fc,Dokshitzer:1991fd} is manifest as a $\tau\to\tau_{\rm min}$ behavior that is less singular than the $\tau\to 0$ limit for massless quark production displayed in Eq.~\eqref{eq:thrustmasslessFO}. However, one can see from the form of the bHQET jet function in Eq.~(\ref{eq:JBoneloop}), that ultra-collinear radiation still involves soft-collinear double-logarithmic singularities which arise from the coherent effect of ultra-collinear gluons physically originating from the associated top quark and its opposite hemisphere~\cite{Fleming:2007qr,Fleming:2007xt}. So, in the context of QCD factorization based on SCET and bHQET the deadcone effect arises from a cancellation of double logarithmic singularities between the ultra-collinear and the large-angle soft radiation (radiated in the collinear direction and called collinear-soft radiation in the following). This can be seen from the expression for the partonic soft function $S^{(\tau)}$ given in Eq.~(\ref{eq:scetsoftfunction}) which exhibits the same double-logarithmic singularity as the bHQET jet function, but with the opposite sign. So the origin of the deadcone effect from the perspective of QCD factorization, which is manifestly gauge invariant, is due to a cancellation of ultra-collinear and collinear-soft radiation. This is somewhat different (but not contradictory) to the conventional and gauge-dependent view that the deadcone originates from  
the suppression of collinear radiation off the boosted top quarks due to the finite top quark mass. 
The relation between these two views is subtle because
in the canonical SCET/bHQET approach (ultra-)collinear jet functions are defined with a zero-bin subtraction~\cite{Manohar:2006nz} to avoid a double counting between (ultra-)collinear and collinear-soft radiation.

Transforming the partonic massive quark thrust distribution to Laplace space with the convention
\begin{align}
\label{eq:sigmaLaplacemassive}
\tilde \sigma(\nu,Q,m^{\rm pole}) \, = \, \int\limits_{\tau_{\rm min}^{\rm pole}} ^\infty  {\rm d}\tau\,e^{-\nu \tau}\,
\frac{1}{\sigma_0}\frac{\mathrm{d}\hat\sigma_s}{\mathrm{d}\tau}(\tau,Q,m^{\rm pole})
\end{align}
the NLL thrust distribution can be written in the condensed form
\begin{align}
\label{eq:Laplacemassive}
&\tilde\sigma(\nu,Q,m^{\rm pole})\,=\,
\mathrm{exp}\biggl[-K(\Gamma_{H_m},\mu_{H,\nu},\mu_{m,\nu})+K(\Gamma_B,\mu_{H,\nu},\mu_{B,\nu})+K(\Gamma_S,\mu_{H,\nu},\mu_{S,\nu})\biggr]\notag \\
&\qquad\times\mathrm{exp}\biggl[\frac{1}{2}\Bigl(\omega(\gamma_{H_m}-\gamma_{H_Q},\mu_{H,\nu},\mu_{m,\nu})+\omega(\gamma_B,\mu_{H,\nu},\mu_{B,\nu})+\omega(\gamma_S,\mu_{H,\nu},\mu_{S,\nu})\Bigr)\biggr]
\end{align}
where the evolution functions $K$ and $\omega$ have been given in Eqs.~(\ref{eq:Komega}) and the cusp and non-cusp anomalous dimensions not already displayed in Eqs.~(\ref{eq:cusp1}) and (\ref{eq:cusp0}) read
\begin{align}
\label{eq:cusp2}
\Gamma_B[\as]&\,=\,-\Gamma_{H_m}[\as]=-\Gamma_{H_Q}[\as]=2\Gcusp[\as]\,,\nonumber\\
\gamma_H[\as]&\, = \, -12C_F \Bigl(\frac{\as}{4\pi}\Bigr)\,,\nonumber\\
\gamma_{H_m}[\as]&\, = \, -8C_F \Bigl(\frac{\as}{4\pi}\Bigr)\,,\nonumber\\
\gamma_{J_B}[\as]&\, = \, 8C_F \Bigl(\frac{\as}{4\pi}\Bigr)
\end{align}
and the scales  $\mu_H$, $\mu_m$, $\mu_{B,\nu}$ and $\mu_{S,\nu}$ are given by 
\begin{align}
\label{eq:salesmassive}
\mu_{H,\nu} = \, Q\,,\qquad
\mu_{m,\nu} = \, m^{\rm pole}\,,\qquad
\mu_{B,\nu}=\, \frac{Q^2}{m^{\rm pole}} (\nu\, \mathrm{e}^{\gamma_E})^{-1} \,,\qquad
\mu_{S,\nu}=\, Q (\nu\, \mathrm{e}^{\gamma_E})^{-1} \,.
\end{align}
As for the case of massless quark production
these scales are fixed to the expressions shown and arise from the combination of the renormalization scale dependent NLL $U$ evolution factors {\it and} the Laplace transformed ${\cal O}(\alpha_s)$ corrections in the hard, mass mode, bHQET jet and soft functions, shown in Eqs.~(\ref{eq:hardfunction}), (\ref{eq:Hmoneloop}), (\ref{eq:JBoneloop}) and (\ref{eq:scetsoftfunction}) respectively, which are logarithmic and plus-distributions. In this combination the dependence on the renormalization scales $\mu_H$, $\mu_m$, $\mu_B$ and $\mu_S$ cancels  and the result shown in Eq.~(\ref{eq:Laplacemassive}) with the physical scales given in Eqs.~(\ref{eq:salesmassive}) emerges. Like in the case of massless quarks, since the structure of these ${\cal O}(\alpha_s)$ corrections is already unambiguously known from the NLL renormalization properties, we consider them part of the NLL logarithmic contributions.
Using in Eq.~(\ref{eq:Laplacemassive}) the renormalization scales $\mu_{i}$ instead of the scales $\mu_{i,\nu}$ ($i=H,m,B,S$) one recovers the renormalization scale dependent results coming from the $U$ evolution factors alone. The mass dependence of the scales in Eq.~(\ref{eq:salesmassive}) and in the rescaled thrust variable $M_\tau$ defined in Eq.~(\ref{eq:Mtaudef}) is subleading and does not generate NLL contributions when the quark mass scheme is changed.

As we show in Sec.~\ref{sec:CBnocutmassive} all terms shown in Eq.~(\ref{eq:Laplacemassive}) are also precisely obtained by the coherent branching formalism at NLL order.

We finally note that all functions and $U$ factors 
that appear in Eqs.~(\ref{eq:thrustmassless2}) and (\ref{eq:thrustmassive2}) have been determined using dimensional regularization to regularize infrared and ultraviolet divergences and the ${\overline{\rm MS}}$ renormalization scheme. At this point the partonic soft function $S^{(\tau)}(k)$ does {\it not} contain any gap subtraction~\cite{Hoang:2007vb} to remove its ${\cal O}(\Lambda_{\rm QCD})$ renormalon ambiguity related to the partonic threshold at $k=0$.

\subsection{Importance of the shape function}
\label{sec:shapefunction}

The soft model shape function $S_{\rm mod}$ appearing in Eqs.~(\ref{eq:thrustmassless1}) and (\ref{eq:thrustmassive1}) represents an essential part of the thrust factorization theorems since it accounts
for the hadronization effects that affect the observable thrust distribution. The shape function leads to a smearing
of the parton level contributions and an additional shift of the peak position since the hadronization effects
increase the hemisphere masses by non-perturbative contributions. It is also essential as far as the shape of the
distribution in the resonance region is concerned where the thrust distribution is peaked.

Since in this work we are mainly interested in the $Q_0$-dependence of the partonic contributions, one may conclude that one should better drop the effects of the shape function $S_{\rm mod}$ in our analysis such that it does not interfere with the perturbative effects. 
However, this is not possible since analyzing the singular partonic corrections of the thrust distribution (and their $Q_0$ dependence) alone without any smearing does not allow for a correct interpretation of their contributions to the observable distribution. This can be easily seen for example from the ${\cal O}(\as)$ fixed-order parton level results for the massless and massive quark thrust distributions shown in Eqs.~(\ref{eq:thrustmasslessFO}) and (\ref{eq:thrustmassiveFO}). Here the partonic contributions to the observable distribution contained in the $\delta$-functions and in the regularized singularity structures of the plus distributions at the partonic thresholds at $\tau=0$ and $\tau=\tau_{\rm min}$, respectively, remain invisible if one simply studies the partonic contributions at a function of $\tau$. One may in particular conclude wrongly, that
the observable peak position is independent of $Q_0$ simply because the partonic threshold always remains at $\tau=0$ and $\tau=\tau_{\rm min}$ for massless and massive quarks, respectively. The essential point is that the complete set of singular structures in the (infinitesimal) vicinity of the threshold contributes in the resonance region and non-trivially affect the observable peak location. Thus, the partonic thresholds alone do not govern the observable peak position and some smearing is crucial to fully resolve the effects of all parton level contributions. 

As a consequence, in our analysis of the partonic effects coming from the shower cut $Q_0$, it is still important that we account for the hadronic smearing of the shape function $S_{\rm mod}$. For the analysis of the partonic effects coming from the shower cut $Q_0$ we therefore include a shape function that is $Q_0$-independent. It has the simple form
\begin{align}
\label{eq:smoddef}
S_{\rm mod}(\ell) \, = \,
\frac{128\, \ell^3}{3\,\Lambda_{\rm m}^4}\exp\Bigl(-\frac{4\ell}{\Lambda_{\rm m}}\Bigr)\,,
\end{align}
and the important properties 
\begin{align}
\int_0^\infty  {\rm d}\ell\,S_{\rm mod}(\ell) \, = \, 1
\quad \mbox{and} \quad
\int_0^\infty  {\rm d}\ell\,\ell\, S_{\rm mod}(\ell) \, = \, \Lambda_{\rm m}\,,
\end{align}
where we consider $\Lambda_{\rm m}$ values between $1$ and $5$~GeV for our conceptual discussions. 
(See also our comment after Eq.~(\ref{eq:thrustmassless1}).)		
We use this shape function for our analytic calculations as well as for the parton level 
numerical results we obtain from the \Herwig\ event generator. This way we can ensure that the 
smearing is precisely equivalent for both types of results.
We note that the exact form of $S_{\rm mod}$ and the size of the smearing scale $\Lambda_{\rm m}$ affect the form and the absolute value of peak location of the distribution in the resonance region. However,
for our analysis only the relative dependence of the peak position on the cut value $Q_0$ is essential, for which the
exact form of the shape function turns out to be irrelevant. 
We further note that for our numerical studies for top quark production we use the smearing due to $S_{\rm mod}$ to also
mimic effects of the top quark width even though the form of $S_{\rm mod}$ does not provide a fully consistent description.

As we show in Sec.~\ref{sec:unreleased}, for making physical predictions the soft model function has to 
compensate for the dependence of the parton level large angle soft radiation on the $Q_0$ cut. 
This is because for large angle soft radiation the shower cut 
represents a factorization scale that separates the parton level and non-perturbative regions. The point of our
examination, however, is not to make physical predictions, but to conceptually quantify the dependence on  
$Q_0$ with the aim to disentangle 
it unambiguously from the effect $Q_0$ has on the mass parameter. 
Along the same lines, we also do not account for the possible effects of a finite experimental resolution. The latter results in an additional smearing of the resonance distribution that, particularly in the context of hadron colliders, may by far exceed the smearing caused by the hadronization effects. While the overall norm still remains irrelevant for the peak position, properties of the theoretical distribution far away from the resonance region could then affect the experimentally observed peak position in a non-negligible way. In such a case the non-singular corrections may have to be included for a reliable description. This is straightforward, but beyond the scope of this work.

\subsection{NLO precision for the resonance location}
\label{sec:peakprecision}

Within quantum field theory a consistent discussion of a quark mass (renormalization) scheme is only meaningful if the theoretical description of the observable of interest has all or at least the dominant ${\cal O}(\alpha_s)$ corrections implemented. In the factorization theorems of Eqs.~(\ref{eq:thrustmassless1}) and (\ref{eq:thrustmassive1}) we can neglect the nonsingular corrections since they are power-suppressed in the resonance region. To be concrete, they lead to negligible shifts in the peak position of order $(1~\mbox{GeV})^2/Q^2$ and $(1~\mbox{GeV})^2/m^2$, respectively, upon including the smearing effects coming from the soft model shape function $S_{\rm mod}$. It is now obvious to ask the question if, apart from the summation of logarithms at NLL order, also the full set of ${\cal O}(\alpha_s)$ non-logarithmic fixed-order corrections contained in the hard, mass mode, jet and soft functions are needed to achieve ${\cal O}(\alpha_s)$ precision in the resonance region. These corrections are either constant (originating from the functions $H_Q$ and $H_m$, see Eqs.~\eqref{eq:hardfunction} and~\eqref{eq:Hmoneloop}, respectively) or proportional to the delta-function (coming from the functions $J^{(\tau)}$, $J_B^{(\tau)}$ and $S^{(\tau)}$, see Eqs.~\eqref{eq:scetjetfunction},\eqref{eq:JBoneloop} and \eqref{eq:scetsoftfunction}, respectively), and their sum is displayed in Eqs.~\eqref{eq:thrustmasslessFO} and \eqref{eq:thrustmassiveFO}.  
If one considers all aspects of the thrust distribution in the resonance region, obviously both, NLL resummation and the full set of ${\cal O}(\alpha_s)$ fixed-order corrections are needed. For example, the one-loop corrections in the hard function lead to ${\cal O}(\alpha_s)$ corrections in the norm of the thrust distributions. This in general favors the so-called "primed" counting scheme~\cite{Abbate:2010xh} where NLL$^\prime$ order refers to the resummation of logarithms at NLL order combined with all additional fixed-order corrections at ${\cal O}(\alpha_s)$. 

However, the mass sensitivity of the thrust distribution in the peak region mainly comes from the location of the resonance peak, $\tau_{\rm peak}$, and properties such as the overall norm of the distribution are less important. For most practical considerations of such kinematic distributions, the norm is even eliminated on purpose by considering distributions that are normalized to a restricted interval in the kinematic variable. Therefore, in our analysis we mainly focus on the resonance peak position of the thrust distribution and do not consider the overall norm. 
Interestingly, as we show in the following, when discussing the peak position with NLO (i.e.\ ${\cal O}(\alpha_s)$) precision, we only have to account for the NLL resummed cross section, and
we can neglect the ${\cal O}(\alpha_s)$ non-logarithmic corrections. 
The reason why these non-logarithmic ${\cal O}(\alpha_s)$ corrections do not contribute to the peak position $\tau_{\rm peak}$ at NLO is that they are represent corrections proportional to the LL cross section.

To see this more explicitly let us rewrite the NLL$+{\cal O}(\alpha_s)$ thrust distributions of Eqs.~(\ref{eq:thrustmassless1}) and (\ref{eq:thrustmassive1}) in the generic form
\begin{align}
\label{eq:generic1}
f_{{\rm NLL}+\alpha_s}(\tau) & \, = \, \int_0^\tau \!\!\mathrm{d}\bar \tau \, \hat f_{{\rm NLL}+\alpha_s}(\bar\tau) \, \bar S_{\rm mod}(\tau-\bar \tau)\,,
\end{align} 
where $f$ and $\hat f$ stand for the hadron and parton level thrust distributions, respectively, and $\bar S_{\rm mod}$ for the hadronization shape function after variable rescaling. The  NLL$+{\cal O}(\alpha_s)$ partonic thrust distribution can then be written in the form 
\begin{align}
\hat f_{{\rm NLL}+\alpha_s}(\tau) \, = \, \hat f_{\rm LL}(\tau) \,+\, \alpha_s\,\Big(\Delta\hat f_{\rm NLL}(\tau) + c \hat f_{\rm LL}(\tau)\Big)
\end{align}
where $\hat f_{\rm LL}$ represents the LL cross section (which provides the complete leading order approximation),
the term $\alpha_s\Delta f_{\rm NLL}$ contains all NLL corrections in the NLL resummed cross section, and
$\alpha_s c$ stands for the non-logarithmic ${\cal O}(\alpha_s)$ corrections mentioned above.
The latter corrections are related to the LL tower of logarithms associated to the term $(2\pi^2/3-2)\delta(\tau)$
in Eq.~(\ref{eq:thrustmasslessFO}) and the term $2\pi^2 \delta(\tau-\tau_{\rm min}^{\rm pole})$
in Eq.~(\ref{eq:thrustmassiveFO}). Note that corrections arising from a change in the quark mass scheme 
are proportional to
derivatives of $\delta(\tau-\tau_{\rm min}^{\rm pole})$ and therefore {\it always} contained
in the term $\alpha_s\Delta f_{\rm NLL}$.
 
The LL peak position $\tau_{\rm peak}^0$ is determined from the equality
\begin{align}
\label{eq:generic2}
0\,\,\stackrel{!}{=}\, & f^\prime_{\rm LL}(\tau_{\rm peak}^0)  \, = \, \int_0^{\tau^0_{\rm peak}} \!\!\mathrm{d}\bar \tau \, \hat f_{\rm LL}(\bar\tau) \, \bar S_{\rm mod}^\prime(\tau^0_{\rm peak}-\bar \tau)\,.
\end{align}
At the NLL level, writing the ${\cal O}(\alpha_s)$ correction to the peak position as $\delta\tau_{\rm peak}$, the corresponding equality reads
\begin{align}
\label{eq:generic3}
0\,\,\stackrel{!}{=}\, &  f^\prime_{{\rm NLL}+\alpha_s}(\tau_{\rm peak}^0+\delta\tau_{\rm peak}) \\[2mm]
\, = \, & 
\int_0^{\tau^0_{\rm peak}+\delta\tau_{\rm peak}} \!\!\mathrm{d}\bar \tau \, 
\Big[ \hat f_{\rm LL}(\bar\tau)  + \alpha_s\,\Big(\Delta\hat f_{\rm NLL}(\bar\tau) + c \hat f_{\rm LL}(\bar\tau)\Big)\Big]\, \bar S_{\rm mod}^\prime(\tau^0_{\rm peak}+\delta\tau_{\rm peak}-\bar \tau)\nonumber \\
\, = \, & 
\delta \tau_{\rm peak}\,
\int_0^{\tau^0_{\rm peak}} \!\!\mathrm{d}\bar \tau \, \hat f_{\rm LL}(\bar\tau) \, \bar S_{\rm mod}^{\prime\prime}(\tau^0_{\rm peak}-\bar \tau) 
\nonumber \\& \quad 
\,+\, \alpha_s\,
\int_0^{\tau^0_{\rm peak}} \!\!\mathrm{d}\bar \tau \, 
\Big[ \Delta\hat f_{\rm NLL}(\bar\tau) + c \hat f_{\rm LL}(\bar\tau)\Big]\, \bar S_{\rm mod}^\prime(\tau^0_{\rm peak}-\bar \tau)
\, + \, {\cal O}(\alpha_s^2)
\nonumber \\
\, = \, & 
\delta \tau_{\rm peak}\,
\int_0^{\tau^0_{\rm peak}} \!\!\mathrm{d}\bar \tau \, \hat f_{\rm LL}(\bar\tau) \, \bar S_{\rm mod}^{\prime\prime}(\tau^0_{\rm peak}-\bar \tau)
\nonumber \\& \quad  
\,+\, \alpha_s
\int_0^{\tau^0_{\rm peak}} \!\!\mathrm{d}\bar \tau \, 
\Delta\hat f_{\rm NLL}(\bar\tau)\, \bar S_{\rm mod}^\prime(\tau^0_{\rm peak}-\bar \tau)
\, + \, {\cal O}(\alpha_s^2)
\nonumber 
\,,
\end{align}
where in the third line we have dropped terms of ${\cal O}(\alpha_s^2)$ and in the fourth we used the LL constraint of Eq.~(\ref{eq:generic2}) for the non-logarithmic ${\cal O}(\alpha_s)$ fixed-order corrections with are proportional to the LL cross section. 

The outcome is that the non-logarithmic ${\cal O}(\alpha_s)$ fixed-order corrections contained in the hard, jet and soft function are not relevant for discussing the peak position $\tau_{\rm peak}$ as far as ${\cal O}(\alpha_s)$ precision is concerned and would only enter when ${\cal O}(\alpha_s^2)$ corrections are considered. Since the peak position represents the dominant characteristics of the thrust distribution entering the mass determination, we can therefore conclude that the resummation of logarithmic correction at the NLL level is sufficient to achieve ${\cal O}(\alpha_s)$ precision for a mass determination based on the resonance peak position.  
Going along the line of arguments we use in the subsequent sections this important result also means that to the extent that parton showers systematically and {\it correctly sum all NLL logarithmic terms}, the peak position of the thrust distribution generated by their evolution is already ${\cal O}(\alpha_s)$ precise, even {\it without} including any additional NLO fixed-order corrections by an NLO matching prescription.

\section{Coherent branching formalism}
\label{sec:coherentbranching}

The coherent branching formalism has proven to be a very powerful tool for
analytic resummation of a large number of observables. Besides the analytic
use, it forms the core rationale behind coherent parton shower algorithms,
notably the angular ordered algorithms of the Herwig family \cite{Bahr:2008pv,Bellm:2015jjp,Bellm:2017bvx} of event
generators. Following earlier work of Refs.~\cite{Catani:1989ne,Catani:1992ua} we use this framework to calculate
the parton level jet mass distributions $J(s,Q^2)$ for massless quarks and $J(s,Q^2,m^2)$ for massive quarks   
originating from successive gluon radiation
off the progenitor quark and anti-quark pair generated by the hard interaction at c.m.\ energy $Q$. Here the variable $s=M_{\rm jet}^2$ stands for the resulting squared jet invariant mass.
This determines the parton level thrust distribution in the peak region as defined in Eq.~(\ref{eq:taudef}) as
\begin{align}\label{eq:sigma_cb}
\frac{{\rm d}\hat{\sigma}^{\rm cb}}{{\rm d}\tau}(\tau,Q) \,= \,&\sigma_0 \int \!{\rm d}s_1 \,{\rm
    d}s_2\,\delta\left(\tau - \frac{s_1+s_2}{Q^2}\right)
  J(s_1,Q^2)J(s_2,Q^2)\,,\notag \\
\frac{{\rm d}\hat{\sigma}^{\rm cb}}{{\rm d}\tau}(\tau,Q,m) \,= \,&\sigma_0 \int \!{\rm d}s_1 \,{\rm
    	d}s_2\,\delta\left(\tau - \frac{s_1+s_2}{Q^2}\right)
    J(s_1,Q^2,m^2)J(s_2,Q^2,m^2)\,,
\end{align}
for massless and massive quark cases, respectively. The jet mass distributions obtained in the 
context of coherent branching incorporate coherently the dynamic effects of soft as well as 
(ultra-)collinear radiation and are UV-finite quantities. 
Thus they differ from the jet functions $J^{(\tau)}$ and $J_B^{(\tau)}$ 
in the QCD factorization approach which describe the factorized collinear and ultra-collinear gluon effects,
respectively,
and are determined from UV-divergent effective theory matrix elements that need to be renormalized.
In order to obtain the observable hadron level thrust distribution, the contributions of the non-perturbative 
effects are accounted for in exactly the same way as for the QCD factorization approach 
by an additional convolution with a soft model shape 
function, as shown in Eqs.~(\ref{eq:thrustmassless1})
and (\ref{eq:thrustmassive1}), see Refs.~\cite{Collins:1985xx,Catani:1989ne,Catani:1992ua}.

We note that in Eqs.~\eqref{eq:sigma_cb} we have used the superscript 'cb' to indicate the cross sections obtained in the coherent branching formalism. We use this notation throughout this paper, when suitable, to distinguish results based on the coherent branching formalism from those obtained in the factorization approach.

While an analytic treatment of the coherent branching formalism in the strict context of perturbation theory does 
not rely on the presence of any infrared cutoff,\footnote{We refer to strict perturbation theory as expanding 
in $\alpha_s$ at a constant renormalization scale such that the evolution is described by higher powers of 
$\alpha_s$ and logarithms only, and that virtual loop and real radiation phase space integrals can be carried out
down to zero momenta.} it is, however, required within the realm 
of an event generator for several reasons. These include the Landau pole singularity of
the strong coupling, which emerges because its renormalization scale is tied to shower
evolution variables, and that the particle multiplicities diverge when the shower 
evolves to infrared scales. In addition, in the limit of small scales the perturbative treatment 
of the parton splitting
breaks down anyway, and it is therefore mandatory to terminate the shower 
at a low scale where the perturbative description is still valid
and hand over
the partonic ensemble generated through the shower emissions 
to a phenomenological model of hadronization.

The variables we consider in the following of this section are used both to derive analytic
results, but we also stress that they precisely correspond to the variables
employed in the angular ordered parton shower of the \Herwig~7 event generator. 
The results obtained from the \Herwig~7 event generator 
only differ from the analytic framework by the implementation of exact momentum
conservation with respect to the momenta of all final state particles that emerge
when the shower has terminated at its infrared cutoff $Q_0$. 
This implementation of momentum conservation shall not change the jet mass distribution and
is explained in more detail in Sec.~\ref{sec:coherentbranchingHerwig}. There we also briefly discuss
some \Herwig~7 (version 7.1.2) specific implementations in its default setting 
that go beyond the coherent branching
formalism and that we do not use in the context of the conceptual studies carried out in this work. 

\subsection{Massless case}
\label{sec:coherentbranchingmassless}

\begin{figure}
  \begin{center}
    \includegraphics[scale=0.8]{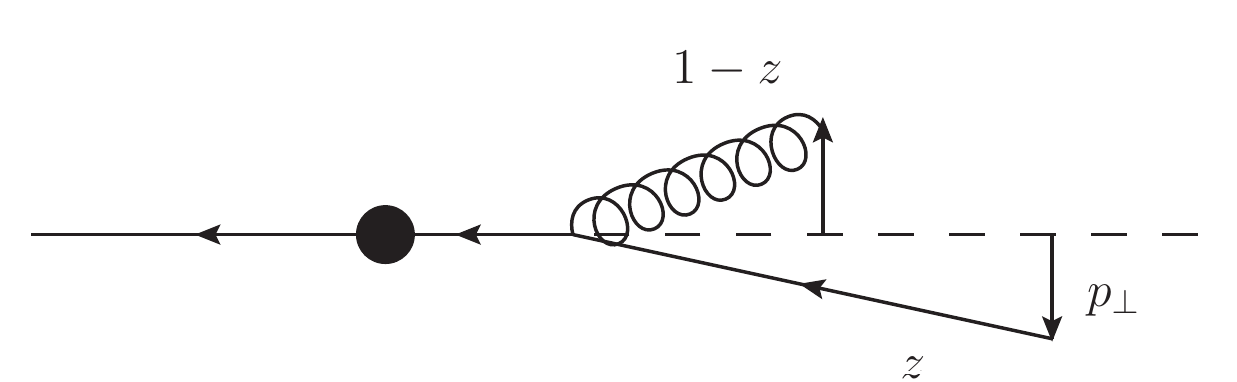}
  \end{center}
  \caption{\label{fig:branching} A gluon branching off a back-to-back
    quark/anti-quark system. The radiated gluon is assumed to carry a fraction
    $1-z$ of the parent's momentum and is emitted at a transverse momentum
    which equals the one acquired by, in this case, the anti-quark after the
    emission.}
\end{figure}

Starting from an initial, color-connected $q\bar{q}$-pair with momenta $p$
and $\bar{p}$, the momenta of the partons emerging from the shower evolution of the quark
carrying the momentum $p$ are parametrized based on
\begin{equation}
\label{eq:globalMomentum}
  k_i^\mu = \alpha_i\ p^\mu + \beta_i\ \bar{n}^\mu + k_{i,\perp}^\mu\,,
\end{equation}
where $k_i$ is the quarks momentum after the $i$-th emission. In the
massless case we use $\bar{n}=\bar{p}$ as the reference direction to specify
the collinear limit, with $k_{i,\perp}\cdot p = k_{i,\perp}\cdot n=0$,
$k_{i,\perp}^2<0$ and $\beta_i$ being determined by the virtualities $k_i\cdot k_i =
k_i^\mu k_{i\mu}=
k_i^2$ as
\begin{equation}
\beta_i = \frac{-k_{i,\perp}^2+k_i^2}{2\alpha_i(p\cdot \bar{n})} \ .
\end{equation}
The radiation off the
anti-quark with momentum $\bar{p}$ is described similarly with a reference
direction $n=p$.  Expressing $k_i^\mu$ in terms of the momentum of the emitter
before the i-th branching we find
\begin{equation}
  k_i^\mu = z_i\ k_{i-1}^\mu + \frac{p_{i,\perp}^2 + k_i^2 - z_i^2 k_{i-1}^2}{2 z_i (k_{i-1}\cdot \bar{n})}\bar{n}^\mu + q_{i,\perp}^\mu
\end{equation}
where the physical splitting variables relative to the quark's momentum $k_{i-1}$ before
the i-th emission relate to the global light-cone decomposition
Eq.~(\ref{eq:globalMomentum}) as
\begin{align}
\label{CBzdef}
z_i &\, =\,  \frac{\alpha_{i}}{\alpha_{i-1}}\,,\\
\label{CBqperpdef}
q_{i,\perp}^\mu &\, =\, k_{i-1,\perp}^\mu- z_i k_{i,\perp}^\mu\,,
\end{align}
where $\alpha_0 = 1$ as well as $q_{0,\perp}^\mu = 0$ are understood.
This means that for the first emission the physical branching variables
coincide with the global parametrization. We have depicted the variables of
one branching in Fig.~\ref{fig:branching}.
Soft gluon coherence is encoded through ordering emissions in an angular
variable~\cite{Catani:1990rr},
\begin{equation}
\label{qtildemassless}
  \tilde{q}_i^2 = \frac{p_{i,\perp}^2}{z_i^2(1-z_i)^2} \ ,
\end{equation}
where $p_{i,\perp}^2=-q_{i,\perp}^2$ is the magnitude of the transverse
momentum, which is purely spacelike and perpendicular to the emitter axis in
the centre-of-mass system of the momenta $k_i$ and $\bar{n}$. The explicit
restrictions of decreasing opening angle of subsequent emissions following a
branching at scale $\tilde{q}_i$ from the evolving quark or anti-quark at
scale $\tilde{q}^2_{i+1}$, and the radiated gluon at scale $\tilde{k}^2_{i}$
are imposed by the conditions
\begin{equation}
\label{eq:angularOrdering}
\tilde{q}^2_{i+1} < z_i^2\ \tilde{q}^2_i\qquad \mbox{and } \qquad 
\tilde{k}^2_{i} < (1-z_i^2)\ \tilde{q}_i^2\,.
\end{equation}
In the context of these variables, the Altarelli-Parisi splitting functions explicitly show
the full Eikonal radiation pattern and the correct collinear limit,
see {\it e.g.} Ref.~\cite{Platzer:2009jq} for an overview and comparison to dipole-type parton
showers. The formalism is appropriate to resum higher order logarithmic corrections for 
observables that are inclusive
concerning the collinear radiation in the same jet {\it and} in the sense that the information
that large-angle soft gluon radiation originates from a particular collinear parton
is unresolved and can hence be described to originate from the net collinear color charge of 
the whole jet. Momentum conservation in the branching $i-1\to i$ implies
\begin{equation}
\label{eq:momentumConservation}
k_{i-1}^2 = \frac{k_{i}^2}{z_i}+\frac{q_i^2}{1-z_i} +z_i(1-z_i)\tilde{q}_i^2
\,,
\end{equation}
where $q_i^2$ is the virtuality of the emitted gluon, the momentum of which is
parametrized in a decomposition similar to Eq.~(\ref{eq:globalMomentum}).

\begin{figure}
  \begin{center}\includegraphics[scale=0.8]{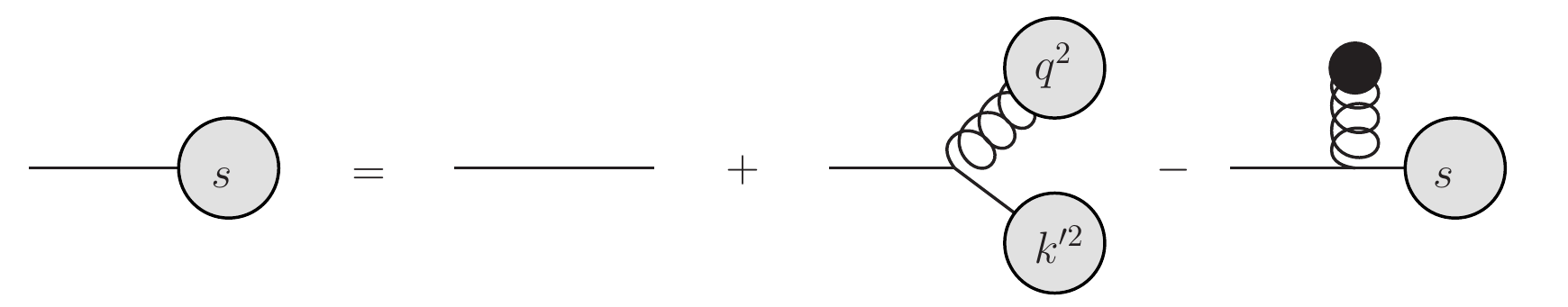}\end{center}
  \caption{\label{fig:jetfunction}Graphical representation of the evolution
    equation Eq.~\ref{jetfunctionmasslessdef}: Grey blobs denote the quark and
    gluon jet function at a given jet mass, a single line implies a
    $\delta$-function at mass zero, while the black dot represents a factor of
    one and implies an unconstrained integration over the gluon's emission
    scale and momentum fraction.}
\end{figure}

We follow Ref.~\cite{Catani:1992ua} and start with an analytic approach for
which the evolution equation for the jet mass distribution starting at
a hard scale $\tilde{q}^2 = Q^2$ has the form
\begin{align}
\label{jetfunctionmasslessdef}
J&(s,Q^2)\,=\,\delta(s)+\int_0^{Q^2}\frac{\mathrm{d}\tilde{q}^2}{\tilde{q}^2}
\int_0^1\mathrm{d}z\,P_{qq}\Bigl[\alpha_s\bigl(z(1-z)\tilde{q}\bigr),z\Bigr]
\\& 
\times \Biggl[\int_0^\infty\mathrm{d}k^{\prime 2}\int_0^\infty
  \mathrm{d}q^2\delta\Bigl(s-\frac{k^{\prime
      2}}{z}-\frac{q^2}{1-z}-z(1-z)\tilde{q}^2\Bigr)
   J(k^{\prime 2},z^2\tilde{q}^2)
   J_g(q^2,(1-z)^2\tilde{q}^2)\notag\\ & \hspace{2cm}
-J(s,\tilde{q}^2)\Biggr]\,,\notag
\end{align}
where $J_g(s,Q^2)$ is the gluon jet mass distribution defined in analogy to the
jet mass distribution $J(s,Q^2)$ for the quarks.
We have illustrated the evolution schematically in Fig.~\ref{fig:jetfunction}.
The splitting function is given by
\begin{equation}
\label{eq:masslessSplittingFunction}
P_{qq}\Bigl[\alpha_s,z\Bigr]\, =\, \frac{\alpha_s C_F}{2\pi}\frac{1+z^2}{1-z} \,=\, 
\frac{\alpha_s C_F}{2\pi}\,\Bigl[ \,\frac{2}{1-z} - (1+z)\,\Bigr]
\end{equation}
where the second equality makes the cusp and non-cusp terms
explicit, which stem from soft ($z\to 1$) and hard collinear emissions,
respectively. 

We note that the evolution equation for the jet mass distribution shown in Eq.~(\ref{jetfunctionmasslessdef}) can
be rendered NLL precise by correctly implementing the analytic form of the two-loop cusp
term in quark splitting function $P_{qq}$. By using the relative transverse momentum of the splitting,
\begin{align}
\label{eq:asscale}
p_{\perp}^2= z^2(1-z)^2\tilde q^2\,,
\end{align}
as the renormalization scale for the strong coupling
the leading $\ln(1-z)/(1-z)$ behavior of the cusp term in the two-loop splitting function
is reproduced exactly. The remaining non-logarithmic term from the two-loop cusp anomalous
dimension and can be incorporated by either
scaling
\begin{equation}
  \label{eq:CMW}
  \alpha_s\to \alpha_s\left(1 + K_g\frac{\alpha_s}{2\pi}\right) \ ,
\end{equation}
or, equivalently, (up to terms of ${\cal O}(\alpha_s^3)$) by adopting a change in renormalization scheme through the 
rescaling
\begin{equation}\label{eq:CMW2}
 \Lambda_{\overline{\text{MS}}}\to
  \Lambda_{\text{MC}}=\Lambda_{\overline{\text{MS}}}\exp\left(\frac{K_g}{\beta_0}\right)
\end{equation}
The constant $K_g$ commonly used in this context relates to the
two-loop cusp anomalous dimension as $\Gcusp_1 = 8 C_F K_g$ shown in
Eqs.~(\ref{eq:cusp0}).  This approach to implement NLL
precision in parton showers is called the CMW
(''Catani-Marchesini-Webber'') or Monte Carlo scheme
\cite{Catani:1990rr}. We note that in the \Herwig\ event generator, the transverse
momentum argument~(\ref{eq:asscale}) is used as the scale of the strong coupling, but that in
the default settings the CMW scheme of Eqs.~\eqref{eq:CMW} and \eqref{eq:CMW2} is not used explicitly.  
Instead the precise value of $\alpha_s$ is obtained from tuning to
LEP data along with the parameters of the hadronization model and the 
shower cut $Q_0$. 
The result, however, numerically resembles the CMW factor in
the relation between $\Lambda_{\overline{\text{MS}}}$ and $\Lambda_{\rm MC}$.
Indeed, for example for a one-loop running the CMW correction implies that
\begin{equation}
\label{eq:MCalphas}
  \alpha_s^{\text{MC}}(M_Z) =
  \frac{\alpha_S^{\overline{\text{MS}}}(M_Z)}{1-\alpha_S^{\overline{\text{MS}}}(M_Z)\frac{K_g}{2\pi}}
  = 0.126\quad\text{at}\quad \alpha^{\overline{\text{MS}}}_S(M_Z) =
  0.118\ ,\quad n_f=5 \ ,
\end{equation}
and the larger value is exactly is the tuned value, with a similar converted
value for $\alpha^{\overline{\text{MS}}}_s(M_Z)$ for the two loop running
actually employed in the \Herwig\ shower.  For our numerical analyses in
Secs.~\ref{sec:herwigmassless} and \ref{sec:herwigmassive}, where we compare
analytic calculations and \Herwig\ results concerning the shower cut $Q_0$
dependence of the thrust peak position, we therefore use the strong coupling
as implemented in \Herwig .

The evolution equation for the jet mass distribution given in Eq.~(\ref{jetfunctionmasslessdef}) 
is an explicit representation of the coherent branching algorithm. Consider 
the distribution of the first emitter's virtuality $k_0^2\equiv k^2$ and one iteration of
the branching algorithm, where one choses  $\tilde{q}^2 \equiv
\tilde{q}_1^2$, $z\equiv z_1$, as well as $k'^2\equiv k_1^2$ and the gluon's
virtuality is denoted by $q^2\equiv q_1^2$ as displayed in Fig.~\ref{fig:jetfunction}.  
There is a
contribution without any branching or virtual effects, encoded in the first $\delta$-function term 
in Eq.~(\ref{jetfunctionmasslessdef}). It describes a
vanishing jet mass that corresponds to the tree-level contribution
and also constitutes the initial condition for the shower evolution at $\tilde{q}^2=Q^2$.
In addition, we need to take into account a resolvable branching at a scale
$\tilde{q}^2$ below the hard scale $Q^2$, which gives rise to a subsequent
evolution of the quark and gluon jet mass distributions at the scales imposed
by the angular ordering criterion of Eq.~(\ref{eq:angularOrdering}). This
contribution is itself constrained by the momentum conservation criterion of
Eq.~(\ref{eq:momentumConservation}). The last contribution originates from an
unresolved emission, which gives rise to an evolution of the quark mass
distribution starting at scale $\tilde{q}^2$ but being unconstrained otherwise. 
Notice that the momentum conservation constraint links the
evolution scale to the specific kinematics that is considered. No further constraints
to the integration over the momenta involved in the emission are present. 

As already mentioned, in the context of an event generator the evolution has to be terminated
by imposing infrared cutoff $Q_0$. This is typically
done by a requiring a minimum transverse momentum for the emissions with respect to the
momentum direction of the emitter. This restricts the integral over $\tilde{q}^2$ and $z$ 
to a region where
\begin{equation}
\label{pperpcut1}
p_{\perp}^2 = \tilde{q}^2\ z^2(1-z)^2 > Q_0^2 \ .
\end{equation}
We note that also other choices are in principle possible and have
been discussed in the context of radiation within the 'dead cone' for
massive quarks \cite{Gieseke:2003rz}. In principle any prescription
that simultaneously cuts off both the collinear $\tilde{q}\to 0$ and
soft $z\to 1$ (as well as $z\to 0$ for a gluon branching) limits, and
also avoids low transverse momenta appearing in the argument of the
strong coupling, is appropriate.

We also note that an analogous evolution equation holds for the gluon jet mass distribution $J_g(s,Q^2)$.
The evolution of the gluon jet is governed by the gluon splitting function, and also describes
gluon branching into a quark/anti-quark pair. However, as
far as the jet mass distributions in the resonance region is are concerned, the contribution of the gluon jet
mass to the quark jet mass is at least at NLL precision suppressed due to the angular ordering
constraint, see e.g.\ Ref.~\cite{Catani:1992ua}. 
Therefore, at NLL several simplifying approximations are in principle possible to solve the evolution
equation for the quark jet mass distribution, which are particularly useful 
for analytic calculations of the jet mass distribution:
(i) we can neglect the contribution
to the jet mass due to the branching of emitted gluons by the replacement 
$J_g((1-z)\tilde{q}^2)\to \delta(q^2)$ for the gluon jet
mass distribution and (ii) we can can take the limit
$z\to 1$ for some terms that do not acquire an enhancement in the soft limit.
Interestingly, this also includes that, once prescription (i) is applied, we can remove the 
remaining, strict angular ordering constraint in the quark jet mass distribution through
modifying the starting scale of the subsequent emission contained in the quark jet mass distribution
by the replacement $J(k^{\prime 2},z^2\tilde{q}^2)\to J(k^{\prime 2},\tilde{q}^2)$.
In Sec.~\ref{sec:herwigthrusttest} we explicitly verify these simplifications 
from numerical simulations using the \Herwig~7 event generator.

\subsection{Massive case}
\label{sec:coherentbranchingmassive}

Moving on to radiation off massive quarks, we consider the
generalizations of coherent branching developed in
Ref.~\cite{Gieseke:2003rz}, based on splitting functions and
factorization {\it in the quasi-collinear limit} for which the emitted
parton's transverse momenta is restricted from above by the mass of
the emitting quark and furthermore small compared to the
scale of the previous emission, $p_{i,\perp}^2\lesssim m_i^2 \ll
2(k_{i-1}\cdot \bar{n})$. In this case we consider a system of a
massive quark and anti-quark, $p^2=\bar{p}^2=m^2$. However we
still use light-like backward directions $\bar{n}$ and $n$
in the momentum parametrization such as~(\ref{eq:globalMomentum}), with
three-momenta pointing along the direction of the massive momenta,
{\it i.e.}  $\bar{n}=(|\vec{p}|,-\vec{p})$ and
$n=(|\vec{p}|,\vec{p})$. This modifies the form of the $\beta_i$
variables to take into account the mass effect,
\begin{equation}
\beta_i = \frac{-k_{i,\perp}^2+k_i^2-\alpha_i^2\ m^2}{2 \alpha_i (p\cdot \bar{n})}\ ,
\end{equation}
while the parametrization of the momenta from the massless case given in
Eq.~(\ref{eq:globalMomentum}) and the relation to the branching variables in
Eqs.~(\ref{CBzdef}) and (\ref{CBqperpdef}) remain unchanged. Following 
Ref.~\cite{Gieseke:2003rz} the evolution variable is generalized to the expression
\begin{equation}
\label{qtildemassive}
\tilde{q}_i^2 = \frac{p_{i,\perp}^2+(1-z_i)^2m^2}{z_i^2(1-z_i)^2} \ .
\end{equation}
Consequently, the generalization of Eq.~(\ref{eq:momentumConservation}) also adopts a 
mass term and reads
\begin{equation}
k_{i-1}^2 = \frac{k_{i}^2-(1-z_i)m^2}{z_i}+\frac{q_i^2}{1-z_i} +z_i(1-z_i)\tilde{q}_i^2 \,.
\end{equation}
The arguments we discussed for the massless quark case concerning the mass of the gluon jet 
apply in the analogous way in the massive quark case. Therefore we do not have to consider
the fully general formalism  for our analytic calculations at NLL order 
and can restrict ourselves to the case of
gluon emission from a massive quark. 
We note that gluon splitting into massive
quarks is also a negligible effect for the jet mass distribution in the resonance region since the
corresponding splitting function is suppressed with respect to the gluon
emission case due to a lack of soft enhancement 
(even in the absence of angular ordering). The variables considered here are
precisely those used in the \Herwig~7 angular ordered shower, which,
in its current version is not relying on a finite $Q_g$ parameter as
quoted in \cite{Gieseke:2003rz}, but is instead using a cutoff on the transverse momentum. 

The evolution equation of the massive quark jet mass
distribution then has the form
\begin{align}
\label{jetfunctionmassivedef}
J&(s,Q^2,m^2)\,=\,\delta(s-m^2)+
\int_{m^2}^{\tilde{Q}^2}\frac{\mathrm{d}\tilde{q}^2}{\tilde{q}^2}\int_0^1\mathrm{d}z\,
P_{QQ}\biggl[\alpha_s(\mu_R^2(\tilde{q}^2,z)),z,\frac{m^2}{\tilde q^2}\biggr]
\\ &
\times
\Biggl[\int_0^\infty\mathrm{d}k^{\prime 2}\int_0^\infty
  \mathrm{d}q^2\delta\Bigl(s-\frac{k^{\prime
      2}-(1-z)m^2}{z}-\frac{q^2}{1-z}-z(1-z)\tilde{q}^2\Bigr)
\notag\\&
\hspace{4cm}\times J(k^{\prime 2},z^2\tilde{q}^2,m^2)J_g(q^2,(1-z)^2\tilde{q}^2)
\notag\\& 
\hspace{2cm}-\,J(s,\tilde{q}^2,m^2)\Biggr] \,.\notag
\end{align}
The initial hard scale of the evolution in $\tilde q^2$ is chosen as
\begin{equation}
\tilde{Q}^2 = \frac{1}{2}Q^2\left(1+\sqrt{1-\frac{4m^2}{Q^2}}\right)
\end{equation}
which amounts to the 'symmetric' phase space choice for the $Q\bar{Q}$
system as suggested in Sec.~3.2 of Ref.~\cite{Gieseke:2003rz}, so that 
the shower evolution off the progenitors $Q$ and $\bar{Q}$ only cover
physically distinct phase space regions.  
For the situation of boosted quarks ($m^2/Q^2\ll 1$) we consider in this paper, however,
we can safely replace $\tilde Q^2\to Q^2$ for all analytic calculations. 
The shower 
cutoff condition in the massive quark case reads
\begin{equation}
\label{pperpcut2}
p_\perp^2 = z^2(1-z)^2\tilde{q}^2 - (1-z)^2m^2 > Q_0^2\,,
\end{equation}
and the splitting function in the quasi-collinear limit generalizes to
\begin{equation}
\label{eq:massiveSplittingFunction}
P_{QQ}\biggl[\alpha_s,z,\frac{m^2}{\tilde q^2}\biggr] \, =\, 
 \frac{\alpha_s C_F}{2\pi}\,\Biggl[\, \frac{1+z^2}{1-z} - \frac{2 m^2}{z(1-z)\tilde q^2}\,\Biggr]
 \,.
\end{equation}
In contrast to the massless quark case where the coherent branching formalism 
has a solid conceptual basis related to the different kinematics of soft and collinear phase 
space regions, the corresponding formalism for massive quarks has in its 
present form higher order ambiguities,
which makes e.g.\ the determination ${\cal O}(\as^2)$ corrections to the
quasi-collinear splitting functions ambiguous. This is related to the more complicated structure 
of collinear, ultra-collinear, mass mode and soft dynamics and phase space regions that emerge
in the presence of the quark mass and which (as we show explicitly in Sec.~\ref{sec:CBnocutmassive}) depends
in addition
on the relation between the jet invariant mass $\sqrt{s}$ and the quark mass $m$. 
This is manifest in the fact that, in contrast to the massless quark case, 
there is no unique choice of the 
renormalization scale of $\alpha_s$
as a function of $z$, $\tilde q$ and the quark mass $m$.
As such, different
choices for $\mu_R^2(\tilde{q}^2,z)$ which reduce to 
Eq.~(\ref{eq:asscale}) in the massless limit
may be considered. The default choice
is the generalized transverse momentum, $\mu_R^2(\tilde{q}^2,z) =
\tilde{q}^2 z^2(1-z)^2$, which adds an additional mass-dependent contribution relative to
the physical transverse momentum given in Eq.~(\ref{pperpcut2}). We demonstrate in Sec.~\ref{sec:CBnocutmassive}
that this choice is fully consistent with the QCD factorization approach for massive quarks at NLL order. 
(See also the power counting shown in Tab.~\ref{tab:counting_massive}: In the soft gluon region the $m^2$ term is 
suppressed and irrelevant, and in the ultra-collinear region the $\tilde q^2$ and the $m^2$ terms are of the same order.)

\subsection{Coherent branching in the Herwig 7 event generator}
\label{sec:coherentbranchingHerwig}

The coherent branching formalis and its variables outlined in the previous two subsections form the core of the
angular ordered parton shower in the \Herwig~7 event generator
\cite{Bahr:2008pv,Bellm:2015jjp,Bellm:2017bvx}, covering the massless 
and the massive quark cases as discussed in 
Secs.~\ref{sec:coherentbranchingmassless} and 
\ref{sec:coherentbranchingmassive}, respectively. 
In the \Herwig~7 parton shower algorithm, a sequence of random values for the variables 
$\tilde{q}$ and $z$ is generated, distributed according to the Sudakov form factor 
that depends on the splitting function. This provides a solution to the
evolution of the jet mass distribution accounting for the branching
and no-branching probabilities in terms of explicit events. 

A major difference to a purely analytic computation of the jet mass
distributions encoded in the evolution equations~(\ref{jetfunctionmasslessdef}) and 
(\ref{jetfunctionmassivedef}),
however, is related to the virtualities, i.e.\ the off-shell invariant
masses of the branching partons. While an analytic calculation of the
jet mass distribution just focuses on the description of the overall
invariant mass of the final state particles produced by the emissions
from the progenitor parton originating from the hard process, event
generators have to face an additional constraint: they have to evolve
the progenitor parton to a final state consisting of partons {\it on
their physical mass shell consistent with overall energy-momentum
conservation} at the point when the shower terminates. This
procedure is called {\it 'kinematic reconstruction'}.  It is the
kinematic reconstruction procedure that fixes the virtualities to the
partons before showering (which are, however, approximated as on-shell
in the splitting function). The kinematic reconstruction is based on the 
information of the entire
evolution tree, the momentum decomposition based on Eq.~(\ref{eq:globalMomentum}),
four-momentum conservation at each vertex, and the knowledge of the
$\tilde{q}$ and $z$ values of each branching to determine explicit
particle momenta and to relate the kinematics of the
subsequent emissions to the associated off-shell invariant masses. 

In this context an
additional important issue the kinematic reconstruction procedure has
to deal with is that the sizes of the physical virtualities are kinematically
limited by the available phase space. However, this phase space constraint
is {\it not} imposed by the parton shower
evolution itself, such that physically inaccessible (i.e.\ too large) invariant masses
can be generated. Given the decomposition of the
momenta based on Eq.~(\ref{eq:globalMomentum}), and a sequence of
$\tilde{q}$ and $z$ values, the kinematic reconstruction algorithms are 
designed such that one single solution for the final state momenta 
is obtained. However, physically, the final state momenta cannot be determined
uniquely such that ambiguities arise in the way how overall
energy-momentum conservation is restored in the event. 

To illustrate the kinematic reconstruction procedure more concretely, 
consider the production of a 
quark/anti-quark progenitor pair produced in $e^+e^-$ annihilation 
carrying on-shell momenta
\begin{equation}
p=\left(\sqrt{\mathbf{p}^2+m^2},\mathbf{p}\right)\qquad\mbox{and}\qquad 
\bar{p}=\left(\sqrt{\mathbf{p}^2+m^2},-\mathbf{p}\right)\,,
\end{equation}
respectively, with the initial tree-level process constraint 
$Q=2\sqrt{\mathbf{p}^2+m^2}$ at the 
starting point of the parton shower evolution.
At the end of the parton shower evolution their showered
counterparts will have gained virtualities $M^2\ge m^2$ and
$\bar{M}^2\ge m^2$ with momenta
\begin{equation}
P = \left(\sqrt{M^2+\mathbf{P}^2},\mathbf{P}\right)\qquad \bar{P} = \left(\sqrt{\bar{M}^2+\bar{\mathbf{P}}^2},\bar{\mathbf{P}}\right)
\end{equation}
and an overall restoration of energy-momentum conservation is
mandatory. The strategy in this case (and similarly its generalizations to
more final and initial state partons) is to transform the
reconstructed momenta of the children coming from the now off-mass-shell shower progenitors
into their common centre-of-mass frame where three-momentum conservation is
guaranteed. Their spatial momentum components will then be re-scaled 
by a common parameter such that the overall invariant mass is consistent with 
energy-momentum conservation, $(P+\bar P)^2=Q^2$. This procedure is
equivalent to specific boosts along the $\mathbf{P}$ and the 
$\bar{\mathbf{P}}$ directions, 
respectively, for the progenitor quark and anti-quark sides. 
In cases that the shower evolution, which -- as we have mentioned before
has no notion of global energy-momentum conservation -- has generated 
virtualities which are inconsistent with the available centre-of-mass energy $Q$, 
the procedure just outlined is not possible.

Different choices for re-interpreting the branching variables when
setting up the full kinematics, with the aim of reducing the occurrence of
unphysically large virtualities have been implemented in \Herwig~7.
The default setting in the released version of \Herwig~7, \texttt{set
ShowerHandler:ReconstructionOption OffShell5}, imposes an additional
constraint in the intermediate evolution by explicitly altering the 
intermediate splitting variables $\tilde{q}$ and $z$ (which are originally 
obtained in the approximation the partons after the splitting are on-shell). 
This scheme absorbs the invariant mass of the children of the branching 
parton~\cite{Reichelt:2017hts} into a redefinition of the splitting variables 
to preserve the originally generated virtuality of the splitting parton. 
This approach, however, 
intrinsically changes the original form of the coherent branching
algorithm as outlined in the previous two subsections, and we
therefore do not consider this default option in the numerical
analyses carried out in Sec.~\ref{sec:herwigcompare}.  Instead, the
setting \texttt{set ShowerHandler:ReconstructionOption CutOff} is
used. It directly uses the variables generated for the splittings, 
and does not redefine the variables used to set up the full kinematics.
Events with unphysically large virtualities are discarded.

An additional difference of the \Herwig~7 parton shower to the analytic 
computation of the jet mass distributions encoded in
Eqs.~(\ref{jetfunctionmasslessdef}) and (\ref{jetfunctionmassivedef}),
is that its default (cluster-type) hadronization model~\cite{Webber:1983if},
imposes, in addition, constituent mass on-shell conditions for all partons that 
emerge when the shower is switched off.
This includes in particular a constituent mass for the gluons of around 
$1\ {\rm GeV}$. 
These parton constituent masses represent tunable parameters of the hadronization model and are thus part of the hadronization model even though they enter the \Herwig~7 parton level output. In particular, the constituent mass allows for a splitting
into quark/anti-quark pairs such that the primary non-perturbative
clusters can be formed.
Within our parton level examination concerning the dependence on the shower cut $Q_0$,
parton constituent masses would represent additional infrared cutoff 
scales that non-trivially interfere with $Q_0$ and in addition may cause 
gauge-invariance issues in higher order perturbative QCD calculations. 
Since we anyway do not use the \Herwig~7 hadronization model in our numerical analyses
of Sec.~\ref{sec:herwigcompare}, as already explained in
Sec.~\ref{sec:shapefunction}, we do not account for these constituent masses in our analytic calculations
and when generating parton level results from \Herwig~7.
We set all quark constituent masses to $m_q^c = 0.01$~MeV, and the gluon mass
parameter to $m_g^c = 2m_q^c$, which is the lower bound dictated by
constraints from the cluster hadronization model.
This effectively eliminates any effect coming from 
the constituent masses.
We note that string hadronization models do not require to assign a mass to
the gluons produced by the shower.

\section{Hemisphere mass distribution from coherent branching without cut}
\label{sec:CBnocut}

In this section we show that -- in the context of strict perturbative computations -- the coherent branching formalism and the factorized QCD 
predictions provide identical results concerning the NLL resummation of logarithmic corrections for the thrust distribution in the absence of any infrared cut, i.e.\ for $Q_0=0$. In the context of our discussions in Sec.~\ref{sec:peakprecision}, this equivalence means that for the thrust distribution the coherent branching formalism with NLL log resummation is already ${\cal O}(\alpha_s)$ precise as far as the peak position is concerned. For the thrust distribution for massive quarks this allows us to identify at ${\cal O}(\alpha_s)$ the coherent branching (CB) mass parameter and the pole mass $m^{\rm pole}$ {\it as long as we consider the resonance peak location as the observable}. We phrase this restricted equivalence by the relation
\begin{align} 
\label{eq:mcbpole1}
m^{\rm CB}(Q_0=0) \, \stackrel{\rm peak}{=} m^{\rm pole} + {\cal O}(\alpha_s^2)
\,.
\end{align}  
We stress that an exact solution for the jet mass distributions in Eqs.~(\ref{jetfunctionmasslessdef}) and (\ref{jetfunctionmassivedef}) (i.e.\ a solution that does not rely on any perturbative expansion or rearrangement of the expressions) is impossible without imposing any infrared cut because of the singularities in the soft and collinear regions of the $(z,\tilde q)$ plane caused by the Landau pole of the strong coupling. So applying the coherent branching formalism without any infrared cut implies (and requires) that the running of the strong coupling is treated strictly perturbatively (see also footnote~2). The equivalence relation~(\ref{eq:mcbpole1}) must therefore be understood strictly in the perturbative sense. From the point of view of an exact solution of the coherent branching formalism the limit $Q_0\to 0$ is impossible to reach. This illustrates the well-known problem of the pole mass being a purely perturbative concept that, however, cannot be associated directly to any physical process at the exact, non-perturbative level.

In the following two subsections we calculate the jet mass distribution in  Eqs.~(\ref{jetfunctionmasslessdef}) and (\ref{jetfunctionmassivedef}) obtained from the coherent branching formalism analytically at NLL order for massless and massive quark, respectively, and show that the results agree identically with those obtained from the factorized QCD calculations for thrust reviewed in Secs.~\ref{sec:factorizationtheoremmassless} and \ref{sec:factorizationtheoremmassive}. For the case of massless quarks this equivalence is well known and has already been studied thoroughly in the literature, see e.g.\ Refs.~\cite{Contopanagos:1996nh,Almeida:2014uva}. We nevertheless lay out the analysis for massless quarks in some detail because it sets the stage for the more complicated discussion for massive quarks in the resonance region, where -- to the best of our knowledge -- such a study has never been carried out before. Moreover, the manipulations are setting the stage for Sec.~\ref{sec:CBunreleased} where we examine the impact of the infrared shower cut $Q_0$ on the resonance location $\tau_{\rm peak}$.  The reader not interested in these computational details may safely skip these two subsection and continue reading with Sec.~\ref{sec:unreleased}.

For simplicity we carry out the bulk of the calculations in Laplace space and define the Laplace transform of the jet mass distributions as
\begin{align}
\label{eq:jetmassLaplace}
\tilde{J}(\bar\nu,Q) &\, = \, \int_{0} ^\infty \!\! {\rm d}s\,e^{-\bar\nu s}\,J(s,Q)\,,\nonumber\\
\tilde{J}(\bar\nu,Q,m) &\, = \, \int_{m^2} ^\infty \!\! {\rm d}s\,e^{-\bar\nu (s-m^2)}\,J(s,Q,m)\,,
\end{align}
such that the Laplace space thrust distributions as defined in Eqs.~(\ref{eq:sigmaLaplacemassless}) and (\ref{eq:sigmaLaplacemassive}) adopt the simple form
\begin{align}
\label{eq:sigmaLaplaceJet}
\tilde \sigma^{\rm cb}(\nu,Q) &\, = \, \Bigl[\tilde{J}\Bigl(\frac{\nu}{Q^2},Q\Bigr)\Bigr]^2\,,\nonumber\\
\tilde \sigma^{\rm cb}(\nu,Q,m) &\, = \, \Bigl[\tilde{J}\Bigl(\frac{\nu}{Q^2},Q,m\Bigr)\Bigr]^2\,.
\end{align}
To keep our notation simple we write the heavy quark mass paramter simply as $m$ instead of $m^{\rm CB}(Q_0=0)$ in the rest of Sec.~\ref{sec:CBnocut}.

\subsection{NLL resummation for massless quarks}
\label{sec:CBnocutmassless}

To analytically determine the NLL jet mass distribution for massless quarks in the peak region from 
Eq.~(\ref{jetfunctionmasslessdef}) we follow Ref.~\cite{Catani:1992ua} 
and replace $z$ by $1$ in all functions that are slowly varying in the limit $z\to 1$, {\it except} in the splitting function. As already discussed at the end of Sec.~\ref{sec:coherentbranchingmassless}, 
this means that the angular ordering constraint can be dropped in the peak region, giving
\begin{align}
\label{eq:jetmass1}
\tilde{J}(\bar\nu,Q)&\, =\,
1+\int_0^{Q^2}\frac{\mathrm{d}\tilde q^2}{\tilde q^2}\int_0^1\mathrm{d}z\; P_{qq}\Bigl[\as\bigl((1-z)\tilde q\bigr),z\Bigr]\Bigl(\mathrm{e}^{-\bar\nu(1-z)\tilde q^2}-1\Bigr)\tilde{J}(\nu,\tilde q)\,.
\end{align}
for the Lapace space integral equation for the jet mass distribution.
From this we find the differential equation
\begin{align}
\frac{\mathrm{d}\tilde{J}(\bar\nu,Q)}{\tilde{J}(\bar\nu,Q)}=\frac{\mathrm{d}Q^2}{Q^2}\int_0^1\mathrm{d}z\; P_{qq}\Bigl[\as\bigl((1-z)Q\bigr),z\Bigr]\Bigl(\mathrm{e}^{-\bar\nu(1-z)Q^2}-1\Bigr)\,,
\end{align}
which gives the solution
\begin{align}
\label{eq:solution1massless}
\ln \tilde{J}(\bar\nu,Q)=\int_0^{Q^2}\frac{\mathrm{d}\tilde q^2}{\tilde q^2}\int_0^1\mathrm{d}z\; P_{qq}\Bigl[\as\bigl((1-z)\tilde q\bigr),z\Bigr]\Bigl(\mathrm{e}^{-\bar\nu(1-z)\tilde q^2}-1\Bigr)\,.
\end{align}
With the substitutions
\begin{align}
\label{eq:qqtildesub}
&\tilde q^2=\frac{q^2}{1-z} \quad\mbox{and}\quad z = 1-\frac{q^{\prime 2}}{q^2}
\end{align}
and using the explicit form of the NLL splitting function in terms of the cusp anomalous dimension of Eq.~(\ref{eq:cusp0}) and a subleading non-cusp term,
\begin{align}
P_{qq}[\as,z] \, = \, \frac{\Gcusp[\as]}{1-z} \, - \, \Bigl(\frac{C_F\as}{2\pi}\Bigr)(1+z)\,,
\end{align}
we arrive at
\begin{align}
\label{eq:pretrick1}
\ln \tilde{J}(\bar\nu,Q)&=\int_0^{Q^2}\frac{\mathrm{d}q^2}{q^2}\;\Bigl(\mathrm{e}^{-\bar\nu q^2}-1\Bigr)\int_{\frac{(q^2)^2}{Q^2}}^{q^2}\frac{\mathrm{d}q^{\prime 2}}{q^{\prime 2}}\;\Bigl[\Gcusp[\as(q^\prime)]
- \, \Bigl(\frac{C_F\as(q^\prime)}{2\pi}\Bigr)\Bigl(2-\frac{q^{\prime 2}}{q^2}\Bigr)\frac{q^{\prime 2}}{q^2}
\Bigr]\,.
\end{align}
For the second non-cusp term we rewrite $\as(q^\prime)$ in terms of $\as(q)$ and powers of $\ln(q^{\prime 2}/q^2)$ and notice that at NLL precision we only have to keep terms that are proportional to $\as^{n+1}(q)\ln^n(q^2/Q^2)$ after the $q^\prime$ integration. Here in turns out that only a single term for $n=0$ has to be kept,
\begin{align}
\label{eq:pretricknoncusp}
-\,\int_{\frac{(q^2)^2}{Q^2}}^{q^2}\frac{\mathrm{d}q^{\prime 2}}{q^{\prime 2}}\;
\Bigl(\frac{C_F\as(q^\prime)}{2\pi}\Bigr)\Bigl(2-\frac{q^{\prime 2}}{q^2}\Bigr)\frac{q^{\prime 2}}{q^2}
\, \stackrel{\rm NLL}{=} \, - 3\,C_F\Bigl(\frac{\as(q)}{4\pi}\Bigr) \, = \, -\frac{1}{4}\gamma_J[\as(q)]\,,
\end{align}
which we have, anticipating the form of the final result, identified with the non-cusp anomalous dimension of the jet function in the factorized QCD cross section. Rewriting in the remaining integral $\as(q)$ in terms of $\as(Q)$ and powers of $\ln(q^2/Q^2)$, we can further simplify the integral by noticing, that for obtaining all NLL logarithmic terms correctly, 
we can use the replacement
\begin{align}
\label{eq:thetarule}
\mathrm{e}^{-\bar\nu q^2}-1 \stackrel{\text{NLL}}{=} -\theta(q^2-w) \qquad \text{with} \qquad w=(\mathrm{e}^{\gamma_E}\bar\nu)^{-1}=Q^2(\mathrm{e}^{\gamma_E}\nu)^{-1}\,.
\end{align}
This replacement technically acts like an infrared cutoff for the $q$ integration. It is, however, not a physical cutoff because it is derived in the context of a strict perturbative expansion (where no infrared Landau Pole singularity arises) and is furthermore not correct beyond NLL order. One should therefore better think of the replacement simply as an algebraic relation that simplifies the perturbative analytic NLL resummation calculation. 

For the remaining double integral with the cusp-anomalous dimension we can now switch the order of integration,
\begin{align}
-\int_{\frac{w^2}{Q^2}}^{w}\frac{\mathrm{d}q^2}{q^2}\;\Gcusp[\as(q^2)]\int_{w}^{qQ}\frac{\mathrm{d}q^{\prime 2}}{q^{\prime 2}}-\int_{w}^{Q^2}\frac{\mathrm{d}q^2}{q^2}\;\Gcusp[\as(q^2)]\int_{q^2}^{qQ}\frac{\mathrm{d}q^{\prime 2}}{q^{\prime 2}}\,,
\end{align}
and reshuffle the $q^{\prime}$ integrations,
\begin{align}
&\int_{w}^{qQ}\frac{\mathrm{d}q^{\prime 2}}{q^{\prime 2}} =\frac{1}{2} \int_{\frac{w^2}{Q^2}}^{q^2}\frac{\mathrm{d}q^{\prime 2}}{q^{\prime 2}}\,, \notag \\
&\int_{q^2}^{qQ}\frac{\mathrm{d}q^{\prime 2}}{q^{\prime 2}} =-\int_{w}^{q^2}\frac{\mathrm{d}q^{\prime 2}}{q^{\prime 2}}+\frac{1}{2} \int_{\frac{w^2}{Q^2}}^{q^2}\frac{\mathrm{d}q^{\prime 2}}{q^{\prime 2}}\,,
\end{align}
to obtain
\begin{align}
-\frac{1}{2}\int_{\frac{w^2}{Q^2}}^{Q^2}\frac{\mathrm{d}q^2}{q^2}\;\Gcusp[\as(q^2)]\int_{\frac{w^2}{Q^2}}^{q^2}\frac{\mathrm{d}q^{\prime 2}}{q^{\prime 2}}+\int_{w}^{Q^2}\frac{\mathrm{d}q^2}{q^2}\;\Gcusp[\as(q^2)]\int_{w}^{q^2}\frac{\mathrm{d}q^{\prime 2}}{q^{\prime 2}}\,.
\end{align}
Noticing the scale identifications
\begin{align}
Q^2=\mu^2_{H,\nu}\,, \quad w=\mu^2_{J,\nu}\,, \quad w^2/Q^2=\mu^2_{S,\nu}
\end{align}
according to Eqs.~(\ref{eq:salesmassless}) and (\ref{eq:thetarule}), we see that at this point we have separated the collinear and soft 
evolution to the hard scale.
The non-cusp term 
\begin{align}
\frac{1}{4}\int_{w}^{Q^2}\frac{\mathrm{d}q^2}{q^2}\;\gamma_J[\as(q)]\,,
\end{align}
on the other hand, describes only a collinear evolution to the hard scale consistent with our assignment in Eq.~(\ref{eq:pretricknoncusp}).
Accounting for Eq.~(\ref{eq:QCDbetafct}) for the QCD beta function and Eq.~(\ref{eq:sigmaLaplaceJet}),
we then arrive at the following form of the NLL Laplace space thrust distribution,
\begin{align}
\ln\tilde\sigma^{\rm cb}(\nu,Q)\,=&\,\,8\int_{\as(\mu_{J,\nu})}^{\as(Q)}\frac{\mathrm{d}\as}{\beta[\as]}\;\Gcusp[\as]\int_{\as(\mu_{J,\nu})}^{\as}\frac{\mathrm{d}\as^{\prime}}{\beta[\as^{\prime}]}  \notag \\
&  -4\int_{\as(\mu_{S,\nu})}^{\as(Q)}\frac{\mathrm{d}\as}{\beta[\as]}\;
\Gcusp[\as]\int_{\as(\mu_{S,\nu})}^{\as}\frac{\mathrm{d}\as^{\prime}}{\beta[\as^{\prime}]}  \notag \\
& +\int_{\as(\mu_{J,\nu})}^{\as(Q)}\frac{\mathrm{d}\as}{\beta[\as]}\;\gamma_J[\as]
\,.
\end{align}
In terms of the $K$ and $\omega$ evolution factors defined in Eq.~\eqref{eq:Komega} this can be rewritten as
\begin{align}
\label{eq:sigmatildemassless}
\tilde \sigma^{\rm cb}(\nu,Q)\,&\,=\,\mathrm{exp}\biggl[4\,K(\Gcusp,\mu_{H,\nu},\mu_{J,\nu})
-2\,K(\Gcusp,\mu_{H,\nu},\mu_{S,\nu})+\frac{1}{2}\omega(\gamma_J,\mu_{H,\nu},\mu_{J,\nu})\biggr] \notag \\
&\,=\,\mathrm{exp}\biggl[K(\Gamma_J,\mu_{H,\nu},\mu_{J,\nu})+K(\Gamma_S,\mu_{H,\nu},\mu_{S,\nu})+
\frac{1}{2}\omega(\gamma_J,\mu_{H,\nu},\mu_{J,\nu})
\biggr]\,,
\end{align}
where in the last line we have used the cusp and non-cusp identities of Eqs.~(\ref{eq:cusp1}).
This agrees identically with the factorized QCD cross section for massless quarks of Eq.~\eqref{eq:Laplacemassless}.

\subsection{NLL resummation for massive quarks}
\label{sec:CBnocutmassive}

To analytically determine the NLL jet mass distribution for massive quarks in the peak region from Eq.~(\ref{jetfunctionmassivedef}) we 
initially proceed in the same way as for the massless quark case. Taking the large $z$ approximation of Ref.~\cite{Catani:1992ua} and in addition
the limit of large boost ($Q^2\gg m^2$), the Laplace space integral equation for the jet mass distribution adopts the form
\begin{align}
\tilde{J}(\bar\nu,Q,m)&\, =\,
1+\int_{m^2}^{Q^2}\frac{\mathrm{d}\tilde q^2}{\tilde q^2}\int_0^1\mathrm{d}z\; 
P_{QQ}\biggl[\as\bigl((1-z)\tilde q\bigr),z,\frac{m^2}{\tilde q^2}\biggr]\Bigl(\mathrm{e}^{-\bar\nu(1-z)\tilde q^2}-1\Bigr)\tilde{J}(\nu,\tilde q)\,,
\end{align}
with
\begin{align}
P_{QQ}\biggl[\as,z,\frac{m^2}{\tilde q^2}\biggr] \, = \, \frac{\Gcusp[\as]}{1-z} \, - \, \Bigl(\frac{C_F\as}{2\pi}\Bigr)(1+z) - \, \Bigl(\frac{C_F\as}{\pi}\Bigr)\frac{m^2}{(1-z)\tilde q^2}\,,
\end{align}
where we have dropped a factor $1/z$ from the mass correction term for a consistent expansion in the $z\to 1$ limit, see also Sec.~\ref{sec:coherentbranching}.
Its solution reads
\begin{align}
\label{eq:solution1massive}
\ln \tilde{J}(\bar\nu,Q,m)&\,=\,\int_{m^2}^{Q^2}\frac{\mathrm{d}\tilde q^2}{\tilde q^2}\int_0^1\mathrm{d}z\; P_{QQ}\Bigl[\as\bigl((1-z)\tilde q\bigr),\tilde q,m,z\Bigr]\Bigl(\mathrm{e}^{-\bar\nu(1-z)\tilde q^2}-1\Bigr)\,,
\end{align}
and with the substitutions of Eq.~(\ref{eq:qqtildesub})
we arrive at
\begin{align}
\label{eq:pretrick1massive}
\ln \tilde{J}(\bar\nu,Q,m)&=\int_{w}^{Q^2}\frac{\mathrm{d}q^2}{q^2}\;\int_{\frac{(q^2)^2}{Q^2}}^{q^2}\frac{\mathrm{d}q^{\prime 2}}{q^{\prime 2}}\;\Bigl[\,-\Gcusp[\as(q^\prime)]\notag\\
&  + \, \Bigl(\frac{C_F\as(q^\prime)}{2\pi}\Bigr)\Bigl(2-\frac{q^{\prime 2}}{q^2}\Bigr)\frac{q^{\prime 2}}{q^2}
+ \, \Bigl(\frac{C_F\as(q^\prime)}{\pi}\Bigr)\frac{m^2 q^{\prime 2}}{(q^2)^2}\,
\Bigr] \,\theta(q^2-q^\prime m)\,
\end{align}
for the NLL resummed Lapace space thrust distribution. 
We have already implemented the NLL relation of Eq.(\ref{eq:thetarule}) to simplify the $q^2$ integration, since it is also valid in the context of massive quarks. 

It it easy to see that the massive quark constraint $q^2>q^\prime m$ is irrelevant for $w=\mu^2_{J,\nu}>m^2$, which refers to the situation where the hemisphere jet masses
are larger than the mass of the quark. In this kinematic situation the mass correction in the splitting function represents the only modification due to the quark mass, but the  
struture of the log resummation is otherwise in complete analogy to the massless quark case. In the context of the factorized QCD calculations, one can then employ usual SCET factorization where the collinear sector of the effective Lagrangian is extended trivially by just accounting for the finite quark mass~\cite{Fleming:2007qr,Fleming:2007xt}.
In this work, however, we are interested in the peak region where the hemisphere jet masses are close to the quark mass, i.e.\ where $w< m^2$. Here the ultra-collinear sector emerges and the QCD factorization requires that the off-shell fluctuations of the massive quark field are integrated out~\cite{Fleming:2007qr,Fleming:2007xt}. So the quark mass effects are much more complicated and lead to a substantial rearrangement of the structure of the resummed logarithms. The physical meaning of $w$ is also modified and the scale identifications read  
\begin{align}
\label{eq:scaleident1}
Q^2=\mu^2_{H,\nu}\,, \quad m^2=\mu^2_{m,\nu}\,, \quad w^2/m^2=\mu^2_{B,\nu}\,, \quad w^2/Q^2=\mu^2_{S,\nu}
\end{align}

Let us now have a closer look at the calculation for the cusp term.
Reversing the order of integration we have to distinguish three integration regions and find
\begin{align}
\label{eq:revseintegration1}
-\int_{\frac{w^2}{Q^2}}^{\frac{w^2}{m^2}}\frac{\mathrm{d}q^2}{q^2}\;\Gcusp[\as(q^2)]& \int_{w}^{qQ}\frac{\mathrm{d}q^{\prime 2}}{q^{\prime 2}} \,-\,\int_{\frac{w^2}{m^2}}^{m^2}\frac{\mathrm{d}q^2}{q^2}\;\Gcusp[\as(q^2)]\int_{mq}^{Q q}\frac{\mathrm{d}q^{\prime 2}}{q^{\prime 2}}
\notag\\ \qquad &
-\int_{m^2}^{Q^2}\frac{\mathrm{d}q^2}{q^2}\;\Gcusp[\as(q^2)]\int_{q^2}^{Q q}\frac{\mathrm{d}q^{\prime 2}}{q^{\prime 2}}\,.
\end{align}
The $q^{\prime}$ integrations can be reshuffled using
\begin{align}
 \int_{w}^{qQ}\frac{\mathrm{d}q^{\prime 2}}{q^{\prime 2}}&=\frac{1}{2}\int_{\frac{w^2}{Q^2}}^{q^2}\frac{\mathrm{d}q^{\prime 2}}{q^{\prime 2}}\,, \\
 \int_{mq}^{Q q}\frac{\mathrm{d}q^{\prime 2}}{q^{\prime 2}}&=-\frac{1}{2}\int_{\frac{w^2}{m^2}}^{q^2}\frac{\mathrm{d}q^{\prime 2}}{q^{\prime 2}}+\frac{1}{2}\int_{\frac{w^2}{Q^2}}^{q^2}\frac{\mathrm{d}q^{\prime 2}}{q^{\prime 2}}\,,\\
 \int_{q^2}^{Q q}\frac{\mathrm{d}q^{\prime 2}}{q^{\prime 2}}&=-\frac{1}{2}\int_{m^2}^{q^2}\frac{\mathrm{d}q^{\prime 2}}{q^{\prime 2}}-\frac{1}{2}\int_{\frac{w^2}{m^2}}^{q^2}\frac{\mathrm{d}q^{\prime 2}}{q^{\prime 2}}+\frac{1}{2}\int_{\frac{w^2}{Q^2}}^{q^2}\frac{\mathrm{d}q^{\prime 2}}{q^{\prime 2}}\,,
\end{align}
such that we get
\begin{align}
\label{eq:cuspmassivefinal}
&\frac{1}{2}\int_{m^2}^{Q^2}\frac{\mathrm{d}q^2}{q^2}\;\Gcusp[\as(q^2)] \int_{m^2}^{q^2}\frac{\mathrm{d}q^{\prime 2}}{q^{\prime 2}}\, +\,\frac{1}{2}\int_{\frac{w^2}{m^2}}^{Q^2}\frac{\mathrm{d}q^2}{q^2}\;\Gcusp[\as(q^2)]\int_{\frac{w^2}{m^2}}^{q^2}\frac{\mathrm{d}q^{\prime 2}}{q^{\prime 2}}
\notag\\ \qquad & \qquad\qquad\qquad
-\frac{1}{2}\int_{\frac{w^2}{Q^2}}^{Q^2}\frac{\mathrm{d}q^2}{q^2}\;\Gcusp[\as(q^2)]\int_{\frac{w^2}{Q^2}}^{q^2}\frac{\mathrm{d}q^{\prime 2}}{q^{\prime 2}}
\notag\\ \qquad &
=\,K(\Gcusp,\mu_{H,\nu},\mu_{m,\nu})+K(\Gcusp,\mu_{H,\nu},\mu_{B,\nu})-K(\Gcusp,\mu_{H,\nu},\mu_{S,\nu})
\,.
\end{align}  
At this point we have separated mass-dependent, ultra-collinear and soft evolution to the hard scale and
have rewritten the result using the evolution function $K$ of Eq.~\eqref{eq:Komega} and the scale identifications of Eq.~(\ref{eq:scaleident1}).

For the non-cusp term, rewriting the constraint $q^2>q^\prime m$ in terms of integration limits,
\begin{align}
\int_{w}^{Q^2}\frac{\mathrm{d}q^2}{q^2}\;\int_{\frac{(q^2)^2}{Q^2}}^{q^2}\frac{\mathrm{d}q^{\prime 2}}{q^{\prime 2}}
\,\theta(q^2-q^\prime m)
\, = \,
\int_{w}^{m^2}\frac{\mathrm{d}q^2}{q^2}\;\int_{\frac{(q^2)^2}{Q^2}}^{\frac{(q^2)^2}{m^2}}\frac{\mathrm{d}q^{\prime 2}}{q^{\prime 2}}
\, + \,
\int_{m^2}^{Q^2}\frac{\mathrm{d}q^2}{q^2}\;\int_{\frac{(q^2)^2}{Q^2}}^{q^2}\frac{\mathrm{d}q^{\prime 2}}{q^{\prime 2}}
\end{align} 
we can use the considerations already applied in the massless quark case and find that only the second integration contributes at NLL order.
This gives 
\begin{align}
\label{eq:noncuspmassive}
\int_{m^2}^{Q^2}\frac{\mathrm{d}q^2}{q^2}\;\Bigl[ 3\,C_F\Bigl(\frac{\as(q)}{4\pi}\Bigr)\Bigr]
\,=\,
\frac{1}{4}\int_{m^2}^{Q^2}\frac{\mathrm{d}q^2}{q^2}\;\Bigl[-\gamma_H[\as(q)]+\gamma_{J_B}[\as(q)]+\gamma_{H_m}[\as(q)]\Bigr]\,,
\end{align}
where we have written the expression in terms of the non-cusp anomalous dimensions of the hard, the mass mode and the bHQET jet functions,
anticipating already the form of the final result.

For the mass correction term in the splitting function we reverse integration order in analogy to our manipulation for the cusp term in Eq.~(\ref{eq:revseintegration1}),
\begin{align}
& \int_{\frac{w^2}{Q^2}}^{\frac{w^2}{m^2}}\frac{\mathrm{d}q^2}{q^2}\;\Bigl(\frac{C_F\as(q^\prime)}{\pi}\Bigr)\int_{w}^{qQ}\frac{\mathrm{d}q^{\prime 2}}{q^{\prime 2}}\frac{m^2q^{2}}{(q^{2\prime})^2} +\int_{\frac{w^2}{m^2}}^{m^2}\frac{\mathrm{d}q^2}{q^2}\;\Bigl(\frac{C_F\as(q^\prime)}{\pi}\Bigr)\int_{mq}^{Q q}\frac{\mathrm{d}q^{\prime 2}}{q^{\prime 2}}\frac{m^2q^{2}}{(q^{2\prime})^2}
\notag\\& 
\qquad\qquad
+\int_{m^2}^{Q^2}\frac{\mathrm{d}q^2}{q^2}\;\Bigl(\frac{C_F\as(q^\prime)}{\pi}\Bigr)\int_{q^2}^{Q q}\frac{\mathrm{d}q^{\prime 2}}{q^{\prime 2}}\frac{m^2q^{2}}{(q^{2\prime})^2}\,.
\end{align}
In the limit of a boosted massive quark $(Q^2\gg m^2)$ only the second term can contribute NLL logarithms.
Using
\begin{align}
&\int_{mq}^{Q q}\frac{\mathrm{d}q^{\prime 2}}{q^{\prime 2}}\frac{m^2q^{2}}{(q^{2\prime})^2}=\frac{1}{2}+\mathcal{O}\Bigl(\frac{m^2}{Q^2}\Bigr) \,,
\end{align}
the contribution from the mass correction term at NLL accuracy then reads
\begin{align}
\label{eq:masscorrmassive}
\int_{\frac{w^2}{m^2}}^{m^2}\frac{\mathrm{d}q^2}{q^2}\;\Bigl[ 2C_F \Bigl(\frac{\as(q)}{4\pi}\Bigr)\Bigr]
&=\frac{1}{4}\int_{\frac{w^2}{m^2}}^{Q^2}\frac{\mathrm{d}q^2}{q^2}\;\gamma_{J_B}[\as(q)]
\,-\,\frac{1}{4}\int_{m^2}^{Q^2}\frac{\mathrm{d}q^2}{q^2}\;\gamma_{J_B}[\as(q)] 
\end{align}

Taking the sum of the NLL contributions from the non-cusp term in Eq.~(\ref{eq:noncuspmassive}) and the mass corrections term in Eq.~(\ref{eq:masscorrmassive})
we obtain 
\begin{align}
\label{eq:massivenoncuspfinal}
&\frac{1}{4}\int_{m^2}^{Q^2}\frac{\mathrm{d}q^2}{q^2}\;\Bigl[-\gamma_H[\as(q)]+\gamma_{H_m}[\as(q)]\Bigr] 
\, + \, \frac{1}{4}\int_{\frac{w^2}{m^2}}^{Q^2}\frac{\mathrm{d}q^2}{q^2}\;\gamma_{J_B}[\as(q)]
\notag\\ & 
\,=\, \frac{1}{4}\Bigl(\omega(\gamma_{H_m}-\gamma_{H_Q},\mu_{H,\nu},\mu_{m,\nu})+\omega(\gamma_{B},\mu_{H,\nu},\mu_{B,\nu})+\omega(\gamma_{S},\mu_{H,\nu},\mu_{S,\nu})\Bigr)
\end{align}
where in the second line we have rewritten the result using the evolution function $\omega$ of Eq.~\eqref{eq:Komega}, the scale identities of Eq.~(\ref{eq:scaleident1}) and that $\gamma_S[\as]=0$ at NLL.
Combining Eqs.~(\ref{eq:cuspmassivefinal}) and (\ref{eq:massivenoncuspfinal}) and using (\ref{eq:sigmaLaplaceJet}) we arrive at the final form for the NLL Laplace space thrust distribution 
\begin{align}
\label{eq:LaplacemassiveCB}
&\tilde\sigma^{\rm cb}(\nu,Q,m^{\rm pole})\,=\,
\mathrm{exp}\biggl[-K(\Gamma_{H_m},\mu_{H,\nu},\mu_{m,\nu})+K(\Gamma_B,\mu_{H,\nu},\mu_{B,\nu})+K(\Gamma_S,\mu_{H,\nu},\mu_{S,\nu})\biggr]\notag \\
&\qquad\times\mathrm{exp}\biggl[\frac{1}{2}\Bigl(\omega(\gamma_{H_m}-\gamma_{H_Q},\mu_{H,\nu},\mu_{m,\nu})+\omega(\gamma_B,\mu_{H,\nu},\mu_{B,\nu})+\omega(\gamma_S,\mu_{H,\nu},\mu_{S,\nu})\Bigr)\biggr]\,,
\end{align}
which agrees identically with the factorized QCD cross section for massive quarks of Eq.~\eqref{eq:Laplacemassive}.

\section{Hemisphere mass distribution with shower cut $Q_0$}
\label{sec:unreleased}

In this section we study analytically the impact of the shower evolution cut $Q_0$ on the thrust distribution in the resonance regions for massless and massive quark production. The main focus is on the effects that cause a dependence of the hemisphere masses that is linear on $Q_0$. The $Q_0$ cut is defined as the restriction $p^2_\perp>Q^2_0$ on the transverse momentum of the emission with respect to the momentum of the emitter. In the context of the coherent branching formalism the dependence of the transverse momentum on the shower evolution variables $\tilde q$ and $z$ for the massless and the massive quark cases are given in Eqs.~(\ref{qtildemassless}) and (\ref{qtildemassive}), respectively, leading to the constraints in Eqs.~(\ref{pperpcut1}) and (\ref{pperpcut2}).

Since {\it in the framework of strict perturbative calculations} the $Q_0$ cut represents an artificial restriction of the radiation generated by the shower, we call the emissions that are allowed by the $Q_0$ cut {\it released} and the emissions that is not allowed by the $Q_0$ cut {\it unreleased}. As we will show, the dominant (linear in $Q_0$) effect of removing the unreleased radiation from the calculation in the resonance region must be {\it reinterpreted as a redefinition of parameters in a perturbative calculation without $Q_0$ cut.}

To elucidate this we compare the effects of the unreleased radiation in the context of the coherent branching formalism for the jet mass distributions as described in Secs.~\ref{sec:coherentbranchingmassless} and \ref{sec:coherentbranchingmassive}, and in the context of QCD factorization using the SCET approach for the thrust distribution as described in Secs.~\ref{sec:factorizationtheoremmassless} and \ref{sec:factorizationtheoremmassive}.
This comparison, together with the facts that in the absence of the $Q_0$ cut coherent branching and QCD factorization provide equivalent results at NLL order and both are ${\cal O}(\alpha_s)$ precise for the resonance peak mass, allows us to relate the quark mass parameter of the coherent branching formalism with $Q_0$ cut (and thus of angular ordered parton showers) to an explicit field theoretic mass renormalization scheme at ${\cal O}(\as)$. 

In subsection~\ref{sec:unreleasedphasespace} we outline the collinear and soft phase space regions and QCD modes relevant for the thrust distribution in the resonance region in the context of coherent branching and QCD factorization, respectively, and we show where a linear $Q_0$ dependence can possibly emerge. In 
subsections \ref{sec:QCDunreleasedmassless} to \ref{sec:CBunreleased} we calculate the effects of the unreleased radiation each for massless and massive quarks for QCD factorization and coherent branching and analyze in detail the effects linear in $Q_0$.

\subsection{Phase space regions with and without $Q_0$ cut}
\label{sec:unreleasedphasespace}

\begin{figure}
\begin{center}
\includegraphics[width=0.65\textwidth]{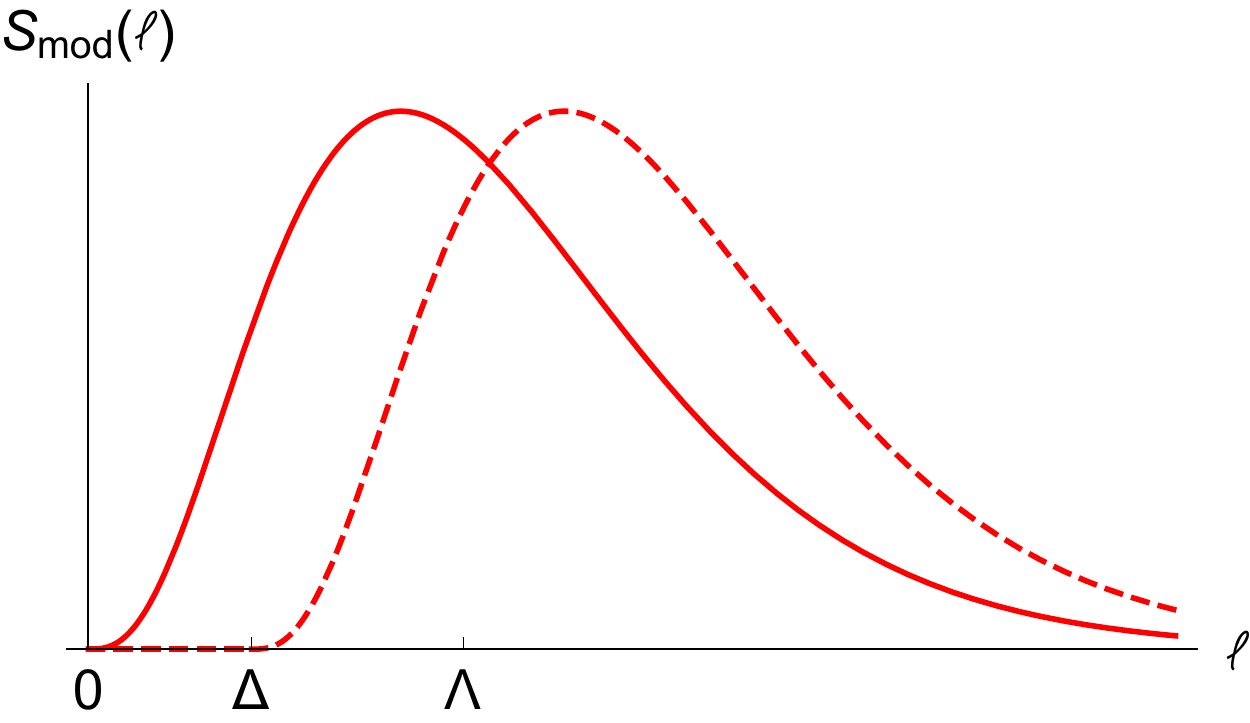}
\caption{
Generic form of the soft model shape function $S_{\rm mod}(\ell)$ in arbitray units for illustration. The original soft model shape function $S_{\rm mod}(\ell)$ is represented by the solid red line and the soft model shape function with a gap shift, $S_{\rm mod}(\ell -\Delta)$ for the case $0<\Delta<\Lambda$, by the dashed red line.
}\label{fig:softmodel}
\end{center}
\end{figure}

To initiate the analytic examinations it is illustrative to first discuss the structure of the phase space and the QCD modes relevant for the resonance region. To define our counting variable we start from the hadron level thrust distributions given in Eqs.~(\ref{eq:thrustmassless1}) and (\ref{eq:thrustmassive1}), where the partonic thrust distribution is convoluted with the soft model shape function $S_{\rm mod}(\ell)$. The parameters of the shape function may be either determined from fits to experimental data or from non-perturbative methods. The shape function has support for positive $\ell$ values and exhibits a peak for $\ell\approx \Lambda$, where $\Lambda$ parametrizes the overall energy the non-perturbative effects add to the parton level hemisphere masses. For larger $\ell$ values the shape function falls quickly and one usually assumes an exponential behavior. A typical generic form for $S_{\rm mod}$ is displayed in Fig.~\ref{fig:softmodel}.
The effect of the shape function on the hadron level prediction is twofold: it smears out the distributive and singular structures of the partonic cross section, and it leads to a shift of the observable resonance peak position in the thrust distribution towards larger values with respect to the partonic thresholds, $\tau_{\rm min}=0$ for massless quarks and $\tau_{\rm min}=2m^2/Q^2$ for massive quarks:
\begin{align}
\tau_{\rm peak}-\tau_{\rm min} \sim  \frac{\Lambda}{Q}\,\ll\, 1\,.
\end{align}
It is therefore natural to adopt $\Lambda/Q$ as the counting parameter in the resonance region.

\begin{figure}
\begin{subfigure}[c]{0.48\textwidth}

\includegraphics[width=1.0\textwidth]{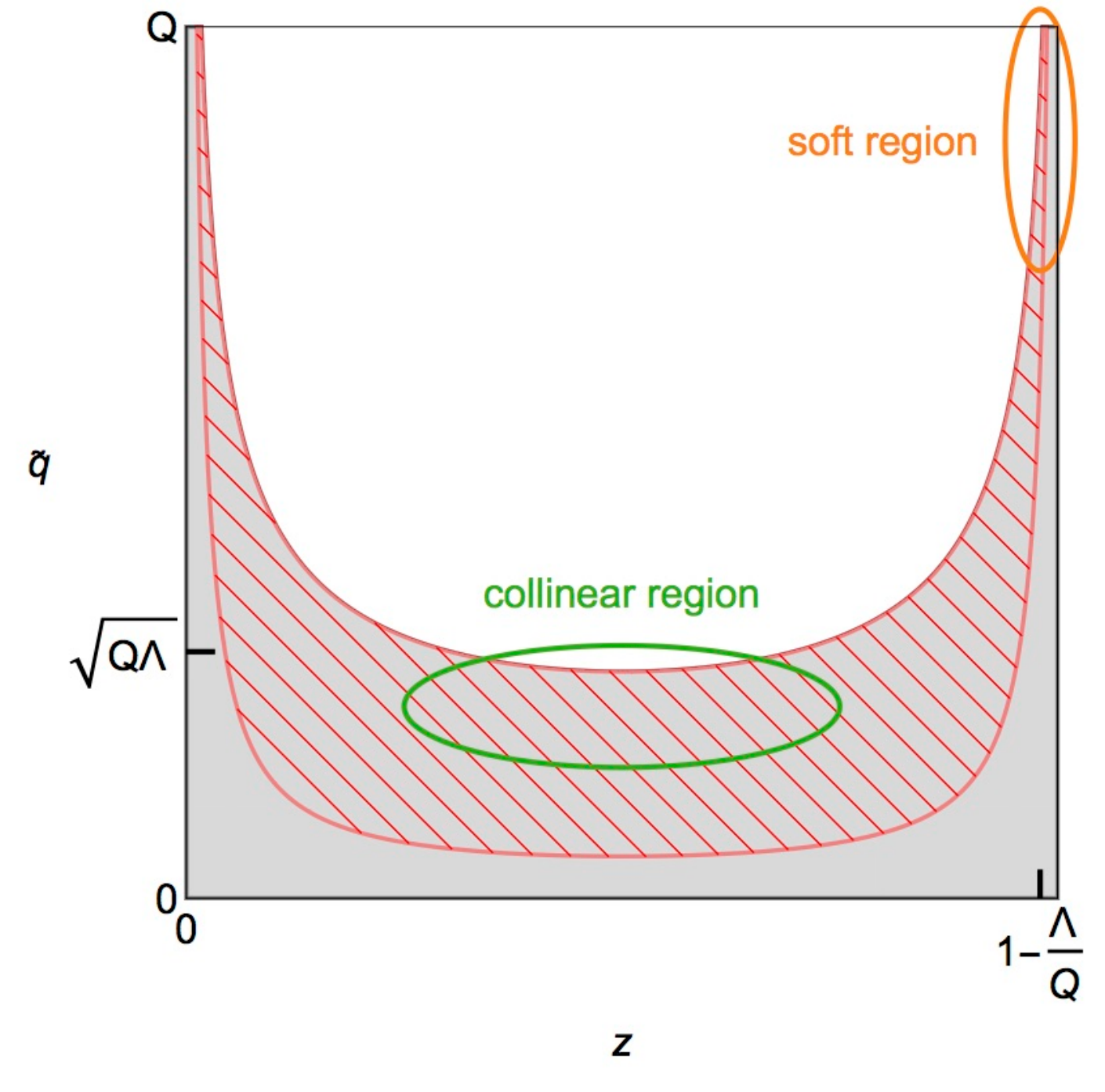}
\subcaption{}\label{fig:phasespace_massless}

\end{subfigure}
\hfill
\begin{subfigure}[c]{0.48\textwidth}
\includegraphics[width=1.0\textwidth]{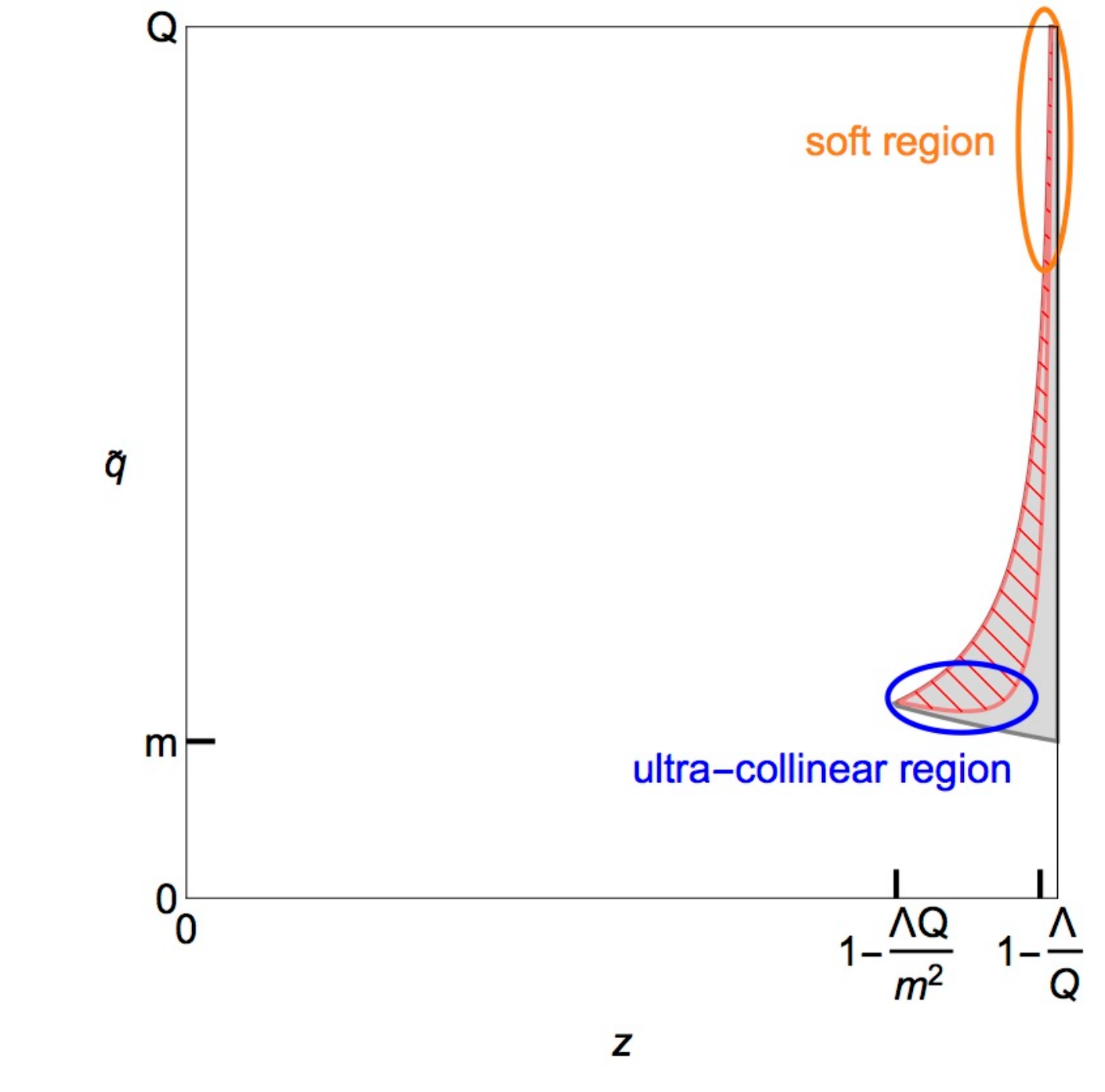}
\subcaption{}\label{fig:phasespace_massive}
\end{subfigure}
\caption{$(z,\tilde q)$ phase space for coherent branching in the (a) massless and (b) massive case, with indication of the relevant soft, collinear and ultra-collinear regions. The hatched area corresponds to the phase space populated in the presence of a shower cut $Q_0$. The soft, collinear and ultra-collinear phase space regions are indicated.}\label{fig:phasespace}
\end{figure}

In Fig.~\ref{fig:phasespace_massless} we show the generic form of the $(z,\tilde q)$ phase space populated by coherent branching for the jet mass distribution for massless quarks, see Eq.~(\ref{jetfunctionmasslessdef}). The gray area represents the phase space without $Q_0$ cut and the hatched area the phase space with $Q_0$ cut. In the peak region the thrust distribution is dominated by soft and collinear gluon radiation, which are also indicated. In QCD factorization the soft and the collinear modes are separated at the operator level by imposing powercounting contraints on the momentum fluctuations these operators can generate. These constraints are most efficiently formulated in the light cone basis where
\begin{align}
\label{QCDfactorizationframe}
p^\mu \, = \, n\cdot p \frac{\bar n^\mu}{2}\, + \, \bar n\cdot p \frac{n^\mu}{2}\, + \, p^\mu_\perp
\end{align}
where $n$ and $\bar n$ are back-to-back light-like vectors than can be directed along the momenta of the progenitor quark-antiquark pair produced by the primary hard scattering. The momentum components in this basis are then denoted by  
$p^\mu = (p^+,p^-,p_\perp)=(n\cdot p,\bar n\cdot p,p_\perp)$ where the momentum square reads $p^2 = p^+ p^- -p_\perp^2$, 
see also Secs.~\ref{sec:coherentbranchingmassless} and \ref{sec:coherentbranchingmassive}. 
The $\Lambda$ counting of the collinear and soft regions formulated in the coherent branching and in the QCD factorization approaches can be connected by the relation 
\begin{align}
(p^+,p^-,p_\perp) \, = \, Q(1-z)\,\Bigl(\frac{\tilde q^2}{Q^2},1,\frac{\tilde q}{Q}\Bigr)
\end{align}
for soft and $n$-collinear modes. For the $\bar n$ collinear modes, the plus and the minus components on the RHS have to be swapped. The momentum power counting for both approaches for massless quark production is summarized in Tab.~\ref{tab:counting_massless}.

\begin{center}
\renewcommand{\arraystretch}{1.5} 
\begin{tabular}{|c|c|c|}
 \hline
 \multicolumn{3}{|c|}{phase space regions for $\displaystyle \tau_{\rm peak}\sim \frac{\Lambda}{Q}\ll 1$, $m=0$} \\
 \hline
 & coherent branching & QCD factorization \\
 \hline
 \multirow{3}{*}{$n$-collinear} & \multicolumn{1}{|l|}{$\displaystyle z\sim(1-z)\sim 1$} & \multirow{3}{*}{$\displaystyle q^\mu\sim (\Lambda,Q,(Q\Lambda)^{\frac{1}{2}})$} \\
 & \multicolumn{1}{|l|}{$\displaystyle \tilde{q}\sim (Q\Lambda)^{\frac{1}{2}}$} &  \\ 
 & \multicolumn{1}{|l|}{$\displaystyle q_\perp \sim (Q\Lambda)^{\frac{1}{2}}$} &  \\ \hline
  \multirow{3}{*}{soft} & \multicolumn{1}{|l|}{$\displaystyle 1-z\sim \frac{\Lambda}{Q},\;z\sim 1$}  & \multirow{3}{*}{$\displaystyle q^\mu\sim (\Lambda,\Lambda,\Lambda)$} \\
 & \multicolumn{1}{|l|}{$\displaystyle \tilde{q}\sim Q$} &  \\ 
 & \multicolumn{1}{|l|}{$\displaystyle q_\perp \sim \Lambda$} &  \\ \hline
\end{tabular}
 \captionof{table}{Power counting for coherent branching and QCD factorization for masseless quarks.}\label{tab:counting_massless}
\end{center}

Imposing the $Q_0$ cut, one has to note that it represents a cut on the transverse momentum of the emission with respect to the momentum of the emitter and not with respect to the momenta of the progenitor quarks. In the coherent branching approach this is automatically taken care of in the definition of the transverse momentum variable $q_{i,\perp}^\mu$ of Eq.~(\ref{CBqperpdef}) which parametrizes the $i$-th branching. 
In QCD factorization, on the other hand, the constraint has a more complicated structure, because the momenta of all radiated partons are usually formulated in one common frame based on Eq.~(\ref{QCDfactorizationframe}). Fortunately at NLL$+{\cal O}(\as)$ precision, the order we consider in this work, only the first emission has to be calculated in the QCD factorization approach to determine the effects linear in $Q_0$. At this level the transverse momentum variable in coherent branching defined in Eq.~(\ref{CBqperpdef}) and the transverse momentum in QCD factorization defined in Eq.~(\ref{QCDfactorizationframe}) agree and can be identified. So at NLL$+{\cal O}(\as)$ precision we can implement the shower cut constraint in the factorized calculation by simply imposing a cut on the transverse momentum in Eq.~(\ref{QCDfactorizationframe}) without further complications. For the rest of this paper we therefore identify the transverse momenta in both apporaches to keep the presentation simple,
and we frequently refer to the shower cut $Q_0$ also as the cut on the transverse momentum $p_\perp$ without further specification. 

From a conceptual point of view the numerical value for $Q_0$ should be chosen such that it is unresolved, i.e.\ it should be smaller than the typical values $p_\perp$ can adopt for the observable of interest. From Tab.~\ref{tab:counting_massless} we can see that soft radiation imposes the strongest constraint and requires that $Q_0$ should in principle be smaller than $\Lambda\sim 1$~GeV.
This is the hierarchy we assume for some of the arguments presented below. For practical parton showers, however, this constraint cannot be satisfied in terms of a strong hierarchy (if at all) due to computational reasons and the proximity to the Landau pole of the strong coupling. As we show in our numerical analysis in Sec.~\ref{sec:herwigcompare} using the approximation $Q_0\ll\Lambda$ in our analytic calculations does very well, even for cases where the both scales are similar in size. In any case, since $Q_0$ represents the smallest scale for the strong coupling, integrations over its Landau pole are prevented as long as $Q_0$ is chosen larger than $\Lambda_{\rm QCD}$, and, moreover, for finite $Q_0$ there are no renormalon ambiguities in perturbative calculations.    

In the context of QCD factorization we can see already at the level of the factorization theorem~(\ref{eq:thrustmassless2}) that a linear dependence on the $p_\perp$ cut $Q_0$ can only arise in the partonic soft function $S$ because it is linear in the light cone momentum $\ell$. In the jet function $J$, however, we expect a quadratic behavior for simple dimensional reasons. This consideration can be confirmed explicitly applying the soft and collinear $(z,\tilde q)$ counting shown Tab.~\ref{tab:counting_massless} to the quark jet mass distribution defined in Eq.~(\ref{jetfunctionmasslessdef}): In the collinear region $z\sim (1-z)\sim 1$ and the cut-dependence arises where $\tilde q^2\sim Q_0^2$.  This leads to changes proportional to $Q_0^2$ on the invariant mass $s$ due to the $\delta$ function constraint. In the soft region we have $\tilde q\sim Q$ and $z\sim 1$, and the cut dependence arises where $(1-z)\sim Q_0/Q$,  and . This then leads to changes in $s$ proportional to $Q Q_0$ due to the $\delta$ function constraint. This simple counting is confirmed by the explicit calculations carried out in Secs.~\ref{sec:CBunreleased} and \ref{sec:QCDunreleasedmassless}. 

In Fig.~\ref{fig:phasespace_massive} we show the generic form of the $(z,\tilde q)$ phase space populated by coherent branching for the jet mass distribution in the resonance region for a massive quark, see Eq.~(\ref{jetfunctionmassivedef}), where the coloring is the same as for the massless quark case. Again the gray region represents the allowed phase space without $Q_0$ cut and the hatched region when the $Q_0$ cut is imposed. We see that the allowed phase space is considerably different from the massless quark case and overall confined to the region of large $z$. This is particular to the resonance region, where $s-m^2\ll m^2$. Here the massive quark thrust distribution is dominated by soft and ultra-collinear gluon radiation, which are also indicated. While the soft region is equivalent to the case of massless quarks, the ultra-collinear region differs substantially from the collinear region for massless quarks because it is related to gluon radiation that is soft in the massive quark rest frame and only becomes collinear due to the massive quark boost. As such the ultra-collinear radiation originating from a boosted massive quark with a given energy is substantially less energetic than the typical collinear radiation originating from a massless quark with the same energy. The resulting power counting is shown in Tab.~\ref{tab:counting_massive}, where we see e.g.\ that ultra-collinear gluons have a typical energy of order $Q^2\Lambda/m^2$, compared to collinear gluons which have a typical energy of order $Q$. Note that if we would consider the situation $s-m^2\gg m^2$ the allowed phase space would look similar to the massless case and we would recover the collinear counting. It is a remarkable fact that, despite its limitations, the coherent branching formalism for massive quarks is capable of describing both limits correctly and provides a smooth connection between them. 
We also note that, since $(p_\perp^2+(1-z)^2m^2)^{1/2}$ is the renormalization scale of the strong coupling,   
integrations over its Landau pole are strictly prevented as long as $Q_0$ is chosen larger than $\Lambda_{\rm QCD}$. Therefore there 
are no renormalon ambiguities in perturbative calculations in the presence of the $p_\perp$ cut $Q_0$. 

\begin{center}
\renewcommand{\arraystretch}{1.5}
\begin{tabular}{|c|c|c|}
 \hline
 \multicolumn{3}{|c|}{phase space regions for $\displaystyle \tau_{\rm peak}-\tau_{\rm min}\sim \frac{\Lambda}{Q}\ll 1$, $m\neq 0$} \\
 \hline
 & coherent branching & QCD factorization \\
 \hline
 \multirow{3}{*}{$n$-ultra-collinear} & \multicolumn{1}{|l|}{$\displaystyle 1-z\sim \frac{Q\Lambda}{m^2},\;z\sim 1$} & \multirow{3}{*}{$\displaystyle q^\mu\sim (\Lambda,\frac{Q^2}{m^2}\Lambda,\frac{Q}{m}\Lambda)$} \\
 & \multicolumn{1}{|l|}{$\displaystyle \tilde{q}\sim m$} &  \\ 
 & \multicolumn{1}{|l|}{$\displaystyle q_\perp \sim \frac{Q}{m}\Lambda$} &  \\ \hline
  \multirow{3}{*}{soft} & \multicolumn{1}{|l|}{$\displaystyle 1-z\sim \frac{\Lambda}{Q},\;z\sim 1$}  & \multirow{3}{*}{$\displaystyle q^\mu\sim (\Lambda,\Lambda,\Lambda)$} \\
 & \multicolumn{1}{|l|}{$\displaystyle \tilde{q}\sim Q$} &  \\ 
 & \multicolumn{1}{|l|}{$\displaystyle q_\perp \sim \Lambda $} &  \\ \hline
\end{tabular}
 \captionof{table}{Power counting for coherent branching and QCD factorization for massive quarks.}\label{tab:counting_massive}
\end{center}

To conclude this section let us also discuss in which sectors we should expect a linear dependence on $Q_0$ for the case of massive quark production.
In the context of QCD factorization, inspecting the factorization theorem~(\ref{eq:thrustmassive2}), we see that a linear dependence on the $p_\perp$ cut $Q_0$ can arise not only in the partonic soft function $S$ but also in the bHQET jet function $J_B$ because it has, in contrast to the massless quark jet function, a linear kinematic dependence on the reduced invariant mass variable $\hat s$, see Eq~(\ref{shatdef}). This simple dimensional argument can again be confirmed applying the ultra-collinear momentum counting shown in Tab.~\ref{tab:counting_massive} to the quark jet mass distribution defined in Eq.~(\ref{jetfunctionmassivedef}): We have $z\sim 1$, $\tilde q\sim m$ and the cut-dependence arises where $(1-z)\sim Q_0/m$. This leads to changes in the squared invariant mass relative to the threshold of order $s-m^2\sim m Q_0$. This simple counting is confirmed by the explicit calculations carried out in Secs.~\ref{sec:CBunreleased} and \ref{sec:QCDunreleasedmassive}.

\subsection{Unreleased radiation: coherent branching}
\label{sec:CBunreleased}

To calculate the effects of the $p_\perp$ cut $Q_0$ on the thrust distribution in the peak region in the context of the coherent branching formalism we can start from the corresponding Lapace space expressions given in Eq.~(\ref{eq:solution1massless}) for massless quarks and Eq.~(\ref{eq:solution1massive}) for massive quarks. In contrast to the calculations we carried out for our examinations concerning 
the summation of logarithms in Secs.~\ref{sec:CBnocutmassless} and \ref{sec:CBnocutmassive}, where the finite quark mass represented a non-trivial modification, we can treat the massless and the massive quark case within the same computation because $Q_0 < m$. We can therefore begin from the Laplace space thrust distribution 
\begin{align}
\ln \tilde{\sigma}^{\rm cb}(\nu,Q,m,Q_0)\,=\,&2\,\int_{m^2}^{Q^2}\frac{\mathrm{d}\tilde{q}^2}{\tilde{q}^2}\int_0^1\mathrm{d}z\,
\theta\bigl(\tilde{q}^2-m^2-\frac{Q_0^2}{(1-z)^2}\bigr)\notag\\
&\qquad\qquad \times P_{QQ}\Bigl[\alpha_s\bigl(\tilde{q}^2(1-z)^2\bigr),z,\frac{m^2}{\tilde{q}^2}\Bigr]
\Bigl(\mathrm{e}^{-\nu\tilde{q}^2(1-z)/Q^2}-1\Bigr) \notag \\
= &\int_{Q_0^2}^{Q^2}\mathrm{d}q_{\perp}^2\int_0^{1-\frac{q_\perp}{Q}}\mathrm{d}z\,\frac{1}{q_\perp^2+m^2(1-z)^2}\notag\\
& \qquad\qquad \times P_{QQ}\Bigl[\alpha_s\bigl(q_\perp^2+m^2(1-z)^2\bigr),z,\frac{m^2(1-z)^2}{q_\perp^2+m^2(1-z)^2}\Bigr]\notag\\
& \qquad\qquad \times\Bigl(\mathrm{e}^{-\nu(q_\perp^2+m^2(1-z)^2)/Q^2(1-z)}-1\Bigr)\,,
\end{align}
where we have implemented the $p_\perp$ cut $Q_0$ according to Eqs.~(\ref{pperpcut1}) and (\ref{pperpcut2}). In the second line we changed the integration variable from $\tilde q$ to $q_\perp$ and used that $m^2/Q^2\ll 1$.  
From the second line one can see that we can write the Laplace space thrust distribution with $Q_0$ cut as
\begin{align}
\label{eq:laplaceshift}
 \tilde{\sigma}^{\rm cb}(\nu,Q,m,Q_0)=\mathrm{e}^{-\mathcal{I}(\nu,Q,m,Q_0)}\times\tilde{\sigma}^{\rm cb}(\nu,Q,m)\,,
\end{align}
where $\tilde{\sigma}^{\rm cb}(\nu,Q,m)$ is the distribution without $Q_0$ cut and the function
\begin{align}\label{eq:I_function}
 \mathcal{I}(\nu,Q, m,Q_0)\,=\,&
2\,\int_{0}^{Q_0^2}\mathrm{d}q_{\perp}^2\int_0^{1-\frac{q_\perp}{Q}}\mathrm{d}z\,\frac{1}{q_\perp^2+m^2(1-z)^2}\notag\\
& \qquad\qquad \times P_{QQ}\Bigl[\alpha_s\bigl(q_\perp^2+m^2(1-z)^2\bigr),z,\frac{m^2(1-z)^2}{q_\perp^2+m^2(1-z)^2}\Bigr]\notag\\
& \qquad\qquad \times\Bigl(\mathrm{e}^{-\nu(q_\perp^2+m^2(1-z)^2)/Q^2(1-z)}-1\Bigr) 
\end{align}
describes the contributions of the unreleased radiation.
Since we are interested in the dominant contribution linear in $Q_0$ we can expand to linear order in $\nu$ and change variables
to $q^2=p_\perp^2+m^2(1-z)^2$ to obtain
\begin{align}
\mathcal{I}(\nu,Q,m,Q_0)\,=\,&\,-\,\frac{4C_F\nu}{\pi Q^2}\biggl\{(Q-2m)\int_0^{Q_0}\mathrm{d}q\,\alpha_s(q)
\\ &\qquad \qquad
+\,m\int_{Q_0}^m\mathrm{d}q\,\alpha_s(q)\,
\frac{\bigl(q-\sqrt{q^2- Q_0^2}\bigr)^2}{q\sqrt{q^2-Q_0^2}}\biggr\}\, + \, {\cal O}\Bigl(\nu^2,Q_0^2,\frac{m^2 Q_0}{Q^3}\Bigr)\,.\notag
\end{align}
where we have dropped terms which are down by additional powers of $Q_0/m$ and $m/Q$.
In addition, to linear order in $Q_0$ we can extend the upper limit of the second integral up to infinity. From this we obtain at ${\cal O}(\as)$ the final result 
\begin{align}
\label{eq:Ifunc}
\mathcal{I}(\nu,Q,m,Q_0)\,=\,&-\Bigl[16 \frac{Q_0}{Q}-8 \pi\frac{Q_0 m}{Q^2}\Bigr] \,\frac{C_F\alpha_s(Q_0)}{4\pi}\,\nu\, + \, {\cal O}(\nu^2,Q_0^2,\frac{m^2 Q_0}{Q^3},\alpha_s^2)\,,
\end{align}
for the unreleased radiation, where we can fix the scale of the strong coupling to $Q_0$ because it represents the only scale the
integral depends. 
For the case of massless quark production the term proportional to $m$ is zero.
A similar calculation for the massless quark case (relevant for an analysis in the effective coupling model) 
was carried out in Ref.~\cite{Davison:2008vx}.

For the thrust distributions obtained from the coherent branching formalism the relations~(\ref{eq:laplaceshift}) and (\ref{eq:Ifunc}) 
in connection with Laplace space identities
imply that up to terms quadratic in $Q_0$, the strong coupling and $m/Q$ the effect of the $p_\perp$ cut is a simple shift in $\tau$ with respect to the thrust distribution without $p_\perp$ cut:   
\begin{align}
\label{eq:CBshiftmassless}
\frac{{\rm d}{\sigma^{\rm cb}}}{{\rm d}\tau}(\tau,Q,Q_0) \,=\, & 
\frac{{\rm d}{\sigma^{\rm cb}}}{{\rm d}\tau}\Bigl(\tau+16 \frac{Q_0}{Q} \,\frac{C_F\alpha_s(Q_0)}{4\pi},Q,Q_0=0\Bigr)
\\
\label{eq:CBshiftmassive}
\frac{{\rm d}{\sigma^{\rm cb}}}{{\rm d}\tau}(\tau,Q,m,Q_0) \,=\, & 
\frac{{\rm d}{\sigma^{\rm cb}}}{{\rm d}\tau}\Bigl(\tau+\Bigl[16 \frac{Q_0}{Q}-8 \pi\frac{Q_0 m}{Q^2}\Bigr] \,\frac{C_F\alpha_s(Q_0)}{4\pi},Q,m,Q_0=0\Bigr)
\end{align}
These shifts are valid for the parton level distributions and through the convolutions of Eqs.~(\ref{eq:thrustmassless1}) and (\ref{eq:thrustmassive1}) also for the hadron level distributions. Numerically, these shifts are far from negligible for $Q_0\sim 1$~GeV, which is the typical size of the shower cut values used in state-of-the-art Monte-Carlo event generators.

Within the coherent branching formalism it is, however, not possible to systematically address the question if these shifts should be interpreted as modifications of QCD parameters such as the mass. This is because the coherent branching formalism provides a convenient computational method to sum cross section contributions that are singular in the soft and collinear limits, but does not provide a field theoretic background where this issue can be discussed conceptually from first principles. We will therefore examine the effects of $p_\perp$ cut $Q_0$ again in the next two subsections in the framework of the factorization theorems~(\ref{eq:thrustmassless2}) and (\ref{eq:thrustmassive2}) for massless and massive quarks, respectively.

\subsection{Unreleased radiation for massless quarks: QCD factorization}
\label{sec:QCDunreleasedmassless}

In the context of QCD factorization the hard, soft and collinear effects are separated at the operator level
and the modifications caused the by the $p_\perp$ cut $Q_0$ can be determined in each sector individually. Possible
cross terms and exponentiation effects are automatically taken care of by the multiplicative structure of the factorization theorem~(\ref{eq:thrustmassless2}). It is then straightforward to see that there is no change in the $U$ factors which sum the large logarithms, since the $p_\perp$ cut acts in the infrared and does not lead to any new types of UV-divergences. As far as the hard function $H_Q$ is concerned, the $p_\perp$ cut contributes only through terms of order $Q_0^2/Q^2$, which are strongly power-suppresed and negligible at the order we are working.
So we only have to analyze the jet function $J^{(\tau)}$ and the soft function $S^{(\tau)}$ as they describe radiation where the $p_\perp$ cut $Q_0$ can leave a non-trivial impact. 

We write the jet function $J^{(\tau)}$ and the soft function $S^{(\tau)}$ in the presence of the $p_\perp$ cut $Q_0$ in the form
\begin{align}
\label{eq:Jfctcut}
J^{(\tau)}(s,\mu_J,Q_0)\, = &\,   J^{(\tau)}(s, \mu_J)\, - \, J_{\mathrm{ur}}^{(\tau)}(s,Q_0)\,,\\
\label{eq:Sfctcut}
S^{(\tau)}(k, \mu_S,Q_0)\, = &\,   S^{(\tau)}(k, \mu_S)\, - \, S_{\mathrm{ur}}^{(\tau)}(k,Q_0)
\end{align}
where $J^{(\tau)}(s, \mu_J)$ and $S^{(\tau)}(k, \mu_S,Q_0)$ are the renormalization scale dependent jet and soft functions from Eq.~(\ref{eq:thrustmassless2}) determined using dimensional regularization for the momentum integrations and defined in the ${\overline{\rm MS}}$ renormalization scheme. Their expressions at ${\cal O}(\as)$ are displayed in Eqs.~(\ref{eq:scetjetfunction}) and (\ref{eq:scetsoftfunction}), respectively. The functions $J_{\mathrm{ur}}^{(\tau)}(s,Q_0)$ and $S_{\mathrm{ur}}^{(\tau)}(k,Q_0)$ represent the unreleased radiation coming from regions {\it below} the $p_\perp$ cut $Q_0$, i.e.\ they describes the perturbative radiation that is prevented if $Q_0$ is finite. Since the $p_\perp$ cut does not lead to any genuine UV divergences, $J_{\mathrm{ur}}^{(\tau)}$ and $S_{\mathrm{ur}}^{(\tau)}$ are renormalization group invariant, which we have indicated by dropping the renormalization scale dependence from their arguments. 
The calculations for $S_{\mathrm{ur}}^{(\tau)}$ and $J_{\mathrm{ur}}^{(\tau)}$ at ${\cal O}(\as)$ are straightforward and described in detail in  App.~\ref{sec:softfctwithcut} and App.~\ref{sec:scetjetfctwithcut}, respectively.

\begin{figure}
	\begin{subfigure}[c]{0.5\textwidth}
		\includegraphics[width=1.0\textwidth]{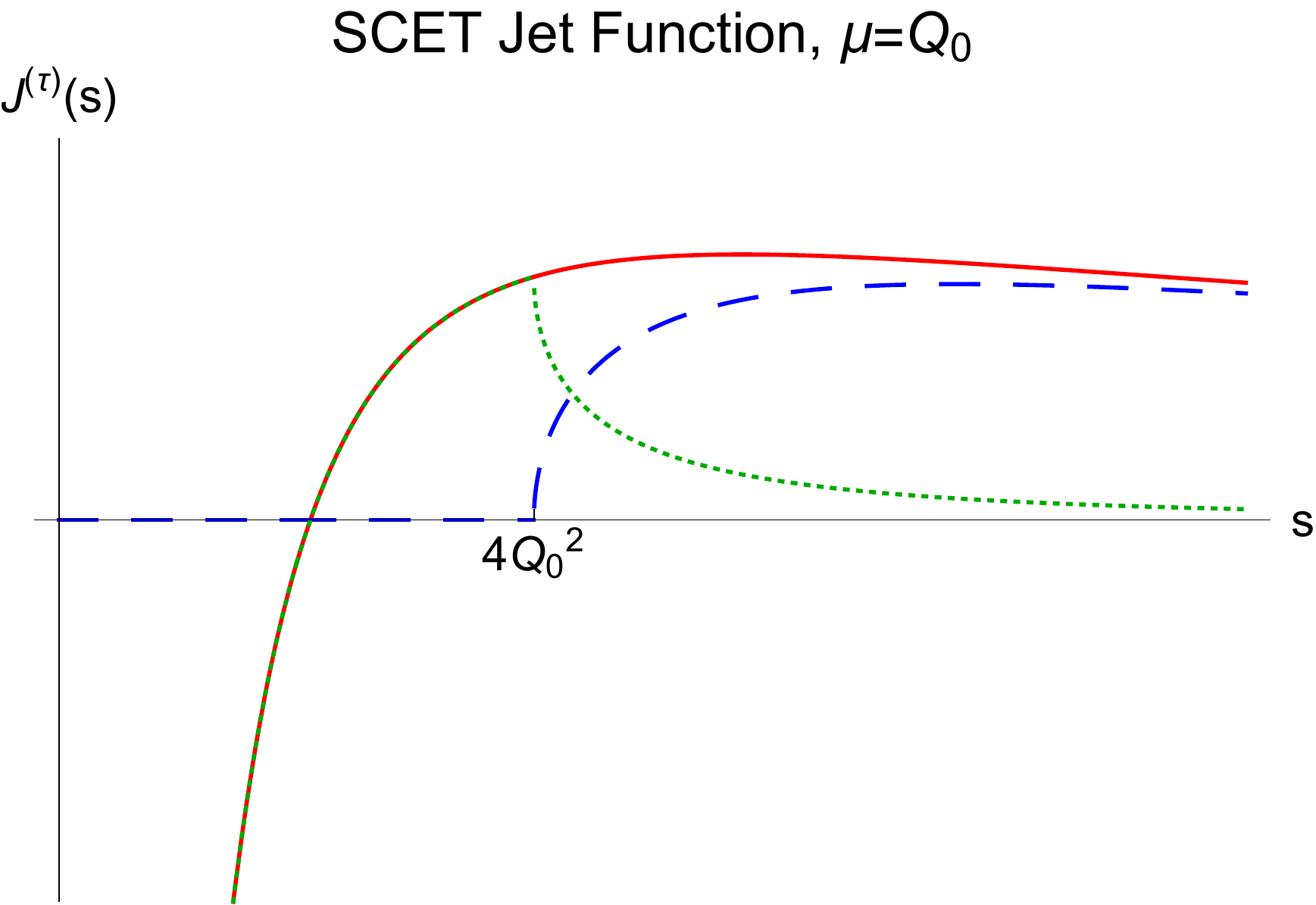}
		\subcaption{}\label{fig:SCETjet}
	\end{subfigure}
	\hfill
	\begin{subfigure}[c]{0.5\textwidth}
		\includegraphics[width=1.0\textwidth]{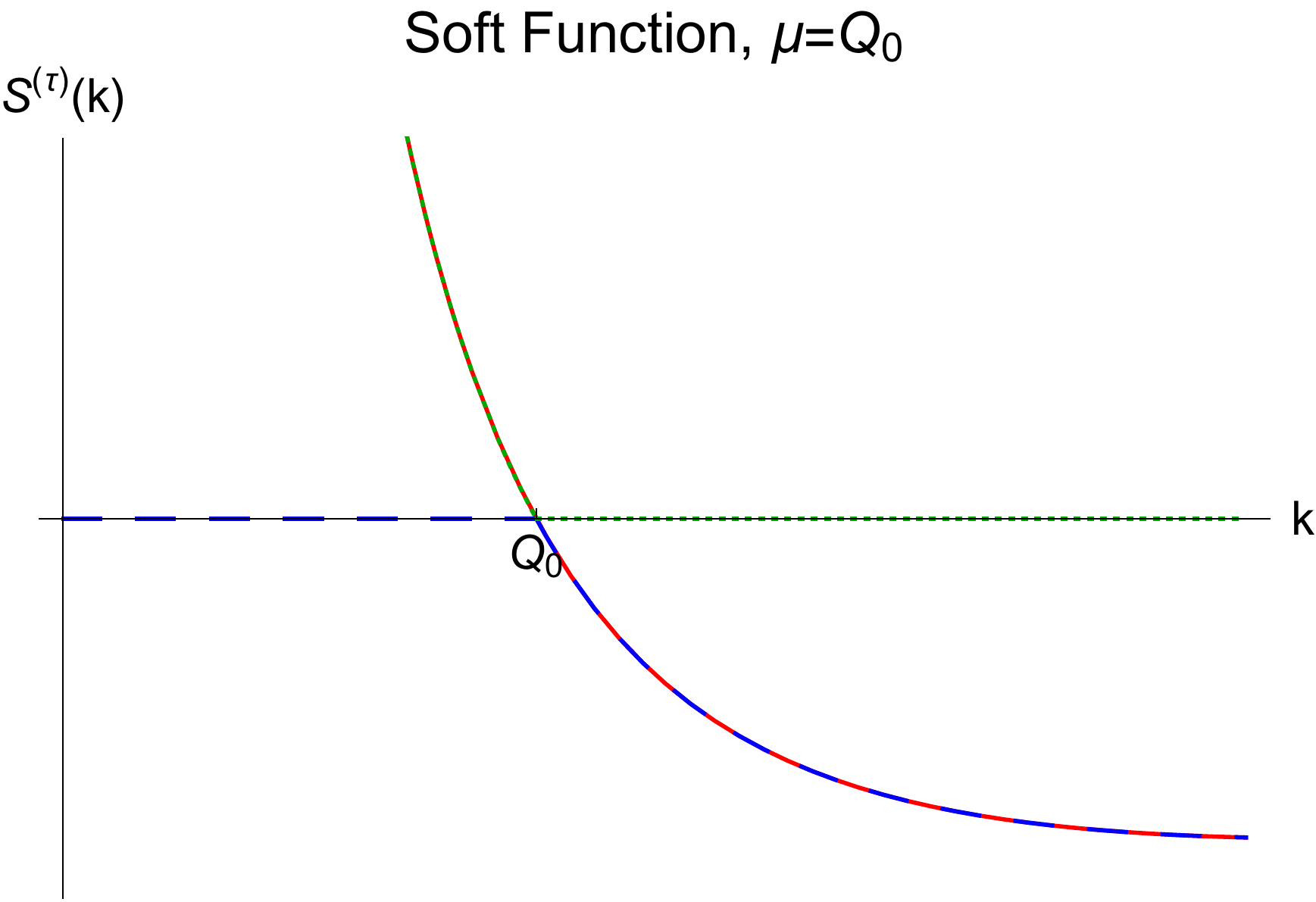}
		\subcaption{}\label{fig:softfunction}
	\end{subfigure}
	\caption{SCET jet function for massless quarks (a) and soft function (b) without cut (solid red), unreleased (dotted green) and with cut (dashed blue) for $\mu=Q_0$.}
\end{figure}

The result for the unreleased jet function has the form [$s^\prime=s/Q_0^2$,  $w(z)=(1-4/z)^{1/2}$]
\begin{align}
\label{eq:Jurresult}
J_{\mathrm{ur}}^{(\tau)}(s,Q_0)&\,=\,\asCF{Q_0}\biggl\{\Bigl(12-\frac{4\pi^2}{3}\Bigr)\delta(s)
 \notag \\
&\,\, +\theta(4Q_0^2-s)\biggl(-\frac{6}{Q_0^2}\plus{s^\prime}+\frac{8}{Q_0^2}\pluss{s^\prime}\biggr) \\
&\,\,+\theta(s-4Q_0^2)\frac{1}{s}\biggl[6(w(s^\prime)-1)-8\Bigl(\ln\Bigl(\frac{1+w(s^\prime)}{1-w(s^\prime)}\Bigr)-\ln\,s^\prime\Bigr)\biggr]\biggr\}+\Oas{2}\,.\notag
\end{align}
In Fig.~\ref{fig:SCETjet} the ${\cal O}(\as)$ corrections for jet function without $p_\perp$ cut, $J^{(\tau)}(s,\mu_J)$ (solid red line), the unreleased jet function $J_{\mathrm{ur}}^{(\tau)}(s,Q_0)$ (dotted green), and the full jet function with   $p_\perp$ cut,  $J^{(\tau)}(s,\mu_J,Q_0)$ (dashed blue line) are shown for $\mu_J=Q_0$ using arbitrary units. For this scale choice the $p_\perp$ cut completely eliminates the plus distributions for $s<4Q_0^2$ in $J^{(\tau)}(s,\mu_J,Q_0)$ and slightly reduces the collinear jet mass distribution for $s$ larger than $4Q_0^2$.
As already argued in Sec.~\ref{sec:unreleasedphasespace}, the unreleased radiation in the collinear sector depends quadratially on $Q_0$ except for the $\delta$-function term, which is however, not affecting the peak location $\tau_{\rm peak}$ at ${\cal O}(\as)$, see the discussion of Sec.~\ref{sec:peakprecision}. The contributions from $s<4 Q_0^2$ as well as from $s>4 Q_0^2$ lead to effects of order $Q_0^2$ in the observable thrust distribution upon integration over the soft model shape function, which corresponds to a smearing in $s$ over an interval of order $Q\Lambda$ which is much larger than $Q_0^2$, see Eq.~(\ref{eq:thrustmassless1}). Since we are interested in effects that are linear in $Q_0$, the unreleased radiation in the collinear sector can thus be ignored in our discussion. 

The result for the unreleased soft function reads [$k^\prime=k/Q_0$]
\begin{align}
\label{eq:Surresult}
S_{\mathrm{ur}}^{(\tau)}(k,Q_0)&=\asCF{Q_0}\,\,16\,\theta(Q_0-k)\,\biggl\{-\frac{1}{Q_0}\pluss{\tilde{k}}\biggr\}+\Oas{2}\,.
\end{align}
In Fig.~\ref{fig:softfunction} 
the ${\cal O}(\as)$ corrections to the soft function without $p_\perp$ cut, $S^{(\tau)}(k,\mu_S)$ (solid red line), the unreleased jet function $S_{\mathrm{ur}}^{(\tau)}(k,Q_0)$ (dotted green), and the full jet function with   $p_\perp$ cut,  $S^{(\tau)}(s,\mu_S,Q_0)$ (dashed blue line) are shown for $\mu_S=Q_0$ for arbitrary units. 
Similar to the case of the jet function, for this scale choice, the $p_\perp$ cut just eliminates the plus distributions for $k<Q_0$ in $S^{(\tau)}(k,\mu_S,Q_0)$, and but has no effects for $k>Q_0$. As already anticipated on general grounds in Sec.~\ref{sec:unreleasedphasespace}, the $p_\perp$ cut indeed leads to a linear dependence on $Q_0$.

As can be seen from the factorization formula~(\ref{eq:thrustmassless1}), the soft model shape function causes a smearing in $k$ over an interval of order $\Lambda$ which we assume to be larger than $Q_0$. Since the unreleased soft function has support only for light cone momenta $k<Q_0$, we can therefore quantify its effect more transparently in terms of a multipole expansion,
\begin{align}
S_{\mathrm{ur}}^{(\tau)}(k,Q_0) \, = & \,
- \Delta_{\rm soft}(Q_0)\,\delta^\prime(k)\,+\,{\cal O}(Q_0^2)\,,
\end{align}
where the term $\Delta_{\rm soft}(Q_0)$ is the first moment of the unreleased soft function,
\begin{align}
\label{eq:Deltasoft}
\Delta_{\rm soft}(Q_0) \, = & \,
\int \mathrm{d}k^\prime\,k^{\prime}\,S_{\mathrm{ur}}^{(\tau)}(k^\prime,Q_0)\,=\,16\, Q_0\,\asCF{Q_0}\,.
\end{align}

Mathematically, this multipole term appears to cause a shift of the partonic soft function threshold by $-\Delta_{\rm soft}(Q_0)$ since it can absorbed into the tree level soft function, $\delta(k)+ \Delta_{\rm soft}(Q_0)\,\delta^\prime(k)\approx \delta(k+ \Delta_{\rm soft}(Q_0))+{\cal O}(\alpha_s^2)$. In the context of the thrust factorization theorem~(\ref{eq:thrustmassless1}) we thus see that this shift agrees identically with the result which we determined from the coherent branching formalism given in Eq.~(\ref{eq:CBshiftmassless}).
However, as we have already mentioned before, in the coherent branching formalism there was no rigorous field theoretical background that strictly enforced this view {\it in the context of perturbation theory} because a perturbative modification of the threshold of a kinematic variable can only be implemented by a renormalization scheme change of a dimensionful parameter. Such a parameter does also not exists for the soft function because is arises from the dynamics of massless gluons and only depends on a light-cone momentum. In the context of the factorization formula~(\ref{eq:thrustmassless1}) 
the correct view is that the linear effect caused by the $p_\perp$ cut $Q_0$ can be reinterpreted as a \emph{shift of the soft model shape function $S_{\rm mod}$~\cite{Collins:1985xx,Catani:1989ne,Catani:1992ua}, called ``gap'' in Ref.~\cite{Hoang:2007vb}}. Following the presentation of Ref.~\cite{Hoang:2007vb} we can write the convolution of the partonic soft function and the non-perturbative shape function as
\begin{align}
\label{eq:softgap}
& \int \mathrm{d}\ell \, S^{(\tau)}(k-\ell, \mu_S,Q_0)\,S_{\rm mod}(\ell) \, = \int \mathrm{d}\ell \, \Bigl[S^{(\tau)}(k-\ell, \mu_S)\, - \, S_{\mathrm{ur}}^{(\tau)}(k-\ell,Q_0)\Bigr]\,S_{\rm mod}(\ell)\,,\notag\\
& \qquad \, = \, \int \mathrm{d}\ell \, S^{(\tau)}\Bigl(k-\ell+\Delta_{\rm soft}(Q_0), \mu_S\Bigr)\,S_{\rm mod}(\ell)
\, + \, {\cal O}(\alpha_s^2,\alpha_s Q_0^2) \notag \\
& \qquad \, = \, 
\int \mathrm{d}\ell \, S^{(\tau)}(k-\ell, \mu_S)\,S_{\rm mod}\Bigl(\ell+\Delta_{\rm soft}(Q_0)\Bigr)
\, + \, {\cal O}(\alpha_s^2,\alpha_s Q_0^2)\,.
\end{align}
This relation shows that the dominant effect of the $p_\perp$ cut $Q_0$ is to modify the interface between perturbation theory and non-perturbative physics and -- from the point of view of a partonic computation carried out without $Q_0$ cut -- acts as a modification of the hadronization contribution from $S_{\rm mod}(\ell)$ to $S_{\rm mod}\bigl(\ell+\Delta_{\rm soft}(Q_0)\bigr)$ as shown in the last line of Eq.~(\ref{eq:softgap}).
As long as the scale $Q_0$ is in the perturbative regime, this scheme change can be described perturbatively. 
This shows that the correct way to deal with a change in $Q_0$ when making physical predictions -- from the point of view of a partonic computation with a $Q_0$ cut -- is to modify the non-perturbative effects by a corresponding change of the shape function gap in order to leave the physical prediction unchanged.

One of the motivations of discussing ''gapped'' soft functions in Ref.~\cite{Hoang:2007vb} was to devise a way consistent with QCD factorization and field theory to eliminate the ${\cal O}(\Lambda_{\rm QCD})$ renormalon from the partonic soft function. This ${\cal O}(\Lambda_{\rm QCD})$ renormalon arises from large factorially growing coefficients in its perturbation series and renders, from the non-perturbative, i.e.\ beyond perturbation theory point of view, the partonic threshold ambiguous to an amount of order of $\Lambda_{\rm QCD}$. While for a massive particle threshold this can be achieved by a modification of the mass scheme, there is no such parameter for gluonic thresholds.
Our argumentation that the effects linear in the $p_\perp$ cut $Q_0$ should be interpreted as a soft function gap are therefore further supported, if the  $p_\perp$ cut eliminates the ${\cal O}(\Lambda_{\rm QCD})$ renormalon behavior of the partonic soft function. To examine this we restrict our discussion to the effects of dressing the gluon propagator with massless fermion bubble chains using the replacement~\cite{Beneke:1992ch}
\begin{align}
\label{eq:dressedgluon}
\frac{1}{q^2+i\epsilon}\,\longrightarrow\, 
\frac{4\pi}{\as(\mu)\beta_0}\Bigl(\frac{e^{5/3}}{\mu^2}\Bigr)^{-u}\,\frac{-1}{(-q^2-i\epsilon)^{1+u}}\,,
\end{align}
to compute the Borel transform,
using the convention $\beta[\as]=-2\beta_0(\as^2/4\pi)+\ldots$ for defining the coefficients of the $\beta$-function (see also Eq.~(\ref{eq:QCDbetafct})),
and focusing on poles in the Borel variable $u$ located a $u=1/2$. The term $e^{5/3}$ is related to using the usual 
$\overline{\mbox{MS}}$ 
renormalization scheme for the strong coupling. In passing we note that using the bubble chain method does not represent a strict all order proof that the $p_\perp$ cut eliminates the ${\cal O}(\Lambda_{\rm QCD})$ renormalon. However, it is sufficient for our discussion that focuses on angular ordered showers which have NLL order precision.  

As was shown in  Ref.~\cite{Hoang:2007vb}, the Laurent expansion of the Borel transform of the partonic soft function $S^{(\tau)}(k,\mu_S)$ around $u=1/2$ reads
\begin{align}
\label{eq:softBorel}
B\Bigl[S^{(\tau)}(k,\mu)\Bigr]\Bigl(u\approx \frac{1}{2}\Bigr)
\, = \, \frac{16 C_F e^{-5/6}}{\pi \beta_0}\,\frac{\mu}{u-\frac{1}{2}}\,\,\delta^\prime(k)\,.
\end{align}
The ${\cal O}(\Lambda_{\rm QCD})$ renormalon is canceled by the $p_\perp$ cut, if the unreleased soft function $S_{\mathrm{ur}}^{(\tau)}$ exhibits also a Borel pole at $u=1/2$ and if the residue agrees with the one shown in Eq.~(\ref{eq:softBorel}).
Some details on the calculation of the Borel transform of the unreleased soft function can be found in App.~\ref{sec:softfctwithcut}.
The result reads
\begin{align}
\label{eq:softurBorel}
B\Bigl[S_{\mathrm{ur}}^{(\tau)}(k,Q_0)\Bigr]\Bigl(u\approx \frac{1}{2}\Bigr)
\, = \, \frac{16 C_F e^{-5/6}}{\pi \beta_0}\,\frac{\mu}{u-\frac{1}{2}}\,\,\delta^\prime(k)\,,
\end{align}
and is identical to Eq.~(\ref{eq:softBorel}) when, consistently, the same scale choice is adopted for the strong coupling. The agreement shows that in the presence of the  $p_\perp$ cut $Q_0$ the ${\cal O}(\Lambda_{\rm QCD})$ renormalon is indeed removed from the partonic soft function due to Eq.~(\ref{eq:Sfctcut}). Thus the $p_\perp$ cut eliminates the ${\cal O}(\Lambda_{\rm QCD})$ renormalon and leads to a more convergent large-order behavior of the partonic soft function. This analysis also reconfirms the view that soft gluon radiation in (at least angular ordered) parton showers used in MC event generators does not suffer from  ${\cal O}(\Lambda_{\rm QCD})$ renormalon ambiguities, in contrast to perturbative calculations without finite infrared cuts.

\subsection{Unreleased radiation for massive quarks: QCD factorization}
\label{sec:QCDunreleasedmassive}

For the massive quark thrust distribution factorization theorem~(\ref{eq:thrustmassless2}) we proceed in a way analogous to the massless quark case. The $p_\perp$ cut does not lead to any modifications for the $U$ factors that sum large logarithms since it does not lead to any new types of UV-divergences. The hard function $H_Q$ is the same as for massless quarks, and the $p_\perp$ cut contributes terms of order $Q_0^2/Q^2$. The mass mode factor $H_m$, which arises from off-shell massive quark fluctuations, obtains modifications of order  $Q_0^2/m^2$. Both effects are strongly power-suppressed and negligible at the order we are working. Since the massive and massless quark factorization theorems contain the same partonic soft function $S^{(\tau)}$ and the same non-perturbative model shape function $S_{\rm mod}$, the effects of the $p_\perp$ cut we have discussed for them in the massless quark case also apply for massive quarks: the $p_\perp$ cut leads to a linear sensitivity to $Q_0$ can can be associated to a gapped soft function, as shown in  Eqs.~(\ref{eq:Deltasoft}) and (\ref{eq:softgap}). This takes care of the $m$-independent shift contribution shown in Eq.~(\ref{eq:CBshiftmassive}).

What remains to be examined is the bHQET jet function $J_B^{(\tau)}$ which contains the dynamics of the ultra-collinear radiation and which, as we have argued in Sec.~\ref{sec:unreleasedphasespace}, can also have a linear sensitivity to the $p_\perp$ cut $Q_0$.
The aim is to show from the field theory perspective that we can associate the $m$-dependent term in Eq.~(\ref{eq:CBshiftmassive}) to a modification of the quark mass scheme different from the pole mass.
This examination of the bHQET jet function represents the central part of our discussion because at NLL$+{\cal O}(\as)$ order the bHQET jet function completely controls the quark mass scheme. We note that the bHQET jet function dominates the mass dependence also at higher orders, while the mass dependence coming from other parts of the factorization formula is subleading.    

We write the bHQET jet function $J_B^{(\tau)}$ in the presence of the $p_\perp$ cut $Q_0$ in the form
\begin{align}
\label{eq:JBfctcut}
J_B^{(\tau)}(\hat s,m^{\rm pole},\mu_B,Q_0)\, = &\,   J_B^{(\tau)}(\hat s,m^{\rm pole},\delta m=0, \mu_B)\, - \, J_{B,\mathrm{ur}}^{(\tau)}(\hat s,Q_0)\,,
\end{align}
where $J_B^{(\tau)}(\hat s,m^{\rm pole},\delta m=0,\mu_B)$ is the renormalization scale dependent bHQET jet from Eq.~(\ref{eq:thrustmassive2}) {\it in the pole mass scheme} determined using dimensional regularization for the momentum integrations and defined in the ${\overline{\rm MS}}$ renormalization scheme. Its expression at ${\cal O}(\as)$ is displayed in Eqs.~(\ref{eq:JBoneloop}). The function $J_{B,\mathrm{ur}}^{(\tau)}(\hat s,Q_0)$ describes the unreleased radiation coming from regions {\it below} the $p_\perp$ cut $Q_0$. The $p_\perp$ cut does not lead to any genuine UV divergences, so $J_{B,\mathrm{ur}}^{(\tau)}$ is renormalization group invariant, which we have indicated by dropping the renormalization scale dependence from its arguments. The calculation of $J_{B,\mathrm{ur}}^{(\tau)}$ is described in detail in App.~\ref{sec:bHQETjetfctwithcut}. 

The result for the unreleased bHQET jet function reads [$\tilde{s}=\hat{s}/Q_0$, $w(z)=(1-4/z)^{1/2}$]
\begin{align}
\label{eq:JBurresult}
m^{\rm pole} J_{B,\mathrm{ur}}^{(\tau)}(\hat{s},Q_0)&\,=\,\asCF{Q_0}\biggl\{-8\pi Q_0\delta^{\prime}(\hat{s})+2\Bigl(4-\frac{\pi^2}{3}\Bigr)\delta(\hat{s}) \notag\\
&\,\,+\theta(2Q_0-\hat{s})\biggl(-\frac{8}{Q_0}\plus{\tilde{s}}+\frac{16}{Q_0}\pluss{\tilde{s}}\biggr) \\
&\,\,+\theta(\hat{s}-2Q_0)\,\frac{8}{\hat{s}}\,\biggl[(w(\tilde{s}^2)-1)-\Bigl(\ln\Bigl(\frac{1+w(\tilde{s}^2)}{1-w(\tilde{s}^2)}\Bigr)-2\ln \tilde{s}\Bigr)\biggr]\biggr\}+\Oas{2} \notag
\end{align}
\begin{figure}
\begin{center}
\includegraphics[width=0.5\textwidth]{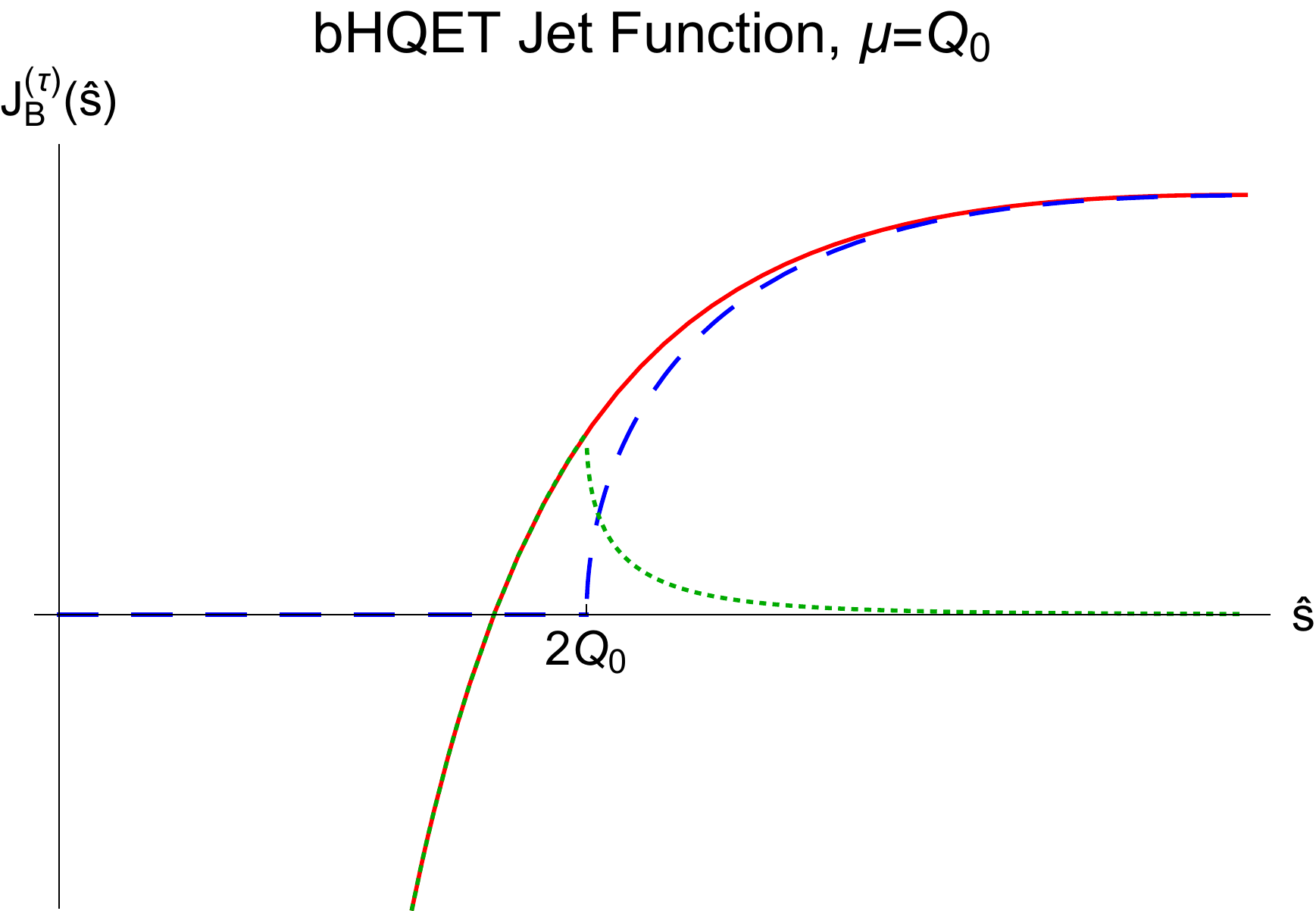}
\caption{bHQET jet function without cut (solid red), unreleased (dotted green) and with cut (dashed blue) for $\mu=Q_0$.}\label{fig:bHQETjet}
\end{center}
\end{figure}
In Fig.~\ref{fig:bHQETjet} 
the ${\cal O}(\as)$ corrections to the bHQET jet function without $p_\perp$ cut, $J_B^{(\tau)}(\hat s,\mu_B)$ (solid red line), the unreleased jet function $J_{B,\mathrm{ur}}^{(\tau)}(\hat s,Q_0)$ 
(dotted green), and the full jet function with the $p_\perp$ cut,  $J_B^{(\tau)}(\hat s,\mu_B,Q_0)$ (dashed blue line) are shown for $\mu_B=Q_0$ for arbitrary units. We see that the effect of the  $p_\perp$ cut has features common to the massless quark jet function: the $p_\perp$ cut eliminates the plus distributions for $\hat s<2Q_0$ and slightly reduces the ultra-collinear jet mass distributions for $\hat s$ larger than $2Q_0$, compare to Fig.~\ref{fig:SCETjet}. However, the difference is that the overall dependence on $Q_0$ is linear, as anticipated in Sec.~\ref{sec:unreleasedphasespace}, and the singular structure at $\hat s=0$ is more complicated due to the appearance of the term proportional to the derivative of the delta function, $\delta^\prime(\hat{s})$. This term arises from the on-shell cuts of the self-energy diagram of the heavy quark with the $p_\perp$ cut $Q_0$, see App.~\ref{sec:bHQETjetfctwithcut} for details.

To understand the result for the unreleased bHQET jet function in Eq.~(\ref{eq:JBurresult}), it is important to recall that for the soft function the interpretation of the effects of the $p_\perp$ cut is related to the interface between partonic cross section and the non-perturbative shape function that describes 
hadronization effects and that there is no partonic parameter involved in the argumentation. This differs from the bHQET jet function which contains the quark mass as a partonic parameter that depends on an explicit decision about its renormalization condition. 
In the expression for the ${\cal O}(\as)$ corrections to the bHQET jet function in Eq.~(\ref{eq:JBoneloop}) this dependence is 
manifest in the term  $-\frac{4\delta m}{m^{\rm pole}}\delta^\prime(\hat{s})$, where $\delta m = m_{\rm pole}-m$ is the difference of the employed mass renormalization scheme to the pole mass. From the structure of the convolutions in the factorization formulae~(\ref{eq:thrustmassive1}) and (\ref{eq:thrustmassive2}), due to the combination $\hat s \,m/Q - k$ appearing in the partonic soft function $S^{(\tau)}$, it is also evident that the effects linear in $Q_0$ contained in Eq.~(\ref{eq:JBurresult}) cannot be associated to a universal (i.e.\ $m/Q$-independent) change of the soft function model gap.
It is therefore mandatory to interpret these contributions from the point of a perturbative mass change alone. 

In the absence of the $p_\perp$ cut, i.e.\ when only dimensional regularization is used to regularize infrared and ultraviolet divergences, the bHQET on-shell heavy quark self energy is a scaleless integral and vanishes to all orders. So in bHQET the quark mass renormalization scheme is automatically the pole mass when we set $\delta m=0$. A change to another scheme is realized by explicitly adopting a finite expressions for $\delta m$ (which is a series that starts at ${\cal O}(\alpha_s)$). In the presence of the $p_\perp$ cut $Q_0$, however, the on-shell self-energy depends on the scale $Q_0$ and does not vanish any more, see App.~\ref{sec:bHQETjetfctwithcut} for details of this calculation. This is the origin of the $\delta^\prime(\hat{s})$ term in Eq.~(\ref{eq:JBurresult}), and it means that {\it in the presence of the $p_\perp$ cut $Q_0$ the pole mass $m^{\rm pole}$, as defined in perturbation theory without any infrared cut, does not any more represent the pole position of the heavy quark propagator}.\footnote{
At this point one may object that in the calculation of the unreleased bHQET jet function one can decide whether one applies the
$p_\perp$ cut $Q_0$ in the on-shell self-energy diagram or not. However, this corresponds to using different infrared regulators for
virtual and real radiation corrections which is inconsistent. In fact, dropping the $p_\perp$ cut $Q_0$ in the on-shell self-energy diagram only and keeping it in the rest of the calculation
is just equivalent to switching from the pole mass scheme to $m^{\rm CB}(Q_0)$. } 
Rather, the pole is located at the $Q_0$-dependent mass
\begin{align}
\label{eq:CBmassschemedef}
m^{\rm CB}(Q_0) \, = \, m^{\rm pole} - \delta m^{\rm CB}(Q_0)\,,
\end{align}
with
\begin{align}
\delta m^{\rm CB}(Q_0) \, = &\, \asCF{Q_0} \,2\pi\,Q_0  \, + \, 
{\cal O}(\as^2)\notag\\
\, = &\, \frac{2}{3}\alpha_s(Q_0)\,Q_0 \, + \, 
{\cal O}(\as^2)\,.
\end{align}
We stress that this means that the pole of the heavy quark propagator is not physical and implicitly depends on the infrared regularization scheme employed. The pole of the heavy quark propagatpr is unique only in the limit of vanishing infrared regulators.
We call $m^{\rm CB}(Q_0)$ the scale-dependent {\it coherent branching (CB) mass}.
It is possible to absorb the $\delta^\prime(\hat{s})$ correction term into the mass scheme (of the
tree-level bHQET jet function) which changes it from $m^{\rm pole}$ to the coherent branching mass $ m^{\rm CB}(Q_0)$.
The essential point is that this scheme change is \emph{implicitly carried out within the coherent branching formalism (and in angular ordered parton showers) because there the $\delta^{\prime}(\hat{s})$ term never arises}. This means that the mass parameter in the coherent branching formalism in the presence of the $p_\perp$ cut $Q_0$ \emph{agrees with the pole of the heavy quark propagator which is the CB mass $m^{\rm CB}(Q_0)$}. As we show in the following, only within this context we find that the result of Eq.~\eqref{eq:JBurresult} is compatible with the mass-dependent shift in Eq.~(\ref{eq:CBshiftmassive})
obtained from the coherent branching formalism in the presence of the $p_\perp$ cut, recalling the definitions of the thrust
variable $\tau$ and the linearized invariant mass variable $\hat s$ given 
in Eqs.~(\ref{eq:taudef}) and (\ref{shatdef}), respectively. 

The subtle issue to fully understand (and appreciate) our conclusion is that all the terms shown in Eq.~\eqref{eq:JBurresult} are required to allow the interpretation that the effects of the $p_\perp$ cut that are linear in $Q_0$ represent a modification of the mass scheme.
The crucial consistency requirement for this interpretation is that the sum of all modifications due to the contributions linear in the cutoff scale $Q_0$ given in Eq.~\eqref{eq:JBurresult} vanish. This is because a change of the quark mass scheme (and of the renormalization scheme of any QCD parameter) leaves the theoretical prediction invariant and essentially represents a mutual exchange
of perturbative corrections between the mass parameter and the dynamical matrix elements. It is therefore mandatory that the contributions linear in $Q_0$ of the remaining corrections (other than the  $\delta^\prime(\hat{s})$ term) 
in the unreleased bHQET jet function given in Eq.~(\ref{eq:JBurresult}) have the same magnitude but the opposite sign as the
contribution coming from the $\delta^\prime(\hat{s})$ term.
Since the soft model in the factorization theorem~(\ref{eq:thrustmassive1}) causes a smearing in $\hat s$ of order $Q \Lambda/m\gg Q_0$, we can - in analogy to our discussion on the unreleased soft function in Sec.~\ref{sec:QCDunreleasedmassless} - use again the multipole expansion to proceed. In contrast to our discussion on the soft function, we do not have to argue about the validity of the multipole expansion because for boosted top quarks we have $Q/m\gg 1$ so that the multipole expansion is well applicable even if $Q_0$ and $\Lambda$ are similar in size. The outcome is that we need to show that {\it the total integral (i.e. the zeroth moment) as well as the first moment of the unreleased bHQET jet function vanish identially}. 
If these conditions are satisfied, we can interpret all effects of the $p_\perp$ cut that are linear in $Q_0$ as a change in the quark mass renormalization scheme. 
   
It is straightforward to check from the result in Eq.~(\ref{eq:JBurresult}) that these properties are indeed satisfied:
\begin{align}
\label{eq:zeromomentzero}
\int \mathrm{d}\hat s\,J_{B,\mathrm{ur}}^{(\tau)}(\hat s,Q_0)& \,=\,  0\,,\\
\label{eq:firstmomentzero}
\int \mathrm{d}\hat s\,\hat s\,J_{B,\mathrm{ur}}^{(\tau)}(\hat s,Q_0)& \,=\,  \Bigl[\asCF{Q_0} \,8\pi\,Q_0\Bigr]_{\delta^\prime} 
- \Bigl[\asCF{Q_0} \,8\pi\,Q_0\Bigr]_{{\rm non}-\delta^\prime}
\notag\\ &
\,=\, \Bigl[ 4\delta m^{\rm CB}(Q_0) \Bigr]_{\delta^\prime}  
\, - \,  \Bigl[ 4\delta m^{\rm CB}(Q_0) \Bigr]_{{\rm non}-\delta^\prime}\, = \, 0\,,
\end{align}
where for the first moment we have indicated by subscripts the contributions from the $\delta^\prime(\hat s)$ term and the rest. Given the complicated structure of the result for the unreleased bHQET jet function in Eq.~(\ref{eq:JBurresult}), the results appear highly non-trivial. From the physical point of view, however, the vanishing zeroth moment is related to the fact that the total ($e^+e^-$ hadronic) cross section is not linearly sensitive to infrared momenta, which is well known. The vanishing of the first moment expresses that, physically, the mass-dependent kinematics threshold generated by the ultra-collinear radiation is not linearly sensitive to infrared momenta either. Linear sensitivity to infrared moments is only introduced by hand when one imposes the pole scheme for the heavy quark mass (defined in the common way by the one-particle irreducible on-shell self energy diagrams in the absence of any infrared regulator)\footnote{In this work we define the pole mass scheme $m^{\rm pole}$ strictly in the generally accepted canonical way, namely in the context of perturbation theory in the limit of vanishing infrared regularization.}. This feature is well known since a long time see e.g. Ref.~\cite{Ball:1995ni}. We can therefore expect that the zeroth and the first moments of the unreleased bHQET jet function vanish to all orders in perturbation theory.

At this point our prove is complete and we have field theoretically shown that -- if one always employs a mass scheme that agrees with the pole of the perturbative heavy quark propagator -- all effects of the $p_\perp$ cut that are linear in $Q_0$ not only can, but rather must
be interpreted as a change of the quark mass scheme from the pole mass to the coherent branching mass: 
\begin{align}
\label{eq:JBshiftrelation}
J_B^{(\tau)}(\hat s,m^{\rm pole},\mu_B,Q_0)\, &= \,   J_B^{(\tau)}(\hat s,m^{\rm pole},\delta m=0, \mu_B)\, - \, J_{B,\mathrm{ur}}^{(\tau)}(\hat s,Q_0)\notag\\[2mm]
&= J_B^{(\tau)}\Big(\hat s,m^{\rm CB}(Q_0),\delta m^{\rm CB}(Q_0), \mu_B\Big)
\, + \, {\cal O}(Q_0^2)
\,,
\end{align}
where at $\mathcal{O}(\alpha_s)$, keeping in mind Eq.~\eqref{eq:JBfctcut} and the form of Eq.~\eqref{eq:JBoneloop}, the term $\delta m^{\rm CB}(Q_0)$ in the 2nd line of Eq.~\eqref{eq:JBshiftrelation} is generated by the non-$\delta^{\prime}$ terms in the unreleased bHQET  jet function of Eq.~\eqref{eq:JBurresult}.
Recalling the definitions of the thrust
variable $\tau$ and the linearized invariant mass variable $\hat s$ given 
in Eqs.~(\ref{eq:taudef}), (\ref{eq:Mtaudef}) and (\ref{shatdef}), we see that the mass dependent $\tau$ shift in Eq.~(\ref{eq:CBshiftmassive}) agrees with the $\tau$ shift generated by $\delta m^{\rm CB}(Q_0)$ in the 2nd line of Eq.~\eqref{eq:JBshiftrelation}. This implies that the mass parameter in the coherent branching formalism (as well as in angular ordered parton showers) in the presence of the $p_\perp$ cut $Q_0$ \emph{is} the CB mass $m^{\rm CB}(Q_0)$.
The result of Eq.~(\ref{eq:JBshiftrelation}) gives us full control over the quark mass scheme in the presence of the $p_\perp$ cut $Q_0$
since, with the help of relation~(\ref{eq:CBmassschemedef}), we can relate the coherent branching mass $m^{\rm CB}(Q_0)$ to any other scheme at ${\cal O}(\alpha_s)$.

It is now natural to ask if the change from the pole mass to the scale-dependent CB mass cures 
the ${\cal O}(\Lambda_{\rm QCD})$ renormalon problem of the thrust distribution in the pole mass scheme. 
We address this question using again the dressed gluon propagator approach of Eq.~(\ref{eq:dressedgluon}) to determine
the Borel transform in the region around $u=1/2$.
As was shown in  Refs.~\cite{Beneke:1994rs,Beneke:1994sw}, the Laurent expansion of the Borel transform of the perturbative series in $\alpha_s(\mu)$ for the pole mass in terms of the $\overline{\mbox{MS}}$ mass around $u=1/2$ reads
\begin{align}
\label{eq:polemassBorel}
B\Bigl[m^{\rm pole}-\overline m(\mu)\Bigr]\Bigl(u\approx \frac{1}{2}\Bigr)
\, = \, -\frac{2 C_F e^{-5/6}}{\beta_0}\,\frac{\mu}{u-\frac{1}{2}}\,.
\end{align}
The corresponding result for the perturbative series  in $\alpha_s(\mu)$ for the pole mass in terms of the CB mass is calculated in 
App.~\ref{sec:bHQETjetfctwithcut} and reads
\begin{align}
\label{eq:CBmassBorel}
B\Bigl[m^{\rm pole}-m^{\rm CB}(Q_0)\Bigr]\Bigl(u\approx \frac{1}{2}\Bigr)
\, = \, -\frac{2 C_F e^{-5/6}}{\beta_0}\,\frac{\mu}{u-\frac{1}{2}}\,.
\end{align}
We see that the result is identical to Eq.~(\ref{eq:polemassBorel}). 
This shows that the scale-dependent CB mass $m^{\rm CB}(Q_0)$ is a low-scale short-distance mass. 
This is  not unexpected, of course, because the CB mass is defined from the bHQET on-shell massive quark self-energy 
with a transverse momentum infrared cut
which prevents the low-virtuality contributions from the evolution of the strong coupling that are responsible for the emergence of
infrared renormalons.
It is also straightforward to check that 
the Borel ambiguities coming from the $\delta^\prime(\hat s)$ self-energy term and the other contributions 
in the unreleased jet function (calculated from the perturbative series in $\alpha_s(\mu)$) cancel exactly:
\begin{align}
B\Bigl[J_{B,\mathrm{ur}}^{(\tau)}(\hat s,Q_0)\Bigr]\Bigl(u\approx \frac{1}{2}\Bigr)&\, = \,
\Bigl[
\frac{8 C_F e^{-5/6}}{\beta_0}\,\frac{\mu}{u-\frac{1}{2}} \,\delta^\prime(\hat s)
\Bigr]_{\delta^\prime}  
\, - \,  \Bigl[
\frac{8 C_F e^{-5/6}}{\beta_0}\,\frac{\mu}{u-\frac{1}{2}} \,\delta^\prime(\hat s)
 \Bigr]_{{\rm non}-\delta^\prime}\notag\\ & \, = \, 0\,.
\end{align}
This reconfirms the relation~(\ref{eq:JBshiftrelation}) also beyond the NLO precision level 
(at least in the large-$\beta_0$ approximation). 
As a consequence, imposing the $p_\perp$ cut $Q_0$ in the massive quark thrust distributions implies that one
uses the CB mass scheme of Eq.~(\ref{eq:CBmassschemedef}) and that all ${\cal O}(\Lambda_{\rm QCD})$ infrared renormalon issues 
are removed.

\section{Summary of all theoretical considerations}
\label{sec:intermediatesummary}

In this section we summarize all theoretical and conceptual results we have obtained  in the previous sections 
in the context of the massless and massive quark thrust distributions (see Eq.~(\ref{eq:taudef}) and (\ref{eq:Mtaudef}) in 
Sec.~\ref{sec:observable}) obtained in the coherent branching formalism and 
the QCD factorization approach. These findings provide the basis of the field theoretic reinterpretation of the effects of
the $p_\perp$ cut $Q_0$ {\it that are linear in $Q_0$} as a modification of hadronization contributions and a redefinition of
the heavy quark mass scheme, valid for boosted massive quarks in the narrow width approximation. We also discuss the meaning of these results in the context
of angular ordered parton showers, which are based on the coherent branching formalism and for which a 
$p_\perp$ cut on the parton shower evolution is mandatory. These considerations set the stage for 
the numerical studies we carry out in Sec.~\ref{sec:herwigcompare} using the \Herwig~7 
event generator~\cite{Bahr:2008pv,Bellm:2015jjp,Bellm:2017bvx}.

Since the QCD factorization approach provides the closest relation to field theory and allows to systematically 
address issues concerning the interpretation of partonic and non-perturbative parameters, the examinations 
in the previous sections
were built around establishing a one-to-one correspondence between the factorized cross sections for thrust and
the corresponding results obtained from the coherent branching formalism. For massive quarks the latter is known to be
valid for quasi-collinear and the former for boosted massive quarks, which here correspond to equivalent kinematic situations.  
Because the peak resonance region of the
thrust distribution, and in particular the peak position, provide the strongest and cleanest top mass sensitivity 
we have focused our considerations on the thrust resonance peak position. 

In Sec.~\ref{sec:peakprecision} we have shown that, for the factorized predictions, resummed results at full NLL order  
(where the dynamical logarithmic terms in the fixed-order matrix elements of the factorized predictions
are understood to be part of full NLL) are
sufficient to describe the peak position with NLO precision, i.e.\ up to higher order terms that enter only at 
${\cal O}(\as^2)$ and beyond. In Secs.~\ref{sec:CBnocutmassless} and \ref{sec:CBnocutmassive} we then 
established for massless and massive quarks, respectively, that in the absence of any infrared cut the NLL 
resummed results provided by the coherent branching formalism and by the usual factorized approach are equivalent. 
Since the massive quark results in the factorized approach we were using for the comparison were determined 
in the strict pole mass scheme $m^{\rm pole}$, we could prove that {\it in the coherent branching formalism  
with NLL resummation of logarithms and in the absence of an infrared cut (i.e.\ for $Q_0=0$) the quark mass 
parameter is equivalent to the pole mass $m^{\rm pole}$ at ${\cal O}(\as)$}:
\begin{align} 
\label{eq:mcbpole2}
m^{\rm CB}(Q_0=0) \,\stackrel{\rm peak}{=}\, m^{\rm pole} + {\cal O}(\alpha_s^2)
\,,
\end{align} 
where $m^{\rm CB}$ is the quark mass parameter in the coherent branching formalism and called the coherent branching (CB) mass.

This relation, however, is only valid in the context of strict QCD perturbation theory, i.e.\ in calculations 
based on expanding in $\alpha_s$ at a constant renormalization scale such that evolution effects are encoded entirely
in powers of logarithms and virtual loop and real radiation 
phase space integrals can be carried out down to zero momenta.
Such a strict perturbative approach, however, 
{\it cannot be applied for angular ordered parton shower algorithms implemented in state-of-the-art MC event generators},
so that it is not possible
to use them without an infrared cut on the parton shower evolution. There are two main reasons for that. The first is 
related to the fact that for the parton showers implemented in multi-purpose MC event generators the renormalization scale of the
strong coupling is a function of kinematic variables that decrease in the course of the shower evolution. In this way
parton showers can account for important subleading NLL information. Without an infrared cut the strong coupling would therefore
run into its Landau pole once the evolution reached virtualities and momenta close to $\Lambda_{\rm QCD}$. 
The second reason is that in the absence of the infrared cut
the particle multiplicities generated by the shower became infinite and made event generation impossible
for pure computational reasons. {\it Thus, relation~(\ref{eq:mcbpole2}) does not apply for parton 
showers that are used in MC event generators.}

In Sec.~\ref{sec:unreleased} we then analyzed the impact of the transverse momentum $p_\perp$ cut $Q_0$ that is imposed 
on angular ordered parton showers. In the evolution described by the coherent branching formalism this cut 
is paraphrased in the 
conditions (\ref{pperpcut1}) and (\ref{pperpcut2}) for the massless and massive quark case, respectively.
In the factorization approach it represents, at NLL$+{\cal O}(\alpha_s)$, a simple cut on the transverse 
momentum of (virtual or real) gluons with respect to the thrust axis in the hard, soft and jet functions. In the presence of the cut $Q_0$ the descriptions
provided by angular ordered parton showers, based on the coherent branching formalism and the one provided by the factorized approach
are all equivalent, and we were thus able to unambiguously track the field theoretic meaning and interpretation of the dominant
contributions \emph{linear} in the
$p_\perp$ cut $Q_0$ through the results obtained in the factorized approach.
At this point we emphasize that our conclusions related to the meaning and reinterpretation of QCD parameters
in the context of computations with the finite $p_\perp$ cut are made {\it from the perspective of computations
without any infrared cut}, since the canonical way how perturbative calculations and the renormalization procedure 
are carried out in collider physics applications is in the
limit of zero infrared cutoff.
Based on our examinations in Secs.~\ref{sec:CBunreleased}, \ref{sec:QCDunreleasedmassless} 
and \ref{sec:QCDunreleasedmassive} we proved the following two statements 
{\it valid in the peak region of the thrust distribution}:
\begin{itemize} 
\item[(1)] 
For massless quark production the dominant linear effects of the shower cut $Q_0$ represent a factorization 
scale\footnote{We adopt the canonical approach of factorization where the factorization scale that separates
perturbative and non-perturbative effects is chosen small, but also sufficiently large such that the interface
can be described within perturbative QCD.} 
at the 
interface of perturbative and non-perturbative large angle soft radiation, and changes in $Q_0$ 
can be reinterpreted as a modification of the non-perturbative contributions in the resonance peak region. 
In the coherent branching formalism and in the QCD factorization approach this modification is related to a shift in the non-perturbative model shape function, called ``gap'' in Ref.~\cite{Hoang:2007vb}\footnote{The name ``gap'' is motivated by the hadronization gap of the hadron mass spectrum.}, that can be 
computed perturbatively. 
For the thrust distribution in the peak region obtained in QCD factorization this is expressed quantitatively by the relation
\begin{align}
\label{eq:thrustmassless3}
\frac{\mathrm{d}\sigma}{\mathrm{d}\tau}&(\tau,Q,Q_0)\,=\,
\int\limits_0^{Q\tau}\!\mathrm{d}\ell\; 
\frac{\mathrm{d}\hat\sigma_s}{\mathrm{d}\tau}\Big(\tau-\frac{\ell}{Q},Q,Q_0\Big)\,\,S_{\rm mod}(\ell) 
\\
&\quad \,=\,
\int\limits_0^{Q\tau}\!\mathrm{d}\ell\; 
\frac{\mathrm{d}\hat\sigma_s}{\mathrm{d}\tau}\Big(\tau-\frac{\ell}{Q},Q,Q_0=0\Big)\,\,S_{\rm mod}(\ell+\Delta_{\rm soft}(Q_0)) 
\,+\,{\cal O}(\as^2,Q_0^2)\notag
\end{align} 
where $\mathrm{d}\hat\sigma/\mathrm{d}\tau$ stands for the partonic and $\mathrm{d}\sigma/\mathrm{d}\tau$ for
the hadron level distribution, $S_{\rm mod}$ is the soft model shape function incorporating the hadronization effects (see Sec.~\ref{sec:shapefunction}),
and having $Q_0$ in the argument of a function refers to a calculation with the $Q_0$ cut imposed.
Here, $\Delta_{\rm soft}(Q_0)$ is the $Q_0$-dependent gap that has the form
\begin{align}
\label{eq:Deltasoftv2}
\Delta_{\rm soft}(Q_0) \, = & \,
16\, Q_0\,\asCF{Q_0}\,+\,{\cal O}(\as^2 Q_0)\,.
\end{align}
The gap function $\Delta_{\rm soft}(Q_0)$ satisfies the
renormalization group equation 
\begin{align}
\label{eq:gapRGE}
R\,\frac{\mathrm{d}}{\mathrm{d}R}\,\Delta_{\rm soft}(R)\, = \, 16\,R\,\asCF{R} \, + \,{\cal O}(\as^2 R)\,,
\end{align}
which, due to the appearance of the scale $R$ on the RHS, describes evolution that is {\it linear in the 
renormalization scale} and is called R-evolution~\cite{Hoang:2008fs,Hoang:2008yj,Hoang:2017suc,Hoang:2017btd}. 
R-evolution differs from usual renormalization group equations such as 
for the strong coupling, which describe logarithmic evolution.  
In the context of multi-purpose MC event generators, where
an angular ordered parton shower is combined with a hadronization model, the relation means that a change of 
the shower cut $Q_0$ needs to be compensated by a retuning of the hadronization model parameters
in order to keep physical predictions effectively unchanged. 
At the level of the hadron level thrust factorization theorem valid in the peak region,
which involves the convolution of the partonic distribution $\frac{\mathrm{d}\hat{\sigma}}{\mathrm{d}\tau}$ with the soft model shape function, 
this feature is quantitatively encoded in the relation
\begin{align}
\label{eq:thrustmassless4}
\frac{\mathrm{d}\sigma}{\mathrm{d}\tau}&(\tau,Q,Q_0)\,=\,
\int\limits_0^{Q\tau}\!\mathrm{d}\ell\; 
\frac{\mathrm{d}\hat\sigma}{\mathrm{d}\tau}\Big(\tau-\frac{\ell}{Q},Q,Q_0^\prime\Big)\,\,
S_{\rm mod}(\ell+\Delta_{\rm soft}(Q_0)-\Delta_{\rm soft}(Q_0^\prime))\,,
\end{align} 
where the difference of the gap functions at the scales $Q_0$ and $Q_0^\prime$ is
\begin{align}
\label{eq:Deltasoftv3}
\Delta_{\rm soft}(Q_0) -\Delta_{\rm soft}(Q_0^\prime) \, = & \,
16\,\int\limits_{Q_0^\prime}^{Q_0}\mathrm{d}R\,\,\Bigl[\,\asCF{R}\,+\,{\cal O}(\as^2)\,\Bigr]\,,
\end{align}
which is manifestly infrared insensitive. Relation \eqref{eq:thrustmassless4} states that the dominant linear effects of a change of the shower cut form $Q_0^\prime$ to $Q_0$ can be compensated, to keep the prediction unchanged, by a modification of the soft model shape function of the form 
\begin{align}
\label{eq:Smodmodified}
S_{\rm mod}(\ell) \to \bar{S}_{\rm mod}(\ell)=S_{\rm mod}\Bigl(\ell -\Delta_{\rm soft}(Q_0)+\Delta_{\rm soft}(Q_0^\prime)\Bigr)\,.
\end{align}
We note that 
relations~(\ref{eq:thrustmassless3}) and (\ref{eq:thrustmassless4}) also have the 
important implication that the size of hadronization corrections for event-shape distributions that are encoded in MC event generators (i.e.\ the difference between parton and hadron level output) depends the value of the shower cut. A discussion of the feature is, however, beyond the scope of this work. We also remark that in practice
a change in the shower cut $Q_0$ may not be entirely compensated by a modification of the gap function alone because of additional non-linear dependence on the shower cut.
\item[(2)]
For massive quark production, the 
dominant linear effects of the shower cut $Q_0$ on the thrust distribution at the resonance peak can be interpreted, from the perspective of a computation in QCD factorization without 
infrared cutoff in the pole mass scheme $m^{\rm pole}$, as a 
modification of the non-perturbative contribution from large angle soft radiation and a 
change of the quark mass scheme from $m^{\rm pole}$ to the
scale-dependent coherent branching (CB) mass scheme $m^{\rm CB}(Q_0)$. 
The modification concerning the non-perturbative effects from large angle soft radiation 
is universal and the same as 
for massless quark production. The modification concerning the quark mass scheme originates from
the restriction the shower cut $Q_0$ imposes on the {\it ultra-collinear radiation}, which corresponds to soft radiation
in the massive quark rest frame and which has to be partly considered as an unresolved contribution to 
the observable top quark state. \emph{The shower cut $Q_0$ changes the position of the pole of the massive quark propagator to $m^{\rm CB}(Q_0)$ and also provides the associated scheme change corrections.}
Starting from a QCD factorization computation  of the thrust distribution in the pole mass scheme, 
this is expressed quantitatively by the relation
\begin{align}
\label{eq:thrustmassive3}
&\frac{\mathrm{d}\sigma}{\mathrm{d}\tau}(\tau,Q,m^{\rm pole},Q_0)\,=\,
\int\limits_0^{Q\tau}\!\mathrm{d}\ell\; 
\frac{\mathrm{d}\hat\sigma_s}{\mathrm{d}\tau}\Big(\tau-\frac{\ell}{Q},Q,m^{\rm pole},Q_0\Big)\,\,S_{\rm mod}(\ell)
\\
&\,=\,
\int\limits_0^{Q\tau}\!\mathrm{d}\ell\; 
\frac{\mathrm{d}\hat\sigma_s}{\mathrm{d}\tau}\Big(\tau-\frac{\ell}{Q},Q,m^{\rm CB}(Q_0),\delta m^{\rm CB}(Q_0),Q_0=0\Big)\,\,
S_{\rm mod}(\ell+\Delta_{\rm soft}(Q_0))\notag\\
&
\hspace{5cm}\,+\,{\cal O}(\as^2,Q_0^2)\notag
\end{align}
where $\mathrm{d}\hat\sigma_s/\mathrm{d}\tau$ stands for the parton level and $\mathrm{d}\sigma/\mathrm{d}\tau$ for
the hadron level distribution, $S_{\rm mod}$ is the soft model shape function incorporating the hadronization effects, having $Q_0$ in the argument of a function refers to a calculation with the $Q_0$ cut imposed coherently in virtual and real radiation calculations, and the argument $\delta m^{\rm CB}(Q_0)$ in $\mathrm{d}\hat\sigma_s/\mathrm{d}\tau$ indicates the modification of the perturbative series due to the scheme change from $m^{\rm pole}$ to $m^{\rm CB}(Q_0)$. 
The soft function gap $\Delta_{\rm soft}(Q_0)$ is given in Eq.~(\ref{eq:Deltasoftv2})
and the scale-dependent CB (coherent branching) mass scheme is defined by
\begin{align}
\label{eq:CBmassschemedef2}
m^{\rm CB}(Q_0) \, = \, m^{\rm pole} - \delta m^{\rm CB}(Q_0)\,,
\end{align}
with
\begin{align}
\label{eq:deltamCB}
\delta m^{\rm CB}(Q_0) \, = &\,
\,\frac{2}{3}\,\alpha_s(Q_0)\,Q_0 \, + \, 
{\cal O}(\as^2 Q_0)\,.
\end{align}
The scale-dependent CB mass $m^{\rm CB}(Q_0)$ is a short-distance mass and thus
does not suffer from the ${\cal O}(\Lambda_{\rm QCD})$ renormalon ambiguity inherent to the
pole mass $m^{\rm pole}$. It satisfies the R-evolution equation~\cite{Hoang:2008yj,Hoang:2017suc,Hoang:2017btd} 
\begin{align}
\label{eq:CBmassRRGE}
R\,\frac{\mathrm{d}}{\mathrm{d}R}\,m^{\rm CB}(R)\, = \, -\,\frac{2}{3}\,R\,\alpha_s(R)\, + \,{\cal O}(\as^2 R)\,,
\end{align}
and evolves linearly in $R$ in the same way as the soft function gap $\Delta_{\rm soft}(R)$.
The difference of the CB masses for the cutoff scales $Q_0^\prime$ and $Q_0$ can then be 
expressed by solving the $R$-evolution equation
\begin{align}
\label{eq:DeltaCBmass}
m^{\rm CB}(Q_0) - m^{\rm CB}(Q_0^\prime)\, = & \,
-\frac{2}{3}\,\int\limits_{Q_0^\prime}^{Q_0}\mathrm{d}R\,\,\Bigl[\,\as(R)\,+\,{\cal O}(\as^2)\,\Bigr]\,,
\end{align}
which is manifestly infrared insensitive.
In the context of angular ordered partons showers with a transverse momentum cut $Q_0$ 
the result implies -- because the parton shower quark mass parameter is implicitly identified with the pole of the quark propagator -- that the parton shower quark mass parameter is the scale-dependent CB mass $m^{\rm CB}(Q_0)$.
In the context of multi-purpose Monte-Carlo event generators, where
an angular ordered parton shower is combined with a hadronization model this means that a change of 
the shower cut from $Q_0^\prime$ to $Q_0$ needs to be compensated by a retuning of the hadronization model parameters compatible with Eq.~\eqref{eq:Smodmodified} {\it and} a
change of the value of the CB mass from $m^{\rm CB}(Q_0^\prime)$ to  $m^{\rm CB}(Q_0)$ according to Eq.~(\ref{eq:DeltaCBmass}) in order to keep physical predictions unchanged.
This puts a stringent field theoretic constraint on properties of the hadronization models, since it is
forbidden that they modify by themselves the mass scheme through the retuning procedure. 
\end{itemize}

Statements (1) and (2) can be cross checked numerically from the side of MC event generators 
by the analysis of the thrust peak position $\tau_{\rm peak}$ as a function of the shower cut $Q_0$
{\it when leaving the hadronization model as well as the numerical value of the generator mass unchanged}. In that case the
sizable linear effects in the $p_\perp$ cut $Q_0$ remain uncompensated and are directly visible in
a characteristic dependence of the thrust peak position, $\tau_{\rm peak}(Q_0)$, on $Q_0$. The resulting $Q_0$-dependence
of $\tau_{\rm peak}(Q_0)$ can be directly read off Eqs.~\eqref{eq:thrustmassless3}, (\ref{eq:Deltasoftv2}), \eqref{eq:thrustmassive3} and (\ref{eq:deltamCB}) giving
the relation
\begin{align}
\label{eq:tauQ0dependence}
 \tau_{\rm peak}(Q_0)\, = \, \tau_{\rm peak}(Q_0^\prime) -
\frac{1}{Q}\Bigl[16 -8 \pi\frac{m}{Q}\Bigr] \,\int\limits_{Q_0^\prime}^{Q_0}\mathrm{d}R\,
\frac{C_F \,\alpha_s(R)}{4\pi}\,,
\end{align}
where $m$ is the generator mass and $Q_0^\prime$ is some reference cutoff scale. Here it is understood that 
only cutoff values $Q_0\ll m$ are employed, and we also remind the reader that the results have been derived in
the limit of boosted massive quarks where $m\ll Q$. 
For the rescaled thrust variable $M_\tau$, see Eq.~(\ref{eq:Mtaudef}), which is suitable for an analysis for
top quarks, the analogous relation reads
\begin{align}
\label{eq:MtauQ0dependence}
M_{\tau,\,{\rm peak}}(Q_0) \, = \, M_{\tau,\,{\rm peak}}(Q_0^\prime) -
\Bigl[8 \frac{Q}{m}-4 \pi\Bigr] \,\int\limits_{Q_0^\prime}^{Q_0}\mathrm{d}R\,
\frac{C_F\, \alpha_s(R)}{4\pi}\,.
\end{align}

We note that in relations~(\ref{eq:tauQ0dependence}) and (\ref{eq:MtauQ0dependence}) the cutoff dependence coming from the large angle soft 
and the ultra-collinear radiation have an opposite sign. This is a characteristic property of these two different
types of effects, which may be used to differentiate between them in the context of quark mass sensitive observables 
which are more exclusive concerning the soft radiation.
In the next section we confront these relations numerically 
with parton-level simulations carried out with the
\Herwig~7 event generator~\cite{Bahr:2008pv,Bellm:2015jjp,Bellm:2017bvx}.

\section{Event generation with Herwig 7}
\label{sec:herwigcompare}

In this section we confront the conceptual and theoretical considerations summarized in Sec.~\ref{sec:intermediatesummary}
and in particular our predictions for the shower cutoff dependence of the peak position of the thrust distributions given in 
Eqs.~(\ref{eq:tauQ0dependence}) and (\ref{eq:MtauQ0dependence})
and our main conclusion that the presence of a shower cutoff $Q_0$ implies that the top quark mass parameter
used in an angular ordered 
parton shower is the scale-dependent CB mass given in
Eq.(\ref{eq:CBmassschemedef2})
with numerical simulations for $e^+e^-$ collisions using the \Herwig~7 event generator
\cite{Bahr:2008pv,Bellm:2015jjp,Bellm:2017bvx} in version 7.1.2. 
The angular ordered parton shower algorithm of \Herwig~7 implements the
coherent branching algorithm outlined in
Secs.~\ref{sec:coherentbranchingmassless} and \ref{sec:coherentbranchingmassive}, 
for massless and massive
quarks, respectively. Since the treatment of the top quark decay goes beyond the coherent branching
formalism outlined in these sections, we provide some more details of 
event generation in \Herwig~7  for top quarks in Sec.~\ref{sec:herwigtopquarks}. 
In Sec.~\ref{sec:herwigsimulationsetup} we explain a number of special setting we use for our \Herwig~7
simulations such that they are precisely in accordance to the coherent branching formalism. 
In Sec.~\ref{sec:herwigthrusttest}, using simulations results obtained with \Herwig~7, 
we reconfirm some approximations 
used in our analytic calculations 
in Secs.~\ref{sec:CBnocutmassless}, \ref{sec:CBnocutmassive} and 
\ref{sec:CBunreleased} within the coherent branching formalism, and 
the insensitivity of thrust to the cut governing the parton shower evolution of the top decay products. 
Our predictions for the shower cutoff dependence of the
thrust peak position for the massless quark and top quark case are then confronted with \Herwig~7
in Secs.~\ref{sec:herwigmassless} and \ref{sec:herwigmassive}, respectively. Here we demonstrate that our 
conceptual predictions for the shower cut dependence of the peak position of the thrust distributions
given in Eqs.~(\ref{eq:tauQ0dependence}) and (\ref{eq:MtauQ0dependence}) are indeed reproduced by the  \Herwig~7
simulations. 
In Sec.~\ref{sec:herwigotherobservables} we address the universality of our findings for thrust 
by discussing the reconstructed ($b$-jet and W boson) top quark invariant mass $m_{b_jW}$ and the endpoint region of the 
$b$-jet and lepton invariant mass $m_{b_j\ell}$.
Finally, in Sec.~\ref{sec:herwigNLO} we comment on the (ir)relevance of NLO-matched simulations 
with respect to the cutoff dependence of the thrust distribution in the resonance region and the kinematic mass sensitivity
of the reconstructed observables $m_{b_jW}$ and $m_{b_j\ell}$.

\subsection{Event generation for top quark production}
\label{sec:herwigtopquarks}

Within \Herwig~7 events with top quarks account for the top quark decay 
in a factorized narrow width approach:
The top quarks are considered stable at the stage of
their production, with momenta $p^\mu$ which satisfy the on-shell
condition $p^2=m_t^2$, where $m_t$ is the \Herwig\ top mass parameter. In the 
default setting  for the \texttt{LEP-Matchbox.in} simulation setup,
no smearing with any Breit-Wigner-type distribution is applied, so that
off-shell effects coming from the finite top quark width are absent.
This default setting is mainly rooted in considerations related to NLO matched 
predictions, where the smearing disrupts the cancellation of virtual and real infrared 
cancellations.
The angular ordered parton shower then attaches radiation to the production process terminated by
the $p_\perp$ cutoff $Q_{0}$, including radiation off the top quarks
(and possibly other colored partons involved in the hard
scattering). After the kinematic reconstruction following the production stage parton shower,
the final state top quarks have definite momenta $p^{\prime\mu}$ which satisfy the on-shell condition
$p^{\prime 2}=m_t^2$, and the
progenitor top quarks, which initiated the showering, have acquired a
virtual mass, see the discussion in
Sec.~\ref{sec:coherentbranchingHerwig}. 
At this point the top quarks decay, where we for simplicity only consider leptonic decays 
of the $W$ bosons coming from the top decays assuming perfect neutrino identification. This is not a restriction for the thrust distributions we examine, but 
simplifies their numerical analyses. The partons originating from the top decays, $t\to b\ W^+$ and $\bar t\to \bar b\ W^-$,
then radiate according to the decay parton shower algorithm from Refs.~\cite{Gieseke:2003rz} and \cite{Hamilton:2006ms}
which is terminated by the $p_\perp$ cutoff $Q_{0,b}$. The radiation from the decay stage parton shower exactly preserves  the  4-momenta of the decaying top and antitop quarks, respectively, and hence their mass shell condition, in a separate kinematic reconstruction procedure.
Within this procedure the $b$-quark shower progenitor that initiates the $b$-jet is allowed to acquire a virtuality according to the decay stage parton shower. 

In the conceptual considerations of the preceding sections we were discussing the effects of the 
production stage parton shower cutoff $Q_0$. The thrust variable is by construction independent of details of the top decay
and therefore also insensitive to the value of the decay state parton shower cutoff $Q_{0,b}$.
In \Herwig~7 the values of $Q_0$ and $Q_{0,b}$ can be chosen independently, which allows us to explicitly check
the insensitivity of thrust to variations of $Q_{0,b}$. This check is carried out in Sec.~\ref{sec:herwigthrusttest}.

\subsection{Settings for MC simulations}
\label{sec:herwigsimulationsetup}

To compare the predictions obtained from analytic examinations of the preceding sections
with the \Herwig\ predictions, which are based on the previously described algorithms and methods,
we use a number of special settings. These settings are used to eliminate default 
features in \Herwig~7 which go beyond the coherent branching formalism as described 
in Sec.~\ref{sec:coherentbranching} or interfere with the $Q_0$ dependence of the {\it parton level
predictions} we aim to analyze. 
We emphasize that the purpose of these settings is to allow for a direct comparison 
of \Herwig\ simulations with our analytic results {\it at the parton level} 
in a conceptually clean and controlled setup.
So these special settings may serve as the starting point of further examinations, 
also accounting for the effects and properties of hadronization models, 
where the impact of 
default settings used in \Herwig\ (or other MC event generators) can be studied in more detail, or
for upcoming releases. 
We emphasize, however, that these special \Herwig~7 settings should be taken with some care since they
are not appropriate to carry out full hadron level simulations. 

As already explained in Sec.~\ref{sec:coherentbranchingHerwig} we set the (constituent) masses
of all quarks and the mass of the gluons that have emerged after the parton showers have terminated to very small 
values to effectively remove their effects in the parton level
results.\footnote{We note that in \Herwig\ all light quarks (i.e.\ up, down and strange quarks) 
	and gluons are treated as exactly massless during the shower evolution and that constituent quark mass and gluon mass conditions
	are only imposed kinematically for the partons that emerge after shower terminations. The constituent quark and gluon masses have to be considered as part of the hadronization model.} 
Zero constituent quark
and gluon masses are required to allow a comparison with our analytic QCD calculations; they are,
however, not compatible with the default \Herwig~7 cluster hadronization model.
Furthermore, in our \Herwig\ simulations we do not include any QED radiation or any
matrix element corrections, except in our discussion of NLO matching carried out in Sec.~\ref{sec:herwigNLO}.
As already discussed in Sec.~\ref{sec:coherentbranchingHerwig} we also
choose the \texttt{CutOff} option for the kinematic reconstruction as
this does not alter the correspondence to the underlying coherent branching algorithm as described in
in Secs.~\ref{sec:coherentbranchingmassless} and \ref{sec:coherentbranchingmassive}.
Finally we note that most analyses we have developed are based on Rivet \cite{Buckley:2010ar}, except those focusing on
particle multiplicities for which an entirely in-house analysis code is used.
In App.~\ref{app:Settings} we give the complete set of input file changes
required to reproduce the parton level results within our special settings, both for 
the massless and massive case. 

\subsection{Monte Carlo tests of approximations for analytic thrust calculations}
\label{sec:herwigthrusttest}

In our analytic calculations of the parton level massless and massive quark 
jet mass distributions at NLL order 
in Secs.~\ref{sec:CBnocutmassless}, \ref{sec:CBnocutmassive} and 
\ref{sec:CBunreleased} within the coherent branching formalism 
we used two approximations which were crucial 
to allow for an analytic all order exponentiation of the computation, see 
e.g.\ Eqs.~(\ref{eq:jetmass1}) to (\ref{eq:solution1massless}). 
In the integral equations for the jet mass distributions 
shown in Eqs.~(\ref{jetfunctionmasslessdef}) and (\ref{jetfunctionmassivedef})
these approximations involve (i)
neglecting the parton branching of the gluon (i.e.\ switching off the $g\to gg$ and $g\to q\bar{q}$ branchings)
and (ii) using the $z\to 1$ limit in the  parts which are slowly varying in the soft limit.
These approximations were
already discussed (and used for analytic calculations) in the seminal coherent branching 
papers for massless quarks, see e.g.\ Ref.~\cite{Catani:1989ne,Catani:1992ua}.
The former approximation is -- {\it for the thrust distribution in the peak region} --  
related to the fact that due to angular ordering the 
showered gluons originating from the progenitor quarks can themselves not radiate to
pick up any significant virtuality. 
The latter approximation implies -- {\it again for the thrust distribution in the peak region} -- 
that once gluon splitting is turned off, also strict 	
angular ordering can be dropped from the
calculations. For simplicity reasons we therefore refer to the latter approximation as ''angular ordering switched off'' 
in the following discussion.

\begin{figure}
	\center
	\begin{subfigure}[c]{0.48\textwidth}
		\includegraphics[width=1.0\textwidth]{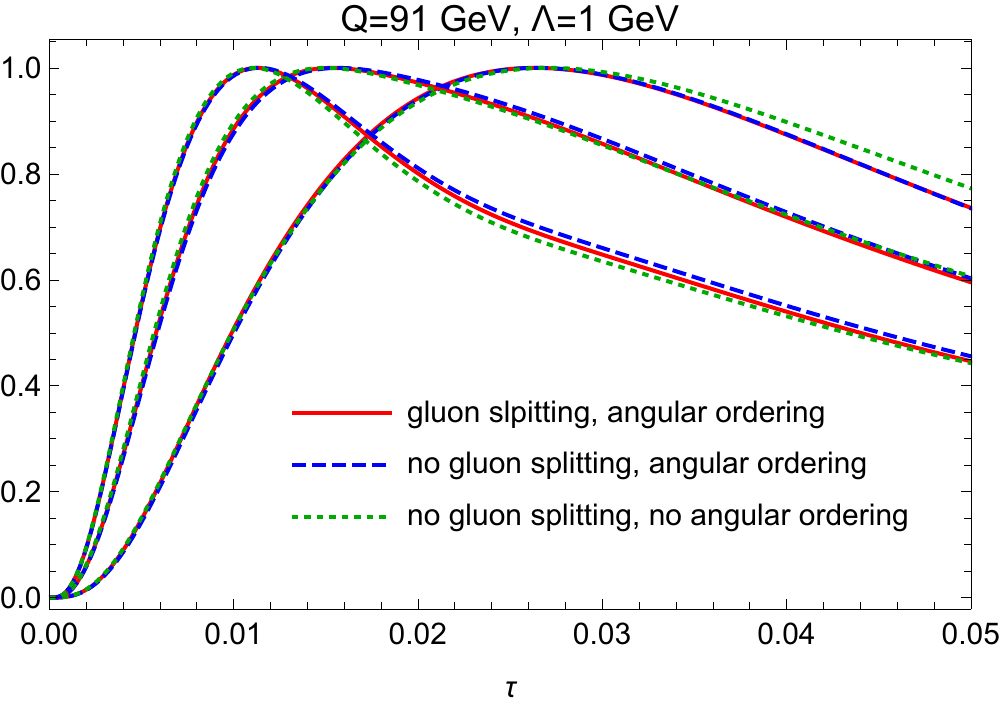}
		\subcaption{\label{fig:checks_a}}
	\end{subfigure}
	\hfill
	\begin{subfigure}[c]{0.48\textwidth}
		\includegraphics[width=1.0\textwidth]{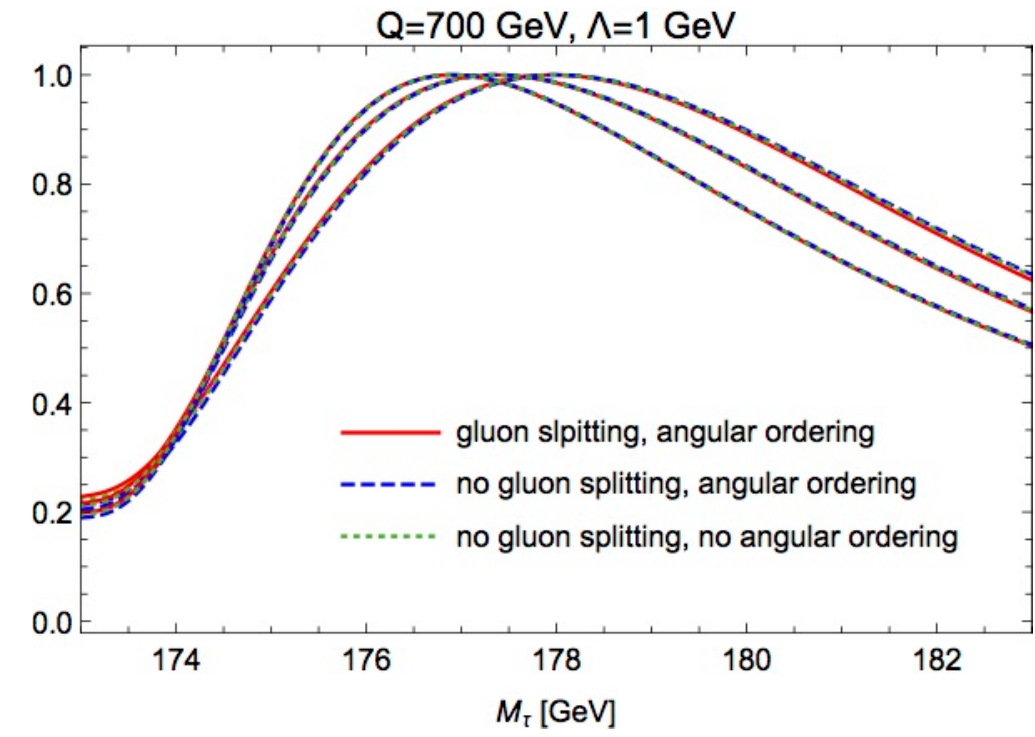}
		\subcaption{\label{fig:checks_b}}
	\end{subfigure}
	
	\caption{\label{fig:checks} Thrust at the parton level in the peak region generated by \Herwig~7 for 
		(a) massless quarks at c.m.\ energy	$Q=91$~GeV and (b) top quarks with mass $m_t=173$~GeV at 
		$Q=700$~GeV. The \Herwig~7 parton level results are smeared with a soft model shape function with smearing parameter 
		$\Lambda=1$~GeV, see Sec.~\ref{sec:shapefunction}. Displayed are simulation results for shower cuts  $Q_0=1$~GeV (right set of curves), 
		$Q_0=1.5$~GeV (middle set of curves) and  $Q_0=2$~GeV (left set of curves) and
		with gluon splitting and angular ordering both turned on (solid red curves), 
		with gluon splitting turned off, but angular ordering turned on (dashed blue curves)
		and with gluon splitting and angular ordering both turned off (dotted green curves).}
\end{figure}

Adopting the settings discussed in Sec.~\ref{sec:herwigsimulationsetup},
these two approximations can be explicitly verified numerically using \Herwig\ 7 
to generate the parton level thrust distribution for massless and massive quark 
production. In Fig.~\ref{fig:checks_a} the parton level thrust distribution,
defined in Eq.~(\ref{eq:taudef}),
obtained from \Herwig\ 7 for massless quarks at c.m.\ energy $Q=91$~GeV
is displayed for shower cuts $Q_0=1$~GeV (right set of curves), 
$Q_0=1.5$~GeV (middle set of curves) and  $Q_0=2$~GeV (left set of curves)
with gluon splitting and angular ordering both turned on (solid red curves), 
with gluon splitting turned off, but angular ordering turned on (dashed blue curves)
and with gluon splitting and angular ordering both turned off (dotted green curves).
All curves are normalized such that at their respective maximum they evaluate to 
unity, which is particularly suitable to discuss the peak region. 
We also remind the reader that
all curves are produced by convolution of the \Herwig~7 parton level results with
the soft model shape function of Eq.~(\ref{eq:smoddef}) for $\Lambda_{\rm m}=\Lambda$ 
with $\Lambda=1$~GeV according to Eq.~\eqref{eq:thrustmassless1}. 
As discussed in Sec.~\ref{sec:shapefunction}, this is essential to obtain a smooth distribution in the peak region that can be interpreted properly. In Fig.~\ref{fig:checks_b}
the parton level rescaled thrust distribution, as defined in Eq.~(\ref{eq:Mtaudef}),
obtained from \Herwig\ 7 for top quarks at c.m.\ energy $Q=700$~GeV and with the 
generator mass set to $m_t=173$~GeV is displayed in the same way and for the same choices for the 
shower cut $Q_0$ and concerning gluon splitting and angular ordering. 
For the top quark case we employed a convolution over the same shape function
according to Eq.~(\ref{eq:Mtau1}) with 
$\Lambda_{\rm m}=\Lambda+4m_t\Gamma_t/Q$ and $\Gamma_t=1.5$~GeV. 
For the top quark case the smearing parameter is larger than for the massless quarks in order
to simulate the additional smearing effects of the top quark width. Note, however, that this
does not represent a systematic treatment of width effects for the top quark.  

From the results in Figs.~\ref{fig:checks_a} and~\ref{fig:checks_b}
we clearly see that the impact of the gluon splitting is very small in the peak region
and, furthermore, that once gluon splitting is turned off the numerical effects of 
angular ordering are very small as well. For the rescaled thrust distribution in the case of
top production these three settings lead to variations in the peak position of less
than $\Delta \tau_{\rm peak} \sim 10^{-3}$ in the massless case and less than $\Delta M_{\tau,\rm{peak}}\sim 100$~MeV in the massive case for $Q_0$ between $1$ and $2$~GeV.
In any case, these variation are considerably smaller than
the variations caused by changes in the shower cut $Q_0$ which we focus on in our 
subsequent examinations. While the validity of the two approximations concerning gluon splitting and
angular ordering for thrust for massless quarks has already been known since Ref.~\cite{Catani:1992ua},
our analysis shows that they are also applicable for the massive quark case, which is new. 
We note that in our analysis of the dependence of the thrust peak position on the shower cut $Q_0$ in 
Secs.~\ref{sec:herwigmassless} and \ref{sec:herwigmassive}, 
we consider \Herwig\ 7 simulations using all three options: (i) full simulation, (i) simulations with 
gluon branchings switched off and (iii) simulations with gluon branchings and angular ordering both 
switched off. The differences of the  \Herwig\ 7 results obtained from these three options should be viewed 
as an illustration of possible subleading effects even though they should not be overinterpreted as a 
systematic error estimate.

\begin{figure}
	\center
	\includegraphics[width=1.0\textwidth]{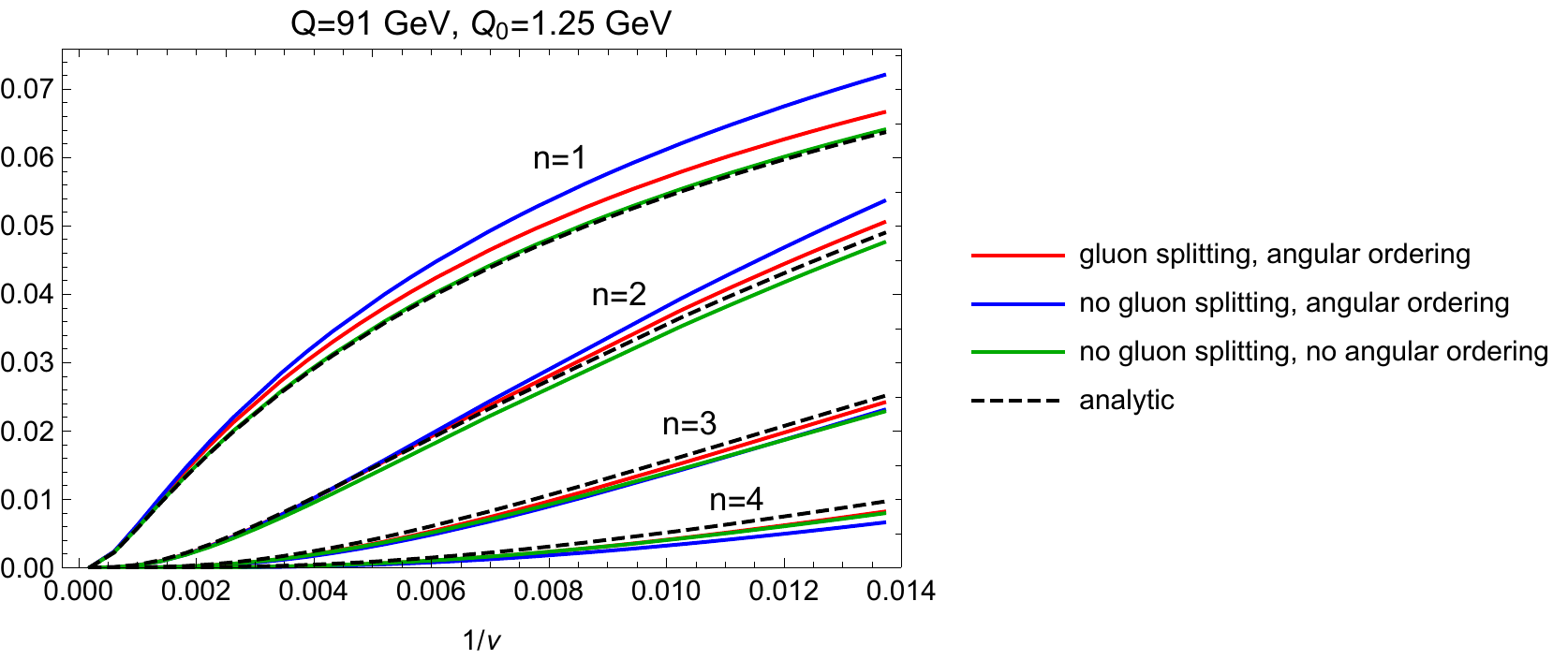}
	\caption{\label{fig:checks-multi} Laplace space parton level thrust distribution over $1/\nu$ in the
		peak region for $Q=91$~GeV and $Q_0=1.25$~GeV shown for the final state parton multiplicities $n=1,2,3,4$.   
		Displayed are the analytic results (dashed black curves) and simulation results 
		with gluon splitting and angular ordering both turned on (red curves), 
		with gluon splitting turned off, but angular ordering turned on (blue curves)
		and with gluon splitting and angular ordering both turned off (green curves).}
\end{figure}

In the context of these results 
an obvious question to ask is whether the suppression of effects coming from the gluon 
branching in the thrust peak region is only a cumulative effect 
visible in the distribution upon accounting for the sum of all emissions, or
whether the suppression takes place literally at the level of the individual parton multiplicities.
To answer this question we can analyze the parton level massless quark thrust distribution for a fixed number 
of final state parton multiplicity, where we define the multiplicity $n$ as the total 
number of partons emitted from the progenitor quark-antiquark pair. 
Interestingly, for the Laplace space parton level distribution~(\ref{eq:sigmaLaplacemassless})
for massless quarks the contribution for a given multiplicity $n$ 
can be determined analytically, in the approximation that gluon splitting and angular ordering
are switched off, simply from Eq.~(\ref{eq:sigmatildemassless}) by taking 
the $n$-th term in the Taylor expansion of the exponential
function. In Fig.~\ref{fig:checks-multi} the Laplace space parton level thrust distribution 
for massless quarks at $Q=91$~GeV with shower cut $Q_0=1.25$~GeV 
is shown as a function of $1/\nu$ in the peak region 
$1/\nu\sim \tau_{\rm peak}\ll 1$ for multiplicities $n=1,2,3,4$. 
Shown are the \Herwig\ 7 full simulation results 
with gluon splitting and angular ordering both turned on (solid red curves), 
with gluon splitting turned off, but angular ordering turned on (solid blue curves)
and with gluon splitting and angular ordering both turned off (solid green curves)
and the analytic result from Eq.~(\ref{eq:sigmatildemassless}), which is calculated 
in the approximation with gluon splitting and angular ordering both turned off  
(dashed black curves). The curves do not include any smearing effects from the
shape function because the Laplace integral of Eq.~(\ref{eq:sigmaLaplacemassless}) already 
provides a sufficient amount of smearing.
We see that \Herwig\ 7 and the analytic results in the various approximations agree very well.
The outcome shows that that the approximations we used in our analytic calculations are also appropriate
at the level of fixed parton multiplicities and may therefore have a more general validity. 

At this point we emphasize that the examination of the effects
of gluon splitting and angular ordering we have just carried out
solely serves as a cross check for the approximations we used in
our analytic calculation for thrust using the coherent branching formalism in Sec.~\ref{sec:CBnocut}
and \ref{sec:CBunreleased} and that these approximation are 
{\it not} a viable option for general phenomenological studies. These approximations
do also not \emph{in any way} constitute conceptual guidelines for predictions based on QCD factorization (or SCET). 
In addition, the consistent use of these approximations for thrust involves that the effects
angular ordering are only small once the gluon branchings are {\it already} switched off.
Indeed, the converse, a simulation with
gluon branchings but strict angular ordering switched off
leads to a dramatic increase of parton radiation and multiplicities and 
to physically meaningless outcomes.

\begin{figure}
		\center
	\begin{subfigure}[c]{0.48\textwidth}
		\includegraphics[width=1.0\textwidth]{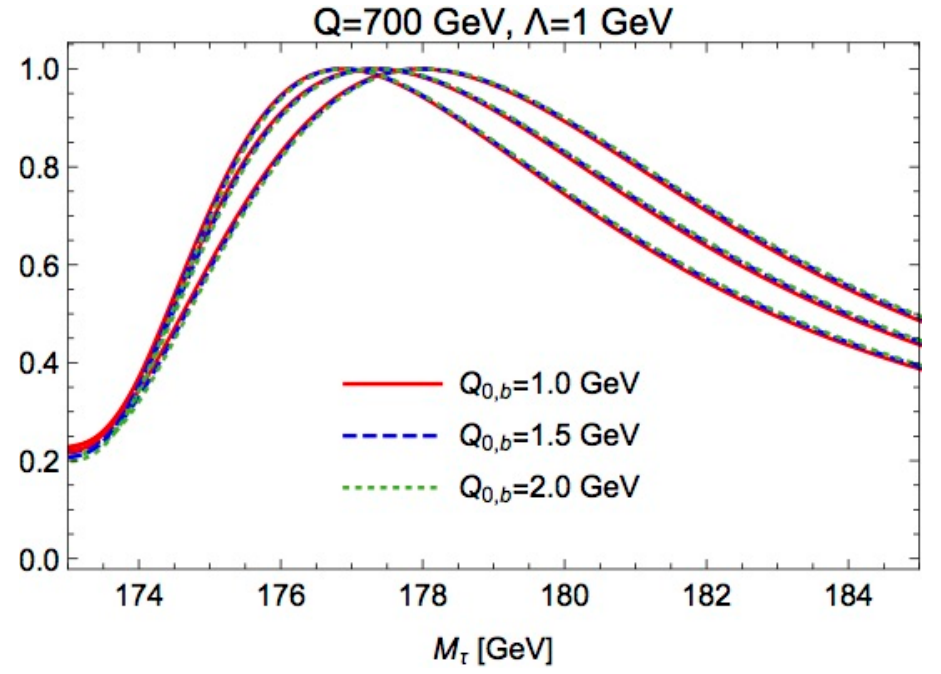}
		\subcaption{\label{fig:checks-b-cut_a}}
	\end{subfigure}
	\begin{subfigure}[c]{0.48\textwidth}
		\includegraphics[width=1.0\textwidth]{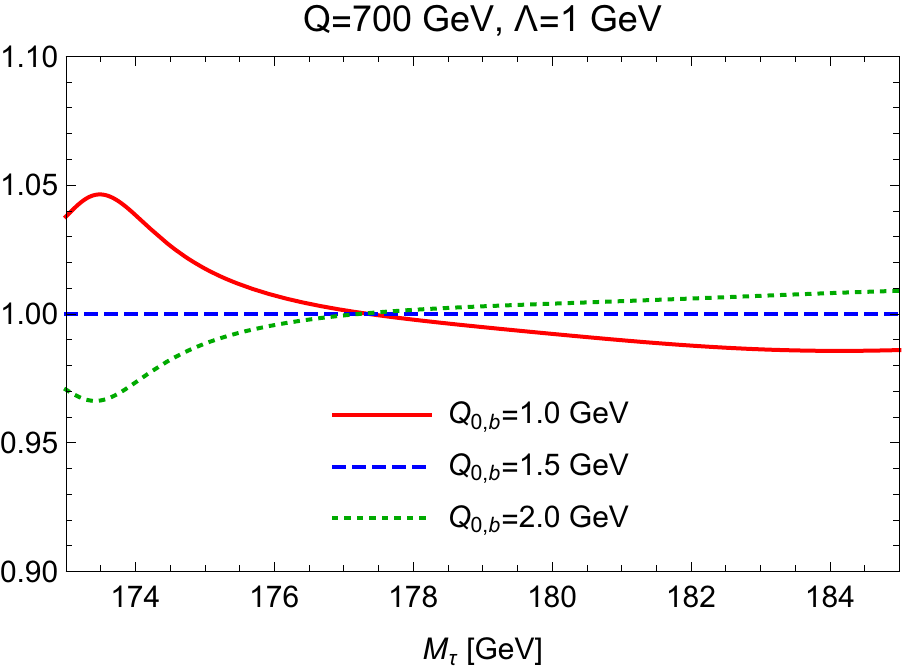}
		\subcaption{\label{fig:checks-b-cut_b}}
	\end{subfigure}
	\caption{\label{fig:checks-b-cut}
	Parton level rescaled thrust for top quarks with $m_t=173$~GeV and $Q=700$~GeV in the peak region 
	generated by \Herwig~7 and smeared with a soft model shape function with smearing parameter 
	$\Lambda=1$~GeV, see Sec.~\ref{sec:shapefunction}. 
	Displayed are simulation results for production stage shower cuts  $Q_0=1$~GeV (right set of curves), 
	$Q_0=1.5$~GeV (middle set of curves) and  $Q_0=2$~GeV (left set of curves) and
	decay stage shower cuts $Q_{0,b} = 1$~GeV (solid red curves), $Q_{0,b} =1.5$~GeV (dashed blue curves)
	and $Q_{0,b} =2$~GeV (dotted green curves).}
\end{figure}

As already elaborated in Sec.~\ref{sec:herwigtopquarks}, for event generation involving 
top quarks \Herwig~7 uses a factorized treatment of production and decay stage 
parton shower evolution. As we argued in Sec.~\ref{sec:observable} the thrust variable
is by construction independent of details of the top decay
and should therefore be insensitive to the value of the decay state parton shower
cutoff $Q_{0,b}$. 
In Fig.~\ref{fig:checks-b-cut_a} the parton level distribution of the rescaled thrust $M_\tau$ 
in the peak region obtained from \Herwig~7
is shown for the c.m.\ energy $Q=700$~GeV and generator mass $m_t=173$~GeV with
production stage shower cuts $Q_0=1$~GeV (right set of curves),
$Q_0=1.5$~GeV (middle set of curves) and $Q_0=2$~GeV (left set of curves),
and decay state shower cuts of $Q_{0,b} = 1$~GeV (solid red curves), $Q_{0,b} =1.5$~GeV (dashed blue curves)
and $Q_{0,b} =2$~GeV (dotted green curves). In Fig~\ref{fig:checks-b-cut_b} a ratio plot for the curves
for the three choices of $Q_{0,b}$ is shown for $Q_0=1$~GeV. For all curves a shape function smearing 
with $\Lambda=1$~GeV has been included following the prescription given above. 
The results confirm that the dependence of the thrust distribution on the decay stage shower cut 
$Q_{0,b}$ is extremely weak and in particular significantly smaller than the corresponding dependence on the 
production stage shower cut $Q_0$. In the resonance region variations due to changes of $Q_{0,b}$ are at the percent level
and negligible as far as the peak position is concerned. 
The results confirm that the thrust variable is ideal to study the production stage shower cutoff dependence
and essentially insensitive to differential details of the top quark decay. 
For our studies of the shower cutoff dependence of the thrust peak position in Secs.~\ref{sec:herwigmassless}
and \ref{sec:herwigmassive} we set $Q_{0,b}=Q_0$, which is the default \Herwig~7 setting.

\subsection{Thrust peak position for massless quarks}
\label{sec:herwigmassless}

In this section we confront our analytic parton level prediction for the $Q_0$ dependence of the 
thrust peak position for massless quarks,
\begin{align}
\label{eq:tauQ0dependencemassless}
\tau_{\rm peak}(Q_0)\, = \, \tau_{\rm peak}(Q_0^\prime) -
\frac{16}{Q} \,\int\limits_{Q_0^\prime}^{Q_0}\mathrm{d}R\,
\frac{C_F \,\alpha_s(R)}{4\pi}\,,
\end{align}
with parton level simulations in \Herwig\ 7 using the specific settings discussed in 
Sec.~\ref{sec:herwigsimulationsetup}. To determine the distribution for a given
c.m.\ energy $Q$ and shower cut $Q_0$ we generated $10^9$ events. The resulting 
binned distribution (with bin size $\Delta\tau =2\times 10^{-4}$) was numerically convoluted 
using a discretized version 
of Eq.~(\ref{eq:thrustmassless1}) with the soft model shape function $S_{\rm mod}$ given in 
Eq.~(\ref{eq:smoddef}) for a given smearing parameter $\Lambda_{\rm m}$. The peak position was then 
determined from fitting a quadratic function to the bin values in the peak region with heights that
differ from the maximum by at most $1$ per mille.
This leads to statistical uncertainties in the peak position well below $10^{-3}$
which is an order of magnitude smaller than the size of the $Q_0$ variations we obtain in our
analysis. The results can thus be considered exact for all practical purposes and we refrain from
quoting any statistical uncertainties in the results we obtain in the simulations.

\begin{figure}
	\center
	\begin{subfigure}[c]{0.49\textwidth}
		\includegraphics[width=1.0\textwidth]{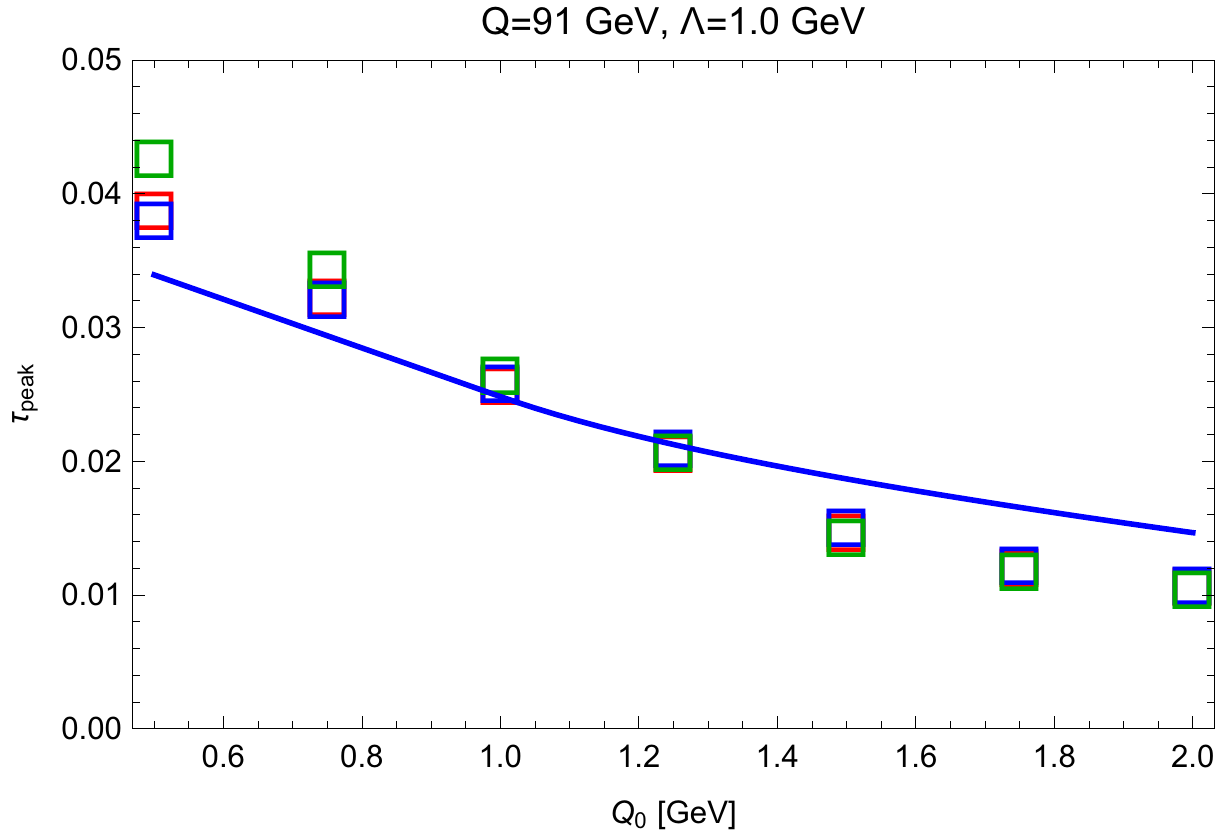}
		\subcaption{}
	\end{subfigure}
	\hfill
	\begin{subfigure}[c]{0.49\textwidth}
		\includegraphics[width=1.0\textwidth]{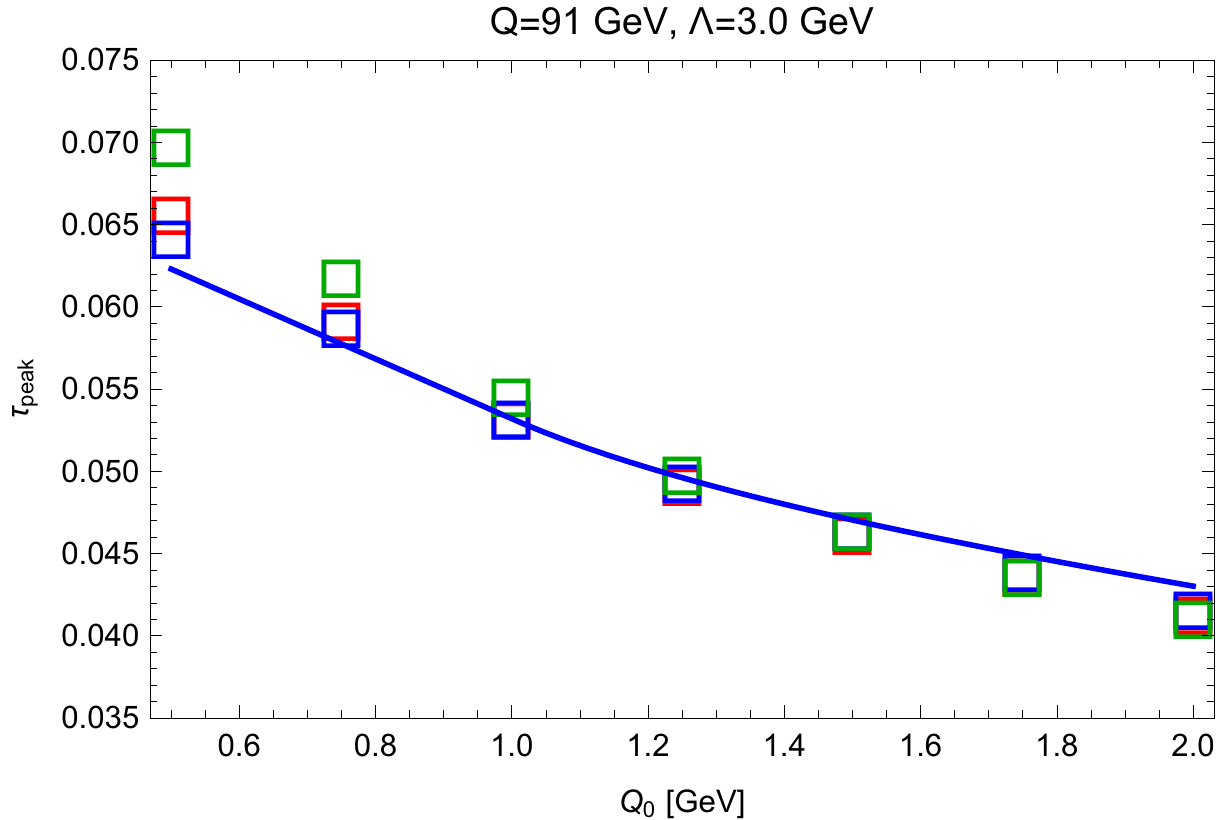}
		\subcaption{}
	\end{subfigure}
	\begin{subfigure}[c]{0.49\textwidth}
		\includegraphics[width=1.0\textwidth]{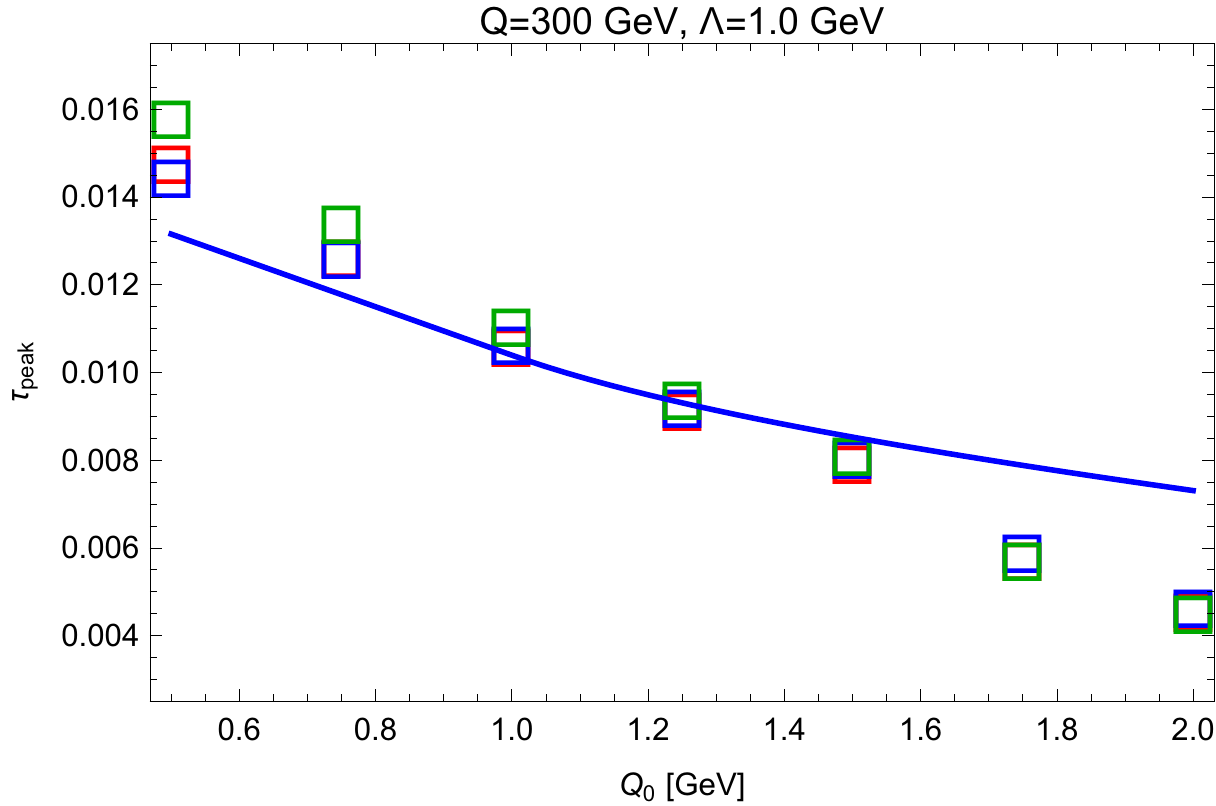}
		\subcaption{}
	\end{subfigure}
	\hfill
	\begin{subfigure}[c]{0.49\textwidth}
		\includegraphics[width=1.0\textwidth]{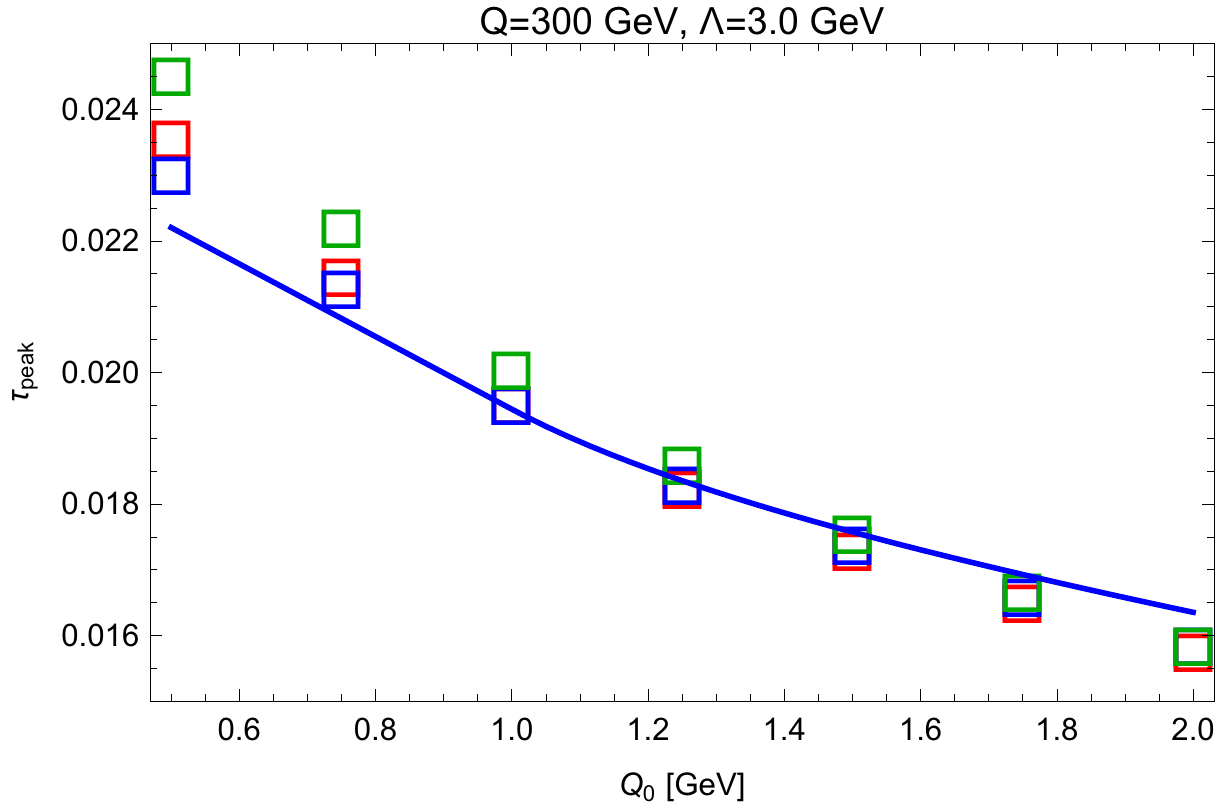}
		\subcaption{}
	\end{subfigure}
	
	\caption{\label{fig:masslesspeak1}
	Peak position $\tau_{\rm peak}$ at the parton level obtained from \Herwig\ 7 as a function of of the shower cut
	$Q_0$ and including a smearing with $\Lambda=1$~GeV (left panels) and $\Lambda=3$~GeV (right panels) for massless quarks and $Q=91$~GeV (upper panels) and $Q=300$~GeV (lower panels). Displayed are the results
	from the full simulation (red squares), 
	with gluon splitting turned off, but angular ordering turned on (blue squares)
	and with gluon splitting and angular ordering both turned off (green squares). The blue solid line 
	is the analytic prediction of Eq.~(\ref{eq:tauQ0dependencemassless}) 
	taking the \Herwig\ 7 result for $Q_0^\prime=1.25$~GeV as the reference.}
\end{figure}

\begin{figure}[t]
	\center
	\begin{subfigure}[c]{0.49\textwidth}
		\includegraphics[width=1.0\textwidth]{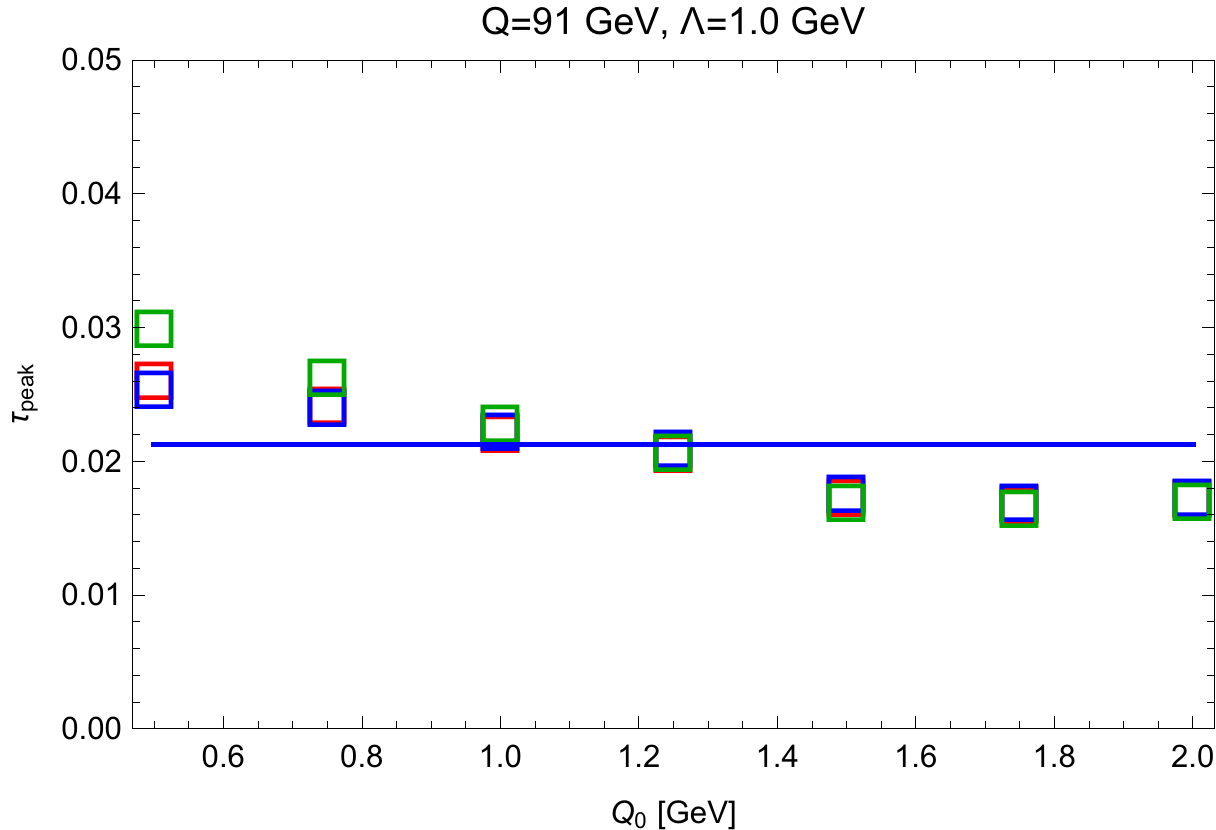}
		\subcaption{}
	\end{subfigure}
	\hfill
	\begin{subfigure}[c]{0.49\textwidth}
		\includegraphics[width=1.0\textwidth]{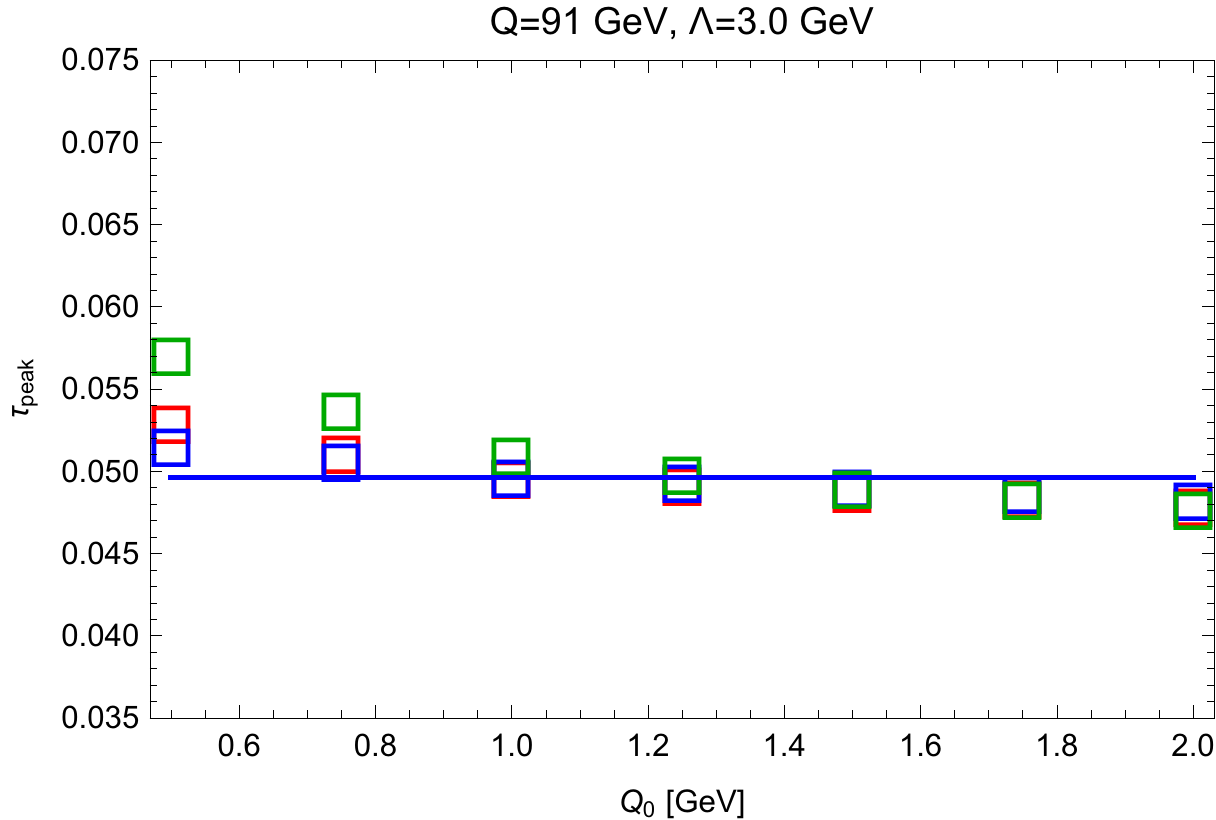}
		\subcaption{}
	\end{subfigure}
	\begin{subfigure}[c]{0.49\textwidth}
		\includegraphics[width=1.0\textwidth]{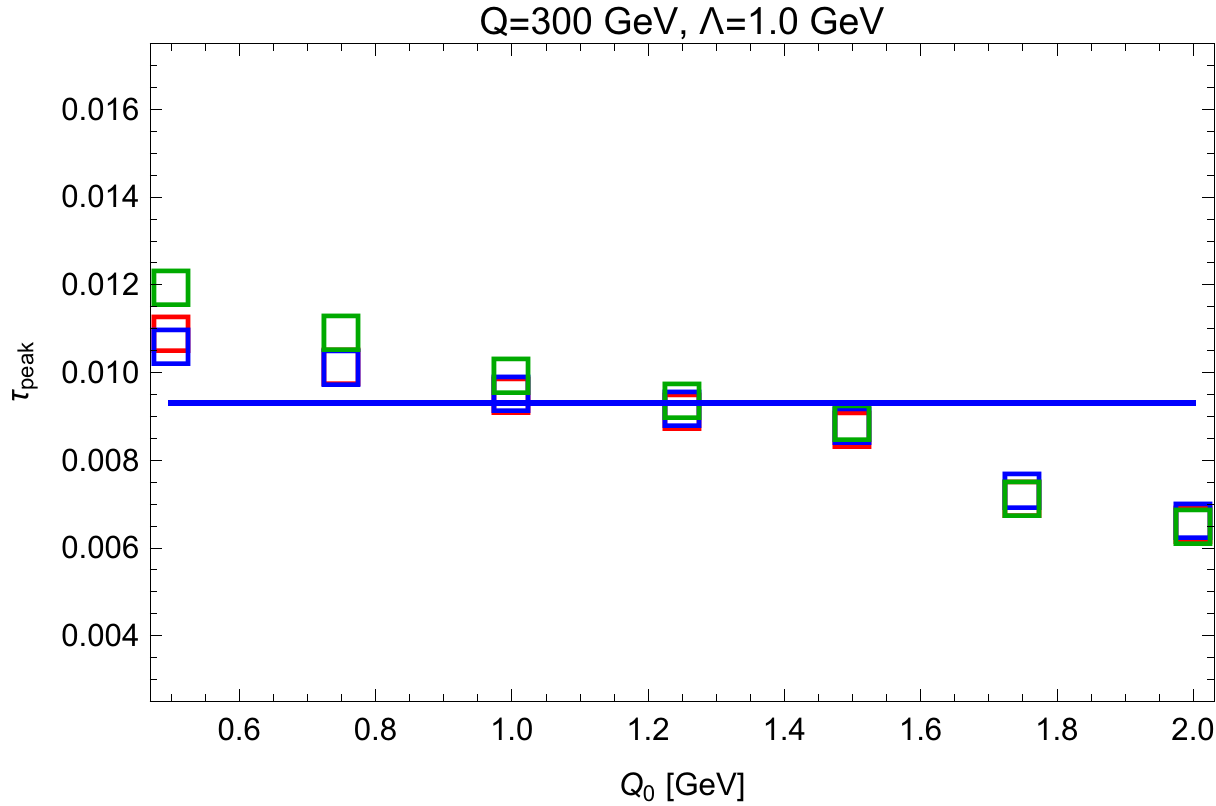}
		\subcaption{}
	\end{subfigure}
	\hfill
	\begin{subfigure}[c]{0.49\textwidth}
		\includegraphics[width=1.0\textwidth]{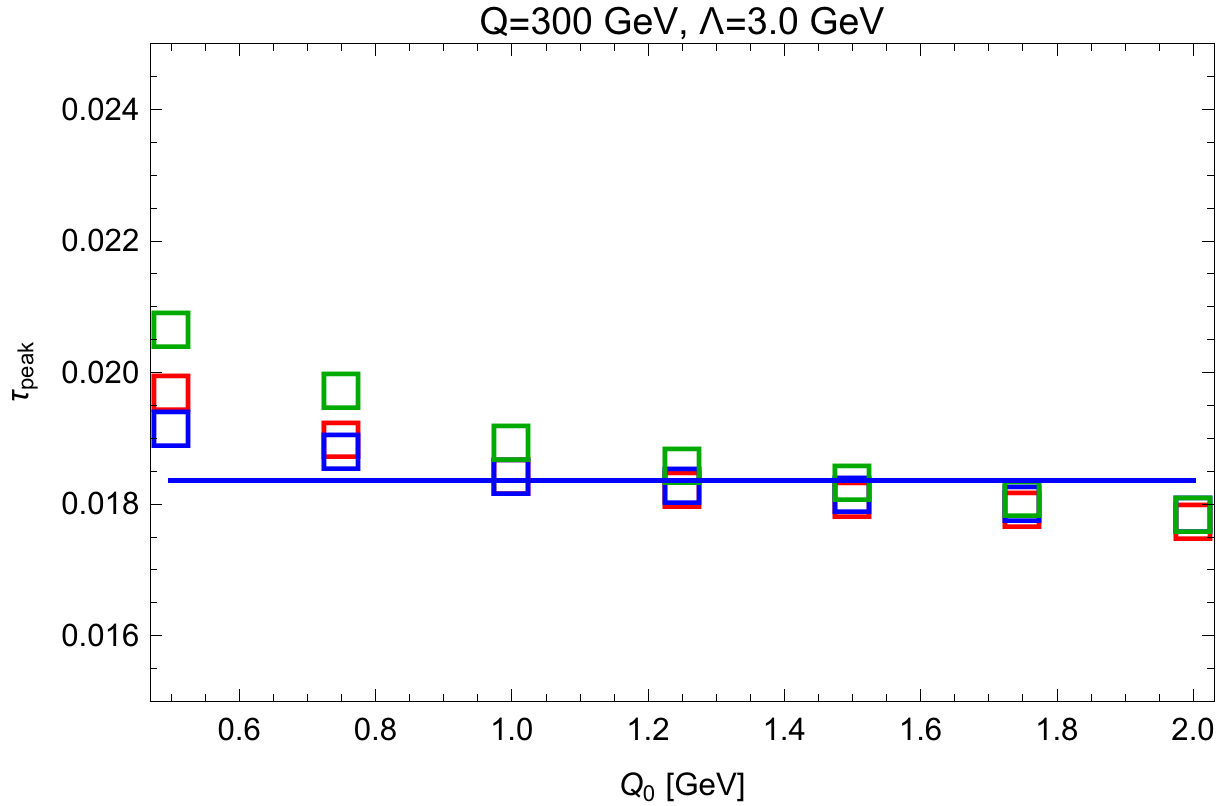}
		\subcaption{}
	\end{subfigure}
	
	\caption{\label{fig:masslesspeak2}
		Peak position $\tau_{\rm peak}$ at the parton level obtained from \Herwig\ 7 for the parameters used 
		in Fig.~\ref{fig:masslesspeak1}, but including the soft function gap calculated analytically to achieve
		results that eliminate the linear dependence on the shower cut $Q_0$ taking $Q_0^\prime=1.25$~GeV as the reference.
		The blue solid line represents the corresponding analytic prediction.}
\end{figure}

In Fig.~\ref{fig:masslesspeak1} the peak position $\tau_{\rm peak}$ obtained from \Herwig\ 7 is shown as a function of
the shower cut $Q_0$ for $Q=91$~GeV (upper panels)
and $Q=300$~GeV (lower panels) for the smearing parameter $\Lambda_{\rm m}=1$~GeV (left panels) and 
$\Lambda_{\rm m}=3$~GeV (right panels). The (center of the) colored squares show the corresponding results 
from the full simulation, i.e.\ with gluon splitting and angular ordering both turned on (red squares), 
with gluon splitting turned off, but angular ordering turned on (blue squares)
and with gluon splitting and angular ordering both turned off (green squares). The solid blue line
represents the analytic prediction of Eq.~(\ref{eq:tauQ0dependencemassless}) with 
$Q_0^\prime=1.25$~GeV as the reference peak position taken form the \Herwig\ 7 simulation
and using the strong coupling employed by the \Herwig\ 7 parton shower to calculate 
$\tau_{\rm peak}$ for $Q_0$ different from $Q_0^\prime$. We have shown the results for 
shower cut values in the range between $(0.5~\mbox{GeV})<Q_0<(2.0~\mbox{GeV})$ even though the 
perturbative treatment is expected to break down for scales below $1$~GeV. Nevertheless,
\Herwig\ 7 can carry out simulations for values of $Q_0$ even below $0.5$~GeV since for scales 
below $1$~GeV the strong coupling used in its parton shower is frozen to the value at $1$~GeV.  
The choice of $Q_0$ in the simulations is in practice limited from below only by computation time since the parton mulplicities
strongly increase for decreasing shower cut. For theoretical considerations, however, only shower cut values
of $1$~GeV and larger can be considered seriously, because $Q_0$ conceptually represents a factorization scale and should be located well
within the regime of perturbation theory. In fact, indications of a breakdown of the perturbative description for $Q_0<1$~GeV are visible in Figs.~\ref{fig:masslesspeak1} (and also in Figs.~\ref{fig:masslesspeak2} and the corresponding results 
for top quarks in the following subsection) as the 
increased spread in the \Herwig\ 7 results arising from 
the different choices concerning gluon branching and angular ordering.

Overall we see quite good agreement between the \Herwig\ 7 simulations and 
the analytic prediction for $\tau_{\rm peak}$ for $\Lambda_{\rm m}=1$~GeV
and excellent agreement for $\Lambda_{\rm m}=3$~GeV. While $\Lambda_{\rm m}=1$~GeV corresponds to the actual size of hadronization effects compatible with experimental data, the choice $\Lambda_{\rm m}=3$~GeV is motivated by the possible size of additional experimental resolution effects.
That we find a much better agreement for larger smearing scale is related to the fact that the
evolution equation~(\ref{eq:tauQ0dependencemassless}) only accounts for the dominant linear dependence on $Q_0$
which was in our analytic calculations in Secs.~\ref{sec:CBunreleased}, \ref{sec:QCDunreleasedmassless} and
\ref{sec:QCDunreleasedmassive} derived by employing the multipole expansion for the contributions of the
unreleased radiation, i.e.\ the radiation originating from below the shower cut. This expansion is formally an expansion
in $Q_0/\Lambda_{\rm m}$, and we see from the agreement between simulation and analytic prediction in
Fig.~\ref{fig:masslesspeak1} that this expansion works already well for $Q_0\sim\Lambda_{\rm m}$ and even better for $\Lambda_{\rm m}>Q_0$. 
Since in realistic simulations and actual experimental
measurements there are additional resolution effects that always lead to a smearing scale that is 
effectively larger than $1$~GeV, we can conclude that the linear dependence on the shower cut $Q_0$ 
expressed by Eq.~(\ref{eq:tauQ0dependencemassless})
represents the dominant effects in all cases and that effects quadratic in $Q_0$ or of even higher power
are small in practice.

At this point it is also illustrative to explicitly show the quality of relation~(\ref{eq:thrustmassless4}) which states that
the observed peak position can be rendered shower cut independent, if the gap of the soft model function used in the
convolution is modified as described in Eq.~(\ref{eq:Smodmodified}). In Fig.~\ref{fig:masslesspeak2} the thrust peak positions obtained from the
\Herwig\ 7 simulations already shown in Fig.~\ref{fig:masslesspeak1} are displayed once again as a function of $Q_0$, however, with 
the corresponding modification of the soft function gap for the reference shower cut value $Q_0^\prime=1.25$~GeV.
As expected, we see that the shower cut dependence is substantially reduced for the smearing scale $\Lambda_{\rm m}=1$~GeV
and almost vanishes for the smearing scale $\Lambda_{\rm m}=3$~GeV in the region $Q_0\ge 1$~GeV, i.e.\ 
where perturbation theory can be employed.

\subsection{Thrust peak position for top quarks}
\label{sec:herwigmassive}

In this section we finally confront our analytic prediction for the $Q_0$ dependence of the 
rescaled thrust peak position for top quarks
\begin{align}
\label{eq:MtauQ0dependencetop}
M_{\tau,\,{\rm peak}}(Q_0) \, = \, M_{\tau,\,{\rm peak}}(Q_0^\prime) -
\Bigl[8 \frac{Q}{m_t}-4 \pi\Bigr] \,\int\limits_{Q_0^\prime}^{Q_0}\mathrm{d}R\,
\frac{C_F\, \alpha_s(R)}{4\pi}\,.
\end{align}
with simulations in \Herwig\ 7. We again used the specific settings discussed in 
Sec.~\ref{sec:herwigsimulationsetup} and generated $10^{9}$ events for a given
c.m.\ energy $Q$ and shower cut $Q_0$. For the convolution with the
soft model shape function $S_{\rm mod}$ given in Eq.~(\ref{eq:smoddef}) 
we employed the discretized version 
of Eq.~(\ref{eq:Mtau1}) with $\Lambda_{\rm m}=\Lambda+4m_t\Gamma_t/Q$ with 
$\Gamma_t=1.5$~GeV for the soft function smearing parameter. It effectively accounts for the  
additional smearing caused by the top quark width. 
Since the resonance region in $\tau$ is substantially more 
narrow compared to the massless case we used a bin size that corresponds to $\Delta\tau =8\times 10^{-6}$
and used the same method to determine the peak position as for the massless quark case.

\begin{figure}
	\center
	\begin{subfigure}[c]{0.49\textwidth}
		\includegraphics[width=1.0\textwidth]{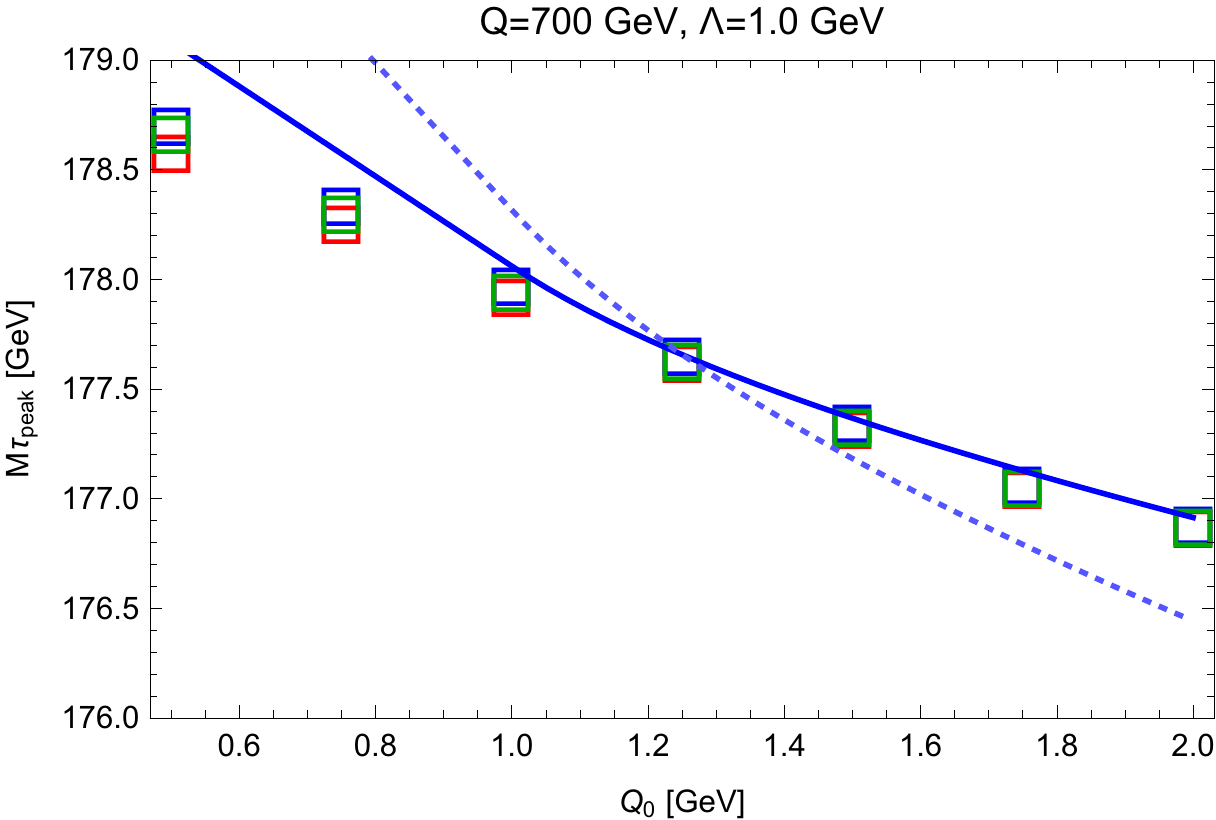}
		\subcaption{}
	\end{subfigure}
	\hfill
	\begin{subfigure}[c]{0.49\textwidth}
		\includegraphics[width=1.0\textwidth]{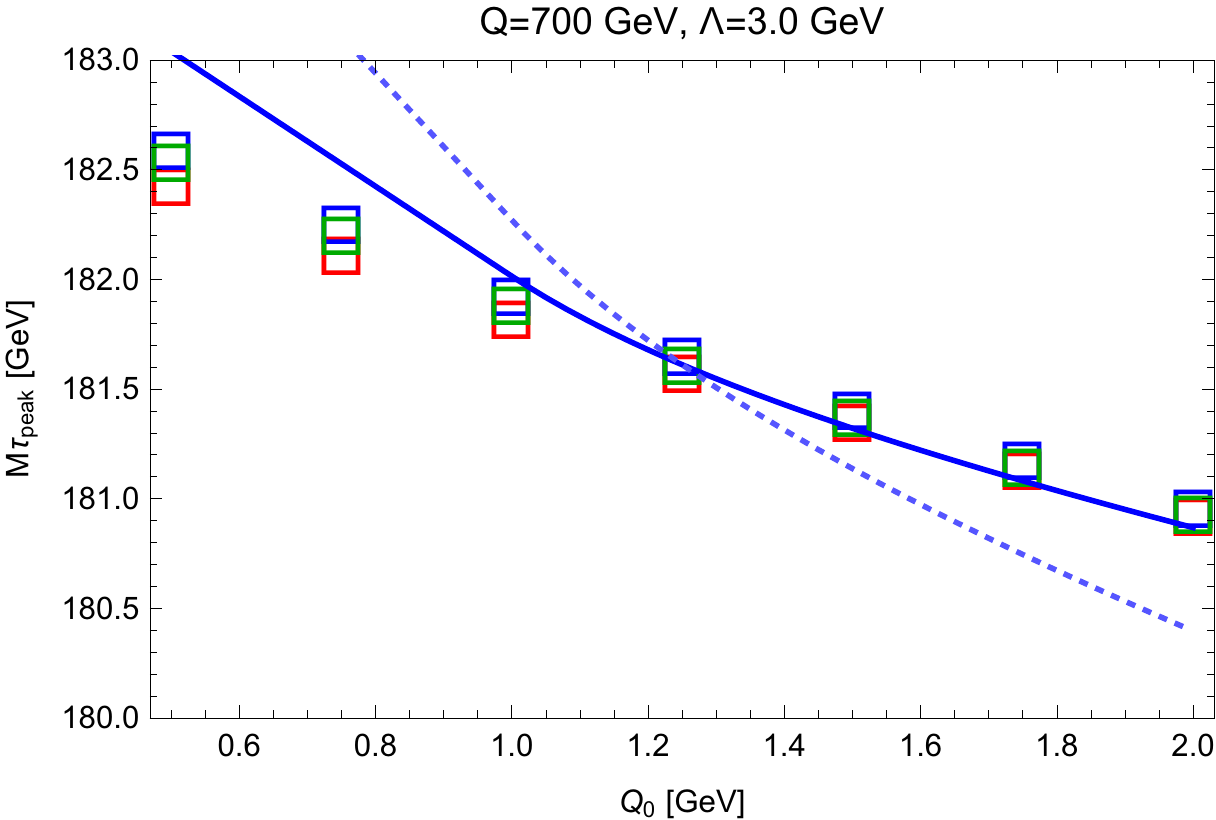}
		\subcaption{}
	\end{subfigure}
	\begin{subfigure}[c]{0.49\textwidth}
		\includegraphics[width=1.0\textwidth]{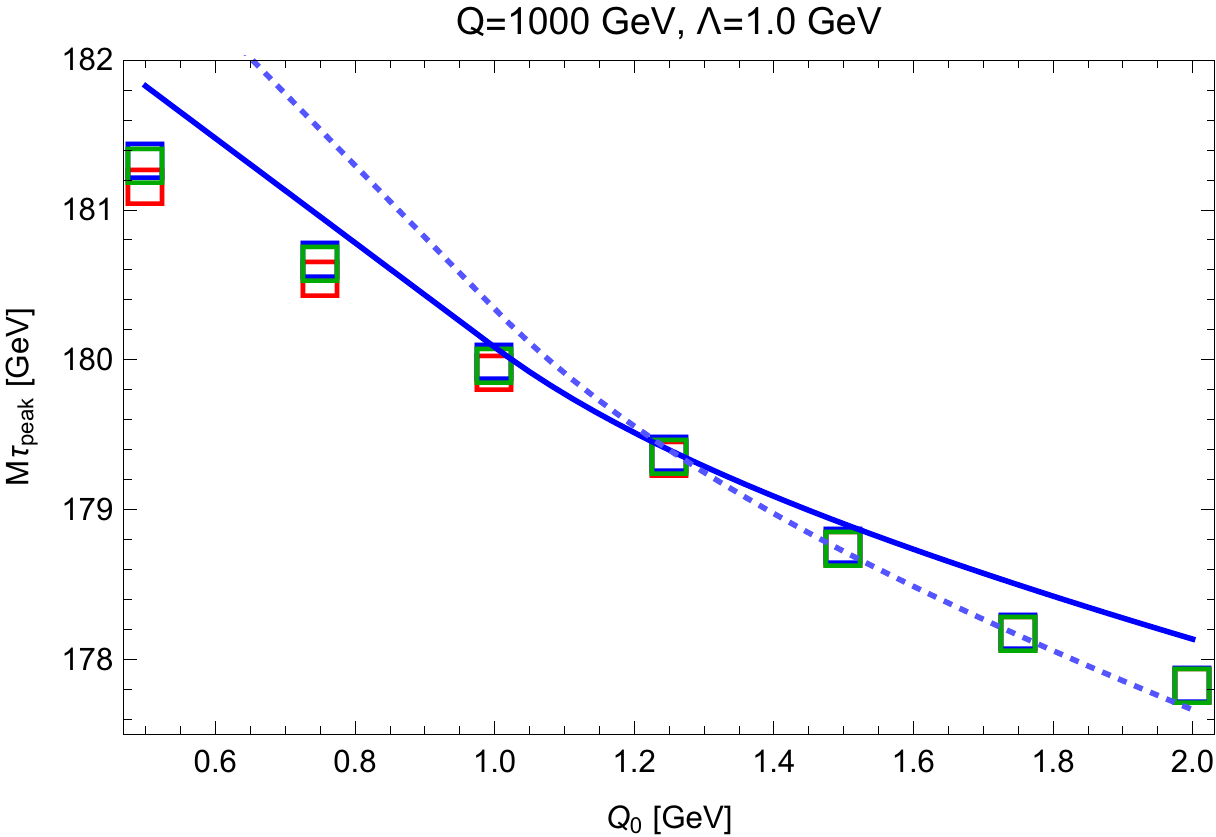}
		\subcaption{}
	\end{subfigure}
	\hfill
	\begin{subfigure}[c]{0.49\textwidth}
		\includegraphics[width=1.0\textwidth]{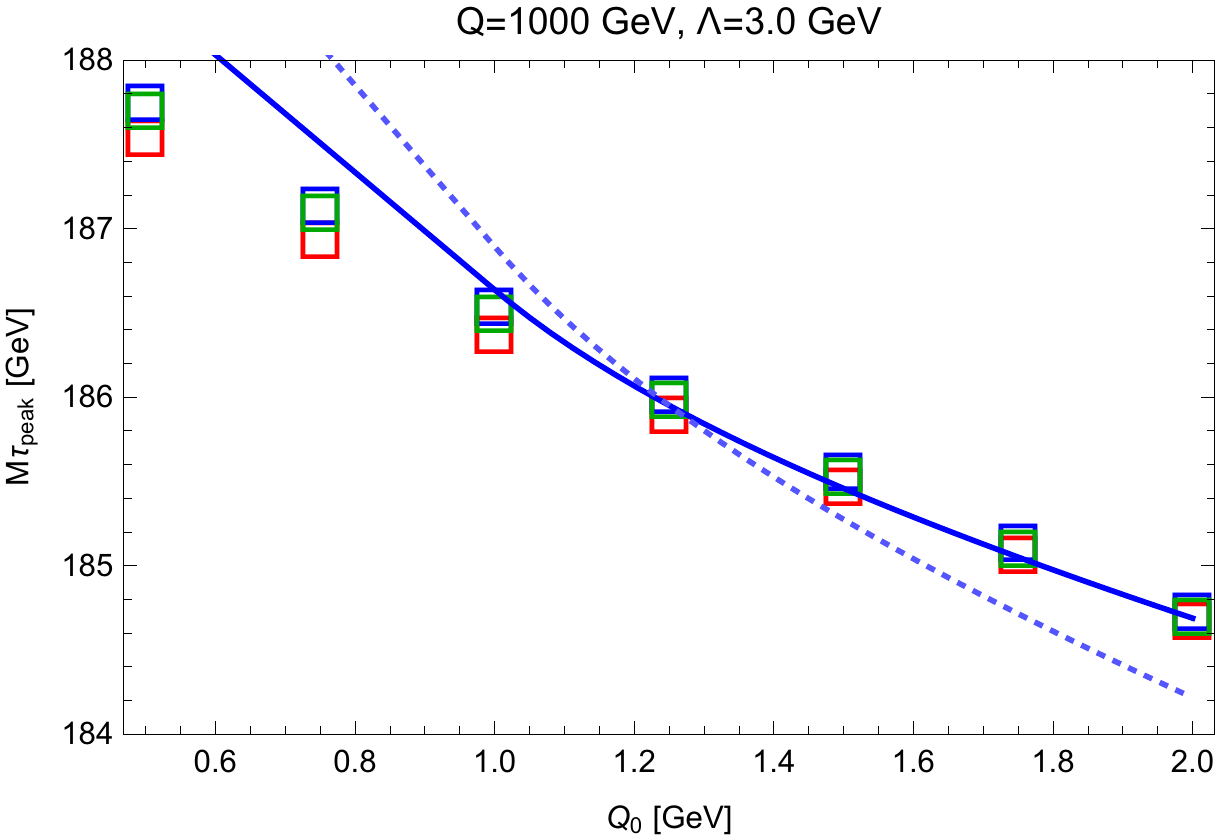}
		\subcaption{}
	\end{subfigure}
	
	\caption{\label{fig:Mtau1}
		Peak position $M_{\tau,{\rm peak}}$ at the parton level obtained from \Herwig\ 7 for the top quark generator mass
		$m_t=173$~GeV as a function of the shower cut
		$Q_0$ for $Q=700$~GeV (upper panels) and $Q=1$~TeV (lower panels) for smearing $\Lambda=1$~GeV (left panels)
		and $\Lambda=3$~GeV (right panels). Displayed are the results
		from the full simulation (red squares), 
		with gluon splitting turned off, but angular ordering turned on (blue squares)
		and with gluon splitting and angular ordering both turned off (green squares). The blue solid line 
		is the analytic prediction of Eq.~(\ref{eq:MtauQ0dependencetop}) 
		taking the \Herwig\ 7 result for $Q_0^\prime=1.25$~GeV as the reference. The dashed blue line 
		is the analytic prediction of Eq.~(\ref{eq:MtauQ0dependencetop}), but only accounting for the large angle
		soft radiation contributions which are multiplied with the $Q/m_t$ factor.}
\end{figure}

In Fig.~\ref{fig:Mtau1} the peak position $M_{\tau,\,{\rm peak}}$ obtained from \Herwig\ 7 with the top quark generator mass
$m_t=173$~GeV is shown as a function of
the shower cut $Q_0$ for $Q=700$~GeV (upper panel)
and $Q=1$~TeV (lower panel) for the smearing parameter $\Lambda=1$~GeV (left column) and 
$\Lambda=3$~GeV (right column). The (center of the) colored squares show the corresponding results 
from the full simulation, i.e.\ with gluon splitting and angular ordering both turned on (red squares), 
with gluon splitting turned off, but angular ordering turned on (blue squares)
and with gluon splitting and angular ordering both turned off (green squares). The solid blue line
represents the analytic prediction of Eq.~(\ref{eq:MtauQ0dependencetop}) with 
$Q_0^\prime=1.25$~GeV as the reference peak position taken form the \Herwig\ 7 simulation
and using the strong coupling employed by \Herwig\ 7 parton shower to calculate 
$M_{\tau,\,{\rm peak}}$ for $Q_0$ different from $Q_0^\prime$. The dashed blue line represents the analytic prediction of Eq.~(\ref{eq:MtauQ0dependencetop}), but only accounting for the large angle
soft radiation contributions which are multiplied with the $Q/m_t$ factor, in order to visualize
the size of the $Q_0$ dependence coming from the ultra-collinear radiation that affects the interpretation of the
mass scheme alone. As in the massless quark case
we have shown the results for 
shower cut values in the range between $(0.5~\mbox{GeV})<Q_0<(2.0~\mbox{GeV})$ and remind the reader that
the results for $Q_0$ below $1$~GeV are only shown for illustration, as already explained 
in Sec.~\ref{sec:herwigmassless}.

We observe that the agreement between the \Herwig\ 7 simulations and the analytic prediction 
for $M_{\tau,\,{\rm peak}}$ is very good for $\Lambda_{\rm m}=1$~GeV as well as for $\Lambda_{\rm m}=3$~GeV.
This shows that for the top quark case, where the width provides an additional irreducible smearing effect,
the linear dependence on the shower cut $Q_0$, which we have determined in our analytic calculations, 
fully captures the complete $Q_0$ dependence and that contributions proportional to higher powers of $Q_0$ are 
negligible for all practical purposes. 
\begin{figure}[t]
	\center
	\begin{subfigure}[c]{0.49\textwidth}
		\includegraphics[width=1.0\textwidth]{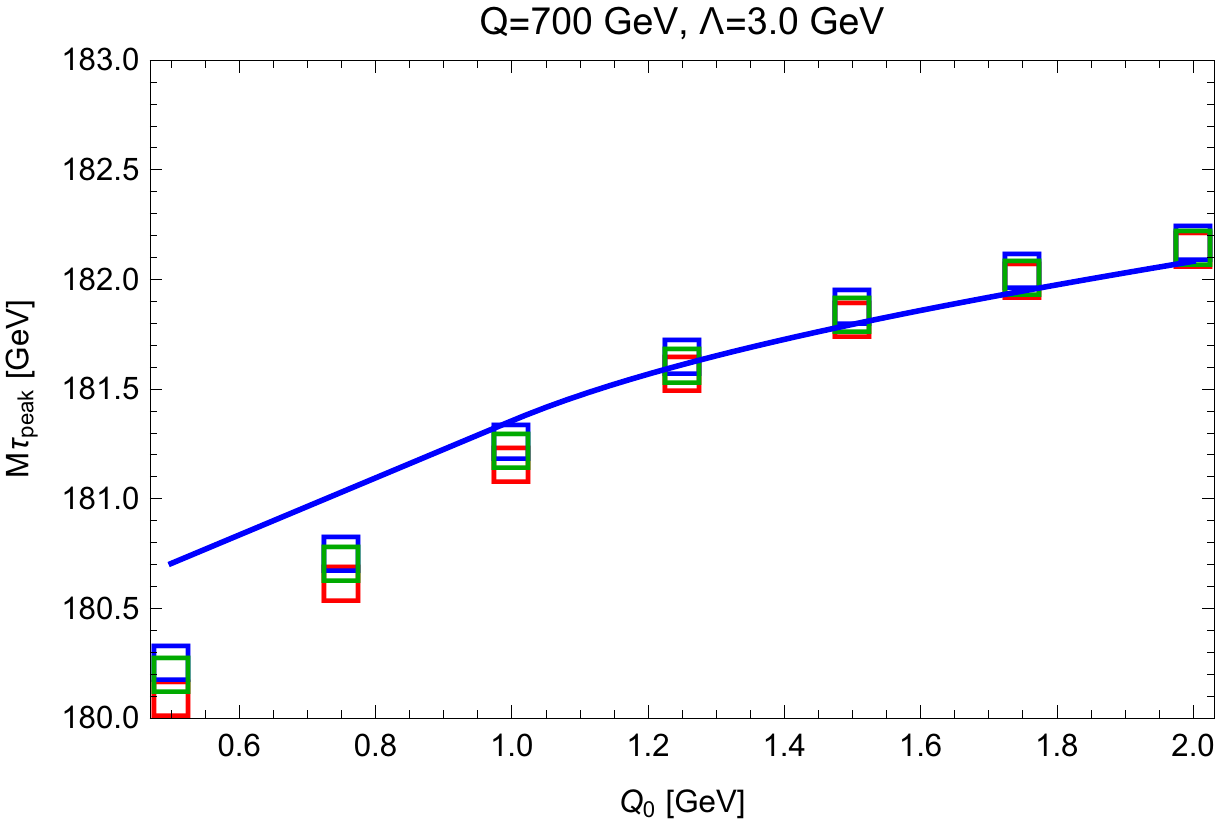}
		\subcaption{}
	\end{subfigure}
	\hfill
	\begin{subfigure}[c]{0.49\textwidth}
		\includegraphics[width=1.0\textwidth]{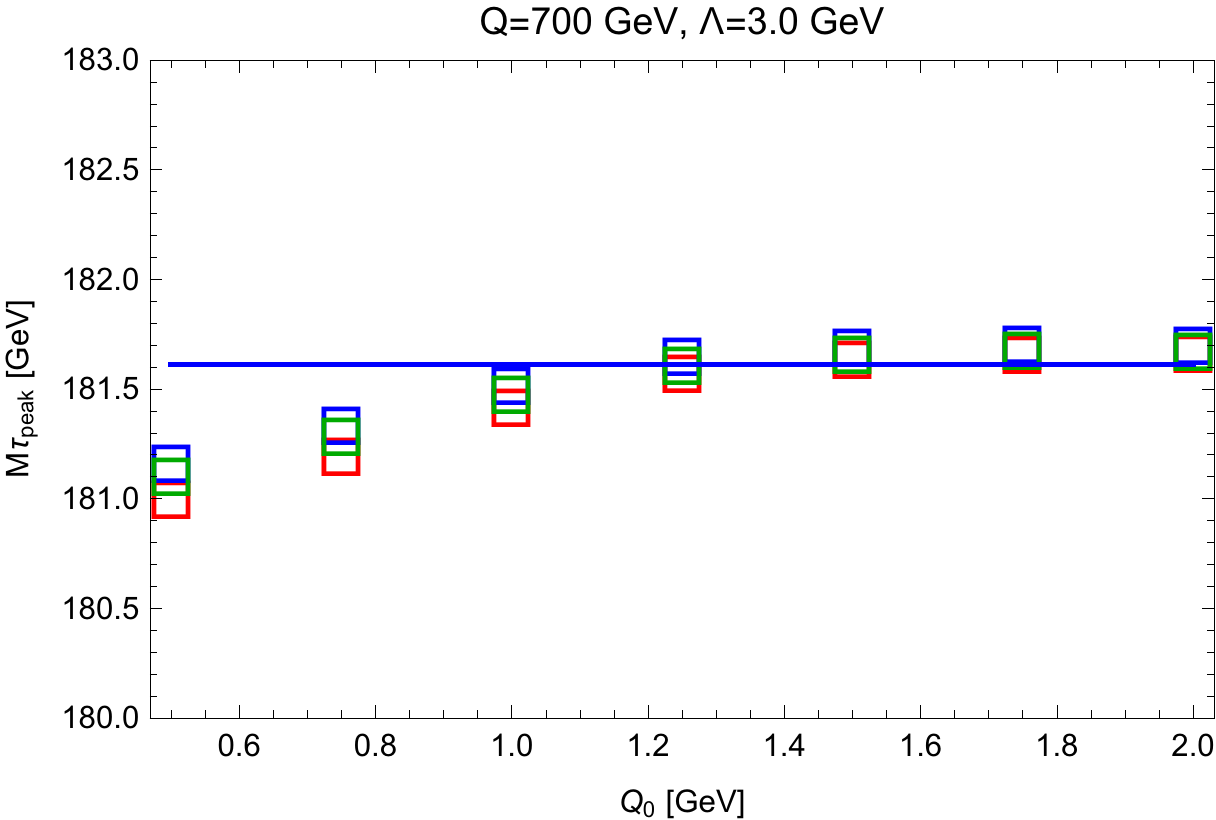}
		\subcaption{}
	\end{subfigure}
	\begin{subfigure}[c]{0.49\textwidth}
		\includegraphics[width=1.0\textwidth]{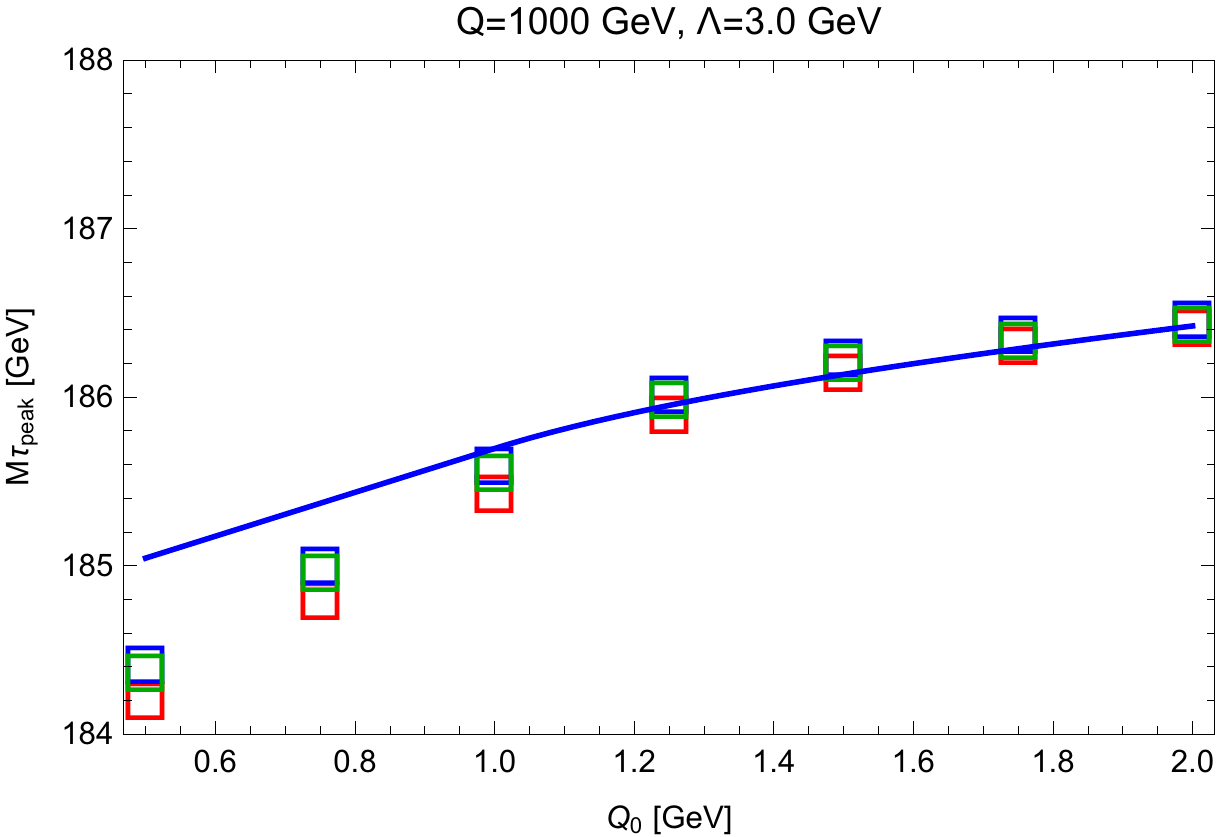}
		\subcaption{}
	\end{subfigure}
	\hfill
	\begin{subfigure}[c]{0.49\textwidth}
		\includegraphics[width=1.0\textwidth]{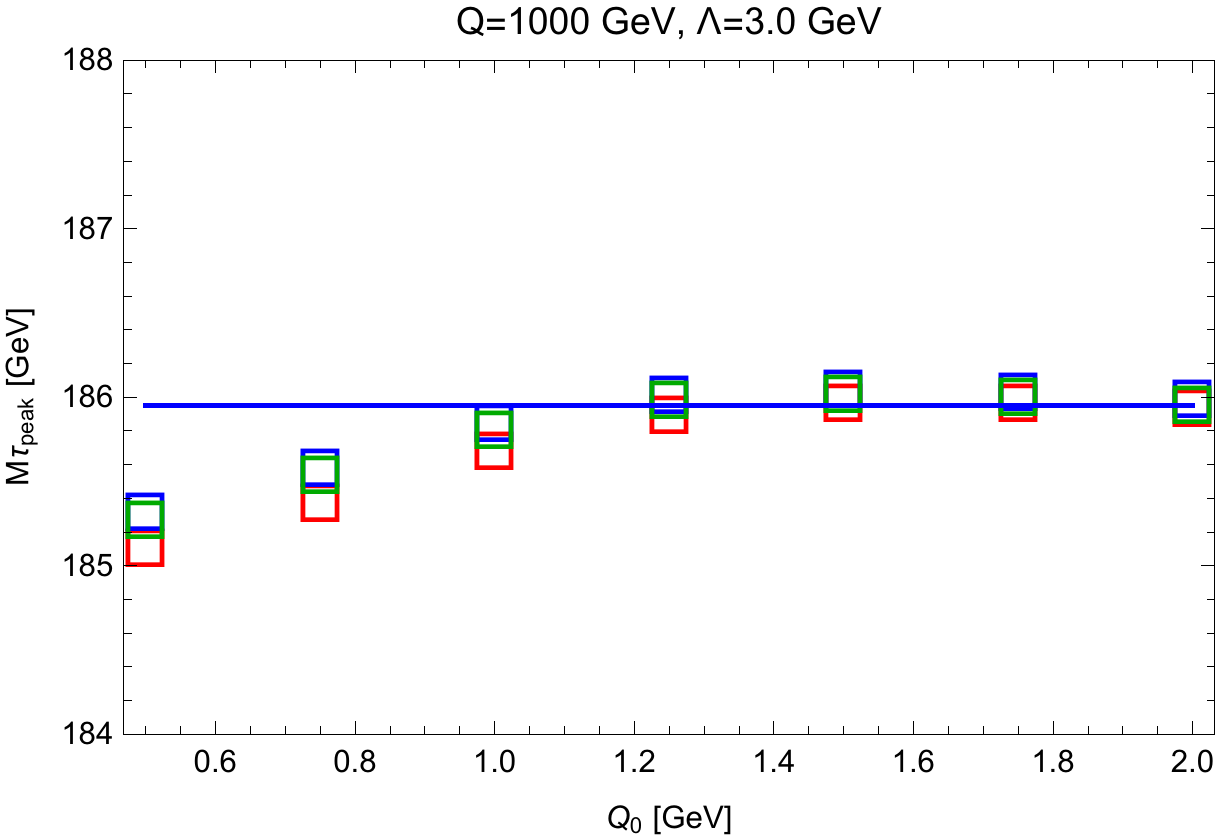}
		\subcaption{}
	\end{subfigure}
	
	\caption{\label{fig:Mtau2}
		Peak position $M_{\tau,{\rm peak}}$ at the parton level obtained from \Herwig\ 7
        for the top quark generator mass
		$m_t=173$~GeV as a function of the shower cut
		$Q_0$ for $Q=700$~GeV (upper panels) and $Q=1$~TeV (lower panels) for smearing with $\Lambda=3$~GeV.
		Left panels: In addition to the results shown 
		in Fig.~\ref{fig:Mtau1} we have included the soft function gap calculated analytically to remove
		the shower cut dependence due to the large angle soft radiation. 
		Right panels: In addition to the results of the left panels 
		we have set the generator mass to $m^{\rm CB}(Q_0)$ such that the peak position becomes
		independent of the shower cut $Q_0$. 
		The blue solid line represents the corresponding analytic prediction for the remaining cutoff scale
		dependence. For all results we used $Q_0^\prime=1.25$~GeV as the reference scale.}
\end{figure}

It is now illustrative to explicitly demonstrate that the observed peak position can be rendered shower cut independent, if - taking $Q_0^\prime=1.25$~GeV as the reference - the gap of the soft model function used in the
convolution is modified according to Eqs.~(\ref{eq:Deltasoftv3}) and \eqref{eq:Smodmodified} and if the generator mass $m_t$ is modified by
\begin{equation}
 \label{eq:generatormass}
 m_t\to m_t+m_t^{\rm CB}(Q_0)-m_t^{\rm CB}(Q_0^\prime)
\end{equation}
according to Eq.~(\ref{eq:DeltaCBmass}).
In Fig.~\ref{fig:Mtau2} the rescaled thrust peak positions obtained from the
\Herwig\ 7 simulations for $Q=700$~GeV (upper panel) and $Q=1$~TeV (lower panel) 
are displayed once again as a function of $Q_0$ for $\Lambda=3$~GeV. In the left panels we have 
in addition to the corresponding curves shown in Fig.~\ref{fig:Mtau1} included  
the corresponding modification of the soft function gap for the reference shower cut value $Q_0^\prime=1.25$~GeV.
This removes the shower cut dependence coming from the large angle soft radiation such that the remaining
$Q_0$ dependence explicitly illustrates the shower cut dependence of the generator mass alone.\footnote{The
rescaled thrust variable $M_\tau$ defined in Eq.~(\ref{eq:Mtaudef}) is normalized such that the $Q_0$ slope 
shown in the left panels of Fig.~\ref{fig:Mtau2} is minus twice the one of the generator mass.} 
Compared to the results shown in Fig.~\ref{fig:Mtau1} we see that the slope in $Q_0$ has an opposite sign
which means that the $Q_0$ dependent CB mass scheme that has to be employed to keep the physical prediction
unchanged is decreasing with $Q_0$ as expressed by the renormalization group equation~(\ref{eq:CBmassRRGE}).  
In the right panels we have then also modified, in addition to the figures in the left column,
the generator mass according to Eq.~\eqref{eq:generatormass} and taking $m^{\rm CB}(Q_0^\prime=1.25\,\text{GeV})=173$~GeV 
as the reference top quark mass. We see that once both modifications are implemented, the shower cut dependence has 
essentially disappeared.

\begin{figure}
	\center
	\begin{subfigure}[c]{0.49\textwidth}
		\includegraphics[width=1.0\textwidth]{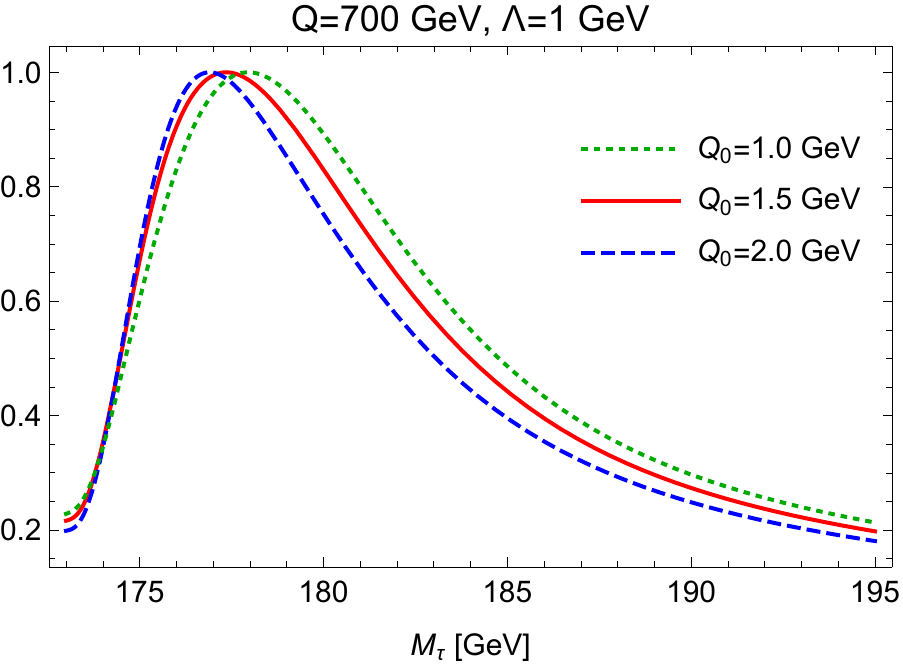}
		\subcaption{}
	\end{subfigure}
	\hfill
	\begin{subfigure}[c]{0.49\textwidth}
		\includegraphics[width=1.0\textwidth]{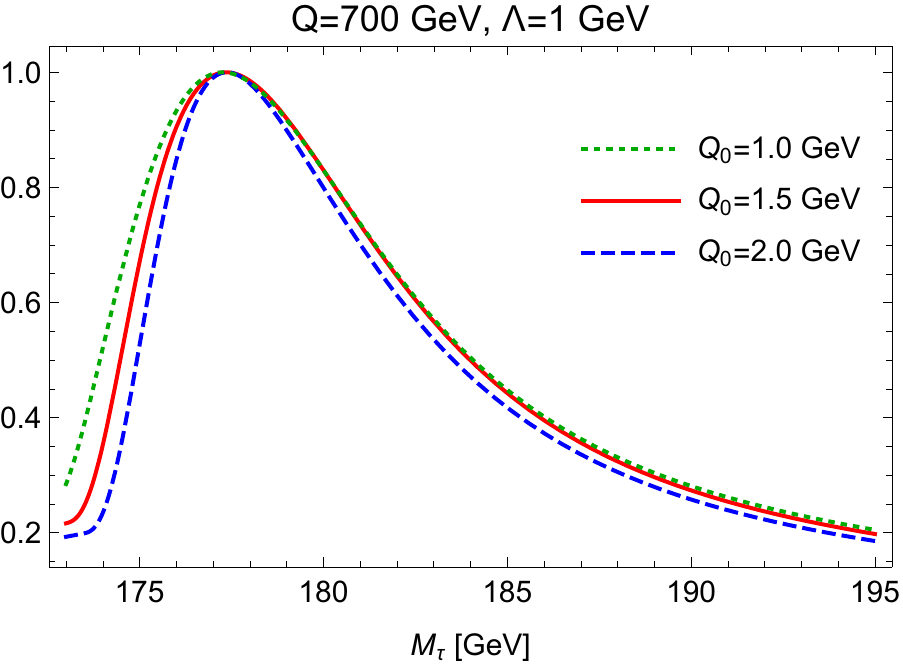}
		\subcaption{}
	\end{subfigure}
	\begin{subfigure}[c]{0.49\textwidth}
		\includegraphics[width=1.0\textwidth]{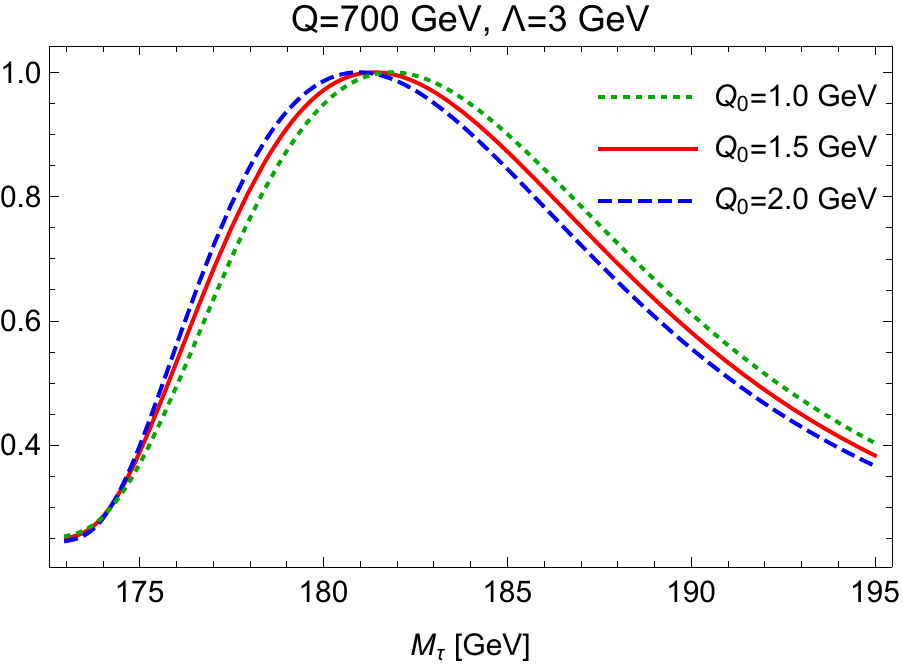}
		\subcaption{}
	\end{subfigure}
	\hfill
	\begin{subfigure}[c]{0.49\textwidth}
		\includegraphics[width=1.0\textwidth]{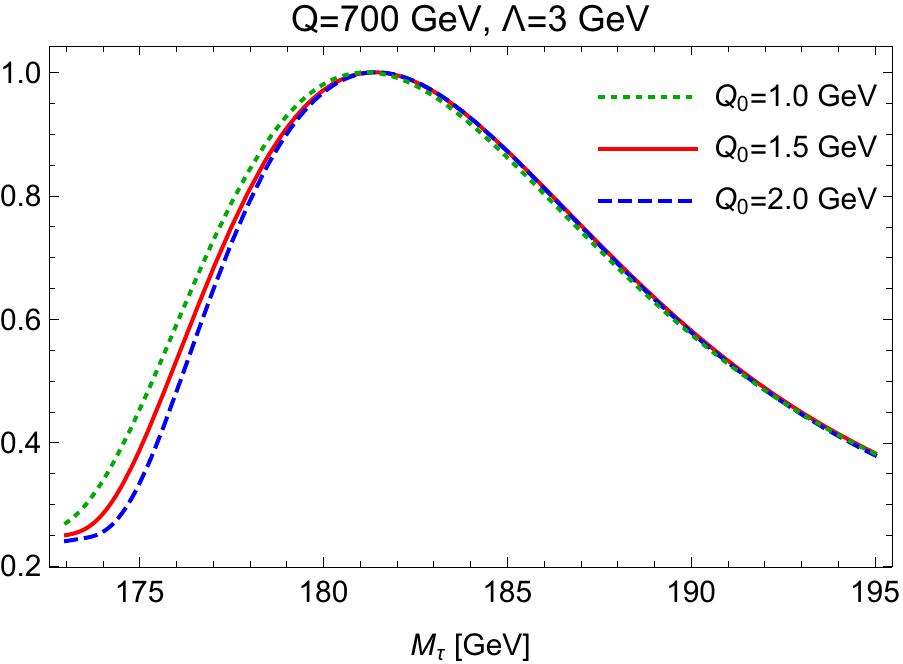}
		\subcaption{}
	\end{subfigure}
	\caption{\label{fig:Mtau3} Parton level rescaled thrust distribution in the peak region obtained from
		\Herwig\ 7 full simulations for $Q=700$~GeV and smearing with $\Lambda=1$~GeV (upper panels) and $\Lambda=3$~GeV
		(lower panels) for shower cut values $Q_0=1$~GeV (dotted green curves), $Q_0=1.5$~GeV (solid red curves) 
		and $Q_0=2$~GeV (dashed blue curves). Left panels: Simulations with generator input mass $m_t=173$~GeV and
		using the same soft model shape function for all shower cut values.
		Right panels: Same distributions, but using a $Q_0$-dependent soft function gap to eliminate the shower cut
		dependence due to large angle soft radiation and using $m^{\rm CB}(Q_0=1.0)=173.22$~GeV (green), 
		$m^{\rm CB}(Q_0=1.5)=173$~GeV (red) and $m^{\rm CB}(Q_0=2)=172.86$~GeV (blue) as the generator masses,
		according to Eq.~(\ref{eq:DeltaCBmass}),
		to render the peak location independent of the shower cut $Q_0$.
		}
\end{figure}

It is interesting to also analyze to which extent the shower cut dependent modifications of the soft function gap and the generator mass we have just discussed for the 
thrust peak position also holds for the whole distribution function in the resonance region. 
This is shown in Figs.~\ref{fig:Mtau3} were the rescaled thrust distributions in the peak region are shown for
$Q=700$~GeV for $Q_0=1$~GeV (dotted green curves), $Q_0=1.5$~GeV (solid red curves) and $Q_0=2$~GeV (dashed blue curves)
obtained from the full simulation.
The left panels show the distributions in the peak region for fixed generators mass $m_t=173$~GeV with smearing parameter $\Lambda=1$~GeV (upper left panel) and $\Lambda=3$~GeV (lower left panel). 
The corresponding right panels show, using $Q_0^\prime=1.5$~GeV as the reference scale, the distributions including the $Q_0$ dependent soft function gap according to Eq~\eqref{eq:Smodmodified} and the $Q_0$ dependent generator mass  according to Eq.~\eqref{eq:generatormass} to keep the peak position cutoff independent.
We see that the resonance distribution tends to be narrower for increasing cutoff $Q_0$, but that this effect is weaker for a
larger smearing. This behavior can be explained from the fact that for increasing cutoff $Q_0$ the no-branching probability (which describes production stage multiplicity $n=0$ events and contributes to the coefficient 
of the tree-level $\delta$-function located at the partonic threshold) is becoming bigger and, correspondingly, the weight of events with branching (which correspond to production stage multiplicities $n>0$ and lead to jet masses above the partonic threshold) is becoming smaller. For a larger smearing this width effects is washed out and therefore less pronounced for $\Lambda=3$~GeV. Thus depending on the size of the experimental resolution the effects that a variation of the
shower cut $Q_0$ has on the whole peak distribution may be more complicated than a simple modification of the soft function gap and the 
generator mass. Since the contributions from ultra-collinear radiation in this context are $m_t/Q$-suppressed, see Eq.~(\ref{eq:MtauQ0dependencetop}), these width effects mostly originate from large angle soft radiation.
One can therefore conclude that these effects may
be properly taken into account during the retuning procedure which has to be carried out upon a change of the shower cut $Q_0$ in MC event generators used for experimental analyses,
and which is substantially more involved than an a simple modification of the soft function gap. 

\subsection{Reconstructed observables and universality}
\label{sec:herwigotherobservables}

After our analysis of the $Q_0$ shower cut dependence of the
MC generator top mass for angular ordered parton showers 
using the thrust distribution in the resonance region there is one 
obvious question to be asked: 
Is our main conclusion concerning the equivalence   
the MC generator top quark mass and the shower cut dependent 
CB mass defined in Eq.~(\ref{eq:CBmassschemedef2}) only valid
for thrust (or very similar event shape type observables),  or is it universal? 
Clearly, examinations at the same level of depth as we carried out for thrust,
where we employed analytic calculations within the coherent branching formalism 
and the QCD factorization approach together with numerical MC 
simulations, will be
difficult for most other observables with strong kinematic quark mass dependence --
most notably because hadron level first principles and factorized predictions (which would allow directly for conclusions at the field theoretic level) are 
not available for them. The question of universality is also made difficult by the fact that the shower cut 
dependence not only affects the meaning of the generator mass for heavy quarks (or potentially 
other QCD parameters), but also modifies the description
of non-perturbative effects through its effects on large angle soft radiation (or other types of
long-range gluon effects), so that the issue may not be resolved completely restricting
the considerations only to partonic cross sections.  

At this point one may also have to define general criteria to prove universality systematically.
Although we hope to address this issue in forthcoming work, at this time such a systematic and
universal approach is lacking.  
However -- \emph{if universality applies} -- the dependence of MC parton level predictions on the shower 
cut $Q_0$, which was one of the main instrument of our thrust examinations, should be visible in 
a predictable, simple and universal way also for other observables and furthermore allow for 
non-trivial consistency checks. While consistency concerning the $Q_0$ dependence among thrust 
and other kinematic observables represents \emph{only a necessary condition} for claiming
universality, it still provides some evidence that universality indeed applies. Furthermore, 
computing the shower cut $Q_0$ dependence analytically for general observables and carrying out 
the corresponding MC simulations as a cross check is a relatively straightforward and easy task
and may even be testable in consistency checks confronting MC generators with experimental data or
in the context of pseudo-data analyses.  
In this section we therefore examine exemplarily two completely different observables with very strong kinematic top 
mass dependence and which are based on a jet clustering procedure acting on the full set of partons 
after production and decay stage parton showers have terminated. In this work we restrict our examinations to
a numerical analysis of the shower cut dependence of these observables, 
and we demonstrate that it can be easily predicted 
and interpreted. Interestingly, we find that the results are compatible with our examinations for the thrust distribution. A more coherent test of consistency in the context of pseudo-data analyses which 
specifically addresses the shower cut dependence of the generator mass shall be
addressed elsewhere. 

\begin{figure}
	\center
	\begin{subfigure}[c]{0.49\textwidth}
		\includegraphics[width=1.0\textwidth]{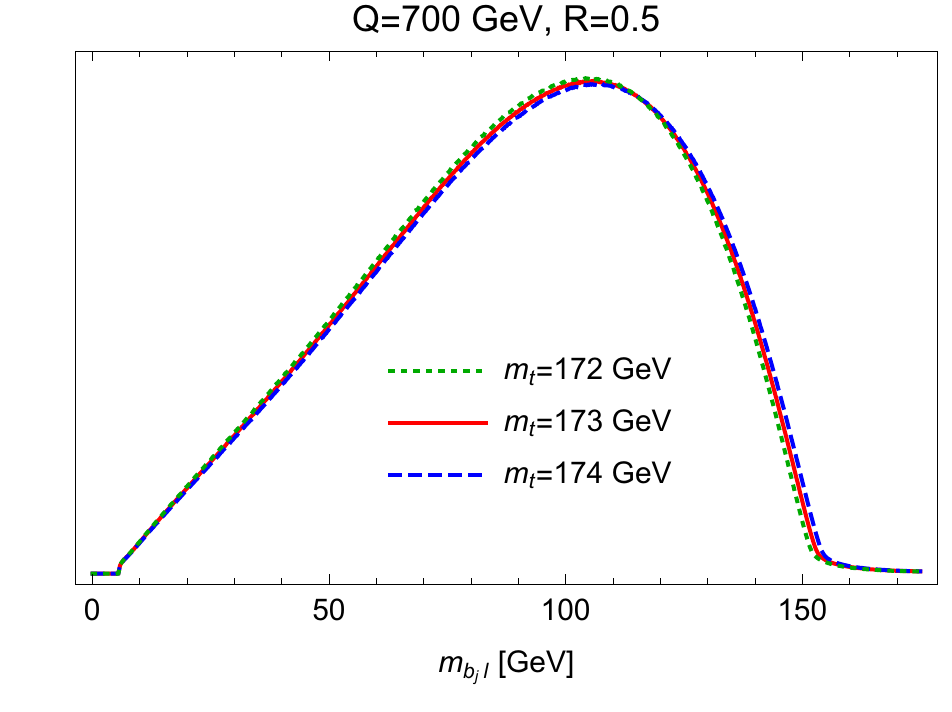}
		\subcaption{\label{fig:mbl}}
	\end{subfigure}
	\hfill
	\begin{subfigure}[c]{0.49\textwidth}
		\includegraphics[width=1.0\textwidth]{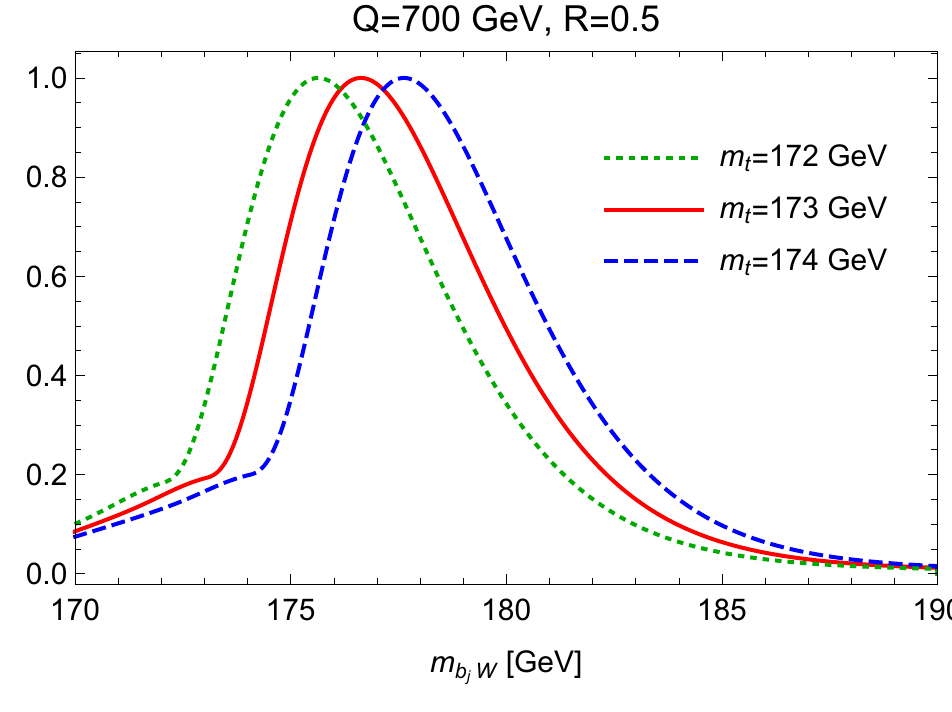}
		\subcaption{\label{fig:mbW}}
	\end{subfigure}
	\caption{\label{fig:mblmbW1}		
		The $m_{b_j\ell}$ (left panel) and $m_{b_jW}$ distributions (right panel) generated by \Herwig\ 7
		for top masses between $m_t=172$ (dotted green), $173$ (solid red) 
		$m_t=174$ GeV (dashed blue) for $Q=700$~GeV, jet radius $R=0.5$ and Cambridge-Aachen-type $b$-jet clustering.
	}
\end{figure}

The first observable is the $b$-jet and lepton invariant mass 
$m_{b_j\ell}$ and the second the reconstructed b-jet and W invariant mass $m_{b_jW}$. Both types of observables 
have been studied intensely in the context of top quark mass measurements at the LHC. 
The kinematic sensitivity of $m_{b_j\ell}$ to the top quark mass $m_t$ arises from the upper endpoint of its distribution, which is, for stable $W$ bosons and at tree-level, located at $(m_t^2-m_W^2)^{1/2}$ neglecting the mass effects of the $b$-jet. But also the bulk of the $m_{b_j\ell}$ distribution has kinematic top mass sensitivity because the region where $m_{b_j\ell}$ is maximal depends on the boost of the $W$ boson in the top rest frame which depends kinematically on the top quark mass. The direct kinematic sensitivity of $m_{b_jW}$ to the top mass arises simply from the kinematic location of the resonance which is tied to $m_t$ in a way very similar to thrust, see Eqs.~(\ref{eq:taudef}). In the following we refer to the top mass sentivities of the endpoint location for $m_{b_j\ell}$ and the peak location for $m_{b_jW}$ simply as 'the kinematic top mass dependence' of these two variables. Typical results for the $m_{b_j\ell}$ and $m_{b_jW}$ distributions using the b-jet clustering described below and generated with \Herwig\ 7 are displayed in Fig.~\ref{fig:mblmbW1} for $Q=700$ GeV and top quark masses 172, 173 and 174 GeV. Overall, we see that $m_{b_jW}$ has a somewhat stronger top mass dependence than $m_{b_j\ell}$.

We consider the production of boosted top quarks at $Q=700$~GeV in $e^+e^-$ annihilation and 
use \Herwig\ 7.1.2 with the same settings as
for the thrust analyses discussed in the previous sections (see Sec.~\ref{sec:herwigsimulationsetup}).
For simplicity we again generate only leptonically decaying $W$ bosons and assume perfect neutrino identification.
Furthermore we neglect any combinatorial background, i.e.\ we assume perfect $b$-jet lepton pairing and perfectly 
reconstructed top or antitop quarks.
While these simplications are not fully realistic, they are, however, fully adequate for our examination of 
the shower cut $Q_0$ dependence. For the $b$-jet clustering we use the FastJet package~\cite{Cacciari:2011ma}
and employ the generalized $k_t$ algorithm for $e^+e^-$ collisions in the inclusive mode with 
the inter-particle and inclusive jet distance measures
\begin{align}
d_{ij}\, = \, &\,\mbox{min}(E_i^{2p},E_j^{2p})\,\frac{1-\cos\theta_{ij}}{1-\cos R}\,,\\
d_{iB}\, = \, &\,E_i^{2p}\notag,
\end{align}
where $E_i$ refers to energy, $R$ is the jet radius\footnote{We note that we use the variable $R$ also for the R-evolution equations \eqref{eq:gapRGE} and \eqref{eq:CBmassRRGE} and the corresponding relations in Eqs.~\eqref{eq:Deltasoftv3}, \eqref{eq:DeltaCBmass}-\eqref{eq:MtauQ0dependence}, \eqref{eq:tauQ0dependencemassless}, \eqref{eq:MtauQ0dependencetop}, \eqref{eq:mtfitQ0dependence} and \eqref{eq:mtfitQ0dependencev2}. Since jet radius and R-evolution are different concepts, the meaning of $R$ should be clear from the context.}, and $\theta_{ij}$ is the relative angle between two momenta. The exponent $p=1$ corresponds to the $k_t$-type generalized clustering algorithm, $p=0$ to the Cambridge-Aachen, and $p=-1$ to the anti-$k_t$-type variant, and we consider all three types of algorithms in our analysis. In Fig.~\ref{fig:mbl} and~\ref{fig:mbW} we show the $m_{b_j\ell}$ distribution and the  $m_{b_jW}$  distribution in the peak region, respectively, generated by \Herwig\ 7 {\it at the parton level} with jet radius $R=0.5$ and Cambridge-Aachen-type jet clustering for generator masses $m_t=172$~GeV (green dashed curves), $m_t=173$~GeV (solid red curves) and $m_t=174$~GeV (dashed blue curves). For $m_{b_jW}$ we have smeared the distribution according to Eq.~(\ref{eq:Mtau1}) using smearing parameter $\Lambda=1 $~GeV as described in Sec.~\ref{sec:herwigmassive}.
Since the $m_{b_j\ell}$ distribution is already smooth by itself at the parton level 
we did not account for any additional smearing. 
For both distributions we see that the top mass dependence is essentially linear and particularly strong in the endpoint region for $m_{b_j\ell}$ and in the peak region for $m_{b_jW}$.

The interesting conceptual aspect of the reconstructed observables $m_{b_j\ell}$ and $m_{b_jW}$ is that, due to the $b$-jet clustering, they
are more exclusive than the hemisphere masses entering the thrust variable of Eq.~(\ref{eq:taudef}).
In particular, $m_{b_j\ell}$ and $m_{b_jW}$ depend on the  $b$-jet radius $R$. For large  $R\sim \pi/2$  we can expect their shower cut dependence
to be very similar to the one for thrust since the ultra-collinear as well as major portions of large angle soft radiation are clustered into the $b$-jet.
On the other hand, due to the boosted top kinematics which confines the top decay products as well as the ultra-collinear radiation inside a cone with angle $\sim m_t/Q$ with respect to the top momentum direction, the clustering should always retain most of the ultra-collinear radiation that is soft in the top rest frame and thus inherently tied to the physical top quark state. Thus for small $R\sim m_t/Q$ we can expect that the majority of the ultra-collinear radiation is still clustered into the $b$-jet while
the majority of the large-angle soft radiation is removed. As a consequence we can expect that the shower cut 
dependence coming from 
the large-angle soft radiation is reduced when $R$ is lowered, while the one 
from ultra-collinear radiation is kept.

To quantify the dependence of $m_{b_j\ell}$ and $m_{b_jW}$ generated from \Herwig~7 on the shower cut we use the following procedure: For a given jet radius $R$ and clustering algorithm (as well as matching scheme for the analysis in Sec.~\ref{sec:herwigNLO}) we take the results for $Q_0=Q_{0,b}=1.5$~GeV as the default and generate $m_{b_j\ell}$ and $m_{b_jW}$ distributions for different generator masses $m_t$ in the range between $172$ and $174$~GeV, which we subsequently use to fit the top quark mass from the distributions generated for $m_t=173$~GeV but with different choices of $Q_0$ or $Q_{0,b}$. The shower cut dependence of the parts of the $m_{b_j\ell}$ and $m_{b_jW}$ distributions used for the fits are then directly transferred into deviations of the fitted top masses with respect to the default mass $m_t=173$~GeV (for $Q_0=Q_{0,b}=1.5$~GeV), which can then be compared with our theoretical expectations. Due to the high number of events we use, statistical uncertainties are negligible and therefore not specified in the following. We emphasize that the shower cut dependence of the fitted top mass we
obtain in this analysis is a representation of the shower cut dependence of  $m_{b_j\ell}$ and $m_{b_jW}$ themselves and not equivalent to the shower cut dependence of the generator mass. 

\begin{figure}
	\center
	\begin{subfigure}[c]{0.49\textwidth}
		\includegraphics[width=1.0\textwidth]{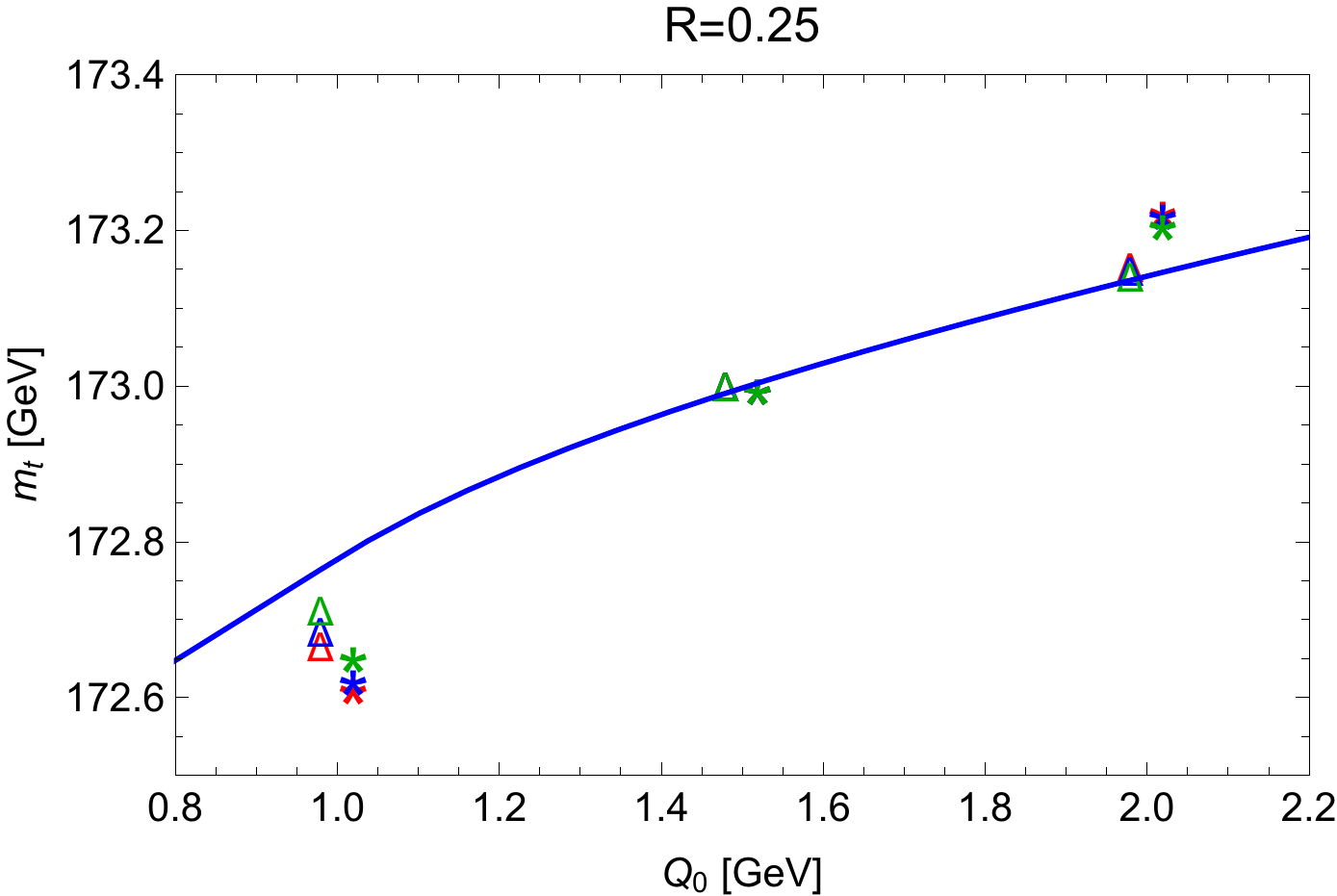}
		\subcaption{}
	\end{subfigure}
	\hfill
	\begin{subfigure}[c]{0.49\textwidth}
		\includegraphics[width=1.0\textwidth]{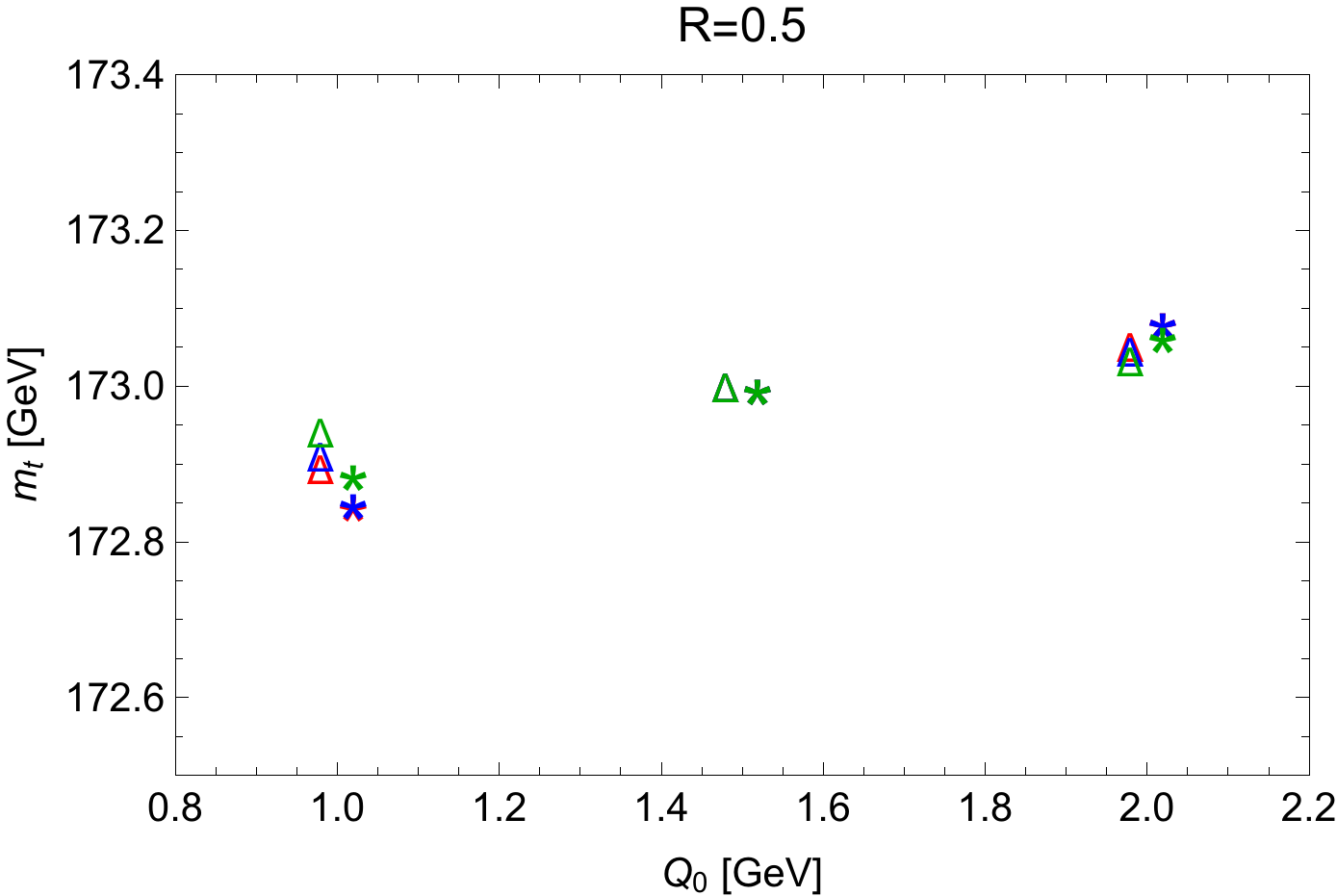}
		\subcaption{}
	\end{subfigure}
	\begin{subfigure}[c]{0.49\textwidth}
		\includegraphics[width=1.0\textwidth]{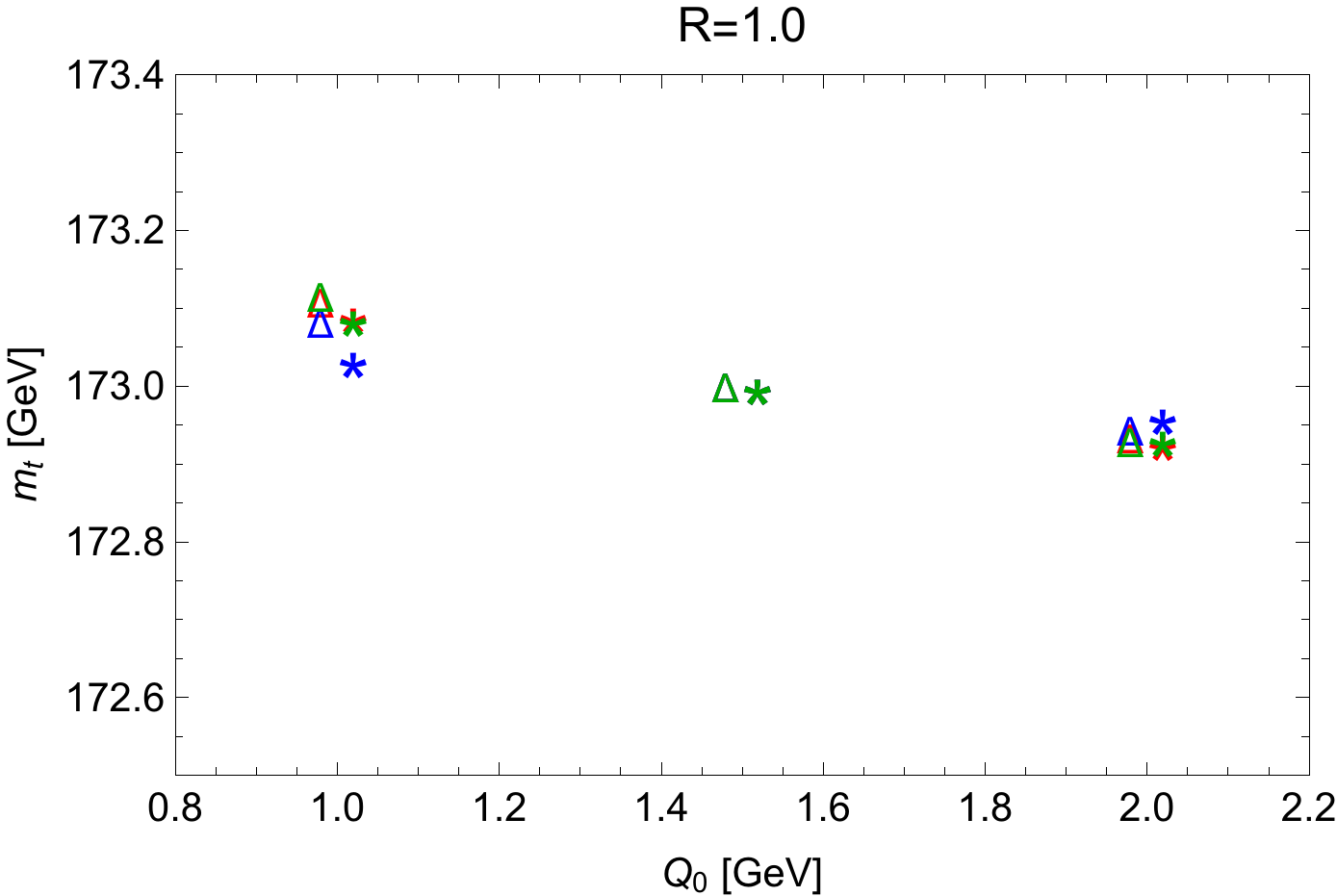}
		\subcaption{}
	\end{subfigure}
	\hfill
	\begin{subfigure}[c]{0.49\textwidth}
		\includegraphics[width=1.0\textwidth]{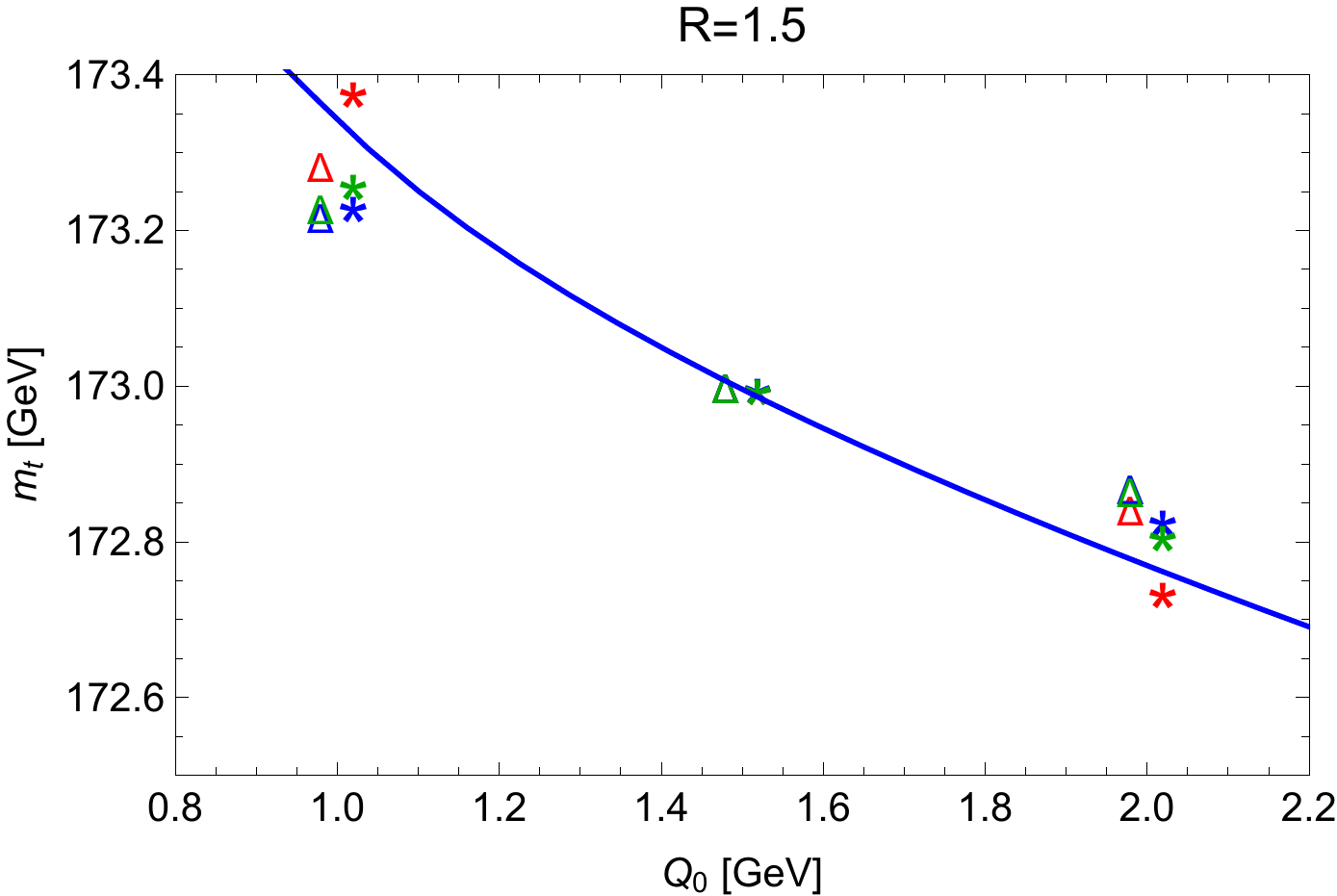}
		\subcaption{}
	\end{subfigure}
	\caption{\label{fig:mblmbWQ0Q0b}
		Fitted top quark mass as a function of the shower cut $Q_0=Q_{0,b}$ for $Q=700$~GeV 
		obtained from the $m_{b_j\ell}$ endpoint (stars) and the $m_{b_jW}$ resonance region (triangles)
		using the $k_t$-type-type algorithm (green), the Cambridge-Aachen-type algorithm (blue) 
		and the anti-$k_t$-type algorithm (red)
		for $b$-jet clustering. Displayed are the results for $b$-jet radii $R=0.25$, $0.5$, $1.0$ and $1.5$. The solid blue 
		line in the lower right panel corresponds to Eq.~(\ref{eq:mtfitQ0dependence}) and the one
		in the upper left panel corresponds to Eq.~(\ref{eq:mtfitQ0dependencev2}) using $Q_0^\prime=1.5$ as the reference
		scale.
	}
\end{figure}

In Figs.~\ref{fig:mblmbWQ0Q0b} the dependence of the fitted top mass on the shower cut
is shown, where production stage and decay stage shower cuts are identified, using for  $Q_{0,b}=Q_{0}$ 
the values $1.0$, $1.5$ and $2.0$~GeV  for jet radii $R=0.25$ (upper left panel),  
$R=0.5$ (upper right panel), $R=1.0$ (lower left panel) and $R=1.5$ (lower right panel).
The top masses obtained from the $m_{b_j\ell}$ endpoint region are shown as colored stars and have been obtained from fits in the  $m_{b_j\ell}$ interval $[150,155]$~GeV. The top masses obtained from the $m_{b_jW}$ resonance region are shown as colored triangles and have been obtained from fits using the highest 20\% of the distribution around to the peak. 
To allow for an easier visual identification we have slightly displaced the stars and the triangles horizontally.  
We have carried out the analyses for
all three jet clustering algorithm where we use green color for the $k_t$-type algorithm ($p=1$), blue color for the Cambridge-Aachen-type algorithm ($p=0$) and red color for the anti-$k_t$-type algorithm ($p=-1$). 
We see that for large hemisphere-type $b$-jet cones the fitted top mass decreases with the shower cut. This means that 
the mass of the reconstructed top quark state, to which $m_{b_j\ell}$ and $m_{b_jW}$ are kinematically sensitive (and which for hemisphere-type b-jets includes the effects of large angle soft radiation),
decreases when $Q_0$ is increased. So the behavior indeed follows the one of thrust we have observed in
Sec.~\ref{sec:herwigmassive}. 
Analytically, the expected $Q_0$ dependence for ideal hemisphere-type $b$-jets has the form
\begin{align}
\label{eq:mtfitQ0dependence}
m_{t,\,{\rm fit}}^{R=\pi/2}(Q_0) \, = \, m_{t,\,{\rm fit}}^{R=\pi/2}(Q_0^\prime) -
\Bigl[4 \frac{Q}{m_t}- 2\pi\Bigr] \,\int\limits_{Q_0^\prime}^{Q_0}\mathrm{d}R\,
\frac{C_F\, \alpha_s(R)}{4\pi}
\end{align}
and is ploted in the lower right panel of Fig.~\ref{fig:mblmbWQ0Q0b}
as the blue solid line using $Q_0^\prime=1.5$~GeV as the reference scale. 
The RHS of Eq.~(\ref{eq:mtfitQ0dependence}) is a factor two smaller than the one for the rescaled thrust $M_{\tau,\,{\rm peak}}$ in Eq.~(\ref{eq:MtauQ0dependence}) since the reconstructed top mass is linear in the top mass while the rescaled thrust variable $M_{\tau,\,{\rm peak}}$ is quadratic in the top mass, see Eqs.~(\ref{eq:taudef}) and (\ref{eq:Mtaudef}).
We see that the expected behavior agrees very well with the results obtained from the fit. The actual fit results for all clustering algorithms
except for anti-$k_t$ tend
to have a slightly smaller slope than Eq.~(\ref{eq:mtfitQ0dependence}), which is mainly due to the fact that even for $R=\pi/2$ the $b$-jets are typically not full hemisphere jets because they are in general not exactly back-to-back and compete with each other in the clustering process.   
For decreasing jet radius $R$, on the other hand, we see that the slope in $Q_0$ of the fitted top mass increases continuously and becomes positive for $R\lsim 0.5$.
This confirms the expectation that the shower cut dependence originating from large-angle soft radiation (which is the contribution proportional to $Q/m$ in Eq.~(\ref{eq:mtfitQ0dependence}) becomes suppressed when $R$ is reduced, while the shower cut-dependence associated to the ultra-collinear radiation is kept. 
For visualization we have plotted  in upper left panel Fig.~\ref{fig:mblmbWQ0Q0b} the relation
\begin{align}
\label{eq:mtfitQ0dependencev2}
m_{t,\,{\rm fit}}^{R\sim m_t/Q}(Q_0) \, = \, m_{t,\,{\rm fit}}^{R\sim m_t/Q}(Q_0^\prime) +
2\pi \,\int\limits_{Q_0^\prime}^{Q_0}\mathrm{d}R\,
\frac{C_F\, \alpha_s(R)}{4\pi}
\end{align}
with $Q_0^\prime=1.5$~GeV as the reference scale as the blue solid line. This is
just Eq.~(\ref{eq:mtfitQ0dependence}) but with the $Q/m_t$ term dropped, that originates from
large angle soft radiation. Again we see excellent agreement between the expected shower 
cut dependence and the actual fit results. It is also conspicuous that the shower cut dependence of 
the fitted top quark masses
we obtain from $m_{b_j\ell}$ and $m_{b_jW}$ for the different jet radii and jet algorithms are essentially equivalent 
and do not exhibit any systematic difference.
This analysis thus fully supports universality concerning the equivalence of
the MC generator top quark mass and the shower cut dependent CB mass defined in Eq.~(\ref{eq:CBmassschemedef2}).

\begin{figure}
	\center
	\begin{subfigure}[c]{0.49\textwidth}
		\includegraphics[width=1.0\textwidth]{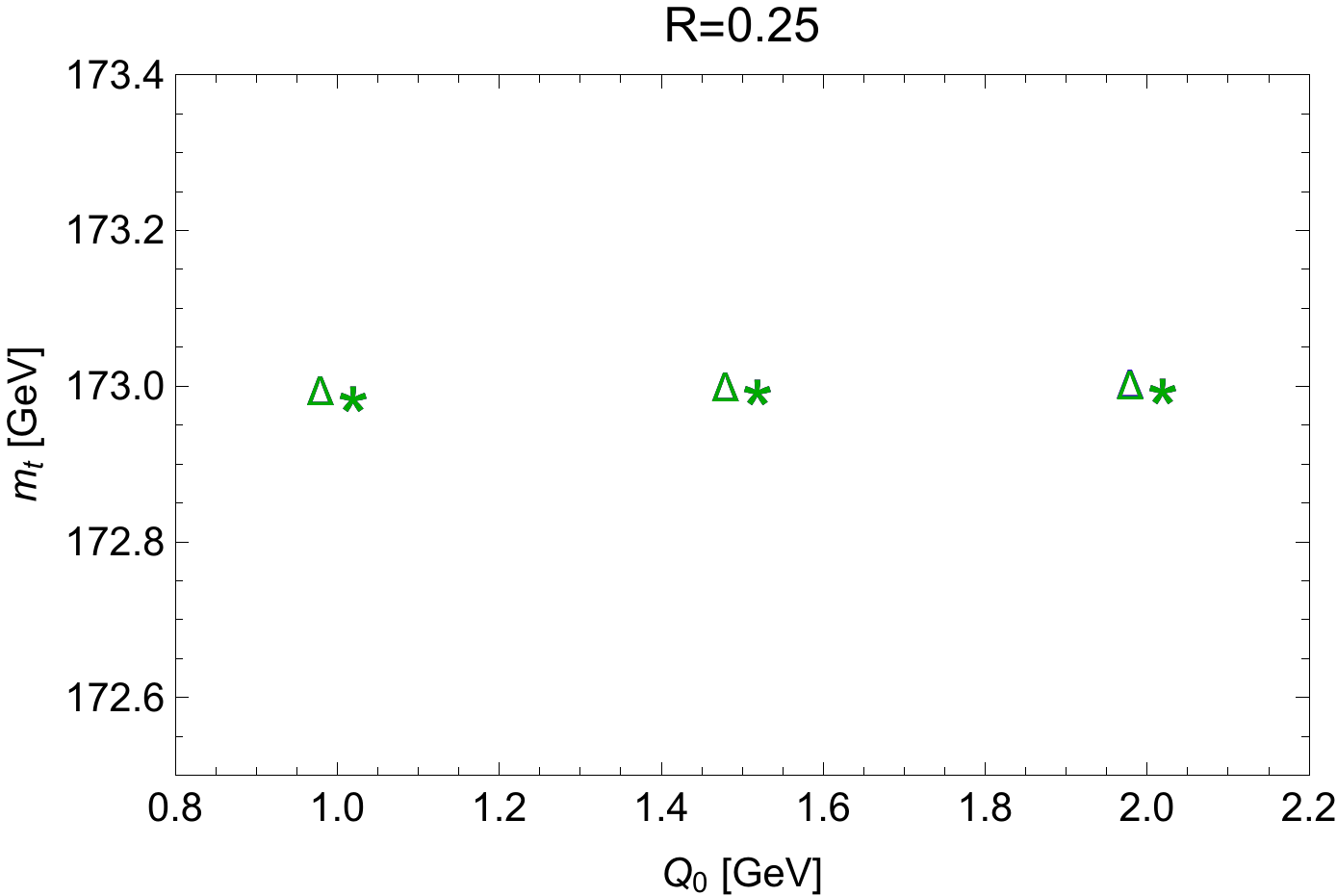}
		\subcaption{}
	\end{subfigure}
	\hfill
	\begin{subfigure}[c]{0.49\textwidth}
		\includegraphics[width=1.0\textwidth]{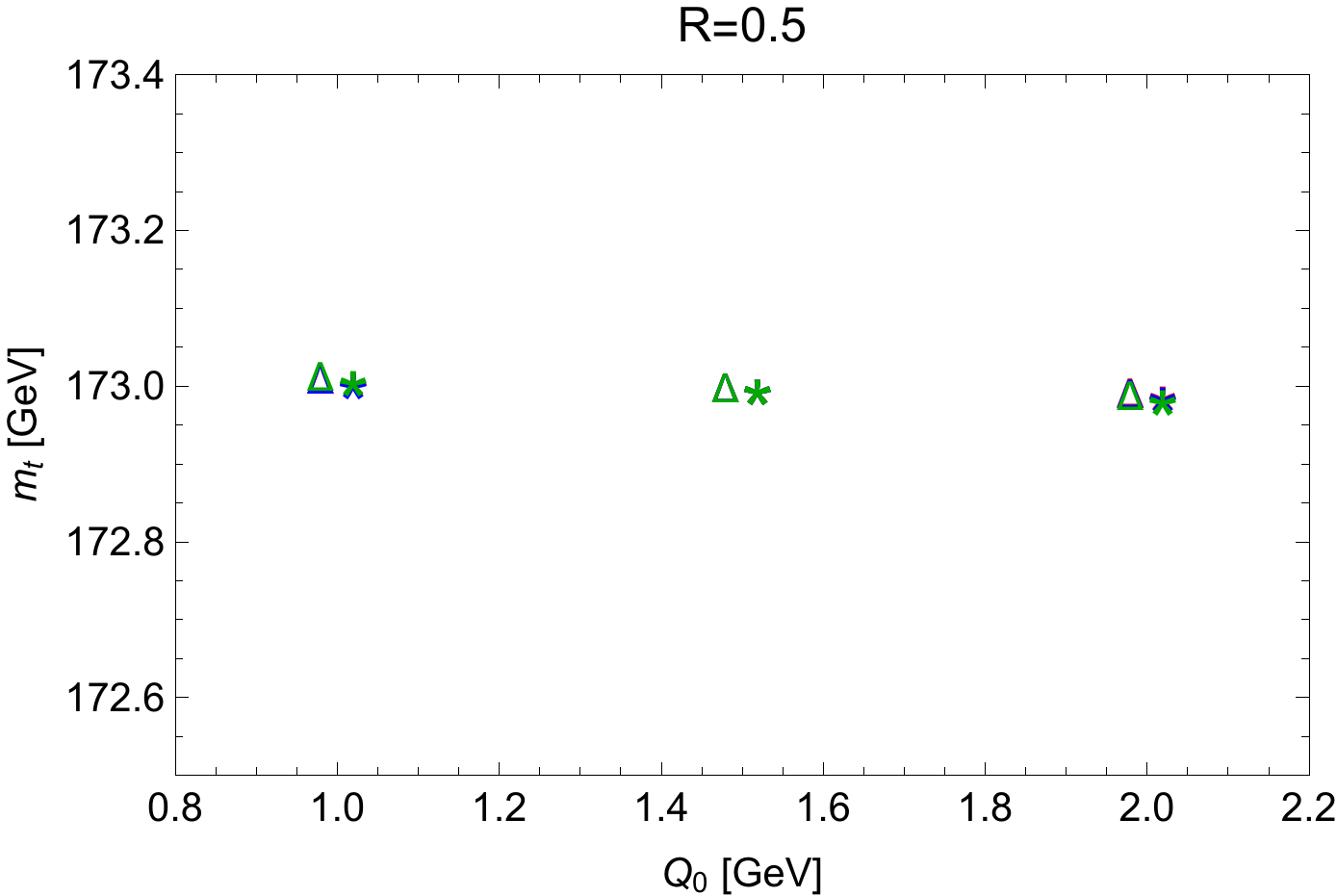}
		\subcaption{}
	\end{subfigure}
	\begin{subfigure}[c]{0.49\textwidth}
		\includegraphics[width=1.0\textwidth]{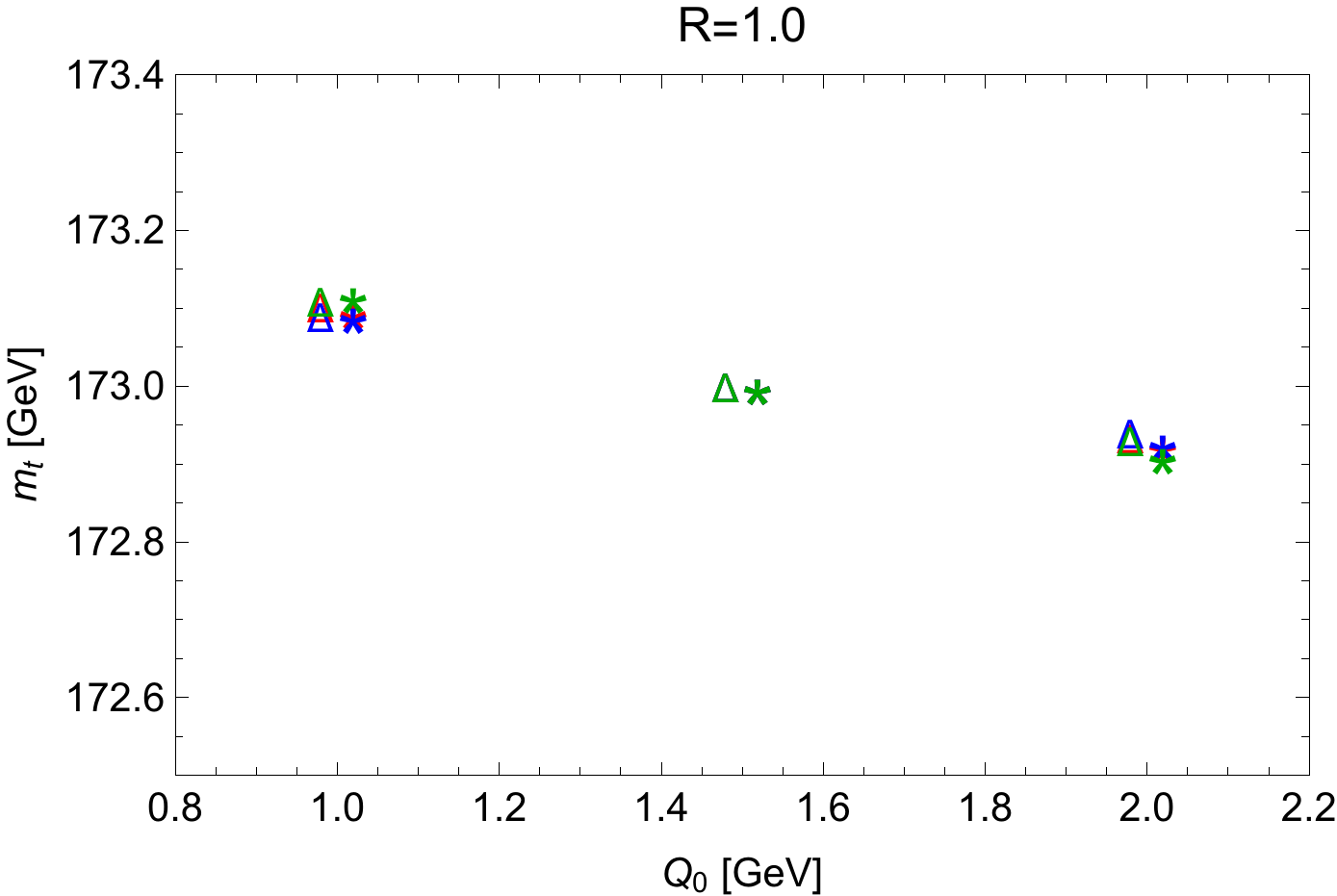}
		\subcaption{}
	\end{subfigure}
	\hfill
	\begin{subfigure}[c]{0.49\textwidth}
		\includegraphics[width=1.0\textwidth]{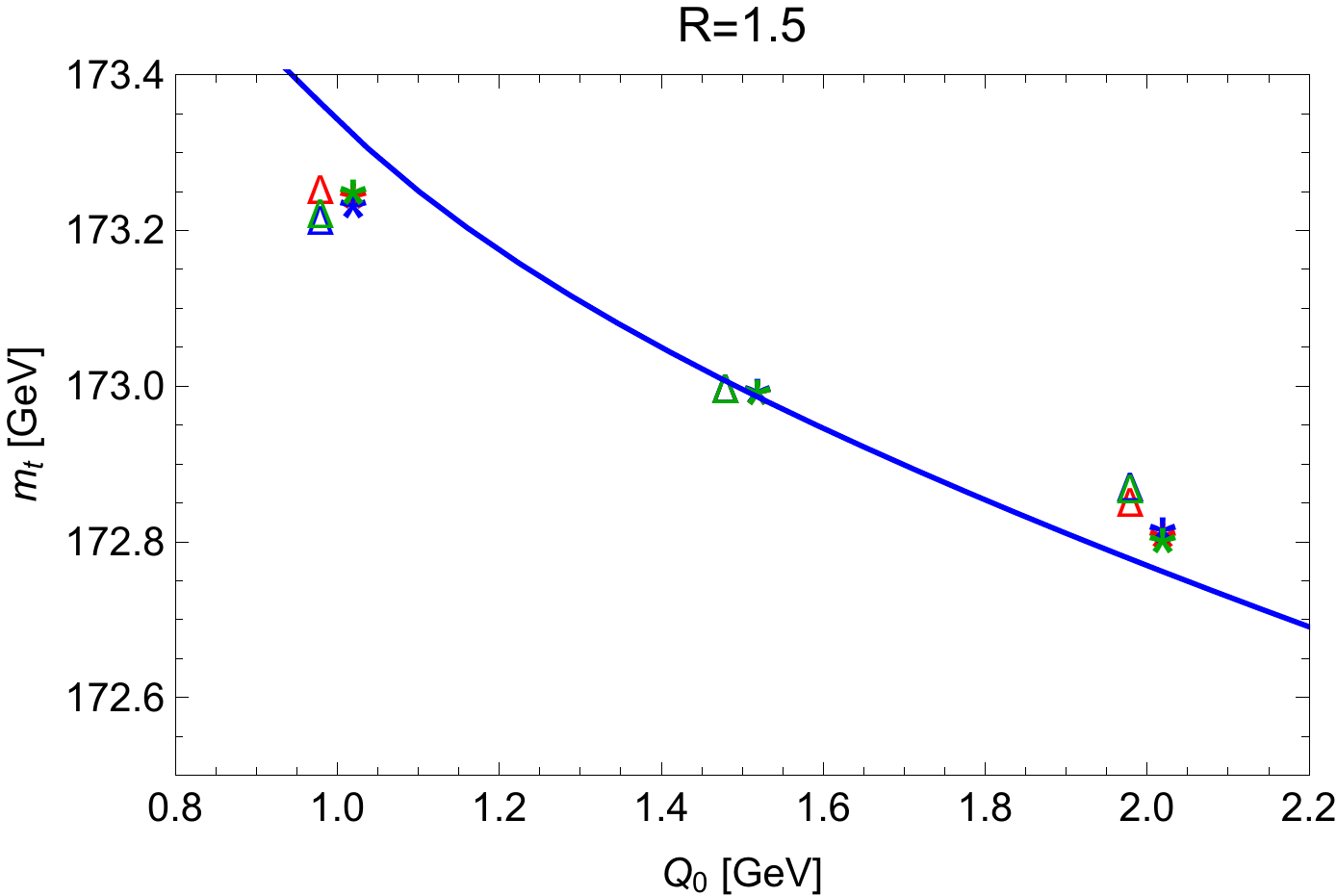}
		\subcaption{}
	\end{subfigure}
	\caption{\label{fig:mblmbWQ0}
		Fitted top quark mass as a function of the production stage shower cut $Q_0$ with decay stage shower cut fixed
		to $Q_{0,b}=1.5$~GeV for $Q=700$~GeV 
		obtained from the $m_{b_j\ell}$ endpoint (stars) and the $m_{b_jW}$ resonance region (triangles)
		using the $k_t$-type algorithm (green), the Cambridge-Aachen algorithm (blue) and the anti-$k_t$-type algorithm (red)
		for $b$-jet clustering. Displayed are the results for $b$-jet radii $R=0.25$, $0.5$, $1.0$ and $1.5$. The solid blue 
		line in the lower right panel corresponds to Eq.~(\ref{eq:mtfitQ0dependence}) using $Q_0^\prime=1.5$ as the reference
		scale. 
	}
\end{figure}

However, in the absence of a systematic factorized analytic approach to the kinematic top mass dependence of the  $m_{b_j\ell}$ and $m_{b_jW}$ this
universality cannot be strictly proven because, in contrast to thrust, $m_{b_j\ell}$ and $m_{b_jW}$ are affected substantially by the MC modelling and the dynamics of the final state and, in particular, by the choice of the decay stage shower cut $Q_{0,b}$. This makes the conceptual background to be examined more involved. In particular, for our {\it parton-level studies} a strict proof would require that we could analytically track the role played by the decay stage shower cut $Q_{0,b}$ for the interpretation of the generator top quark mass in a systematic manner.  

To visualize the relevance of the decay stage shower cut $Q_{0,b}$ for small $b$-jet radii we show in
Figs.~\ref{fig:mblmbWQ0} again the dependence of the fitted top mass on the production stage shower cut $Q_0$ for the same cases displayed 
in Figs.~\ref{fig:mblmbWQ0Q0b},
but using a fixed decay stage shower cut $Q_{0,b}=1.5$~GeV.  
We see that for a large hemisphere-type $b$-jet radius $R=1.5$ the results are equivalent to the corresponding ones for $Q_{0,b}=Q_{0}$ shown in 
lower right panel of Figs.~\ref{fig:mblmbWQ0Q0b}.
For decreasing jet radii we see that the dependence of the fitted top mass on $Q_0$  decreases continuously remains  essentially flat for $R<0.5$ in contrast to Figs.~\ref{fig:mblmbWQ0Q0b} where a positive slope in $Q_0$ was
emerging.   
This shows that for small jet radii the shower cut dependence of the kinematic top mass sensitivity of $m_{b_j\ell}$ and $m_{b_jW}$ arises from the decay stage shower cut $Q_{0,b}$. Even though it appears hard to believe that the good agreement we observed for small $b$-jet radii and $Q_{0,b}=Q_0$ between the fit results and the naive expectations 
is purely accidental, the case of small jet radii is strictly speaking not covered by the conceptual considerations we have carried out for thrust. 

To conclude the question of universality, at the present stage, we can say that the shower cut dependence 
we observe for the kinematic 
top mass dependence of $m_{b_j\ell}$ and $m_{b_jW}$ is compatible with the one we have proven for thrust and thus supports universality. 
This is quite encouraging and motivates further systematic and more general consistency studies that may be carried out with MC simulations 
and relatively simple analytical computations alone. 
However, a strict conceptual proof would also involve a precise quantification of the role of the decay 
stage shower cut $Q_{0,b}$ (and maybe other issues relevant
for exclusive observables with strong kinematic top sensitivity), 
preferably in the context of a factorized approach where the types of radiation relevant for the interpretation of the top quark mass can be ambiguously 
separated from other types of radiation and discussed at the field theoretic level.
This strongly motivates the development of factorized predictions for reconstructed and exclusive observables 
such as $m_{b_j\ell}$ and $m_{b_jW}$.

At this point we would also like to remind the reader that all our examinations above have been carried out for boosted top quarks. The direct reconstruction top mass measurements at the LHC are, on the other hand, based on top quarks with $p_T$ values in the range of 50 to 100~GeV, which corresponds predominantly to unboosted top quarks. We stress that for unboosted top quarks a classification of the radiation modes relevant for a systematic discussion of the meaning of the generator mass is currently lacking and that, in particular, the concepts of large angle soft and ultra-collinear radiation do not apply. Therefore, none of the above considerations or argumentations are applicable for the reconstructed observables $m_{b_j\ell}$ and $m_{b_jW}$  for unboosted top quarks.

\subsection{Impact of NLO matching}
\label{sec:herwigNLO}

A crucial precondition of our examinations on 
the shower cut dependence was that NLL precise angular ordered parton showers, 
based on the coherent branching formalism described in Secs.~\ref{sec:coherentbranchingmassless} and 
\ref{sec:coherentbranchingmassive}, are already NLO precise \emph{as far as the dominant linear 
shower cut dependence of the thrust peak position 
is concerned}, see Sec.~\ref{sec:peakprecision}. 
This means in turn that the ${\cal O}(\alpha_s)$ 
QCD corrections added to simulations in NLO matched MC setups should show very small or even
negligible effects in the numerical studies that we have carried out in 
Secs.~\ref{sec:herwigthrusttest}, \ref{sec:herwigmassless} and \ref{sec:herwigmassive}. 
It is the purpose of this section to demonstrate this 
explicitly by comparing \Herwig\ 7 simulations with and 
without NLO matching. Furthermore we show by general considerations that NLO matched MC simulation can a priori 
not modify the shower cut dependence present in MC simulations at NLL for which 
NLO matching is not accounted for. Since we believe that this discussion may add to a more refined 
understanding of NLO matching for the general reader, we explain some generic features  
of NLO matching in the following with a special focus on the role of the shower cut. 
The reader familiar with the details of NLO matched MC simulations may skip 
this conceptual discussion and jump directly to the numerical discussion 
starting after Eq.~(\ref{eq:powheg}). 

Matching parton showers to NLO QCD corrections for the hard process
has by now become a default requirement for event
generation in LHC data analysis. 
In the following we review the prototypical structure of a parton shower
NLO matching algorithm with particular focus on the effects related
to the parton shower cutoff. The core of general matching algorithms for parton showers
is based on a careful analysis of a single
shower emission in order to avoid double counting with the 
corresponding NLO cross section prediction. At this point, one has to keep in mind that
the main aim is to improve the hardest emission, which in general may
not be the first one to occur, particularly in the case of
angular ordered parton showers. However, to first subleading order in
$\alpha_s$, the first and the hardest emission
are the same. At this level, global recoil schemes like the kinematic reconstruction 
of the angular ordered shower tree discussed in Sec.~\ref{sec:coherentbranchingHerwig}, 
as well as local recoil schemes which
restore the kinematics after each successive emission can 
be discussed in a unified fashion. Given the $n$-parton
phase space point 
\begin{align}
\phi_n\equiv\{p_1,...,p_n\}
\end{align} 
associated to the initial hard process, a phase space point with one
additional emission off the progenitor leg $i$ can be parametrized in terms of
the momentum scale $\tilde{q}$ of the emission and the momentum fraction $z$,
where for simplicity throughout our discussion we suppress an additional azimuthal angle variable
required to set up a complete momentum for the emission:
\begin{equation}
(\phi_{n},\tilde{q},z)\,\to\, \Phi^{(i)}_{n+1}(\phi_n,\tilde{q},z) \equiv
\{q^{(i)}_1(\phi_{n},\tilde{q},z),...,q^{(i)}_{n}(\phi_{n},\tilde{q},z),q^{(i)}_{n+1}(\phi_{n},\tilde{q},z)\}\,.
\end{equation}
At this point it is useful to also introduce 
the inverse mapping from the $(n+1)$-parton phase space point 
\begin{align}
\phi_{n+1} \equiv\{q_1,...,q_{n+1}\}
\end{align} 
to the $n$-parton phase space
$\Phi^{(i)}_n(\phi_{n+1})$ and associated evolution
variables $\tilde{Q}^{(i)}(\phi_{n+1})$, $Z^{(i)}(\phi_{n+1})$ such that
\begin{equation}
\label{eq:invmap}
\Phi^{(i)}_{n+1}(\Phi^{(i)}_n(\phi_{n+1}),\tilde{Q}^{(i)}(\phi_{n+1}),Z^{(i)}(\phi_{n+1})) = \phi_{n+1} \ .
\end{equation}
Defining the general infinitesimal $m$-parton phase space volume element for a total momentum $p_{e^+}+p_{e^-}$ as
\begin{equation}
  {\rm d}\text{PSP}_{m}(k_1,...,k_m) \equiv
  (2\pi)^4\delta^{(4)}\left(\sum_{i=1}^{m} k_i - (p_{e^+}+p_{e^-})\right) 
\prod_{i=1}^{m} \frac{{\rm d}^4 k_i}{(2\pi)^{3}}\delta(k_i^2-m_i^2) \theta(k_i^0)
\end{equation}
and using abbreviations
\begin{equation}
  {\rm d}\phi_{n} \equiv {\rm d}\text{PSP}_{n}(p_1,...,p_n)\,,\qquad
  {\rm d}\phi_{n+1} \equiv {\rm d}\text{PSP}_{n+1}(q_1,...,q_{n+1}) \,,
\end{equation}
the kinematic mapping implies a factorization of the $(n+1)$-parton phase space volume element
of the form
\begin{eqnarray}
{\rm d}\phi_{n+1}|_{\phi_{n+1}=\Phi^{(i)}_{n+1}(\phi_n,\tilde{q},z)}&\equiv& {\rm d}\text{PSP}_{n+1}(q^{(i)}_1(\phi_n,\tilde{q},z),...,
q^{(i)}_{n+1}(\phi_n,\tilde{q},z))\\\nonumber
&=& J^{(i)}(\phi_n,q,z){\rm d}\phi_n{\rm d}\tilde{q}{\rm d}z \,,
\end{eqnarray}
where the term $J^{(i)}$ is the associated Jacobian factor. 
At the cross section level the parton shower splitting rate is then given by combining the 
$(n+1)$-parton phase space element
with the propagator factor and the corresponding splitting function $P_i$:
\begin{equation}
\label{eq:dPJdpdz}
{\rm d}P^{(i)}(\phi_n,\tilde{q},z) = \frac{16\pi^2}{(q_i+q_{n+1})^2-m_i^2} P_i\biggl[\alpha_s,z,\frac{m_i^2}{\tilde{q}^2}\biggr] 
J^{(i)}(\phi_n,\tilde{q},z){\rm d}\tilde{q}{\rm d}z \ ,
\end{equation}
and the arguments of the mapped momenta $q_{i}=q_{i}(\phi_n,\tilde{q},z)$ are
understood implicitly.
The exponent of the Sudakov form factor, which quantifies the no-branching probability for possible emissions
between the scale $\tilde{Q}$ and $\tilde{q}$, is
then simply given by 
\begin{equation}
\Delta(\phi_n,\tilde{q},\tilde{Q}) = \exp\biggl[\,-\int_{\tilde{q}}^{\tilde{Q}}
{\rm d}\tilde{k} \int{\rm d}z \sum_i \frac{{\rm
    d}P^{(i)}(\phi_n,\tilde{k},z)}{{\rm d}\tilde{k}{\rm
    d}z}\theta\left(P^{(i)}_\perp(\tilde{k},z)-Q_0\right)\theta(\tilde{k}-m_i)\,\biggr]\,,
\end{equation}
with a cut on the transverse momentum $P_\perp^{(i)}(\tilde{q},z)$ of the splitting
expressed as a function of $\tilde{q}$ and $z$ (see Eq.~(\ref{pperpcut2})),
and where the sum is over all possible splittings off the progenitor partons.
We can now define the one-emission action of the parton shower on an observable $\eta=f_{\eta,m}(\phi_m)$,
where $f_{\eta,m}$ is the value of the observable for an $m$-parton final state,  by
\begin{multline}
\text{PS}[u](\phi_n) = \Delta(\phi_n,0,\tilde{Q}) u(\phi_n) +\\ \sum_i \int_{0}^{\tilde{Q}} {\rm d}\tilde{q} \int {\rm d}z
\frac{{\rm d}P^{(i)}(\phi_n,\tilde{q},z)}{{\rm d}\tilde{q} {\rm d}z}
\Delta(\phi_n,\tilde{q},\tilde{Q})\ \times\\\theta\left(P^{(i)}_\perp(\tilde{q},z)-Q_0\right)\theta(\tilde{q}-m_i)
u(\Phi_{n+1}^{(i)}(\phi_n,\tilde{q},z))\,,
\end{multline}
where $u(\phi_m)=\delta(\eta-f_{\eta,m}(\phi_m))$ is the measurement function. 
Starting from a given $n$-parton (LO) cross section 
\begin{equation}
\sigma_{\text{LO}}[u] \equiv \int {\rm d}\sigma_{\text{LO}}(\phi_n)u(\phi_n)
\end{equation}
the one-emission action of the parton shower can then be expressed as
\begin{equation}
\sigma_{\text{LO+PS}}[u] = \sigma_{\text{LO}}[\text{PS}[u]]\,.
\end{equation}
Expanding the one-emission
action to first subleading order in $\alpha_s$ we find
\begin{multline}
\sigma_{\text{LO+PS}}[u] = \int {\rm d}\sigma_{\text{LO}}(\phi_n)\,u(\phi_n)\\
+\sum_i \int_{0}^{\tilde{Q}} {\rm d}\tilde{q} \int {\rm d}z \int
{\rm d}\sigma_{\text{LO}}(\phi_n)\frac{{\rm d}P^{(i)}(\phi_n,\tilde{q},z)}{{\rm d}\tilde{q}
  {\rm d}z}\ \times\\
\theta\left(P^{(i)}_\perp(\tilde{q},z)-Q_0\right)\theta(\tilde{q}-m_i)\left(u(\Phi_{n+1}^{(i)}(\phi_n,\tilde{q},z)) - 
u(\phi_n)\right)\,,
\end{multline}
which, with the help of the inverse mapping~(\ref{eq:invmap}), can be
cast into the form
\begin{multline}
\label{eq:showerexpanded}
\sigma_{\text{LO+PS}}[u] = \int {\rm d}\sigma_{\text{LO}}(\phi_n) u(\phi_n)
+\sum_i \int
{\rm d}\sigma^{(i)}_{\text{PS}}(\phi_{n+1})\ \times\\
\theta\left(\tilde{Q}-\tilde{Q}^{(i)}(\phi_{n+1}))\right)\theta\left(P_\perp^{(i)}(\phi_{n+1})-Q_0\right)
\left(u(\phi_{n+1}) - u(\Phi^{(i)}_n(\phi_{n+1}))\right) \,,
\end{multline}
with
$P_\perp^{(i)}(\phi_{n+1})=P_\perp^{(i)}(\tilde{Q}^{(i)}(\phi_{n+1}),Z^{(i)}(\phi_{n+1}))$,
and where the starting point of the evolution $\tilde{Q}$ and the infrared shower cut
$Q_0$ are now made explicit in terms of $\theta$-functions at the level of the
squared matrix element, and we have introduced the parton shower approximation
to the NLO fixed-order correction to the cross section:
\begin{multline}
{\rm d}\sigma^{(i)}_{\text{PS}}(\phi_{n+1}) \equiv {\rm
  d}\sigma_{\text{LO}}(\Phi^{(i)}_{n}(\phi_{n+1})) \ \times\\
{\rm
  d}P^{(i)}(\Phi^{(i)}_{n}(\phi_{n+1}),\tilde{Q}^{(i)}(\phi_{n+1}),Z^{(i)}(\phi_{n+1}))\theta\left(\tilde{Q}^{(i)}(\phi_{n+1})
- m_i\right) \ .
\end{multline}
Expression~(\ref{eq:showerexpanded}) is particularly useful to formulate NLO matching since it is
very close to the generic form of the corresponding NLO fixed-order cross section obtained in full QCD
using the subtraction approach:\footnote{For simplicity we assume here that the kinematic mapping
	used in the subtraction formalism is the same as the one used for
	shower emissions, which is, however, not mandatory. In the general case one has to account 
	for additional Jacobian factors that, however, do not alter the
	line of reasoning.}
\begin{align}
\label{eq:NLOFO}
\sigma_{\text{NLO}}[u] \equiv \,&\,\sigma_{\text{LO}}[u] + \int {\rm d}\sigma_{V+I}(\phi_n)u(\phi_n)\notag\\
&+ \int\Bigl[ {\rm d}\sigma_R(\phi_{n+1})u(\phi_{n+1}) -\sum_i {\rm d}\sigma_A^{(i)}(\phi_{n+1})u(\Phi_n^{(i)}(\phi_{n+1}))\Bigr]\,.
\end{align}
Here
the terms ${\rm d}\sigma_A^{(i)}$ are the subtraction cross sections to
cancel the infrared divergences of the real emission cross section 
${\rm d}\sigma_R$ coming from progenitor leg $i$, and 
${\rm d}\sigma_{V+I}$ denotes the combination of the NLO virtual corrections
and the subtraction cross sections integrated over the emission phase spaces, which
is free of poles in $\epsilon$ in dimensional regularization. This concludes our notation to
discuss the generic formalism of NLO matched parton showers.

NLO matching, see
\cite{Frixione:2002ik,Frixione:2007vw,Platzer:2011bc,Hoeche:2011fd,Nason:2012pr}
for initial development concerning multi-purpose event generators as
well as a general review, including the modified hardest emission
approach as employed in the POWHEG formalism~\cite{Frixione:2007vw},
is then performed by subtracting the ${\cal O}(\alpha_s)$ contribution
of Eq.~(\ref{eq:showerexpanded}) from the NLO fixed-order cross
section. To be specific, one sets up a subtracted (or 'matched') NLO
cross section $\sigma_{\text{NLO}-\text{PS}}$, such that
\begin{equation}
\sigma_{\text{NLO}-\text{PS}}[\text{PS}[u]] = \sigma_{\text{NLO}}[u]\,\Big[1 + {\cal O}(\alpha_s)\Big]
\end{equation}
having the important condition in mind that
the total NLO inclusive cross section precisely agrees with
the NLO fixed-order calculation:
\begin{equation}
\label{eq:matchedconsist1}
\sigma_{\text{NLO}-\text{PS}}[\text{PS}[1]] = \sigma_{\text{NLO}-\text{PS}}[1] = \sigma_{\text{NLO}}[1]\,,
\end{equation}
where the first equality arises from the unitarity property of the parton shower.
This can be achieved by
\begin{equation}
\label{eq:subtractedNLOcs}
\sigma_{\text{NLO}-\text{PS}}[u] = \sigma_{\text{LO}}[u] + \sigma_{V+I}[u] + \sigma_{R-A-\text{PS}}[u]\,,
\end{equation}
where $\sigma_{\text{LO}}$ and $\sigma_{V+I}$ are directly taken from Eq.~(\ref{eq:NLOFO}) 
and the matching subtraction term is of the form
\begin{multline}
\label{eq:showermatched}
\sigma_{R-A-\text{PS}}[u] = \\
\sum_i \int \left(
\theta(\tilde{Q}-\tilde{Q}^{(i)}_{n+1})\theta(P_{\perp,n+1}^{(i)}-Q_0)
      {\rm d}\sigma^{(i)}_{\text{PS}}(\phi_{n+1}) -
      {\rm d}\sigma^{(i)}_{A}(\phi_{n+1}) \right)
      u(\Phi_n^{(i)}(\phi_{n+1})) +\\ \int \left({\rm
        d}\sigma_R(\phi_{n+1})-\sum_i
      \theta(\tilde{Q}-\tilde{Q}^{(i)}_{n+1})\theta(P_{\perp,n+1}^{(i)}-Q_0)
            {\rm d}\sigma^{(i)}_{\text{PS}}(\phi_{n+1})
            \right)u(\phi_{n+1}) \ ,
\end{multline}
where we have introduced the shorthand notations 
$\tilde{Q}^{(i)}_{n+1}\equiv\tilde{Q}^{(i)}(\phi_{n+1})$ and 
$P_{\perp,n+1}^{(i)}\equiv P_{\perp}^{(i)}(\phi_{n+1})$. 

The important point of Eq.~(\ref{eq:showermatched}) is that within the NLO matched
parton shower evolution algorithm the expression in the
first line constitutes together with $\sigma_{\text{LO}}[u] + \sigma_{V+I}[u]$ 
the tree-level $n$-parton cross section with $n$ progenitor 
partons, while the second line represents a new $(n+1)$-parton tree-level configuration
with $n+1$ progenitors. Both types of configuration are therefore showered separately
and need to be independently infrared safe and numerically stable without relying on
any cross talk between both contributions. However, there are by construction infrared divergences
that only cancel in the combination of $ {\rm d}\sigma_R$ and the ${\rm d}\sigma^{(i)}_A$
because of the presence of the infrared cuts contained in the parton shower contributions.  
As a consequence the coefficients of the individual
$u(\Phi_{n}^{(i)}(\phi_{n+1}))$ and $u(\phi_{n+1})$ diverge for
emissions at scales below the cutoff, and 
$\sigma_{R-A-\text{PS}}$ cannot be used in this form for event generation. 

Therefore, at this point, {\it all NLO matching algorithms} supplement the matched cross section of 
Eq.~(\ref{eq:showermatched}) by an additional contribution which avoids
the divergences individually contained in the parton shower evolution starting from
the $n$- and the $(n+1)$-parton configurations. This modified version of the matching subtraction
term has the generic form
\begin{multline}
\label{eq:showermatchedmodified}
\tilde{\sigma}_{R-A-\text{PS}}[u] = \sigma_{R-A-\text{PS}}[u] \\
-\sum_i \int{\rm d}\sigma^{(i)}_X(\phi_{n+1})\left(u(\phi_{n+1})-u(\Phi_{n}^{(i)}(\phi_{n+1})\right)
\theta(Q_0-P_{\perp,n+1}^{(i)}) \ .
\end{multline}
The term ${\rm d}\sigma^{(i)}_X(\phi_{n+1})$ is an additional subtraction (auxiliary)
cross section that is designed to reproduce locally in phase phase the singularities of 
the subtraction terms ${\rm d}\sigma^{(i)}_{A}$,
and the real emission contribution ${\rm d}\sigma_R$. 
We note that the formalism could also be implemented in a different way
by simply removing the shower cut $\theta$-functions in Eq.~(\ref{eq:showermatched})
if care is taken that the parton shower
approximation precisely reproduces, locally in the phase space, 
all singularities in the fixed-order real radiation and subtraction 
cross sections.\footnote{This is
the case for the splitting functions employed in the Powheg formalism~\cite{Frixione:2007vw},
as well as for parton showers featuring full color matrix element
corrections \cite{Platzer:2012np,Hoeche:2011fd} and spin correlations \cite{Collins:1987cp,Knowles:1987cu,Richardson:2001df}.}
This approach, however, just corresponds a particular choice of
${\rm d}\sigma_X$ in the  matching subtraction
term already shown in Eq.~(\ref{eq:showermatchedmodified}).

Only after the modified subtracted (or 'matched') NLO cross section is constructed,
events can be generated with finite weights and leading to finite cross section.
We stress that depending on the construction of the additional subtraction
in the modified matched NLO cross section it is in principle possible that the parton shower  
{\it allows emissions below the cutoff $Q_0$}. However, the weights in these regions of phase space
are without any logarithmic enhancement. Their contributions are typically very small, but
depend on the form of the auxiliary cross section ${\rm d}\sigma_X$. An important consequence is
that the consistency relation for the total inclusive cross section of  
Eq.~(\ref{eq:matchedconsist1}) is still satisfied, but that
for differential cross sections in dynamical kinematic variables $\eta$ such as thrust, 
where $\eta$ refers to the difference to a threshold or an endpoint where linear sensitivity to 
the shower cut can arise,
the relation between parton shower approximation to the NLO cross section and full fixed-order
NLO cross section reads
\begin{equation}
\label{eq:matchedconsist3}
\sigma_{\text{NLO}-\text{PS}}[\text{PS}[u]] = \sigma_{\text{NLO}}[u]\,\Big[1 + {\cal O}\Big((Q_0/\eta),\alpha_s\Big)\Big],
\end{equation}
and the cutoff dependence is linear and can be significant compared to achievable experimental precision. 
It is these contributions which were the focus in the preceedings parts of this paper, and the bottom line is
that in NLO matched partons showers they are still present and do not modify the principle precison with respect to
the unmatched NLL parton showers. 

Within the \Herwig\ 7 event generator's \Matchbox\ module \cite{Platzer:2011bc} 
subtractive (which call MC@NLO-type~\cite{Frixione:2002ik}) as well as multiplicative (which we call 
POWHEG-type~\cite{Frixione:2007vw}) matching can be performed, which both are particular 
incarnations of the matching principles just described above. In the
latter case, however,  an additional matrix element correction is employed in 
Eq.~(\ref{eq:showermatched}) to change the hardest
emission to be described by a splitting function given by the ratio of exact real
emission and Born matrix elements,
\begin{equation}
  \label{eq:powheg}
  P^{(i)}(\phi_n,\tilde{q},z)\to
  \frac{w^{(i)}(\Phi^{(i)}_{n+1}(\phi_n,\tilde{q},z))}{\sum_j
    w^{(j)}(\Phi^{(j)}_{n+1}(\phi_n,\tilde{q},z))} \frac{|{\cal
      M}_R(\Phi^{(i)}_{n+1}(\phi_n,\tilde{q},z)|^2}{|{\cal M}_B(\phi_n)|^2}\,.
\end{equation}
The terms $w^{(i)}$ are weight functions that partition the phase space into
different emitter regions, for which in practice we choose dipole-type
factors, each of which has a collinear divergence only if the emission becomes
collinear to the emitter $i$. Both types of matching schemes employ the NLO subtraction
cross section ${\rm d}\sigma_A^{(i)}$ as the auxiliary cross section ${\rm
  d}\sigma_X^{(i)}$, i.e.\ we have ${\rm d}\sigma_X^{(i)}={\rm
  d}\sigma_A^{(i)}$. This provides a more transparent and stable
implementation of the matched cross section. Furthermore, for the Powheg-type
matching, the first emission off the $n$-parton Born configuration is
generated using the splitting kernel and Sudakov form factor determined with
Eq.~(\ref{eq:powheg}) and the transverse momentum of all subsequent emissions
(with respect to the parent parton momentum) is vetoed not to exceed the transverse
momentum of the first. At this point emissions with larger angles but
transverse momenta smaller than the emission generated according to
Eq.~(\ref{eq:powheg}) are included using in addition a so-called vetoed, truncated
shower \cite{Frixione:2007vw,Hoche:2010kg}.

\begin{figure}
	\center
	\begin{subfigure}[c]{0.49\textwidth}
		\includegraphics[width=1.0\textwidth]{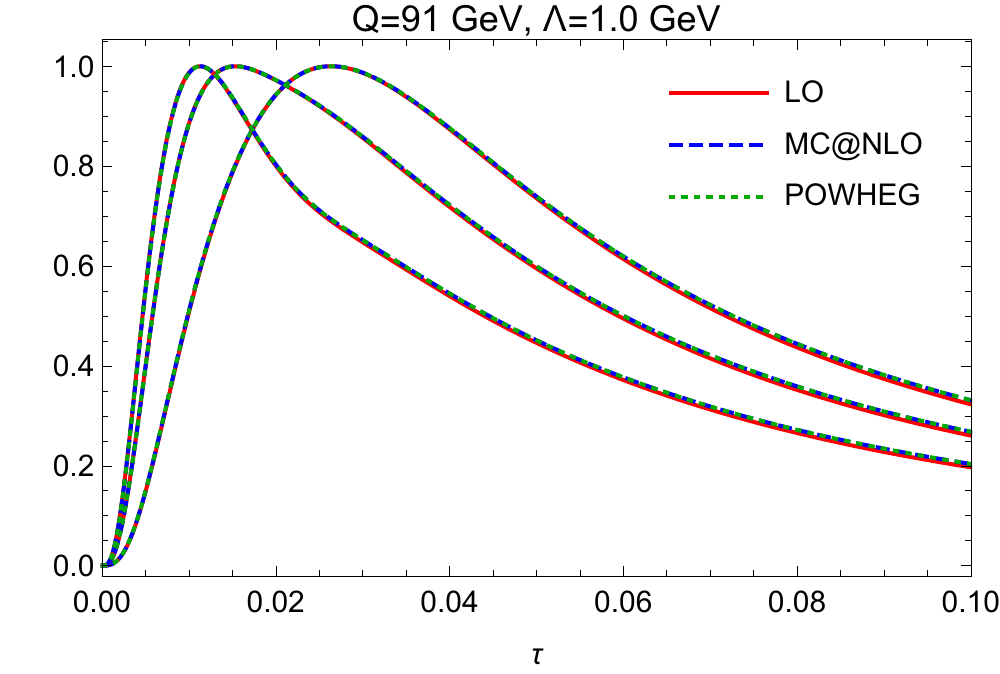}
		\subcaption{}
	\end{subfigure}
	\hfill
	\begin{subfigure}[c]{0.49\textwidth}
		\includegraphics[width=1.03\textwidth]{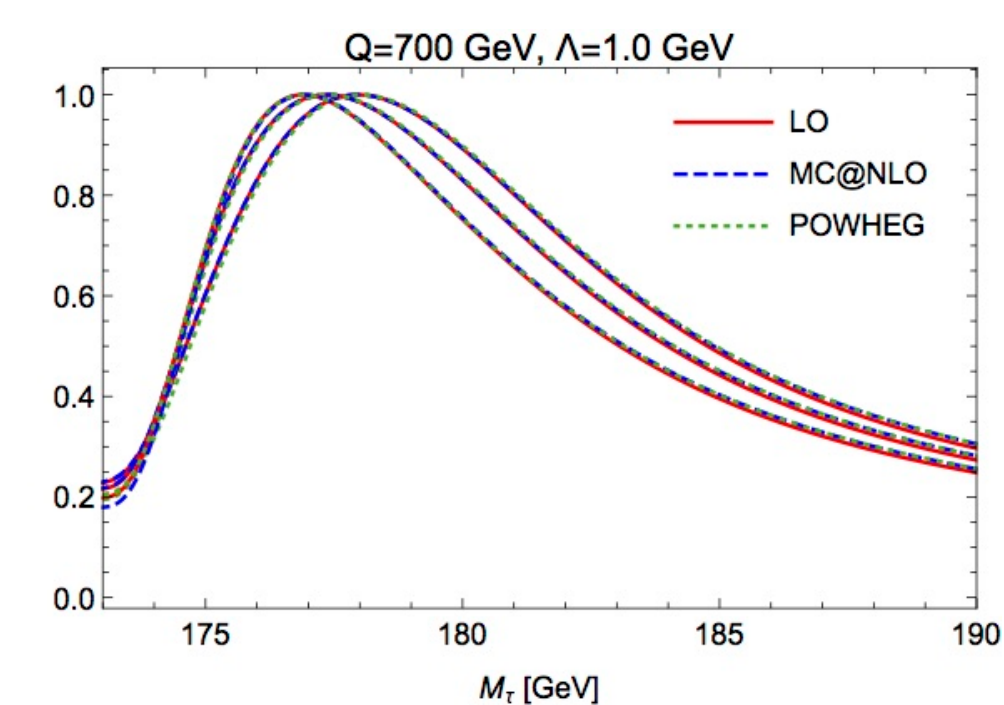}
		\subcaption{}
	\end{subfigure}
	\caption{\label{fig:nlopeakshift}
	 Thrust at the parton level in the peak region generated with \Herwig~7 full simulations for 
	 (a) massless quarks at c.m.\ energy	$Q=91$~GeV and (b) top quarks with mass $m_t=173$~GeV at 
	 $Q=700$~GeV. The parton level results are smeared with a soft model shape function with smearing parameter 
	 $\Lambda=1$~GeV, see Sec.~\ref{sec:shapefunction}. Displayed are simulation results for shower cuts  
	 $Q_0=1.0$~GeV (right set of curves), 
	 $Q_0=1.5$~GeV (middle set of curves) and  $Q_0=2$~GeV (left set of curves)
	 at LO (i.e.\ without any NLO matching, solid red curves), with MC@NLO-type matching (dashed blue curves)
	 and POWHEG-type matching (dotted green curves).
	}
\end{figure}

Let us now compare numerical results obtained with \Herwig\ 7 without NLO matching -- referred to as 'LO' ('leading-order') for
the rest of this section --
(which is the setup we have used for our simulation studies in Secs.~\ref{sec:herwigthrusttest},
\ref{sec:herwigmassless}, \ref{sec:herwigmassive} and \ref{sec:herwigotherobservables})
and with NLO matching using the MC@NLO-type and the POWHEG-type matching.
In Figs.~\ref{fig:nlopeakshift} we show the thrust distribution for massless quark production at $Q=91$~GeV
(left panel) and the rescaled thrust distribution for top quark production with $m_t=173 $~GeV at $Q=700$~GeV 
(right panel) for $Q_0=Q_{0,b}=1.0$~GeV (right set of curves), $1.5$~GeV (middle set of curves) and $2.0$~GeV 
(left set of curves) at LO (solid red), with MC@NLO-type matching (dashed blue curves) and with POWHEG-type matching 
(dotted green curves). All curves are normalized to unity 
at the peak position. We hardly see any difference between the LO and NLO matched simulations in the resonance
regions. Visible effects arise only in the tails away from the resonances, which can be understood from the
fact that the hardest gluon emission, which is improved to full NLO precision by the matching procedure, 
only obtains sizable NLO corrections away from the singular resonance region. In the resonance region, however,
the NLL splitting function approach already provides a fully adequate description and the genuine non-singular
NLO corrections are very small. For the cases
shown in  Figs.~\ref{fig:nlopeakshift} the peak shifts due to NLO effects are typically less than
$\Delta \tau_{\rm peak} \sim 10^{-4}$ in the massless case and less than
$\Delta M_{\tau,\rm{peak}}\sim 100$~MeV in the massive case which is considerably 
smaller than the effects of the shower cut dependence we consider.
We have checked that this property is generic and valid for all energies and shower cut values we have examined
in our earlier studies. The results
confirm that NLO matched parton shower simulations do not add more precision in the thrust resonance region and,
in particular, do not modify the shower cut dependence of the simulations without NLO matching. 

\begin{figure}
	\center
	\begin{subfigure}[c]{0.49\textwidth}
		\includegraphics[width=1.0\textwidth]{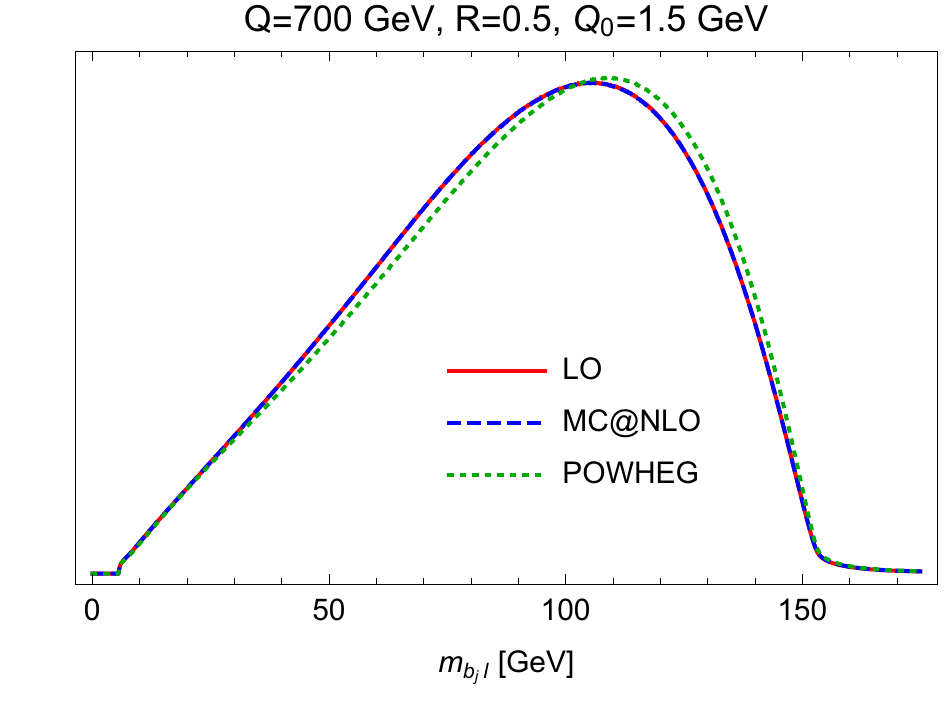}
		\subcaption{}
	\end{subfigure}
	\hfill
	\begin{subfigure}[c]{0.49\textwidth}
		\includegraphics[width=1.0\textwidth]{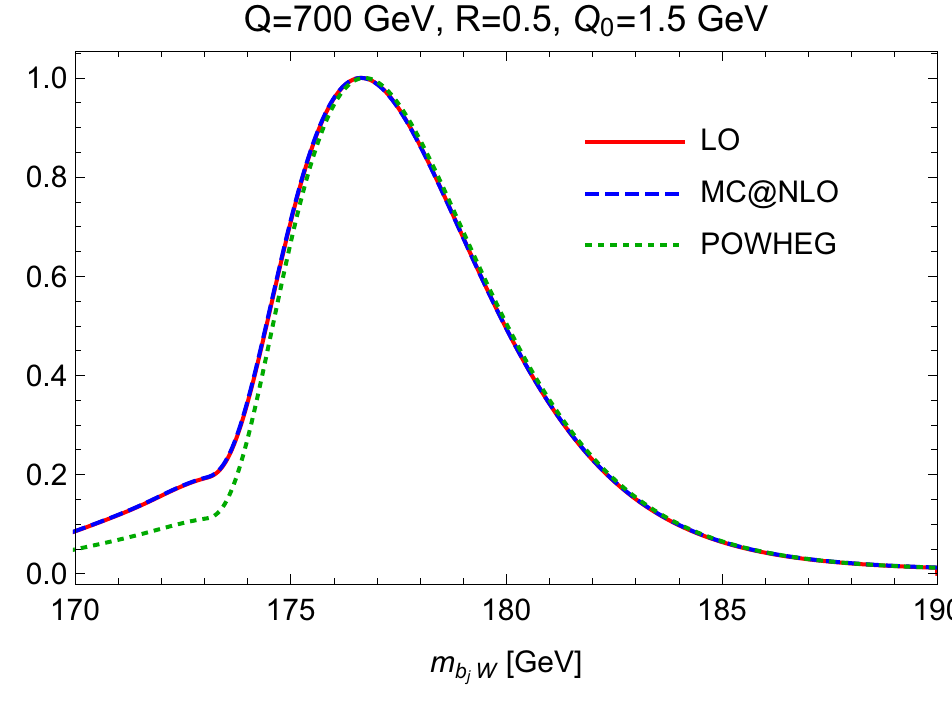}
		\subcaption{}
	\end{subfigure}
	
	\caption{\label{fig:NLOmblmbW}
		The $m_{b_j\ell}$ (left panel) and $m_{b_jW}$ distributions (right panel) generated with \Herwig\ 7
		full simulations
		for $m_t=173$~GeV for $Q=700$~GeV, jet radius $R=0.5$ and Cambridge-Aachen-type $b$-jet clustering.
		Show are results at LO (solid red curves), with MC@NLO-type matching (dashed blue curves) and
		POWHEG-type matching (dotted green curves).
	}
\end{figure}

At this point it is also instructive to examine the impact of NLO matching to the reconstructed observables
$m_{b_j\ell}$ and $m_{b_jW}$, which we have already examined at LO in 
Sec.~\ref{sec:herwigotherobservables}. Within \Herwig\ 7, concerning the description of top quarks, 
the MC@NLO-type matching provides only
NLO improved simulations concerning the production of the top quarks while the POWHEG-type matching  
provides NLO improved simulations concerning the production and the decay of the top quarks, where we
refer to Ref.~\cite{Hamilton:2006ms} for more details. In our LO examination in Sec.~\ref{sec:herwigotherobservables} 
we have already seen that $m_{b_j\ell}$ and $m_{b_jW}$ are quite sensitive to the modeling of the decay for 
$b$-jet clustering for small jet radii as they are used in experimental reconstruction analyses.
At the same time, for small jet radii $m_{b_j\ell}$ and $m_{b_jW}$ are by construction insensitive to details of the top quark 
production. We can therefore expect that the LO and MC@NLO-type simulation results are very similar, while
the POWHEG-type results may receive notable NLO corrections. 
This is shown in Figs.~\ref{fig:NLOmblmbW} where the $m_{b_j\ell}$ (left panel)
and the $m_{b_jW}$ distributions are displayed for $Q=700$~GeV, $m_t=173$~GeV, $R=0.5$ and $Q_0=Q_{0,b}=1.5$~GeV 
at LO (solid red curve), with MC@NLO-type
matching (dashed blue curve) and POWHEG-type matching (dotted green curve). As expected, we see that the MC@NLO-type
matching for top production has essentially no impact, while we find visible effects in the distribution 
for POWHEG-type matching.
However, in the top mass sensitive regions these are substantially smaller for $m_{b_jW}$ than for $m_{b_j\ell}$, 
which is particularly conspicuous when comparing the curves in Figs.~\ref{fig:NLOmblmbW} to the corresponding ones 
shown in Figs.~\ref{fig:mblmbW1}, where the dependence on the top quark mass was illustrated.

\begin{figure}
	\center
	\begin{subfigure}[c]{0.49\textwidth}
		\includegraphics[width=1.0\textwidth]{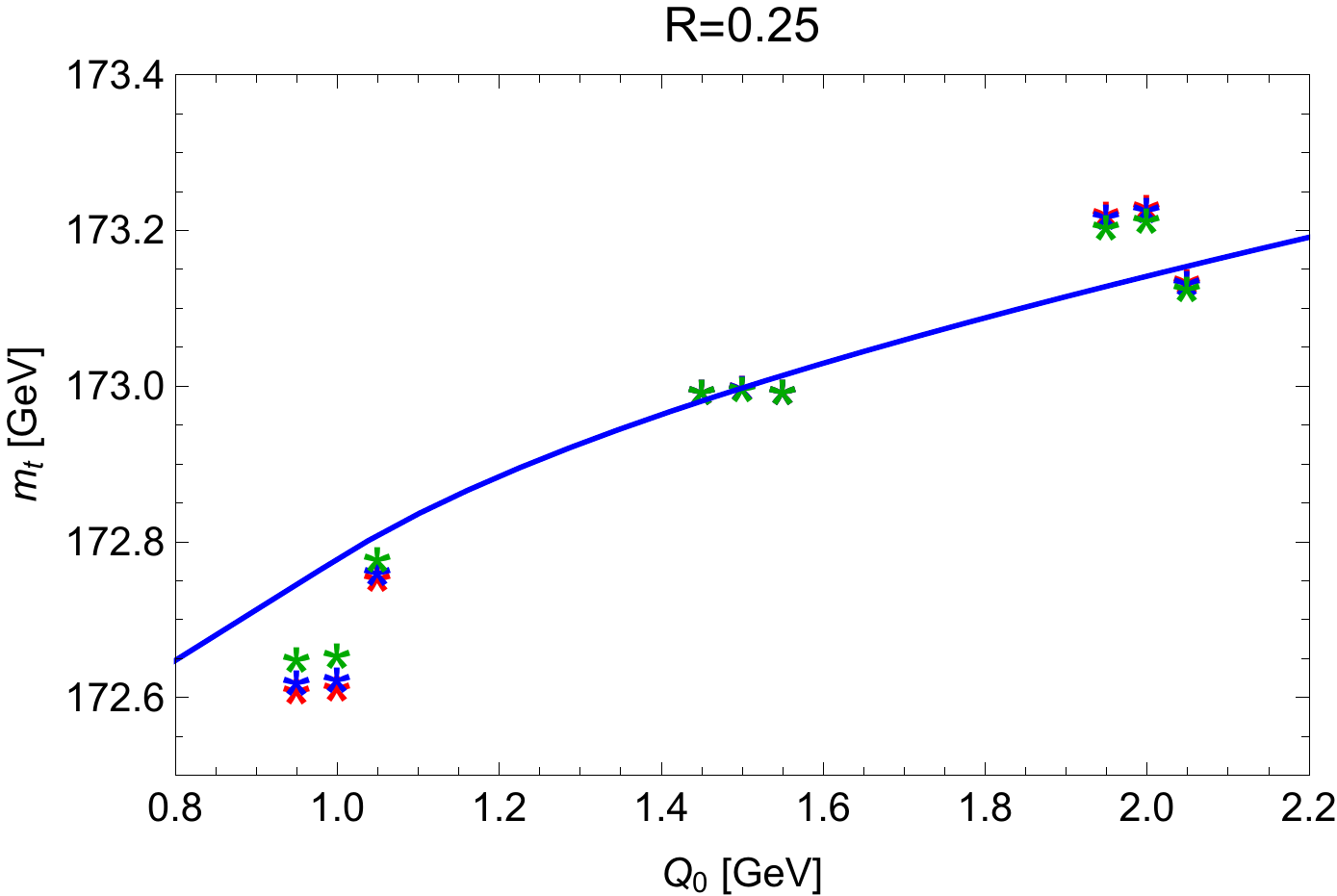}
		\subcaption{}
	\end{subfigure}
	\hfill
	\begin{subfigure}[c]{0.49\textwidth}
		\includegraphics[width=1.0\textwidth]{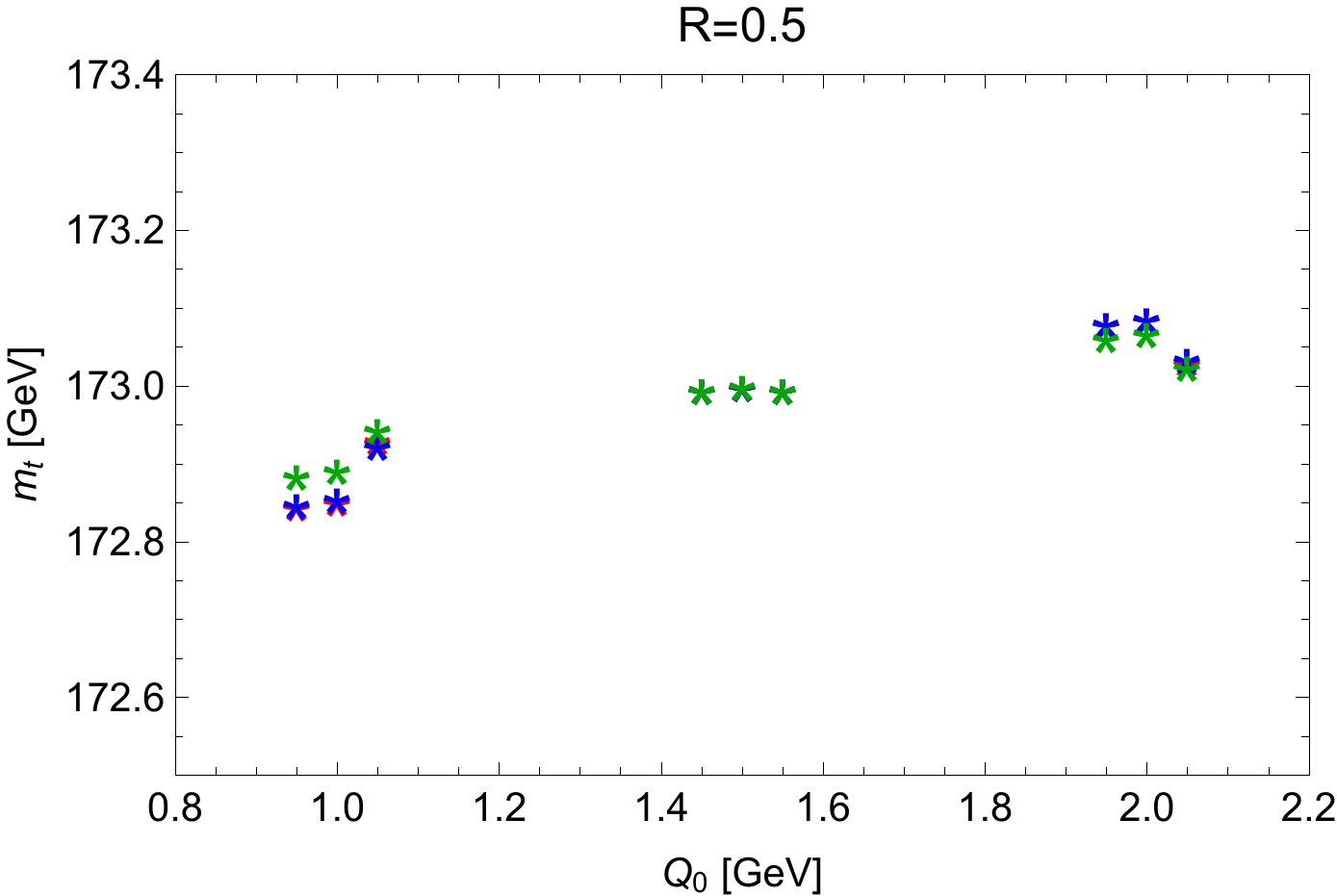}
		\subcaption{}
	\end{subfigure}
	\begin{subfigure}[c]{0.49\textwidth}
		\includegraphics[width=1.0\textwidth]{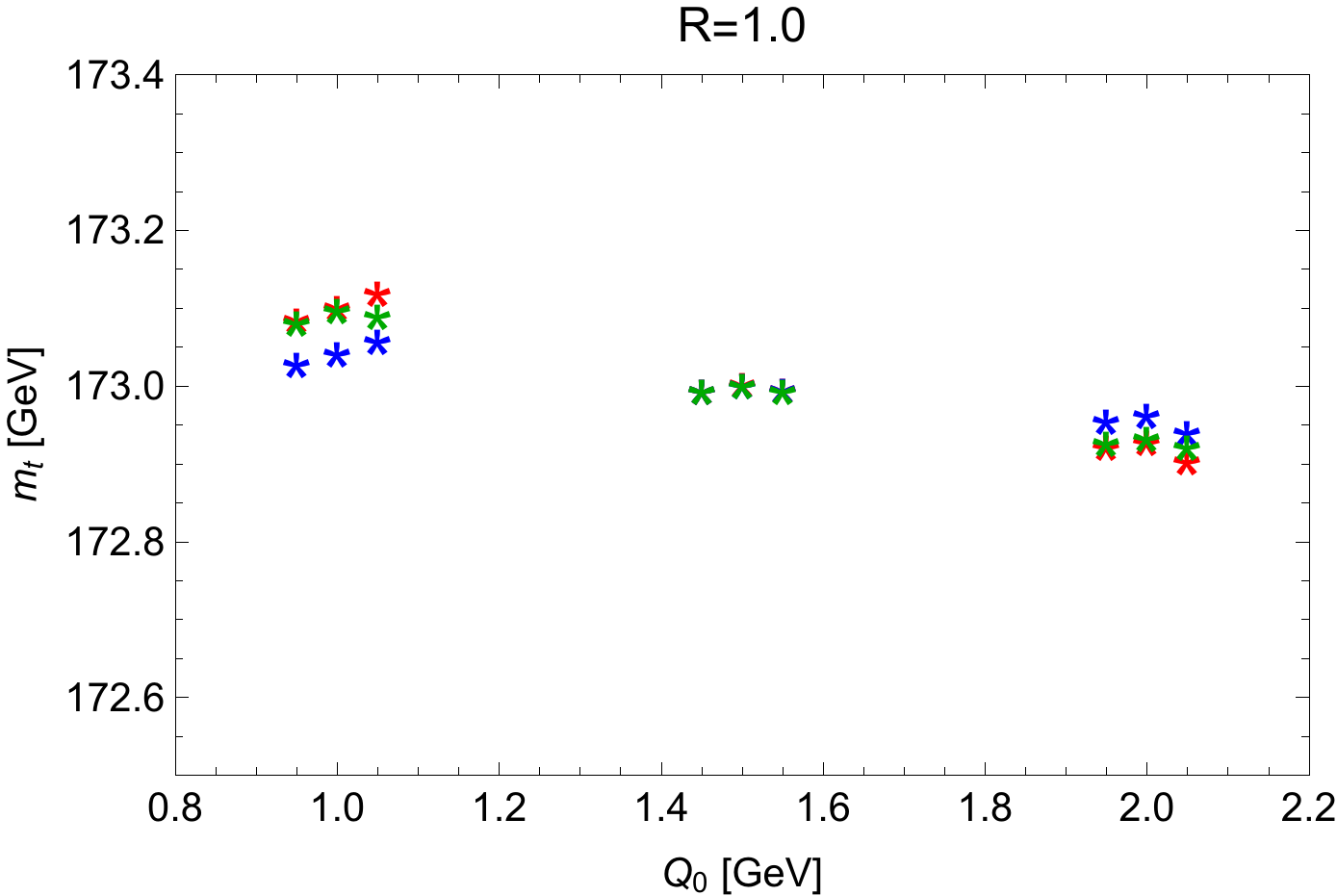}
		\subcaption{}
	\end{subfigure}
	\hfill
	\begin{subfigure}[c]{0.49\textwidth}
		\includegraphics[width=1.0\textwidth]{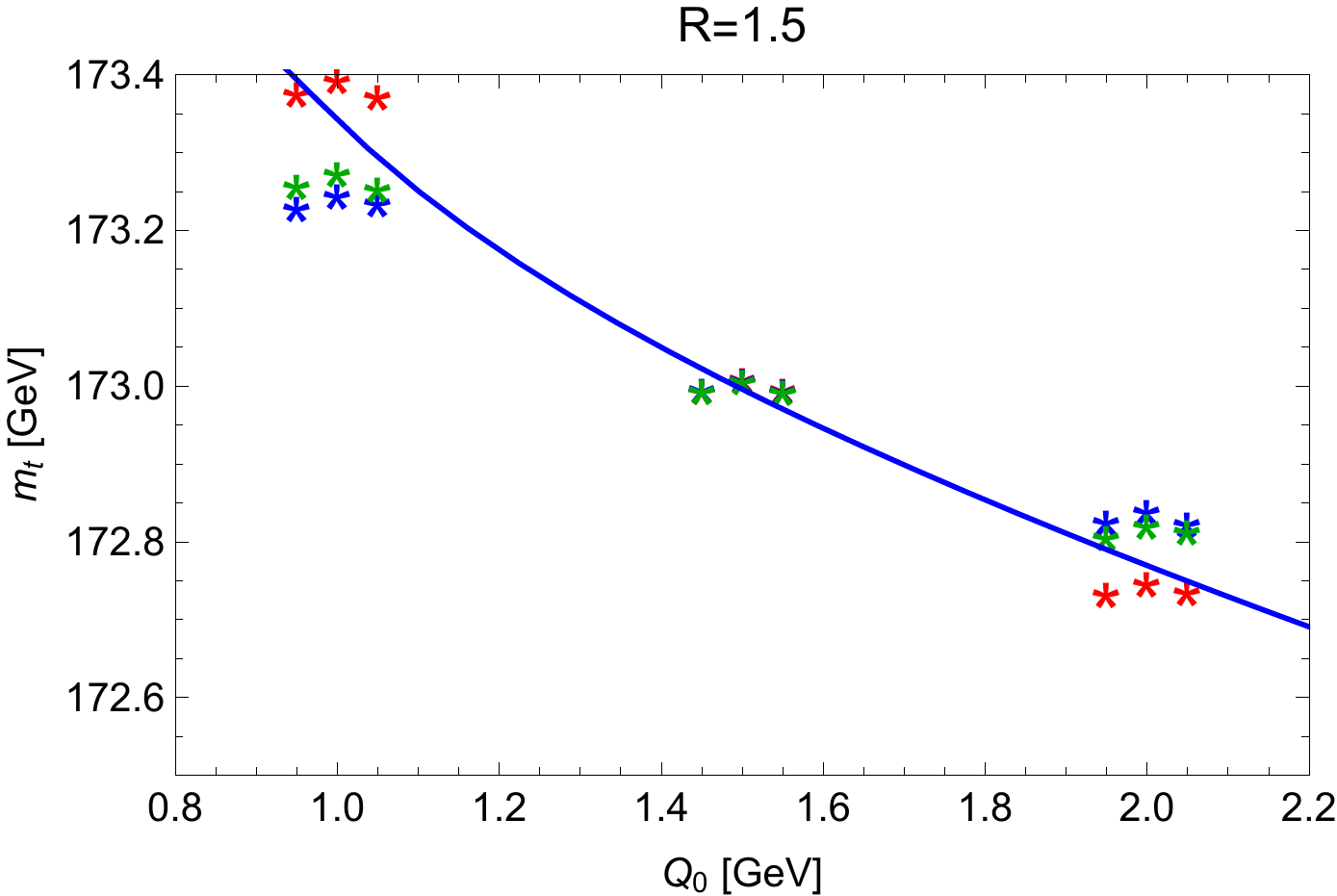}
		\subcaption{}
	\end{subfigure}
	\caption{
		Fitted top quark mass obtained from the $m_{b_j\ell}$ endpoint region for shower cut values
		$Q_0=Q_{0,b}=1.0$, $1.5$ and $2.0$~GeV for $Q=700$~GeV 
		using the $k_t$-type algorithm (green), the Cambridge-Aachen algorithm (blue) and the anti-$k_t$-type algorithm (red)
		for $b$-jet clustering.  For each $Q_0$ value the respective left set of symbols come from the LO simulations, 
		the respective middle set of symbols come from simulations with MC@NLO-type 
		matching and the respective right set of symbols from simulations with POWHEG-type matching. 
		Displayed are the results for $b$-jet radii $R=0.25$, $0.5$, $1.0$ and $1.5$. The solid blue 
		line in the lower right panel corresponds to Eq.~(\ref{eq:mtfitQ0dependence}) and the one
		in the upper left panel corresponds to Eq.~(\ref{eq:mtfitQ0dependencev2}) using $Q_0^\prime=1.5$ as the reference
		scale.
		\label{fig:mblNLOQ0Q0b}
	}
\end{figure}

\begin{figure}
	\center
	\begin{subfigure}[c]{0.49\textwidth}
		\includegraphics[width=1.0\textwidth]{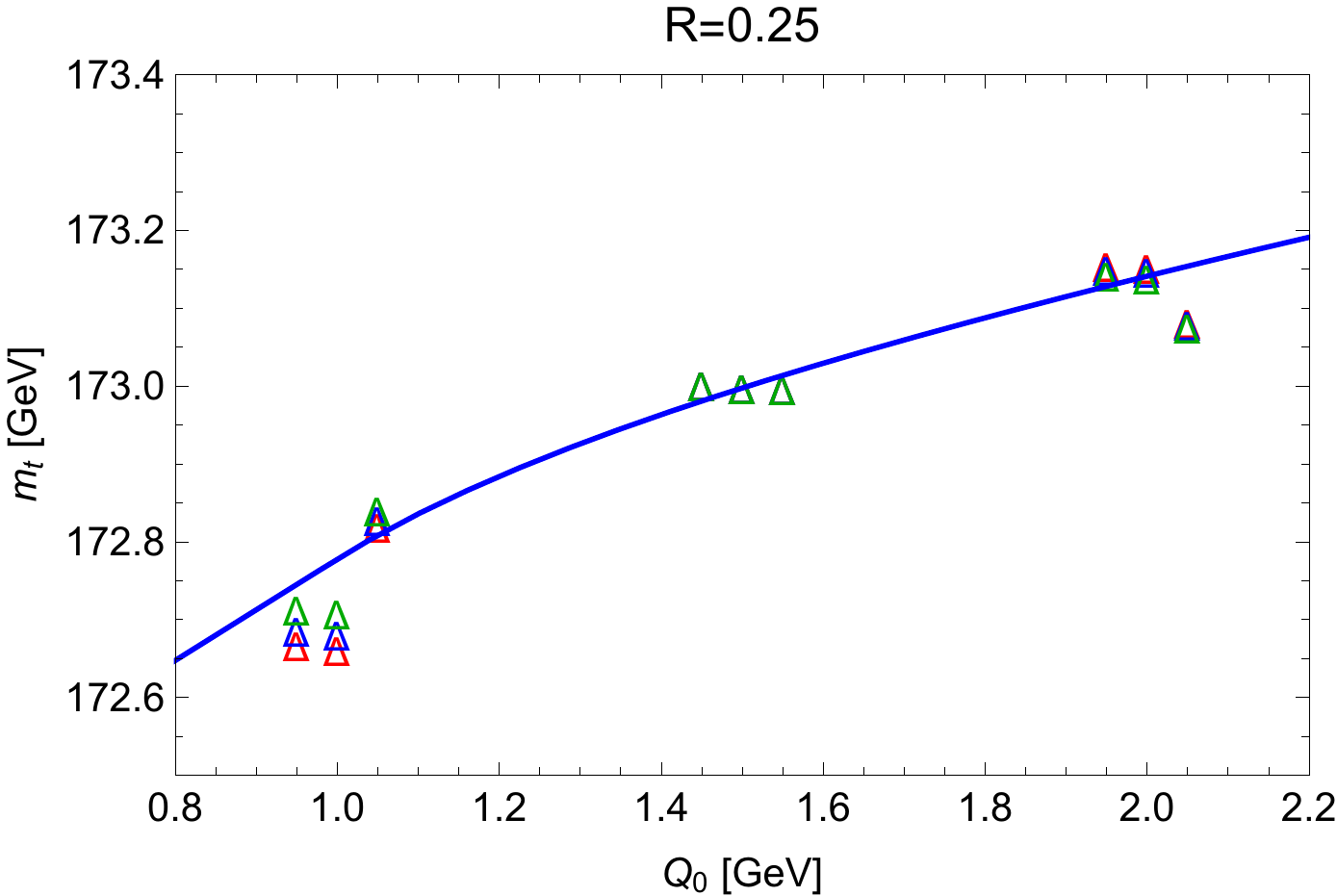}
		\subcaption{}
	\end{subfigure}
	\hfill
	\begin{subfigure}[c]{0.49\textwidth}
		\includegraphics[width=1.0\textwidth]{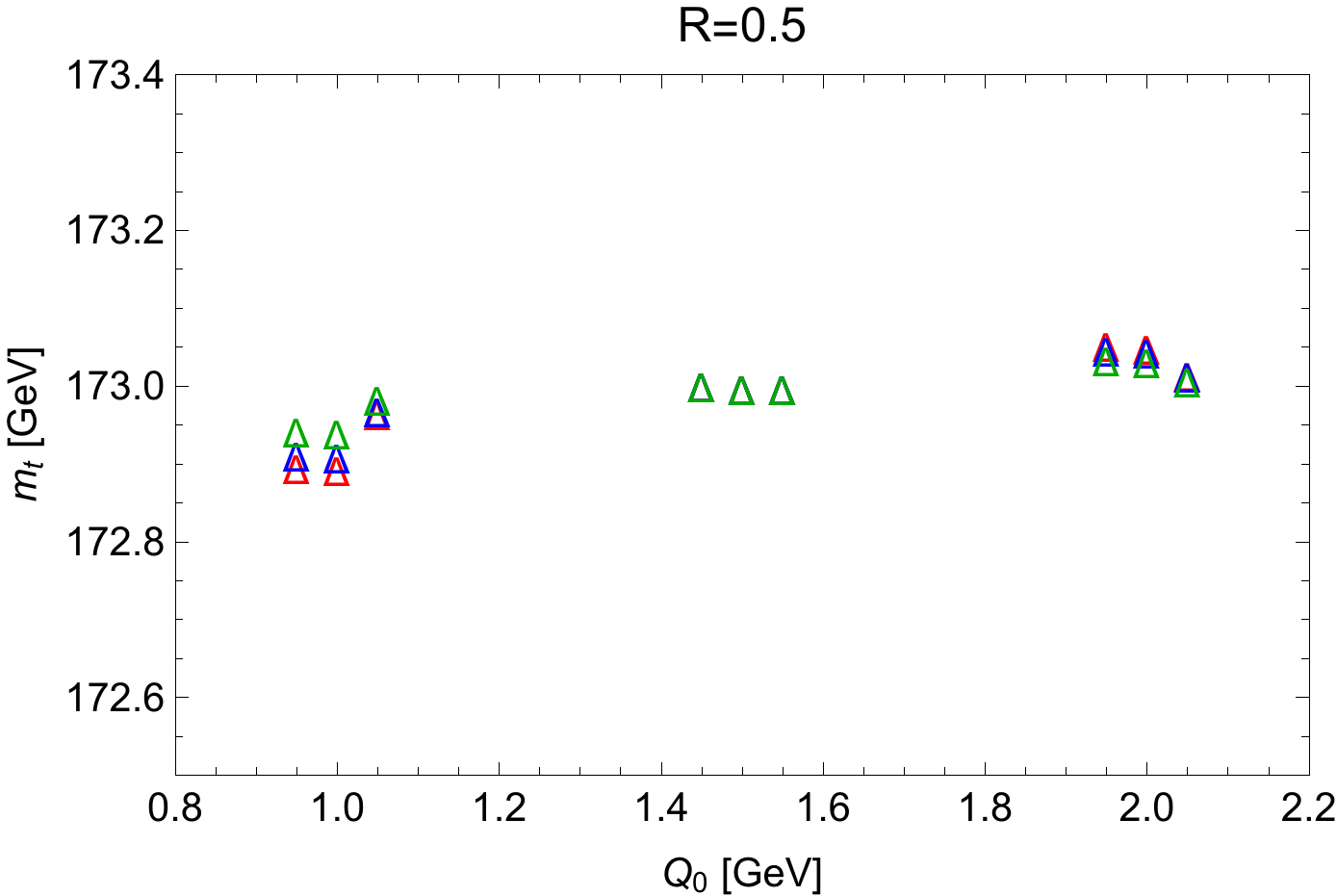}
		\subcaption{}
	\end{subfigure}
	\begin{subfigure}[c]{0.49\textwidth}
		\includegraphics[width=1.0\textwidth]{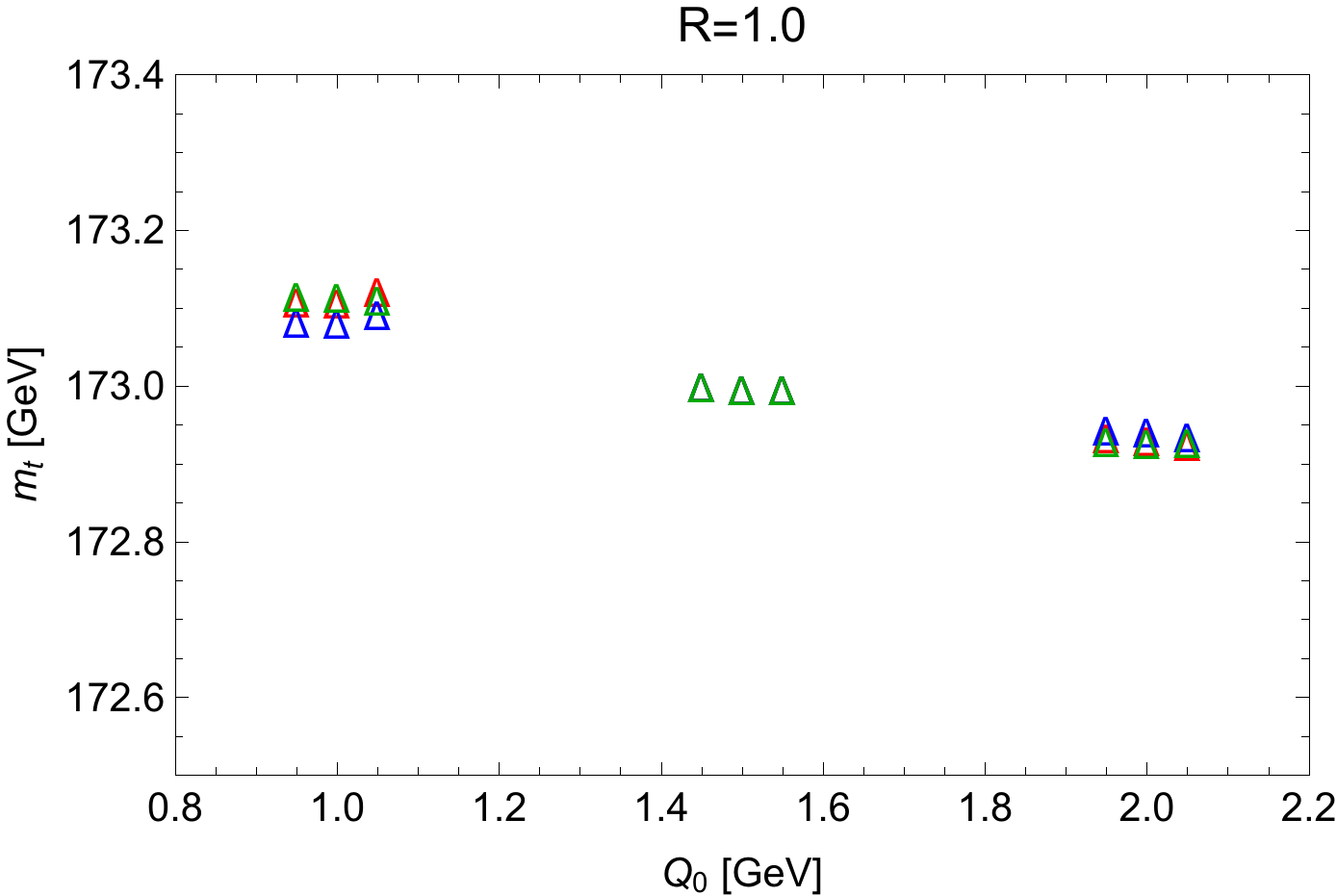}
		\subcaption{}
	\end{subfigure}
	\hfill
	\begin{subfigure}[c]{0.49\textwidth}
		\includegraphics[width=1.0\textwidth]{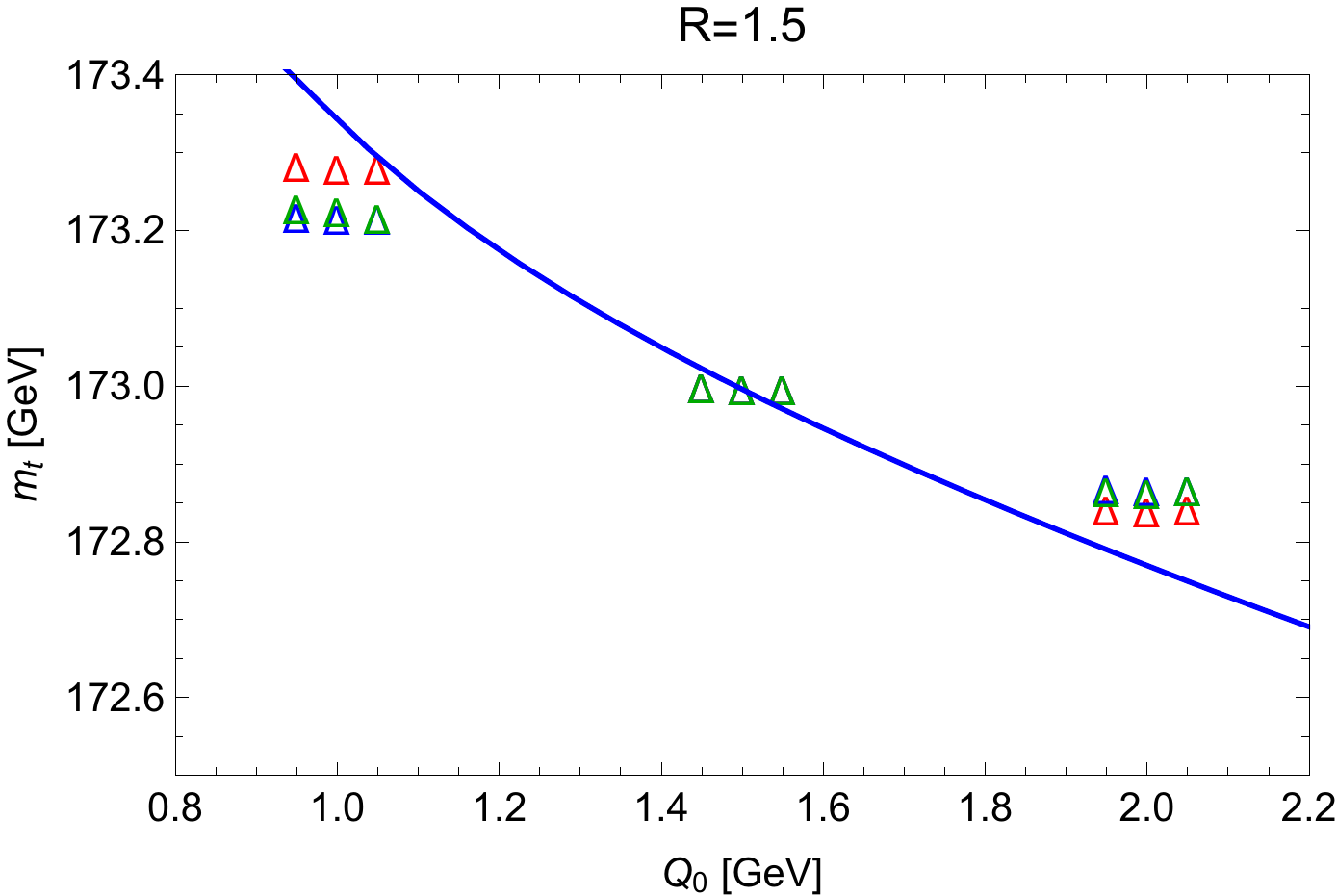}
		\subcaption{}
	\end{subfigure}
	\caption{
		Fitted top quark mass obtained from the $m_{b_jW}$ resonance region for shower cut values
		$Q_0=Q_{0,b}=1.0$, $1.5$ and $2.0$~GeV for $Q=700$~GeV 
		using the $k_t$-type algorithm (green), the Cambridge-Aachen algorithm (blue) and the anti-$k_t$-type algorithm (red)
		for $b$-jet clustering.  For each $Q_0$ value the respective left set of symbols come from the LO simulations, 
		the respective middle set of symbols come from simulations with MC@NLO-type 
		matching and the respective right set of symbols from simulations with POWHEG-type matching. 
		Displayed are the results for $b$-jet radii $R=0.25$, $0.5$, $1.0$ and $1.5$. The solid blue 
		line in the lower right panel corresponds to Eq.~(\ref{eq:mtfitQ0dependence}) and the one
		in the upper left panel corresponds to Eq.~(\ref{eq:mtfitQ0dependencev2}) using $Q_0^\prime=1.5$ as the reference
		scale.
		\label{fig:mbWNLOQ0Q0b}
	}
\end{figure}

Focusing on the shower cut dependence of the kinematic top mass sensitivity of $m_{b_j\ell}$ and $m_{b_jW}$ we again
use the approach of Sec.~\ref{sec:herwigotherobservables} by fits of the top quark mass with respect to the default
shower cut setting $Q_0=Q_{0,b}=1.5$~GeV (see the paragraph prior to Eq.~(\ref{eq:mtfitQ0dependence}) 
in Sec.~\ref{sec:herwigotherobservables} for the description of the fitting approach). 
In Figs.~\ref{fig:mblNLOQ0Q0b} and Figs.~\ref{fig:mbWNLOQ0Q0b}
the dependence of the fitted top mass obtained from the $m_{b_j\ell}$ endpoint region (stars)
and from the $m_{b_jW}$ resonance region (triangles), respectively, is displayed at LO and with NLO matching using the  
same settings as in Figs.~\ref{fig:mblmbWQ0Q0b} where we only displayed the LO results. 
We again show the results for shower cuts $Q_{0,b}=Q_{0}=1.0$, $1.5$ and $2.0$~GeV for jet radii 
$R=0.25$ (upper left panels), $R=0.5$ (upper right panels), $R=1.0$ (lower left panels) and $R=1.5$ (lower right panels),
and we have carried out the analyses for $b$-jet clustering using the
$k_t$-type algorithm (green symbols), the Cambridge-Aachen-type algorithm (blue symbols) and the anti-$k_t$-type algorithm (red symbols). 
To allow for an easier visual identification of the results we have slightly displaced the symbols horizontally,
where for each $Q_0$ value the respective left set of symbols come from the LO simulations (already 
displayed in Figs.~\ref{fig:mblmbWQ0Q0b}), the respective middle set of symbols come from simulations with MC@NLO-type 
matching and the respective right set of symbols from simulations with POWHEG-type matching. 

We see that the NLO matching has essentially 
no impact on the fitted top mass for large jet radii and the cutoff dependence agrees again very well with 
Eq.~(\ref{eq:mtfitQ0dependence}), which is displayed in the lower right panel ($R=1.5$) as the solid blue line with 
$Q_0^\prime=1.5$~GeV as the reference scale. This is expected since $m_{b_j\ell}$ and $m_{b_jW}$ with large 
$b$-jet clustering radius are by construction neither sensitive to the top production mechanism and nor to 
details of the top quark decay. 
It is conspicuous, however, that there is also very good agreement between the LO and NLO fitted top masses 
for small jet radii. For comparison we have displayed again Eq.~(\ref{eq:mtfitQ0dependencev2}) with
$Q_0^\prime=1.5$~GeV as the reference scale in the upper left panel ($R=0.25$). We recall that Eq.~(\ref{eq:mtfitQ0dependencev2})
describes the expected shower cut dependence for $R\sim m_t/Q$ with the contributions coming from large angle soft radiation being 
removed while those from the ultra-collinear radiation being kept. So we see that, even though the POWHEG-type matching
has sizable nominal effects on the distributions for the reconstructed observables, particularly for $m_{b_j\ell}$, the relative shower cut 
dependence itself it essentially unchanged. 

This outcome 
again fully supports the idea of universality of the shower cut dependence and its independence concerning
NLO matched predictions, and it
is precisely what is to be expected if the equivalence of the
MC generator top mass and the shower cut dependent CB mass of Eq.~(\ref{eq:CBmassschemedef2}) is universal. 
However, as already noted in Sec.~\ref{sec:herwigotherobservables}, a strict proof would require a
a thorough quantitative (and preferably analytic) understanding of the $b$-jet clustering 
for exclusive observables such as $m_{b_j\ell}$ and $m_{b_jW}$ to unambiguously track the shower cut dependence.
We emphasize again, that such quantitative understanding should at best be achieved in the context of
a QCD factorization approach as it allows for a direct, clean and unambiguous field theoretical association of the different types of 
radiation concerning dynamical physical effects and contributions affecting the interpretation of QCD parameters such
as the top quark mass.

\section{Conclusions}
\label{sec:conclusions}

The emergence of infrared divergences and their proper treatment to achieve meaningful physics predictions
represents one of the major conceptual and technical issues in modern applications of perturbative QCD in 
the context of collider physics. These divergences emerge in partonic computations in the (unphysical) 
limit of infinitesimally small
resolution concerning infrared energies and momenta and are resolved by treating partonic configurations 
below the resolution scale as contributions to the same observable configuration. Within this approach, 
infrared cuts used to regulate the infrared divergences at the intermediate steps of the perturbative 
calculations, can then be sent to zero, where the limit of this procedure typically defines what is commonly 
perceived as the perturbative
component of cross section predictions. In the context of multi-purpose MC event generators, where the parton 
showers are responsible for the description of the parton dynamics below the hard interaction scale, the same 
principles are applied. However, the infrared shower cut $Q_0$ which terminates the parton shower evolution is 
finite, typically in the range of $1$~GeV, and leads to a power-like dependence of the parton-level predictions on $Q_0$
depending on the mass dimension and the infrared sensitivity of the observable. As we have discussed in this work for observables with kinematic top mass
sensitivity this dependence on the shower cut $Q_0$ turns out to be linear and non-negligible 
given that the current experimental precision in top quark mass determinations based on direct reconstruction methods 
already reached the level of $0.5$~GeV.

In this work we analyzed in detail the role of the shower cut $Q_0$ in angular ordered parton showers based 
on the coherent branching formalism \emph{for quasi-collinear, i.e. boosted, massive quarks} at NLL. We have demonstrated, using an 
eventshape-type observable based on hemisphere masses and closely related to thrust (see Eqs.~(\ref{eq:taudef}) 
and (\ref{eq:Mtaudef})) in the resonance region where the highest kinematic top mass sensitivity is located, that 
the finite shower cut
automatically implies that the generator top quark mass is the $Q_0$-dependent 
{\it coherent branching} (CB) mass, $m_t^{\rm MC}=m_t^{\rm CB}(Q_0)$, even though the underlying analytic expressions that go into the formulation of 
the parton shower are derived in the pole mass scheme.  
The CB mass is a low-scale short-distance mass and free of an ${\cal O}(\Lambda_{\rm QCD})$ renormalon ambiguity.
At ${\cal O}(\alpha_s)$ its relation to the pole mass $m_t^{\rm pole}$ reads
\begin{equation}
\label{eq:conclusions1}
m_t^{\rm CB}(Q_0) - m_t^{\rm pole} \, = \, - \frac{2}{3}\,\alpha_s(Q_0)\,Q_0 \, + \,{\cal O}(\alpha_s^2 Q_0)\,.
\end{equation}
The inclusion of NLO corrections in the context of NLO-matched parton showers does not add more precision 
to this relation as the additional NLO information improves the perturbative description of configurations that 
are located outside the resonance region, i.e. outside the region where the main kinematic top mass sensitivity arises. 
In Sec.~\ref{sec:intermediatesummary} we have provided a detailed
summary of all our theoretical findings and in Sec.~\ref{sec:herwigcompare} we have confronted them with parton-level
simulations carried out using the \Herwig\ 7 event generator. The simulation results fully confirm our
conceptual conclusions concerning the equivalence of the generator top mass and the shower-cut-dependent
CB mass, and we also gathered evidence that the equivalence is universal and also applies for other
more exclusive observables such as the $b$-jet and lepton invariant mass $m_{b_j\ell}$ and the
reconstructed top invariant mass $m_{b_jW}$ in the limit of boosted top quarks. In the course of our examinations we also analyzed in detail the
shower cut dependence coming from large-angle soft radiation which is universal for the
production of massless quarks and boosted top quarks and which represents an interface to the hadronization model
used in the MC event generator. These results have implications for the hadronization 
corrections in event-shape distribution and the extraction of $\alpha_s$ which we have, however, 
not addressed in this work and will be discussed elsewhere.

To conclude this work we address two important questions which have not been addressed in the main body of the paper. The first 
is about the remaining conceptual issues that have to be resolved to universally explore the meaning of the
MC generator top quark mass in the context of state-of-the-art MC event generators that are used in the 
experimental analyses. The second 
is about how to best
convert $m_t^{\rm CB}(Q_0)$ to other top quark mass schemes. This issue gains particular importance if one assumes that
the MC top quark mass $m_t^{\rm MC}$ determined in direct reconstruction methods at hadron colliders
is indeed equal to the coherent branching mass $m_t^{\rm CB}(Q_0)$.

Before we address these issues we would like to emphasize that the proper field theoretic 
specification of the generator top quark mass $m_t^{\rm MC}$ as a particular mass renormalization scheme 
does not touch in any way the important questions how MC modeling uncertainties such as
for the description of multi parton interactions or relevant for the event selection and
the description of hadronization effects in the final states such as color reconnection
affect the top mass measurements. 
These uncertainties are and shall continue to be under scrutiny, and their study may lead to improved  
MC generators in the future. The focus of the present work, on the other hand, 
is that the principle field theoretic meaning of the 
cutoff-dependence of the generator top quark mass can be studied and resolved independently of 
these issues and thus deserves particular attention by itself. Associated dedicated studies cover subtle effects
that are, however, already relevant in view of the current experimental uncertainties in top quark mass
determinations and may in a complementary way contribute to improved MC generators. 

Let us now address the first issue.
The basic simplifications for the examinations carried out in this work were that we used
(i) parton level studies, 
(ii) the narrow width approximation, 
(iii) boosted (quasi-collinear) top quark kinematics
and (iv) hemiphere masses in $e^+e^-$ collisions closely related to the thrust/2-jettiness event-shape.
In the context of hemisphere mass studies, the extension to MC hadron level studies is straightforward and shall
be carried out in forthcoming works. Here the main question to be addressed is how well the MC hadronization models
are compatible with the parton-hadron level factorization of Eqs.~(\ref{eq:thrustmassless1}) and (\ref{eq:thrustmassive1}),
which is an intrinsic property of QCD. The main point then to be clarified is, whether MC hadronization models have the 
capability to retune the top quark mass -- a property that would make the MC top quark generator mass a hadronization
parameter und mean that there are additional MC dependent non-perturbative contributions that have to be accounted for in the relation between the MC generator top mass $m_t^{\rm MC}$ and the coherent branching mass $m_t^{\rm CB}(Q_0)$. 
Concerning the narrow width approximation, we note that state-of-the-art
parton showers for massive quarks do not have the capability to describe unstable particle effects from first principles. 
These unstable particle effects include the top quark intrinsic Breit-Wigner smearing of its 
invariant mass as well as interference effects connecting
top and non-top processes through equivalent top decay final states. Accounting for unstable particle effects may
be possible in the context of MC generators matched to calculations including the full top production and decay 
process or in the context of new MC generators which incorporate unstable particle effects in terms of systematic expansions
that are more general than the narrow width approximation. Such studies are in reach and may be addressed in the near future.
Concerning the approximation of boosted top quark kinematics, imagining systematic studies for slow top quarks 
comparable to the examinations carried out in this work, 
the prospects are far less clear. This is because the existing parton shower formalisms for massive quarks
are by construction designed to be valid in the quasi-collinear limit - even though the bulk of the top quarks entering
current experimental analyses have relatively low transverse momenta between $50$ and $100$~GeV and cannot be considered to
be quasi-collinear. So progressing into this 
direction involves general studies of the MC modeling for top quarks that in principle go beyond the problem of the MC 
top quark generator mass. Finally, concerning the extension of examinations at the level of those carried out in this work 
to other types of observables covering also hadronic collisions at the LHC, such studies require the development of 
new types of factorization theorems. For groomed fat jet masses for boosted top quark production at the LHC a factorization
approach was recently developed~\cite{Hoang:2017kmk}, but factorization theorems for more exclusive variables 
such as $m_{b_j\ell}$ or $m_{b_jW}$, which are currently absent, are desirable as well. 
Furthermore, pushing the existing factorization approach for 
thrust and the description of the shower cut dependence to one higher order would be useful as well since it would allow for
an explicit check of the ${\cal O}(\alpha_s^2)$ corrections to relation (\ref{eq:conclusions1}).

Let us now address the second issue.
Assuming that the currently most precise top quark mass measurements of $m_t^{\rm MC}$ can be identified with
a measurement of the CB mass $m_t^{\rm CB}(Q_0)$ defined in relation~(\ref{eq:conclusions1}), how well can it be converted
to other mass schemes? Given that most theoretical predictions for top quark physics at the LHC are 
carried out in the pole mass scheme, one may simply convert the CB mass to the pole mass using 
Eq.~(\ref{eq:conclusions1}). The \Herwig\ 7 event generator uses $Q_0=1.25$~GeV as the default value for the shower
cut, and using the $\overline{\text{MS}}$ scheme for the strong coupling 
with $\alpha_s^{\overline{\text{MS}},(n_f=5)}(M_Z)=0.118$ we obtain 
$m_t^{\rm pole}-m_t^{\rm CB}(Q_0=1.25~\mbox{GeV})=330$~MeV, where we have evaluated the strong coupling 
in the 3-flavor scheme using $\alpha_s^{(n_f=5)}(M_Z)=0.118$ as the input.
On the other hand, using the Monte Carlo (MC) scheme for $\alpha_s$, 
which accounts for the two-loop cusp anomalous dimension 
contained in the NLL quark splitting function and which is effectively used in \Herwig\ 7 
(see Eq.~(\ref{eq:MCalphas}) in Sec.~\ref{sec:coherentbranchingmassless} on the MC scheme for the strong coupling), 
we obtain $m_t^{\rm pole}-m_t^{\rm CB}(Q_0=1.25~\mbox{GeV})=520$~MeV. The difference of about $200$~MeV 
between both conversions can be viewed as an illustration of the currently unknown 
${\cal O}(\alpha_s^2)$ corrections and indicates that the convergence is not particularly good. This is, 
however, expected since the pole mass has an
${\cal O}(\Lambda_{\rm QCD})$ renormalon ambiguity. From the analysis of Ref.~\cite{Hoang:2017btd}, where
a determination of the pole mass from a short-distance mass at the scale $1.3$~GeV was studied in detail, 
we can expect that ${\cal O}(\alpha_s^2)$ and ${\cal O}(\alpha_s^3)$ corrections are also needed to determine the
pole mass value and that at  ${\cal O}(\alpha_s^3)$ there is a remaining irreducible uncertainty of around 
$250$~MeV due to the ${\cal O}(\Lambda_{\rm QCD})$ renormalon ambiguity of the pole mass 
(see also \cite{Beneke:2016cbu} for an alternative view on the size of the renormalon ambiguity of 
the pole mass). Thus the determination of the currently unknown ${\cal O}(\alpha_s^2)$ and 
${\cal O}(\alpha_s^3)$ corrections to Eq.~(\ref{eq:conclusions1}) is important to reliably determine the pole mass.
To determine the ${\cal O}(\alpha_s^2)$ corrections in the factorization approach the effects of the 
shower cut need to be implemented into the bHQET jet function at ${\cal O}(\alpha_s^2)$. To determine the 
${\cal O}(\alpha_s^2)$ corrections in the context of coherent branching formalism (or angular ordered parton showers)
the effects of the shower cut have to be analyzed in the context of a fully consistent 
next-to-next-to-leading order evolution. 
The overall conclusion is that the difference between the pole mass and the CB mass,
$m_t^{\rm pole}-m_t^{\rm CB}(Q_0=1.25~\mbox{GeV})$, is likely at least as large as the current uncertainties 
in top quark mass measurements from direct reconstruction of around $500$~MeV (see Sec.~\ref{sec:prelude}) and
requires the determination of two and three-loop corrections. Even when these corrections become available,
there will be is an irreducible uncertainty of $250$~MeV. So, for a reliable determination of the pole mass the unknown
higher order corrections to Eq.~(\ref{eq:conclusions1}) are very important, and the ultimate uncertainty in the pole mass is 
at the same level as the precision of $200$~MeV that may be achieved for measurements of $m_t^{\rm MC}$
in the future high-luminosity run of the LHC~\cite{CMS:2013wfa,CMS:2017gvo}.

Alternatively, since physical observables are not tied conceptually to the pole mass scheme in any way and its
${\cal O}(\Lambda_{\rm QCD})$ ambiguity is a pure artefact of the pole mass definition, one 
can as well parametrize calculations using suitable short-distance top quark mass schemes. In this approach the sizeable 
corrections and the renormalon ambiguity associated to the pole mass scheme -- as well as any controversial discussion
on the actual size of this ambiguity -- can be avoided entirely. To illustrate this approach
let us consider a determination of the MSR mass $m_t^{\rm MSR}(Q_0)$
from a given value of the CB mass $m_t^{\rm CB}(Q_0)$. At ${\cal O}(\alpha_s)$ the relation between the 
scale-dependent MSR mass~\cite{Hoang:2008yj,Hoang:2017suc}  and the pole mass
reads $m_t^{\rm MSR}(Q_0) - m_t^{\rm pole}= - 4\,\alpha_s(Q_0)Q_0/(3\pi)$. For the strong coupling in the
$\overline{\text{MS}}$ scheme this gives $m_t^{\rm MSR}(Q_0) -m_t^{\rm CB}(Q_0=1.25~\mbox{GeV})=120$~MeV compared
to $190$~MeV in the Monte Carlo scheme. As expected from the fact that the difference of MSR and CB masses does
not contain any ${\cal O}(\Lambda_{\rm QCD})$ renormalon ambiguity, the scheme corrections to obtain the MSR mass are 
small, and one can also expect that they exhibit good convergence because MSR and CB masses are both short-distance mass
schemes. The difference of $70$~MeV can be viewed as an illustration of the currently unknown 
${\cal O}(\alpha_s^2)$ corrections and indicates that the knowledge of the two-loop corrections in Eq.~(\ref{eq:conclusions1})
may be sufficient to convert the CB mass to the MSR with a precision of better than $50$~MeV.
Compared to the current uncertainties 
in top quark mass measurements from direct reconstruction of around $500$~MeV these corrections are small and
the knowledge of these two-loop corrections is not required.
Furthermore, as was shown in Ref.~\cite{Hoang:2017suc}, one can convert the MSR mass to all other commonly used
short-distance mass schemes, such as the 1S~\cite{Hoang:1998ng,Hoang:1998hm,Hoang:1999ye}, 
the PS~\cite{Beneke:1998rk} or the $\overline{\text{MS}}$ schemes, with a
precision of $10$~MeV. The overall conclusion is that, when using only short-distance mass schemes, the achievable precision 
in converting the MC/CB mass to other mass schemes is already at this stage 
substantially higher than the current experimental uncertainties
and also than extrapolations concerning the future high-luminosity run of the LHC which indicate that a precision of 
$200$~MeV~\cite{CMS:2013wfa,CMS:2017gvo} for a determination of the top quark mass can be reached.

\section*{Acknowledgments}

We acknowledge partial support by the FWF Austrian Science Fund under the Doctoral
Program (``Particles and Interactions'') No.\ W1252-N27 and by the European Union under the
COST action (``Unraveling new physics at the LHC through the precision
frontier'') No.\ CA16201, and under the Horizon 2020 research and innovation
programme as part of the Marie Sklodowska-Curie Innovative Training Network
MCnetITN3 (grant agreement no. 722104).
We also thank the Erwin Schr\"odinger International Institute for Mathematics and Physics 
for partial support through the 2016 program ``Challenges and Concepts for Field Theory and Applications 
in the Era of LHC Run-2''. We appreciate valuable comments to our manuscript from Marcel Vos as well as from  
Stefan Kluth and Martijn Mulders.

\appendix

\section{Ingredients of for 2-jettiness at NLL from effective field theory}
\label{app:thrustformulae}

\subsection{Resummation for massless quarks in SCET}
\label{app:masslessthrust}

The ${\cal O}(\as)$ the hard, jet and soft functions appearing in the dominant singular partonic contributions
of the 2-jettiness factorization theorem obtained in SCET for massless quarks have the form~\cite{Schwartz:2007ib} 
(see also Refs.~\cite{Becher:2008cf,Abbate:2010xh})
\begin{align}
\label{eq:hardfunction}
H_{Q}(Q,\mu)&=1+\frac{\as(\mu) C_F}{4\pi}\biggl(-2\ln^2\frac{Q^2}{\mu^2}+6\ln\frac{Q^2}{\mu^2}-16+\frac{7\pi^2}{3}\biggr)+\mathcal{O}(\as^2) \,, \\%
\label{eq:scetjetfunction}
J^{(\tau)}(s,\mu)&=\delta(s)+\frac{\as(\mu) C_F}{4\pi}\biggl(\frac{8}{\mu^2}\biggl[\frac{\mu^2\ln\frac{s}{\mu^2}}{s}\biggr]_+-\frac{6}{\mu^2}\biggl[\frac{\mu^2}{s}\biggr]_++(14-2\pi^2)\delta(s)\biggr)+\mathcal{O}(\as^2) \,,\\
\label{eq:scetsoftfunction}
S^{(\tau)}(k,\mu)&=\delta(k)+\frac{\as(\mu) C_F}{4\pi}\biggl(-\frac{16}{\mu}\biggl[\frac{\mu\ln\frac{k}{\mu}}{k}\biggr]_++\frac{\pi^2}{3}\delta(k)\biggr)+\mathcal{O}(\as^2) \,, 
\end{align}
where we have used the notations of Ref.~\cite{Fleming:2007xt}. 
Their respective anomalous dimensions can (to all orders) be written in the form
\begin{align}
\label{eq:hardfctanomdim}
\mu\frac{d}{d\mu} U_H(Q,\mu_H,\mu)\,  = \,&\,\Big( \Gamma_H[\alpha_s(\mu)]\ln\Big(\frac{\mu^2}{Q^2}\Big)+
\gamma_H[\alpha_s(\mu)]\Big)\,U_H(Q,\mu_H,\mu)\,, \\
\label{eq:jetfctanomdim}
\mu\frac{d}{d\mu} U_J(s,\mu,\mu_J) \, = \,&\,\int\!\! d s^\prime\Big(\! -\frac{\Gamma_S[\alpha_s(\mu)]}{\mu^2}\Big[\frac{\mu^2\theta(s-s^\prime)}{s-s^\prime}\Big]_+
+\gamma_J[\alpha_s(\mu)]\delta(s-s^\prime)\Big)\notag\\ &
\hspace{1cm} \times U_s(s-s^\prime,\mu,\mu_J)\,, \\
\label{eq:softfctanomdim}
\mu\frac{d}{d\mu} U_S(k,\mu,\mu_J)\,  = \,&\,\int\!\! d k^\prime\Big(\! -\frac{2\Gamma_S[\alpha_s(\mu)]}{\mu}\Big[\frac{\mu\theta(k-k^\prime)}{k-k^\prime}\Big]+
\gamma_S[\alpha_s(\mu)]_+\delta(k-k^\prime)\Big)\\ &
\hspace{1cm} \times U_s(k-k^\prime,\mu,\mu_S)\,,\notag
\end{align}
where the coeffients at NLL precision needed for discussions are given in Eqs.~(\ref{eq:cusp1}), (\ref{eq:cusp0})
and (\ref{eq:cusp2}).
These results have been obtained using dimensional regularization to regulate ultraviolet as well as infrared divergences
and do not account for any other infrared cutoff. Ultraviolet renormalization has been carried out in the ${\overline{\rm MS}}$ scheme.

\subsection{Resummation for massive quarks in SCET and bHQET}
\label{app:massivethrust}

As shown in Ref.~\cite{Fleming:2007qr,Fleming:2007xt} using SCET and bHQET,
for the dominant singular partonic contributions of the 2-jettiness distribution 
in the peak resonance region for massive quarks, the hard and soft functions, $H_Q$ and $S^{(\tau)}$, are the same
as for massless quarks. Their expressions at ${\cal O}(\as)$ and the general form for their anomalous dimensions 
are given in Eqs.~(\ref{eq:hardfunction}), (\ref{eq:scetsoftfunction}), (\ref{eq:hardfctanomdim}) and (\ref{eq:softfctanomdim}),
respectively. The ${\cal O}(\as)$ results for the mass mode and the bHQET jet functions read~\cite{Fleming:2007xt}
\begin{align}
\label{eq:Hmoneloop}
H_m(m,\mu)&\, =\, 1+\frac{\as(\mu) C_F}{4\pi}\biggl(2\ln^2\frac{m^2}{\mu^2}-2\ln\frac{m^2}{\mu^2}
+ 8+\frac{\pi^2}{3}\biggr)+\mathcal{O}(\as^2) \,, \\
\label{eq:JBoneloop}
m\,J_B(\hat{s},m,\delta m,\mu)&\,=\,\delta(\hat{s})-4\,\delta m\,\delta^\prime(\hat{s})+\frac{\as(\mu) C_F}{4\pi}\biggl(\frac{16}{\mu}\biggl[\frac{\mu\ln\frac{\hat{s}}{\mu}}{\hat{s}}\biggr]_+
-\frac{8}{\mu}\biggl[\frac{\mu}{\hat{s}}\biggr]_+\notag \\ &\hspace{2cm}
+(8-\pi^2)\delta(\hat{s})\biggr)+\mathcal{O}(\as^2) \,,
\end{align}
where the result for the bHQET jet function has been displayed for a general quark mass renormalization scheme $m$ which is related to 
the pole mass scheme by the relation $\delta m = m_{\rm pole}-m$. 
So in Eq.~(\ref{eq:JBoneloop}) we have $\hat s=(s-m^2)/m$.
Their respective anomalous dimensions can (to all orders) written in the form
\begin{align}
\label{eq:mmfctanomdim}
\mu\frac{d}{d\mu} U_m\Big(\frac{Q}{m},\mu_m,\mu\Big)  \,  = \,&\, \Big( \Gamma_m[\alpha_s(\mu)]\ln\Big(\frac{m^2}{Q^2}\Big)+\gamma_m[\alpha_s(\mu)]\Big)\,U_m\Big(\frac{Q}{m},\mu_m,\mu\Big)\\
\label{eq:bhqetjetfctanomdim}
\mu\frac{d}{d\mu} U_{J_B}(\hat s,\mu,\mu_B)  \,  = \,&\, \int\!\! {\rm d} \hat s^\prime\Big(\! -\frac{\Gamma_{J_B}[\alpha_s(\mu)]}{\mu}\Big[\frac{\mu\theta(\hat s-\hat s^\prime)}{\hat s-\hat s^\prime}\Big]_++\gamma_J[\alpha_s(\mu)]\delta(\hat s-\hat s^\prime)\Big)\\ &
\hspace{1cm} \times U_{J_B}(\hat s-\hat s^\prime,\mu,\mu_{J_B})\,,\notag
\end{align}
where the coeffients at NLL precision are given in Eqs.~(\ref{eq:cusp2}), see also Eqs.~(\ref{eq:cusp1}) and (\ref{eq:cusp0}).
These results have been obtained using dimensional regularization to regulate ultraviolet as well as infrared divergences
and do not account for any other infrared cutoff. Ultraviolet renormalization has been carried out in the ${\overline{\rm MS}}$ scheme.

\section{Jet and soft functions in SCET and bHQET with a $p_\perp$ cut at ${\cal O}(\as)$}
\label{app:thrustformulaewithcut}

\subsection{Unreleased soft function for thrust}
\label{sec:softfctwithcut}

\begin{figure}
	\center
	\begin{subfigure}[c]{0.30\textwidth}
		\includegraphics[width=1.0\textwidth]{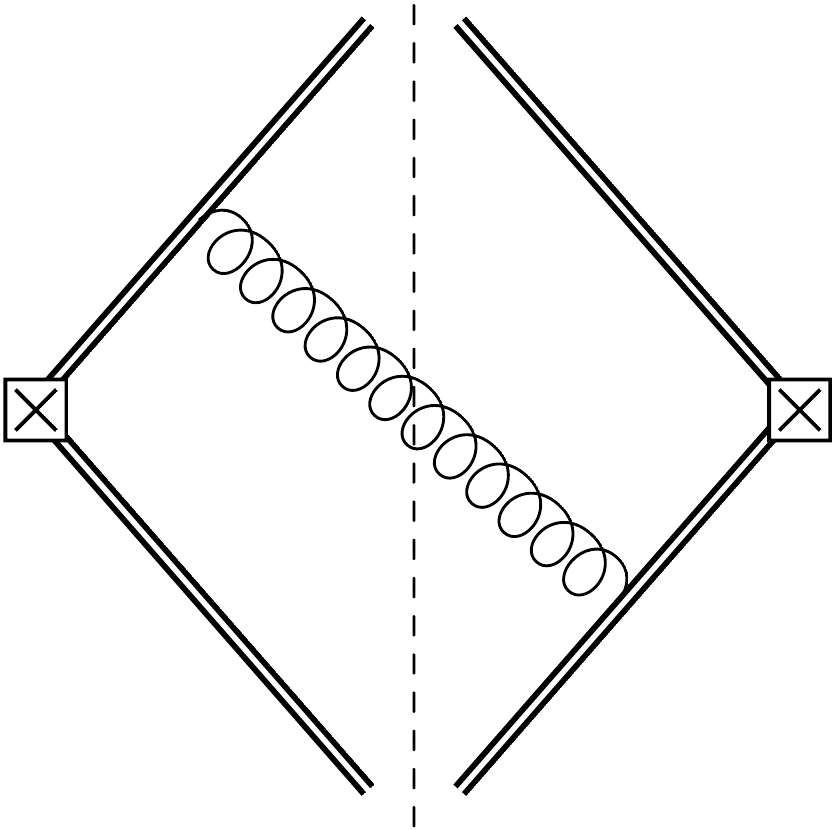}
		\subcaption{\label{fig:softfunctiondiagrams_real}}
	\end{subfigure}
	\hspace{2.0cm}
	\begin{subfigure}[c]{0.30\textwidth}
		\includegraphics[width=1.0\textwidth]{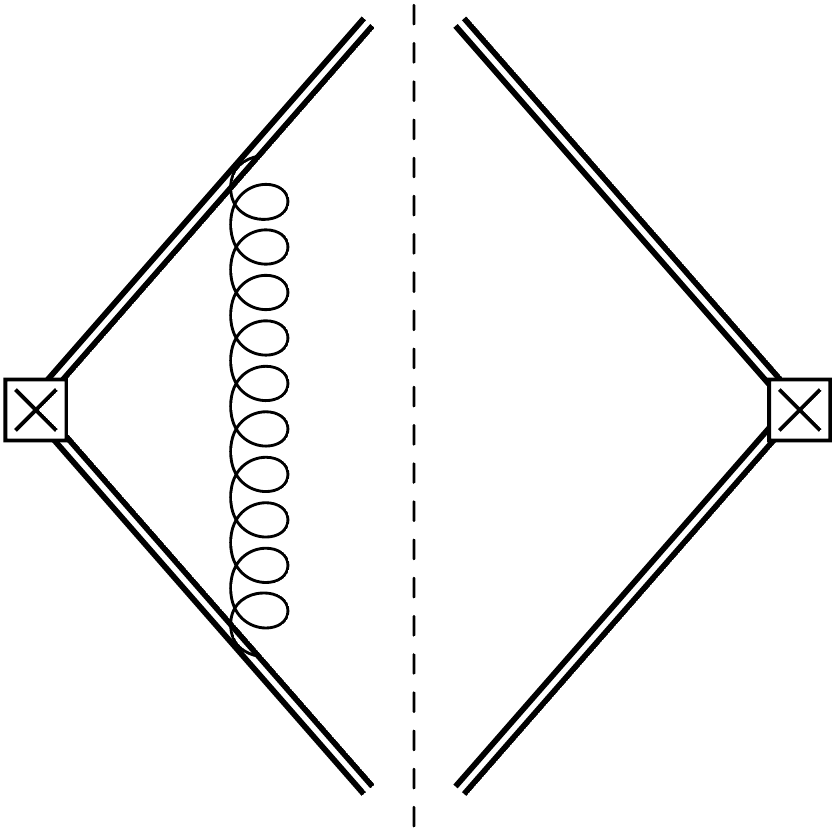}
		\subcaption{\label{fig:softfunctiondiagrams_virt}}
	\end{subfigure}
	\caption{\label{fig:softfunctiondiagrams}
		Diagrams relevant for computation of the partonic soft function at ${\cal O}(\alpha_s)$.
		}
\end{figure}

In this section we provide details on the calculation of the unreleased 
thrust soft function $S^{(\tau)}_{\rm ur}$ at ${\cal O}(\as)$. 
The unreleased soft function describes
large angle soft radiation originating {\it from below} the $p_\perp$ cut $Q_0$. 
We carry out the calculation using the dressed gluon propagator of Eq.~(\ref{eq:dressedgluon})
which is suitable to obtain the soft function
in Borel space (accounting for fermion bubble resummation to all orders).  
From this we can easily identify the ${\cal O}(\Lambda_{\rm QCD})$ renormalon pole
located at $u=1/2$. 
To obtain the usual one-loop result one can take the limit $u\to 0 $ in the end and
multiply back the factor $(\alpha_s \beta_0)/(4\pi)$ effectively removed by the
dressed gluon propagator in this limit. We note that at ${\cal O}(\as)$ all integrations
can be readily carried out in $4$ dimensions because the unreleased radiation does not result 
in any ultraviolet divergences. However, in contrast to the calculations without
any $p_\perp$ cut we also have to consider the contributions from the virtual diagrams, because the
scale $Q_0$ constitutes an additional scale such that the virtual diagrams may lead to finite
contributions. Interestingly, as we show below, the virtual diagrams lead to vanishing results 
even for finite $Q_0$. 

The Borel space contribution from the real radiation diagrams (including the mirror diagram) 
shown in Fig.~\ref{fig:softfunctiondiagrams_real} reads
\begin{align}
\label{eq:softfunction_diagram}
B\Bigl[S^{(\tau,{\rm real})}_{\rm ur}(k,Q_0)\Bigr]=&\frac{4 C_F}{\beta_0}\bigl(\mu^2 \mathrm{e}^{-c}\bigr)^u\theta(k)\int\frac{\mathrm{d}q^+\mathrm{d}q^-}{q^+(q^-)^{1+\alpha}}\notag\\
& \,\times\Bigl(\theta(q^--q^+)\delta(k-q^+)+\theta(q^+-q^-)\delta(k-q^-)\Bigr) \notag \\
&\, \times \frac{1}{\pi}\,\rm{Im}\biggl[\int_{0}^{Q_0^2}\mathrm{d}q_{\perp}^2\,\frac{1}{\bigl(q_\perp^2-(q^+q^-+i0)\bigr)^{1+u}}\biggr]\,,
\end{align}
with $c=5/3$ in the $\overline{\rm{MS}}$ renormalization scheme for the strong coupling. 
In Eq.~\eqref{eq:softfunction_diagram} we introduced the rapidity regulator $\alpha$ on the $q^-$ light-cone component. 
This is regulator is useful because the upper bound for the transverse momentum integration leads
to intermediate $1/\alpha$ rapidity divergences, which, however, cancel when summing the contributions from the two hemispheres
(defined by the contributions associated to the two $\theta$ step functions).
So, overall there are no rapidity divergences in the ${\cal O}(\as)$ thrust soft function with an upper $p_\perp$ cutoff.
Doing the trivial delta function integrations gives
\begin{align}
&\frac{4 C_F}{\beta_0}\bigl(\mu^2 \mathrm{e}^{-c}\bigr)^u\frac{\theta(k)}{k}\int_{k}^{\infty}\frac{\mathrm{d}q}{q}\,\Bigl(q^{-\alpha}+k^{-\alpha}\Bigr) 
\times \frac{1}{\pi}\,\rm{Im}\biggl[\int_{0}^{Q_0^2}\mathrm{d}q_{\perp}^2\,\frac{1}{\bigl(q_\perp^2-(qk+i0)\bigr)^{1+u}}\biggr]\,.
\end{align}
Next we can calculate the $q_\perp$ integral and take the imaginary part employing the relation
\begin{align}\label{eq:soft_qperp_integral}
&\frac{1}{\pi}\,\rm{Im}\biggl[\int_{0}^{Q_0^2}\mathrm{d}q_{\perp}^2\,\frac{1}{\bigl(q_\perp^2-(qk+i0)\bigr)^{1+u}}\biggr] 
\,=\,\frac{1}{u\pi}\,\rm{Im}\Bigl[(-qk-i0)^{-u}-(Q_0^2-qk-i0)^{-u}\Bigr] 
\notag \\ &
\hspace{2cm} \,=\,
\frac{1}{\Gamma(1-u)\Gamma(1+u)}\Bigl((qk)^{-u}-\theta(qk-Q_0^2)(qk-Q_0^2)^{-u}\Bigr)\,,
\end{align}
where in the last line we have used the fact that $qk>0$. This leaves us with the sum of three integrals
\begin{align}
&\frac{4C_F}{\beta_0}\frac{\bigl(\mu^2\mathrm{e}^{-c}\bigr)^u}{\Gamma(1-u)\Gamma(1+u)}\frac{1}{k^{1+u}}\biggl[2\int_{k}^{\infty}\frac{\mathrm{d}q}{q^{1+u}}-2\,\theta(k-Q_0)\int_{k}^{\infty}\frac{\mathrm{d}q}{q}\bigl(q-\frac{Q_0^2}{k}\bigr)^{-u} \notag \\
&\hspace{4.5cm}-\theta(Q_0-k)\int_{\frac{Q_0^2}{k}}^{\infty}\frac{\mathrm{d}q}{q}\bigl(q-\frac{Q_0^2}{k}\bigr)^{-u}\bigl(q^{-\alpha}+k^{-\alpha}\bigr)\biggr]  \\
&\,=\,\frac{2C_F}{\beta_0}\frac{\bigl(\mu^2\mathrm{e}^{-c}\bigr)^u}{\Gamma(1-u)\Gamma(1+u)}\biggl[\frac{2}{u\,k^{1+2u}}-\theta(k-Q_0)\frac{2\,Q_0^{-2u}}{k}\rm{B}\Bigl[\frac{Q_0^2}{k^2};u,1-u\Bigr] \notag \\
&\hspace{3.7cm}-\theta(Q_0-k)Q_0^{-2u-\alpha}\Gamma(1-u)\Bigl(\frac{Q_0^\alpha\Gamma(u)}{k^{1+\alpha}}+\frac{Q_0^{-\alpha}\Gamma(u+\alpha)}{\Gamma(1+\alpha)\,k^{1-\alpha}}\Bigr)\biggr]\,,
\notag
\end{align}
where we have already taken the limit $\alpha\to 0$ in the first two terms since they are finite for $\alpha\to 0$, and $\rm{B}[z;a,b]$ is the incomplete beta function.
For the third term the $\alpha\to 0$ limit has to be taken more carefully, using 
\begin{align}
\frac{Q_0^{\pm\alpha}}{k^{1\pm\alpha}}=\mp\frac{1}{\alpha}\delta(k)+\frac{1}{Q_0}\plus{k/Q_0}+\mathcal{O}(\alpha)\,.
\end{align}
With this we finally arive at
\begin{align}\label{eq:softfunction_fullresult}
B\Bigl[S^{(\tau)}_{\rm ur}(k,Q_0)\Bigr]\,&\,=\,B\Bigl[S^{(\tau,{\rm real})}_{\rm ur}(k,Q_0)\Bigr]\notag\\
&\,=\,
\frac{8C_F}{\beta_0}\frac{\Bigl(\frac{\mu^2\mathrm{e}^{-c}}{Q_0^2}\Bigr)^u}{\Gamma(1-u)\Gamma(1+u)}
\biggl[\frac{Q_0^{2u}}{u\,k^{1+2u}}-\theta(k-Q_0)\frac{1}{k}\,\rm{B}\Bigl[\frac{Q_0^2}{k^2},u,1-u\Bigr] \\
&-\theta(Q_0-k)\Gamma(u)\Gamma(1-u)\Bigl(\delta(k)\frac{\rm{H}_{u-1}}{2}+\frac{1}{Q_0}\plus{k/Q_0}\Bigr)\biggr]\,,\notag
\end{align}
where $\rm{H}_{n}=\psi(n+1)+\gamma_E$ is the harmonic number function and $\rm{B}\bigl[z;a,b\bigr]$ 
the incomplete Beta function.
As already discussed before, there are no rapidity divergences in the thrust soft function and all $\frac{1}{\alpha}$ poles cancel in the final result. Since the virtual diagrams turn out to vanish (see below), this represents already the full result for the unreleased thrust soft function.

To identify the leading renormalon pole we Laurent expand Eq.~\eqref{eq:softfunction_fullresult} around $u= 1/2$. Using the relation
\begin{align}
\frac{\mu^{2u}}{k^{1+2u}}=\frac{\mu}{2(u-\frac{1}{2})}\,\delta^\prime(k)+\mathcal{O}\Bigl((u-\frac{1}{2})^0\Bigr)\,,
\end{align}
we find the pole contribution
\begin{align}
B\Bigl[S^{(\tau)}_{\rm ur}(k,Q_0)\Bigr]\Bigl(u\approx \frac{1}{2}\Bigr)=\frac{16C_F\mathrm{e}^{-\frac{c}{2}}}{\pi\beta_0}\frac{\mu}{u-\frac{1}{2}}\,\delta^{\prime}(k)\,.
\end{align}
To obtain the ${\cal O}(\as)$ unreleased soft function one has to take the limit $u\to 0$ of Eq.~\eqref{eq:softfunction_fullresult} 
and include back again the factor $(\alpha_s \beta_0)/(4\pi)$. The result is
\begin{align}\label{eq:unrleased_soft_oneloop}
S^{(\tau)}_{\rm ur}(k,Q_0)=\frac{\alpha_sC_F}{4\pi}\,16\,\theta(Q_0-k)\biggl\{-\frac{1}{Q_0}\pluss{\tilde{k}}\biggr\}\,,
\end{align}
with $\tilde{k}=k/Q_0$.

We will now show that the virtual contributions to unreleased soft function vanish even in the presence of a $p_\perp$ cut. 
The Borel space contribution from the sum of the virtual diagrams shown in Fig.~\ref{fig:softfunctiondiagrams_virt} reads 
\begin{align}
B\Bigl[S^{(\tau,{\rm virt})}_{\rm ur}(k,Q_0)\Bigr]=\frac{i\,64\,C_F\pi^2}{\beta_0}\bigl(\mu^2\mathrm{e}^{-c}\bigr)^u\delta(k)\int\frac{\mathrm{d}^4q}{(2\pi)^4}\,\frac{\theta(Q_0-q_\perp)}{(-q^2)^{1+u}(n\cdot q)(\bar{n}\cdot q)^{1+\alpha}}\,,
\end{align}
where we have again introduced the $\alpha$ regulator and used $n^2=\bar{n}^2=0$ and $n\cdot \bar{n}=2$. This integral
is  most conveniently solved by using Feynman parameters of the form
\begin{align}\label{eq:FeynmanParameter}
\frac{1}{a^{\alpha}b^{\beta}c^{\gamma}}=\frac{\Gamma(\alpha+\beta+\gamma)}{\Gamma(\alpha)\Gamma(\beta)\Gamma(\gamma)}\int_{0}^{\infty}\mathrm{d}\lambda_1\mathrm{d}\lambda_2\,\frac{\lambda_1^{\beta-1}\lambda_2^{\gamma-1}}{(a+\lambda_1b+\lambda_2c)^{\alpha+\beta+\gamma}}\,,
\end{align}
such that one finds
\begin{align}
& \frac{-i\,64\,C_F}{\beta_0}\bigl(\mu^2\mathrm{e}^{-c}\bigr)^u\frac{\Gamma(3+u+\alpha)}{\Gamma(1+u)
\Gamma(1+\alpha)}(-1)^{u+\alpha}\delta(k)
\notag \\ &\qquad \times
\int_0^{\infty}\mathrm{d}\lambda_1\,\mathrm{d}\lambda_2\,\,\lambda_2^\alpha\int\frac{\mathrm{d}^4q}{(2\pi)^4}\,\frac{\theta(Q_0-q_\perp)}{(q^2-\lambda_1\lambda_2)^{3+u+\alpha}}\,.
\end{align}
The $q$ integral is solved by using Eq.~\eqref{eq:unreleased_integral} and leads to
\begin{align}
B\Bigl[S^{(\rm virt)}_{\rm ur}(k,Q_0)\Bigr]\,=\,&
\frac{-4C_F}{\beta_0}\bigl(\mu^2\mathrm{e}^{-c}\bigr)^u\frac{\Gamma(1+u+\alpha)}{\Gamma(1+u)
	\Gamma(1+\alpha)}\delta(k)\notag \\
&\hspace{2.0cm}\times\int_0^{\infty}\mathrm{d}\lambda_2\,\lambda_2^\alpha\int_0^\infty\mathrm{d}\lambda_1\,
\Bigl((\lambda_1\lambda_2)^{-1-u-\alpha}-(Q_0^2+\lambda_1\lambda_2)^{-1-u-\alpha}\Bigr) \notag \\
\,=\,&
\frac{4C_F}{\beta_0}\Bigl(\frac{\mu^2\mathrm{e}^{-c}}{Q_0^2}\Bigr)^u
\frac{\Gamma(u+\alpha)}{\Gamma(1+u)\Gamma(1+\alpha)}\delta(k)\,Q_0^{-2\alpha}
\int_0^{\infty}\mathrm{d}\lambda_2\,\lambda_2^{-1+\alpha} \notag \\
\, =\, &\,0\,.
\end{align}
The integral is scaleless and thus vanishes. Thus virtual diagrams do not contribute to the unreleased soft function 
at ${\cal O}(\as)$.

\subsection{Unreleased soft function for  angularities and C-parameter}
\label{sec:othersoftfctswithcut}

It is straightforward to determine soft functions for other event shape variables using the method described in
Sec.~\ref{sec:softfctwithcut}. In the following we provide the corresponding results for the angularities $\tau_\alpha$
and the $C$-parameter for future use.

For the angularities the measurement function shown in the second line of Eq.~(\ref{eq:softfunction_diagram}) for thrust
reads
\begin{align}
 \theta(q^--q^+)\delta\Bigl(k-(q^+)^{1-\frac{a}{2}}(q^-)^{\frac{a}{2}}\Bigr)\;+\; \bigl[\;q^- \leftrightarrow q^+\;\bigl]\,.
\end{align}
and the resulting unreleased soft function at ${\cal O}(\as)$ has the form 
\begin{align}
S_{\mathrm{ur}}^{(\tau_a)}(k,Q_0)&=\asCF{Q_0}\,\frac{16\,\theta(Q_0-k)}{1-a}\biggl\{-\frac{1}{Q_0}\pluss{\tilde{k}}\biggr\}\,.
\end{align}
The pole of the Borel transform at $u=1/2$ reads
\begin{align}
B\Bigl[S^{(\tau_a)}_{\rm ur}(k,Q_0)\Bigr]\Bigl(u\approx \frac{1}{2}\Bigr)&=\frac{16C_F\mathrm{e}^{-\frac{c}{2}}}{\pi\beta_0(1-a)}\frac{\mu}{u-\frac{1}{2}}\,\delta^{\prime}(k)\,,
\end{align}
and the first moment has the form
\begin{align}
\int \mathrm{d}k\,k\,S_{\mathrm{ur}}^{(\tau_a)}(k,Q_0)=\asCF{Q_0}\,\frac{16 \,Q_0}{1-a}\,.
\end{align}

For the $C$-parameter the measurement function shown in the second line of Eq.~(\ref{eq:softfunction_diagram}) for thrust
reads
\begin{align}
 \delta\Bigl(k-\frac{q^-q^+}{q^-+q^+}\Bigr)
\end{align}
and the resulting unreleased soft function at ${\cal O}(\as)$ has the form [$w(z)=(1-4/z)^{1/2}$]
\begin{align}
S_{\mathrm{ur}}^{(C)}(k,Q_0)&=\asCF{Q_0}\,16\,\theta\bigl(\frac{Q_0}{2}-k\bigr)\biggl\{-\frac{1}{Q_0}\pluss{\tilde{k}}+
\frac{\pi^2}{24}\delta(k)+\frac{\ln\Bigl(\frac{1+w(1/\tilde{k}^2)}{2}\Bigr)}{k}\biggr\}\,.
\end{align}
The pole of the Borel transform at $u=1/2$ reads
\begin{align}
B\Bigl[S^{(C)}_{\rm ur}(k,Q_0)\Bigr]\Bigl(u\approx \frac{1}{2}\Bigr)&=\frac{4C_F\mathrm{e}^{-\frac{c}{2}}}{\beta_0}\frac{\mu}{u-\frac{1}{2}}\,\delta^{\prime}(k)\,.
\end{align}
and the first moment has the form
\begin{align}
\int \mathrm{d}\ell\,\ell\,S_{\mathrm{ur}}^{(C)}(\ell,Q_0)=\asCF{Q_0}\,4\pi Q_0\,.
\end{align}
For all results we have $c=5/3$ in the $\overline{\rm MS}$ scheme and $\tilde{k}\equiv k/Q_0$.

\subsection{Unreleased bHQET jet function}
\label{sec:bHQETjetfctwithcut}

\begin{figure}
	\center
	\begin{subfigure}[c]{0.45\textwidth}
		\includegraphics[width=1.0\textwidth]{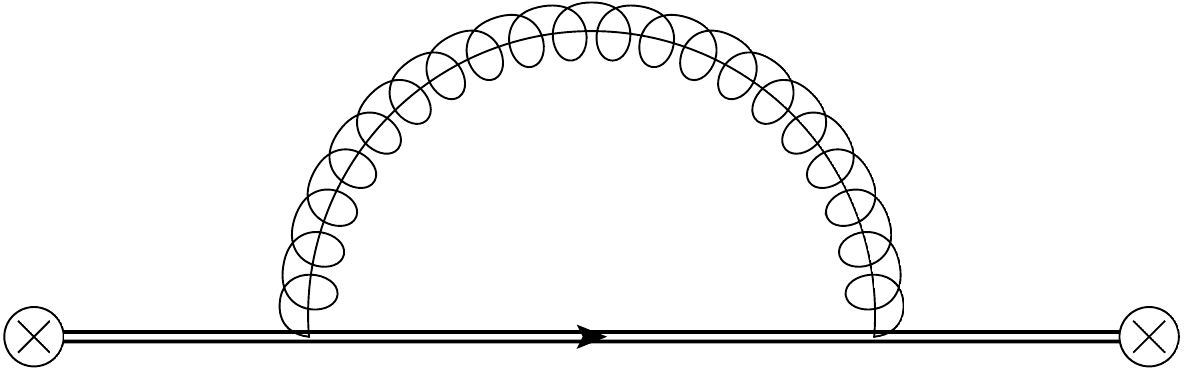}
		\subcaption{\label{fig:bHQETdiagrams_selfenergy}}
	\end{subfigure}
	\hfill
	\begin{subfigure}[c]{0.45\textwidth}
		\includegraphics[width=1.0\textwidth]{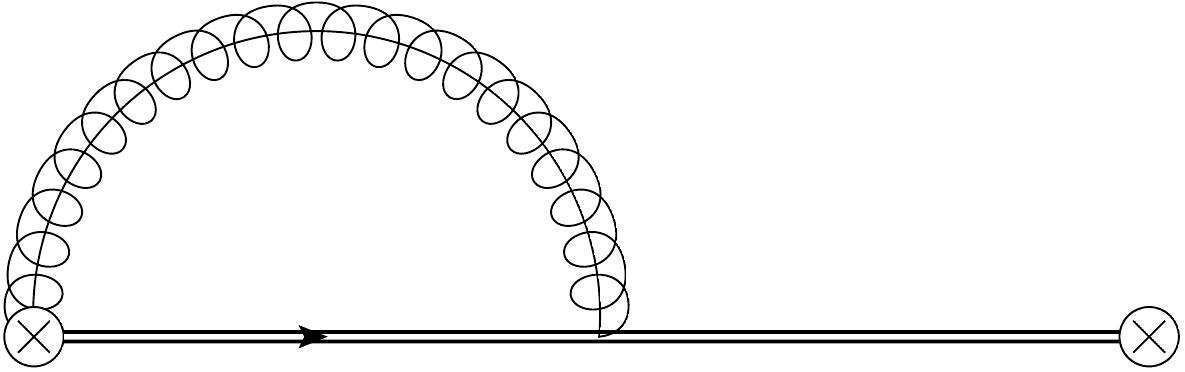}
		\subcaption{\label{fig:bHQETdiagrams_Wilsonline}}
	\end{subfigure}
	\caption{\label{fig:bHQETdiagrams}
			Diagrams relevant for computation of the bHQET jet function at ${\cal O}(\alpha_s)$.
		}
\end{figure}

In this section we calculate the unreleased bHQET jet function at ${\cal O}(\as)$. 
The unreleased bHQET jet function arises from ultra-collinear radiation off the massive quark 
coming {\it from below} the $p_\perp$ cut $Q_0$. As for the unreleased soft function we carry out the calculation
using the dressed gluon propagator of Eq.~(\ref{eq:dressedgluon}), which is suitable to obtain the 
Borel transform accounting for fermion bubble resummation to all orders and to obtain the usual
${\cal O}(\as)$ result from the limit $u\to 0$ and accounting for the additional factor  
$(\alpha_s \beta_0)/(4\pi)$. All integrals can again be carried out in $4$ dimensions since 
in the presence of the $p_\perp$ cut they are ultraviolet finite. In contrast to the calculation for the
soft function, there are no rapidity divergences at intermediate steps of the calculation.
We note that in the following we determine the ${\cal O}(\alpha_s)$ corrections to the unreleased bHQET jet function matrix element 
$\mathcal{J}_{B,\rm ur}^{(\tau)}(\hat{s},Q_0)$ with and $\hat{s}=\hat{s}+i0$ following the conventions from Ref.~\cite{Fleming:2007xt}.
The actual unreleased bHQET jet function $J_{B,\rm ur}(\hat{s},Q_0)$ appearing in the
factorization theorem of Eq.~(\ref{eq:thrustmassive2}) is then obtained by taking the imaginary part: 
\begin{align}
\label{eq:takeimpart}
J_{B,\rm ur}^{(\tau)}(\hat{s},Q_0) \, = \, \mbox{Im}\Big[ \mathcal{J}_{B,\rm ur}^{(\tau)}(\hat{s}+i0,Q_0)\Big]\,.
\end{align}

The self energy diagram Fig.~\ref{fig:bHQETdiagrams_selfenergy} in Borel space reads (already including a factor two because the jet function in the factorization theorem Eq.~\eqref{eq:thrustmassive2} accounts for both hemispheres)
\begin{align}
 B\Bigl[\mathcal{J}_{B,\rm ur}^{(a)}(\hat{s},Q_0)\Bigr]=\frac{i\,64\pi\,C_F}{m\hat{s}^2\beta_0}\bigl(\mu^2\mathrm{e}^{-c}\bigr)^u\theta(\hat{s})\int\dd{4}{q}\,\frac{\theta(Q_0-q_\perp)}{(-q^2)^{1+u}(-v\cdot q -\frac{\hat{s}}{2})}\,,
\end{align}
with $v^2=1$ and $c=5/3$ in the $\overline{\rm{MS}}$ renormalization scheme. The integral is evaluated in 4 dimensions because in the unreleased contributions there are no divergences that need to be regularized by dimensional regularization. It can be calculated by using Feynman paramters of the form
\begin{align}
 \frac{1}{a^{\alpha}b^{\beta}}=\frac{\Gamma(\alpha+\beta)}{\Gamma(\alpha)\Gamma(\beta)}\int_{0}^{\infty}\mathrm{d}\lambda\,\frac{\lambda^{\beta-1}}{(a+\lambda b)^{\alpha+\beta}}\,,
\end{align}
such that
\begin{align}
 B\Bigl[\mathcal{J}_{B,\rm ur}^{(a)}(\hat{s},Q_0)\Bigr]=\frac{i\,64\pi\,C_F}{m\hat{s}^2\beta_0}&\bigl(\mu^2\mathrm{e}^{-c}\bigr)^u(1+u)(-1)^u\theta(\hat{s})\notag\\
 &\times\,\int_0^\infty\mathrm{d}\lambda\,\int\dd{4}{q}\,\frac{\theta(Q_0-q_\perp)}{\bigl(q^2-\frac{\lambda}{2}(\frac{\lambda}{2}-\hat{s})\bigr)^{2+u}}\,.
\end{align}
The $q$ integral is solved by using Eq.~\eqref{eq:unreleased_integral} and leads to (after doing the additional substitution $\lambda\to 2Q_0\lambda$ and $\tilde{s}=\hat{s}/Q_0$)
\begin{align}\label{eq:bHQETjet_integral}
 B\Bigl[\mathcal{J}_{B,\rm ur}^{(a)}(\hat{s},Q_0)\Bigr]=\frac{-8C_F\,Q_0}{m\hat{s}^2\pi\beta_0}\Bigl(\frac{\mu^2\mathrm{e}^{-c}}{Q_0^2}\Bigr)^u\theta(\hat{s})\frac{1}{u}\int_0^\infty\mathrm{d}\lambda\,\Bigl[\bigl(\lambda(\lambda-\tilde{s})\bigr)^{-u}-\bigl(1+\lambda(\lambda-\tilde{s})\bigr)^{-u}\Bigr]\,.
\end{align}
Let us first look at the on-shell (os) self-energy contribution of this diagram. Due to the cutoff $Q_0$ it does not vanish and is therefore relevant for the mass renormalization scheme. It is obtained by 
setting $\hat{s}\to 0$ under the integral. This yields
\begin{align}
 B\Bigl[\mathcal{J}_{B,\rm ur}^{(a,\rm os)}(\hat{s},Q_0)\Bigr]=&\frac{-8C_F\,Q_0}{m\hat{s}^2\pi\beta_0}\Bigl(\frac{\mu^2\mathrm{e}^{-c}}{Q_0^2}\Bigr)^u\theta(\hat{s})\frac{1}{u}\int_0^\infty\mathrm{d}\lambda\,\Bigl[\lambda^{-2u}-(1+\lambda^2)^{-u}\Bigr] \notag \\
 =&\frac{4C_FQ_0}{m\hat{s}^2\pi\beta_0}\Bigl(\frac{\mu^2\mathrm{e}^{-c}}{Q_0^2}\Bigr)^u\theta(\hat{s})\frac{\sqrt{\pi}\,\Gamma\bigl(u-\frac{1}{2}\bigr)}{\Gamma(1+u)}\,.\label{eq:bHQETjet_onshell_fullresult}
\end{align}
The first term under the integral in the first line 
is scaleless and can be dropped. To identify the leading renormalon pole we expand Eq.~\eqref{eq:bHQETjet_onshell_fullresult} for $u\to 1/2$ to obtain
\begin{align}\label{eq:onshell_Borel}
 B\Bigl[\mathcal{J}_{B,\rm ur}^{(a,\rm os)}(\hat{s},Q_0)\Bigr]\Bigl(u\approx\frac{1}{2}\Bigr)=\frac{8C_F\mathrm{e}^{-\frac{c}{2}}}{m\hat{s}^2\pi\beta_0}\theta(\hat{s})\frac{\mu}{u-\frac{1}{2}}\,.
\end{align}
To get the one-loop result one has to take the limit $u\to 0$ of Eq.~\eqref{eq:bHQETjet_onshell_fullresult} and include again the factor $(\alpha_s\beta_0)/(4\pi)$:
\begin{align}\label{eq:onshell_oneloop}
 \mathcal{J}_{B,\rm ur}^{(a,\rm os)}(\hat{s},Q_0)= 
- \frac{1}{m}\frac{\alpha_sC_F}{4\pi}\frac{\theta(\hat{s})}{\pi\hat{s}^2}\,8\pi Q_0\,.
\end{align}

To get the result of the integral in Eq.~\eqref{eq:bHQETjet_integral} for finite $\hat{s}$ we first note that this does not give a pole for $u\to 1/2$. This can be seen by setting $u=1/2$ and investigating the behavior of the integrand for small and large values of $\lambda$:
\begin{align}
 \lambda^{-\frac{1}{2}} \quad &\text{for} \quad \lambda \to 0\,,\\
 \lambda^{-3}\quad &\text{for}\quad \lambda \to \infty\,.
\end{align}
This implies that the integral converges and that there is no renormalon pole at $u=1/2$ in the off-shell case. To get the corresponding one-loop result we take the limit $u\to 0$ and multiply back the factor $(\alpha_s\beta_0)/(4\pi)$ and get [$w(z)=(1-4/z)^{1/2}$]

\begin{align}
 \mathcal{J}_{B,\rm ur}^{(a)}(\hat{s},Q_0)&=\frac{1}{m}\frac{\as C_F}{4\pi}\frac{\theta(\hat{s})}{\pi\hat{s}^2}8Q_0\int_0^\infty\mathrm{d}\lambda\,\ln\biggl(\frac{\lambda(\lambda-\tilde{s})}{1+\lambda(\lambda-\tilde{s})}\biggr)  \\
 &=\frac{1}{m}\frac{\as C_F}{4\pi}\frac{\theta(\hat{s})}{\pi\hat{s}}8\Biggl[\theta(2Q_0-\hat{s})\biggl(-\frac{w(\frac{16}{\tilde{s}^2})\bigl(2\rm{Tan}^{-1}\bigl(\frac{\tilde{s}}{2w(\frac{16}{\tilde{s}^2})}\bigr)+\pi\bigr)}{\tilde{s}}+\ln(-\tilde{s})\biggr) \notag \\
 &\hspace{1.5cm}+\theta(\hat{s}-2Q_0)\biggl(i\pi(w(\tilde{s}^2)-1)+\ln \tilde{s}-\frac{w(\tilde{s}^2)}{2}\ln\Bigl(\frac{1+w(\tilde{s}^2)}{1-w(\tilde{s}^2)}\Bigr))\biggr)\Biggr]\,.\notag
\end{align}

The diagram in Fig.~\ref{fig:bHQETdiagrams_Wilsonline} in Borel space reads (including a factor two
to account for both hemispheres and a factor two for the mirror diagram)
\begin{align}
 B\Bigl[\mathcal{J}_{B,\rm ur}^{(b)}&(\hat{s},Q_0)\Bigr]=\frac{-i\,64\pi\,C_F}{m\hat{s}\beta_0}\bigl(\mu^2\mathrm{e}^{-c}\bigr)^u\theta(\hat{s})\frac{Q}{m} \\
 &\times \int\dd{4}{q}\,\biggl[\frac{\theta(Q_0-q_\perp)}{(-q^2)^{1+u}(-v\cdot q-\frac{\hat{s}}{2})(-n\cdot q)}-\frac{\theta(Q_0-q_\perp)}{(-q^2)^{1+u}(-\frac{Q}{2m}\bar{n}\cdot q-\frac{\hat{s}}{2})(-n\cdot q)}\biggr]\,,\notag
\end{align}
with $v^2=1,\,n^2=\bar{n}^2=0,\,n\cdot\bar{n}=2\;\text{and}\;n\cdot v=Q/m$. Again the prescription $\hat{s}=\hat{s}+i0$ is implied.
The second term under the integral is the 0-bin that needs to be subratcted to avoid a double counting between the soft and the collinear regions. We can again use Feynman parameters of the form Eq.~\eqref{eq:FeynmanParameter} to obtain
\begin{align}
  B\Bigl[\mathcal{J}_{B,\rm ur}^{(b)}&(\hat{s},Q_0)\Bigr]=\frac{i\,64\pi\,C_F}{m\hat{s}\beta_0}\bigl(\mu^2\mathrm{e}^{-c}\bigr)^u\frac{(-1)^u\Gamma(3+u)}{\Gamma(1+u)}\theta(\hat{s})\frac{Q}{m} \\
  &\times\int_0^\infty\mathrm{d}\lambda_1\mathrm{d}\lambda_2\int\dd{4}{q}\biggl[\frac{\theta(Q_0-q_\perp)}{\bigl(q^2-\frac{\lambda_1}{2}(\frac{\lambda_1}{2}+\frac{Q\lambda_2}{m}-\hat{s})\bigr)^{3+u}}-\frac{\theta(Q_0-q_\perp)}{\bigl(q^2-\frac{\lambda_1}{2}(\frac{Q\lambda_2}{m}-\hat{s})\bigr)^{3+u}}\biggr]\,.\notag
\end{align}
The $q$ integral is solved by using Eq.~\eqref{eq:unreleased_integral} and leads to (after doing the additional substitution $\lambda_1\to 2Q_0\lambda_1$, $\lambda_2\to Q_0m\lambda_2/Q$ and $\tilde{s}=\hat{s}/Q_0$)
\begin{align}
 B\Bigl[\mathcal{J}_{B,\rm ur}^{(b)}(\hat{s},Q_0)\Bigr]=&\frac{8\,C_F}{m\hat{s}\pi\beta_0}\Bigl(\frac{\mu^2\mathrm{e}^{-c}}{Q_0^2}\Bigr)^u\theta(\hat{s}) \\
 &\times\int_0^\infty\mathrm{d}\lambda_1\mathrm{d}\lambda_2\,\biggl[\bigl(\lambda_1(\lambda_1+\lambda_2-\hat{s})\bigr)^{-1-u}-\bigl(1+\lambda_1(\lambda_1+\lambda_2-\hat{s})\bigr)^{-1-u}\notag \\
 &\hspace{2.5cm}-\bigl(\lambda_1(\lambda_2-\hat{s})\bigr)^{-1-u}+\bigl(1+\lambda_1(\lambda_2-\hat{s}))\bigr)^{-1-u}\biggr] \notag \\
 =&\frac{8\,C_F}{m\hat{s}\pi\beta_0}\Bigl(\frac{\mu^2\mathrm{e}^{-c}}{Q_0^2}\Bigr)^u\theta(\hat{s})\frac{1}{u}\int_0^\infty\frac{\mathrm{d}\lambda_1}{\lambda_1}\,\biggl[\bigl(\lambda_1(\lambda_1-\hat{s})\bigr)^{-u}-\bigl(1+\lambda_1(\lambda_1-\hat{s})\bigr)^{-u}\notag \\
 &\hspace{5.8cm}-\bigl(-\lambda_1\hat{s}\bigr)^{-u}+\bigl(1-\lambda_1\hat{s})\bigr)^{-u}\biggr]\,.\notag
\end{align}
We note that the integral does not lead to a pole at $u=1/2$, because the $\lambda$ integral is finite. 
This can be seen by investigating the behavior of the integrand for the small and larger $\lambda$:
\begin{align}
 \lambda^{-\frac{1}{2}} \quad &\text{for} \quad \lambda \to 0\,,\\
 \lambda^{-\frac{3}{2}}\quad &\text{for}\quad \lambda \to \infty\,.
\end{align}
To get the one-loop result we take the limit $u\to 0$ and multiply back the factor $(\alpha_s\beta_0)/(4\pi)$ to
obtain [$w(z)=(1-4/z)^{1/2}$]
\begin{align}
&\mathcal{J}_{B,\rm ur}^{(b)}(\hat{s},Q_0)= -\frac{1}{m}\frac{\as C_F}{4\pi}\frac{\theta(\hat{s})}{\pi\hat{s}}8\int_0^\infty\frac{\mathrm{d}\lambda_1}{\lambda_1}\,\ln\biggl(\frac{-(\lambda_1-\hat{s})(1-\lambda_1\hat{s})}{\hat{s}\bigl(1+\lambda_1(\lambda_1-\hat{s})\bigr)}\biggr)  \\
&=\frac{1}{m}\frac{\as C_F}{4\pi}\frac{\theta(\hat{s})}{\pi\hat{s}}8\Biggl[\theta(2Q_0-\hat{s})\biggl(-\ln^2(-\tilde{s})-\frac{\pi^2}{4}-\Bigl(\rm{Tan}^{-1}\Bigl(\frac{\tilde{s}}{2w(\frac{16}{\tilde{s}^2})}\Bigr)+\pi\Bigr)\rm{Tan}^{-1}\Bigl(\frac{\tilde{s}}{2w(\frac{16}{\tilde{s}^2})}\Bigr)\biggr)\notag \\
&\hspace{1.0cm}+\theta(\hat{s}-2Q_0)\biggl(i\pi\Bigl(2\ln\tilde{s}-\ln\Bigl(\frac{1+w(\tilde{s}^2)}{1-w(\tilde{s}^2)}\Bigr)\Bigr)-\ln\Bigl(\frac{1+w(\tilde{s}^2)}{2}\Bigr)\ln\Bigl(\frac{1-w(\tilde{s}^2)}{2}\Bigr)\biggr)\Biggr]\,.\notag
\end{align}

The complete sum of all off-shell ${\cal O}(\alpha_s)$ corrections, 
defined as the sum of all contributions for finite $\hat{s}$ minus the on-shell diagram of 
Eq.~\eqref{eq:onshell_oneloop}, reads
\begin{align}\label{eq:offshell_oneloop}
\mathcal{J}^{(\rm{off})}_{B,\rm{ur}}(\hat{s},Q_0)=&\mathcal{J}^{(a)}_{B,\rm{ur}}(\hat{s},Q_0)+\mathcal{J}^{(b)}_{B,\rm{ur}}(\hat{s},Q_0)-\mathcal{J}^{(a,\rm{os})}_{B,\rm{ur}}(\hat{s},Q_0) \notag \\
=&\frac{1}{m}\frac{\alpha_sC_F}{4\pi}\frac{1}{\pi\hat{s}}\biggl\{\theta(2Q_0-\hat{s})\biggl[-8\ln^2(-\tilde{s})+8\ln(-\tilde{s})+\frac{8\pi(1-w(\frac{16}{\tilde{s}^2}))}{\tilde{s}}\notag \\
 &-\frac{16w(\frac{16}{\tilde{s}^2})\rm{Tan}^{-1}\Bigl(\frac{\tilde{s}}{2w(\frac{16}{\tilde{s}^2})}\Bigr)}{\tilde{s}}-8\rm{Tan}^{-1}\Bigl(\frac{\tilde{s}}{2w(\frac{16}{\tilde{s}^2})}\Bigr)\Bigl(\rm{Tan}^{-1}\Bigl(\frac{\tilde{s}}{2w(\frac{16}{\tilde{s}^2})}\Bigr)+\pi\Bigr)-2\pi^2\biggr] \notag \\
 &+\theta(\hat{s}-2Q_0)\biggl[i\pi\Bigl(8w(\tilde{s}^2)-8+16\ln\tilde{s}-8\ln\Bigl(\frac{1+w(\tilde{s}^2)}{1-w(\tilde{s}^2)}\Bigr)\Bigr)+8\ln\tilde{s}\notag \\
 &-4w(\tilde{s}^2)\ln\Bigl(\frac{1+w(\tilde{s}^2)}{1-w(\tilde{s}^2)}\Bigr)-8\ln\Bigl(\frac{1-w(\tilde{s}^2)}{2}\Bigr)\ln\Bigl(\frac{1+w(\tilde{s}^2)}{2}\Bigr)+\frac{8\pi}{\tilde{s}}\biggr]\biggr\}\,,
\end{align}
and the corresponding Borel space result at the pole at $u=1/2$ has the form
\begin{align}\label{eq:offshell_Borel}
 B\Bigl[\mathcal{J}_{B,\rm ur}^{(\rm off)}(\hat{s},Q_0)\Bigr]\Bigl(u\approx\frac{1}{2}\Bigr)=\frac{-8C_F\mathrm{e}^{-\frac{c}{2}}}{m\hat{s}^2\pi\beta_0}\theta(\hat{s})\frac{\mu}{u-\frac{1}{2}}\,.
\end{align}

To obtain the ${\cal O}(\alpha_s)$ results for the bHQET jet funtion we take the imaginary part of
Eqs.~\eqref{eq:onshell_oneloop} and \eqref{eq:offshell_oneloop} following Eq.~(\ref{eq:takeimpart}).
Writing the jet function as a sum of the on-shell (os) self-energy contribution and the remaining 
off-shell contributions (off)
\begin{align}
J_{B,\rm ur}^{(\tau)}(\hat{s},Q_0) \, = \, J_{B,\rm ur}^{(\rm os)}(\hat{s},Q_0) +  J_{B,\rm ur}^{(\rm off)}(\hat{s},Q_0)
\end{align}
we obtain [$w(z)=(1-4/z)^{1/2}$, $\tilde{s}=\hat{s}/Q_0$]
\begin{align}
\label{eq:JBos}
J_{B,\rm ur}^{(\rm os)}(\hat{s},Q_0)&=\frac{1}{m}\frac{\alpha_sC_F}{4\pi}\Bigl(-8\pi Q_0\,\delta^\prime(\hat{s})\Bigr)\,,\\
\label{eq:JBoff}
J_{B,\rm ur}^{(\rm off)}(\hat{s},Q_0)&=\frac{1}{m}\frac{\alpha_sC_F}{4\pi}\biggl\{2\Bigl(4-\frac{\pi^2}{3}\Bigr)\delta(\hat{s}) \\
&\,\,+\theta(2Q_0-\hat{s})\biggl(-\frac{8}{Q_0}\plus{\tilde{s}}+\frac{16}{Q_0}\pluss{\tilde{s}}\biggr) \notag\\
&\,\,+\theta(\hat{s}-2Q_0)\,\frac{8}{\hat{s}}\,\biggl[\Bigl(w(\tilde{s}^2)-1\Bigr)-\Bigl(\ln\Bigl(\frac{1+w(\tilde{s}^2)}{1-w(\tilde{s}^2)}\Bigr)-2\ln \tilde{s}\Bigr)\biggr]\biggr\}\,.\notag
\end{align}
The corresponding Borel space result at the pole at $u=1/2$ are obtained 
by taking the imaginary parts of  Eqs.~\eqref{eq:onshell_Borel} and~\eqref{eq:offshell_Borel} giving
\begin{align}
\label{eq:JBonBorel}
 B\Bigl[J_{B,\rm ur}^{(\rm os)}(\hat{s},Q_0)\Bigr]\Bigl(u\approx\frac{1}{2}\Bigr)&=\frac{8C_F\mathrm{e}^{-\frac{c}{2}}}{m\beta_0}\frac{\mu}{u-\frac{1}{2}}\,\delta^\prime(\hat{s})\,,\\
\label{eq:JBoffBorel}
 B\Bigl[J_{B,\rm ur}^{(\rm off)}(\hat{s},Q_0)\Bigr]\Bigl(u\approx\frac{1}{2}\Bigr)&=\frac{-8C_F\mathrm{e}^{-\frac{c}{2}}}{m\beta_0}\frac{\mu}{u-\frac{1}{2}}\,\delta^\prime(\hat{s})\,.
\end{align}

\subsection{Unreleased SCET jet function}
\label{sec:scetjetfctwithcut}

The calculation for the unreleased SCET jet function for massless quark production 
can be carried out in close analogy to Sec.~\ref{sec:bHQETjetfctwithcut}.
The full result (accounting for the contributions arising from two hemispheres) 
at ${\cal O}(\as)$ reads [$s^\prime=s/Q_0^2$, $w(z)=(1-4/z)^{1/2}$]
\begin{align}
J_{\mathrm{ur}}^{(\tau)}(s,Q_0)&=\asCF{Q_0}\biggl\{\Bigl(12-\frac{4\pi^2}{3}\Bigr)\delta(s)+\theta(4Q_0^2-s)\biggl(-\frac{6}{Q_0^2}\plus{s^\prime}+\frac{8}{Q_0^2}\pluss{s^\prime}\biggr) \notag \\
&\qquad +\theta(s-4Q_0^2)\frac{1}{s}\biggl[6(w(s^\prime)-1)-8\Bigl(\ln\Bigl(\frac{1+w(s^\prime)}{1-w(s^\prime)}\Bigr)-\ln\,s^\prime\Bigr)\biggr]\biggr\}\,.
\end{align}
For future reference we also provide some useful intermediate results the 
SCET jet function in Feynman gauge. The self-energy diagrams yields
\begin{align}
\asCF{\mu}\biggl\{-\delta(s)+\frac{1}{\mu^2}\plus{s/\mu^2}\biggr\} 
\end{align}
without any cut and
\begin{align}
\asCF{Q_0}\biggl\{-2\delta(s)+\theta(4Q_0^2-s)\frac{1}{Q_0^2}\plus{s/Q_0^2}+\frac{\theta(s-4Q_0^2)}{s}\bigl(1-w(s/Q_0^2)\bigr)\biggr\}
\end{align}
for the unreleased contribution. 
The Wilson-line diagrams (containing the Eikonal propagator) yield
\begin{align}
\asCF{\mu}\biggl\{\bigl(4-\frac{\pi^2}{2}\bigr)\delta(s)-\frac{2}{\mu^2}\plus{s/\mu^2}+\frac{2}{\mu^2}\pluss{s/\mu^2}\biggr\}
\end{align}
without any cut and
\begin{align}
\asCF{Q_0}&\biggl\{\bigl(4-\frac{\pi^2}{3}\bigr)\delta(s)+\theta(4Q_0^2-s)\biggl(-\frac{2}{Q_0^2}\plus{s/Q_0^2}+\frac{2}{Q_0^2}\pluss{s/Q_0^2}\biggr) \notag \\
&\hspace{1.0cm}+\theta(s-4Q_0^2)\frac{2}{s}\biggl[w(s/Q_0^2)-1+\ln\frac{s}{Q_0^2}-\ln\Bigl(\frac{1+w(s/Q_0^2)}{1-w(s/Q_0^2)}\Bigr)\biggr]\biggr\}
\end{align}
for the unreleased contribution.

\section{Integrals in d-dimensions  with $p_{\perp}$ cut}
\label{app:integrals}

For the calculation of the jet and soft functions with and without a $q_{\perp}$-cut we need to solve d-dimensional integrals of the form
\begin{align}
 \int\dd{d}{q}\,\frac{f_{Q_0}(q_\perp)}{(q^2+\Delta)^n}\;,
\end{align}
for the cases $f_{Q_0}(q_\perp)=1$ (no cut), $f_{Q_0}(q_\perp)=\theta(q_\perp-Q_0)$ (only above cut) and $f_{Q_0}(q_\perp)=\theta(Q_0-q_\perp)$ (unreleased). After a Wick-rotation and doing the energy and angular integrals that are not affected by the cutoff, one arrives at
\begin{align}
 \frac{2i(-1)^n}{(4\pi)^{\frac{d}{2}}(n-1)\Gamma\bigl(\frac{d}{2}-1\bigr)}\int_0^\infty \mathrm{d}q_\perp\,f_{Q_0}(q_\perp)\frac{q_\perp^{d-3}}{(q_\perp^2-\Delta)^{n-1}}\;.
\end{align}
This can be solved for the three different cases and gives
\begin{align}
 \int\dd{d}{q}\,\frac{1}{(q^2+\Delta)^n}&=\frac{i(-1)^n\,\Gamma\bigl(n-\frac{d}{2}\bigr)}{(4\pi)^{\frac{d}{2}}\Gamma(n)}(-\Delta)^{\frac{d}{2}-n} \;,\\
  \int\dd{d}{q}\,\frac{\theta(q_{\perp}-Q_0)}{(q^2+\Delta)^n}&=\frac{i(-1)^n\,\mathrm{B}\bigl(\frac{\Delta}{Q_0^2},n-\frac{d}{2},2-n\bigr)}{(4\pi)^{\frac{d}{2}}(n-1)\Gamma\bigl(\frac{d}{2}-1\bigr)}(\Delta)^{\frac{d}{2}-n}\;,\\
   \int\dd{d}{q}\,\frac{\theta(Q_0-q_{\perp})}{(q^2+\Delta)^n}&=\frac{i(-1)^n\,{}_2\mathrm{F}_1\bigl(\frac{d}{2}-1,n-1,\frac{d}{2},\frac{Q_0^2}{\Delta}\bigr)}{(4\pi)^{\frac{d}{2}}(n-1)\Gamma(\frac{d}{2})}(Q_0^2)^{\frac{d}{2}-1}(-\Delta)^{1-n}\\
   &\hspace{-0.2cm}\xrightarrow[d\to 4]{}\;\frac{i(-1)^n}{16\pi^2(n-1)(n-2)}\Bigl((-\Delta)^{2-n}-(Q_0^2-\Delta)^{2-n}\Bigr)\;,\label{eq:unreleased_integral}
\end{align}
where in the last line we took the limit $d \to 4$ since there are no divergences in the unreleased contributions that need to be regularized by dimensional regularization.

\section{Simulation settings}
\label{app:Settings}

In this appendix we document the changes relative to the default
settings in \Herwig\ version 7.1.2 \cite{herwig-web} with which these
studies have been carried out. All of the results are parton level
simulations, with special settings to make contact with the analytic
approach and are not advocated to be used in a full simulation.  All
simulation is based on the default \texttt{LEP-Matchbox.in} input
file, which is prepared to generate both leading order and
next-to-leading order matched simulations.

\subsection{Common settings}

In all of the simulations we consider, we use light quarks $u,d,s,c,b$ by
setting their nominal mass to zero, and their consitutent masses, as well as
the gluon's constituent mass to be effectively zero,
\begin{verbatim}
set /Herwig/Particles/x:NominalMass 0*GeV
set /Herwig/Particles/x:ConstituentMass 0.00001*GeV
set /Herwig/Particles/g:ConstituentMass 0.000021*GeV
\end{verbatim}
where \texttt{x} = \texttt{u,d,s,c,b}. We also switch off QED initial state
radiation,
\begin{verbatim}
set /Herwig/Particles/e+:PDF /Herwig/Partons/NoPDF
set /Herwig/Particles/e-:PDF /Herwig/Partons/NoPDF
\end{verbatim}
and do consider the kinematic reconstruction option employed in earlier
\Herwig\ versions,
\begin{verbatim}
set /Herwig/Shower/ShowerHandler:ReconstructionOption CutOff
\end{verbatim}
We always consider parton level results
\begin{verbatim}
read Matchbox/PQCDLevel.in
\end{verbatim}
The parton shower cutoff is changed via
\begin{verbatim}
set /Herwig/Shower/QtoQGSudakov:pTmin X*GeV
set /Herwig/Shower/GtoGGSudakov:pTmin X*GeV
set /Herwig/Shower/GtoQQbarSudakov:pTmin X*GeV
\end{verbatim}
where \texttt{X}=$Q_0/{\rm GeV}$. If gluon branchings are desired to be
switched off, we use
\begin{verbatim}
cd /Herwig/Shower
do SplittingGenerator:DeleteFinalSplitting g->g,g; GtoGGSudakov
do SplittingGenerator:DeleteFinalSplitting g->x,xbar; GtoQQbarSudakov
\end{verbatim}
where \texttt{x} again runs over the different quark flavors. We always
switch off $g\to t\bar{t}$ branchings by an according statement.
If additionally we qant to quantify the remaining impact of angular ordering,
we choose
\begin{verbatim}
cd /Herwig/Shower
set QtoQGSplitFn:AngularOrdered No
set GtoGGSplitFn:AngularOrdered No
set QtoQGSplitFn:ScaleChoice pT
set GtoGGSplitFn:ScaleChoice pT
\end{verbatim}
As far as the calculation of the production process is concerned, we either
use the leading order, subtractive or multiplicative matched simulation,
\begin{verbatim}
read Matchbox/MCatLO-DefaultShower.in
read Matchbox/MCatNLO-DefaultShower.in
read Matchbox/Powheg-DefaultShower.in
\end{verbatim}
respectively. The \Matchbox\ build-in matrix elements for $e^+e^-\to
\text{jets}$ at leading and next-to-leading order are employed in our
simulation. Unless stated otherwise, matrix elment corrections are switched
off,
\begin{verbatim}
set /Herwig/Shower/ShowerHandler:HardEmission None
\end{verbatim}

\subsection{Massless case}

In the massless case we generate two-jet events using light flavours only
\begin{verbatim}
set Factory:OrderInAlphaS 0
set Factory:OrderInAlphaEW 2
do Factory:Process e+ e- -> j j
\end{verbatim}
No other special settings are applied.

\subsection{Massive case}

In the massive case, we produce top quark pairs on-shell,
\begin{verbatim}
read Matchbox/OnShellTopProduction.in
set Factory:OrderInAlphaS 0
set Factory:OrderInAlphaEW 2
do Factory:Process e+ e- -> t tbar
\end{verbatim}
We exclusively select leptonic decays,
\begin{verbatim}
do /Herwig/Particles/t:SelectDecayModes t->nu_mu,mu+,b; t->nu_e,e+,b;
do /Herwig/Particles/tbar:SelectDecayModes tbar->nu_mubar,mu-,bbar; \
 tbar->nu_ebar,e-,bbar; 
\end{verbatim}
If an independent cutoff on the top quark shower is desired, we use
\begin{verbatim}
cd /Herwig/Shower
do SplittingGenerator:DeleteFinalSplitting t->t,g; QtoQGSudakov
cp QtoQGSudakov TtoTGSudakov
do SplittingGenerator:AddFinalSplitting t->t,g; TtoTGSudakov
set /Herwig/Shower/TtoTGSudakov:pTmin X*GeV
\end{verbatim}

\bibliography{shower-top-mass}

\providecommand{\href}[2]{#2}\begingroup\raggedright\begin{thebibliography}{100}

\bibitem{Khachatryan:2015hba}
{\scshape CMS} collaboration, V.~Khachatryan et~al., \emph{{Measurement of the
  top quark mass using proton-proton data at ${\sqrt{(s)}}$ = 7 and 8 TeV}},
  \href{https://doi.org/10.1103/PhysRevD.93.072004}{\emph{Phys. Rev.}
  {\bfseries D93} (2016) 072004},
  [\href{https://arxiv.org/abs/1509.04044}{{\ttfamily 1509.04044}}].

\bibitem{Aaboud:2016igd}
{\scshape ATLAS} collaboration, M.~Aaboud et~al., \emph{{Measurement of the top
  quark mass in the $t\bar{t}\to$ dilepton channel from ${\sqrt{{s}}}$ = 8 TeV
  ATLAS data}},
  \href{https://doi.org/10.1016/j.physletb.2016.08.042}{\emph{Phys. Lett.}
  {\bfseries B761} (2016) 350--371},
  [\href{https://arxiv.org/abs/1606.02179}{{\ttfamily 1606.02179}}].

\bibitem{Tevatron:2014cka}
{\scshape CDF, D0} collaboration, T.~E.~W. Group, \emph{{Combination of CDF and
  D0 results on the mass of the top quark using up to 9.7 fb$^{-1}$ at the
  Tevatron}},  \href{https://arxiv.org/abs/1407.2682}{{\ttfamily 1407.2682}}.

\bibitem{Bellm:2016rhh}
J.~Bellm, G.~Nail, S.~Pl{\"a}tzer, P.~Schichtel and A.~Si{\'o}dmok,
  \emph{{Parton Shower Uncertainties with Herwig 7: Benchmarks at Leading
  Order}}, \href{https://doi.org/10.1140/epjc/s10052-016-4506-x}{\emph{Eur.
  Phys. J.} {\bfseries C76} (2016) 665},
  [\href{https://arxiv.org/abs/1605.01338}{{\ttfamily 1605.01338}}].

\bibitem{Bendavid:2018nar}
J.~R. Andersen et~al., \emph{{Les Houches 2017: Physics at TeV Colliders
  Standard Model Working Group Report}},  in \emph{{10th Les Houches Workshop
  on Physics at TeV Colliders (PhysTeV 2017) Les Houches, France, June 5-23,
  2017}}, 2018, \href{https://arxiv.org/abs/1803.07977}{{\ttfamily
  1803.07977}},
  \href{http://inspirehep.net/record/1663483/files/1803.07977.pdf}{http://inspirehep.net/record/1663483/files/1803.07977.pdf}.

\bibitem{Dasgupta:2018nvj}
M.~Dasgupta, F.~A. Dreyer, K.~Hamilton, P.~F. Monni and G.~P. Salam,
  \emph{{Logarithmic accuracy of parton showers: a fixed-order study}},
  \href{https://arxiv.org/abs/1805.09327}{{\ttfamily 1805.09327}}.

\bibitem{Baer:2013cma}
H.~Baer, T.~Barklow, K.~Fujii, Y.~Gao, A.~Hoang, S.~Kanemura et~al., \emph{{The
  International Linear Collider Technical Design Report - Volume 2: Physics}},
  \href{https://arxiv.org/abs/1306.6352}{{\ttfamily 1306.6352}}.

\bibitem{Asner:2013hla}
D.~Asner, A.~Hoang, Y.~Kiyo, R.~P{\"o}schl, Y.~Sumino and M.~Vos, \emph{{Top
  quark precision physics at the International Linear Collider}},  in
  \emph{{Proceedings, 2013 Community Summer Study on the Future of U.S.
  Particle Physics: Snowmass on the Mississippi (CSS2013): Minneapolis, MN,
  USA, July 29-August 6, 2013}}, 2013,
  \href{https://arxiv.org/abs/1307.8265}{{\ttfamily 1307.8265}},
  \href{http://inspirehep.net/record/1245485/files/arXiv:1307.8265.pdf}{http://inspirehep.net/record/1245485/files/arXiv:1307.8265.pdf}.

\bibitem{Vos:2016til}
M.~Vos et~al., \emph{{Top physics at high-energy lepton colliders}},
  \href{https://arxiv.org/abs/1604.08122}{{\ttfamily 1604.08122}}.

\bibitem{Jain:2008gb}
A.~Jain, I.~Scimemi and I.~W. Stewart, \emph{{Two-loop Jet-Function and
  Jet-Mass for Top Quarks}},
  \href{https://doi.org/10.1103/PhysRevD.77.094008}{\emph{Phys. Rev.}
  {\bfseries D77} (2008) 094008},
  [\href{https://arxiv.org/abs/0801.0743}{{\ttfamily 0801.0743}}].

\bibitem{Butenschoen:2016lpz}
M.~Butenschoen, B.~Dehnadi, A.~H. Hoang, V.~Mateu, M.~Preisser and I.~W.
  Stewart, \emph{{Top Quark Mass Calibration for Monte Carlo Event
  Generators}}, {\emph{Phys. Rev. Lett.} {\bfseries 117} (2016) 232001},
  [\href{https://arxiv.org/abs/1608.01318}{{\ttfamily 1608.01318}}].

\bibitem{Hoang:2000yr}
A.~H. Hoang et~al., \emph{{Top - anti-top pair production close to threshold:
  Synopsis of recent NNLO results}},
  \href{https://doi.org/10.1007/s1010500c0003}{\emph{Eur. Phys. J.direct}
  {\bfseries 2} (2000) 3},
  [\href{https://arxiv.org/abs/hep-ph/0001286}{{\ttfamily hep-ph/0001286}}].

\bibitem{Hoang:2008xm}
A.~H. Hoang and I.~W. Stewart, \emph{{Top Mass Measurements from Jets and the
  Tevatron Top-Quark Mass}},
  \href{https://doi.org/10.1016/j.nuclphysbps.2008.10.028}{\emph{Nucl. Phys.
  Proc. Suppl.} {\bfseries 185} (2008) 220--226},
  [\href{https://arxiv.org/abs/0808.0222}{{\ttfamily 0808.0222}}].

\bibitem{Fleming:2007qr}
S.~Fleming, A.~H. Hoang, S.~Mantry and I.~W. Stewart, \emph{{Jets from massive
  unstable particles: Top-mass determination}},
  \href{https://doi.org/10.1103/PhysRevD.77.074010}{\emph{Phys. Rev.}
  {\bfseries D77} (2008) 074010},
  [\href{https://arxiv.org/abs/hep-ph/0703207}{{\ttfamily hep-ph/0703207}}].

\bibitem{Fleming:2007xt}
S.~Fleming, A.~H. Hoang, S.~Mantry and I.~W. Stewart, \emph{{Top Jets in the
  Peak Region: Factorization Analysis with NLL Resummation}},
  \href{https://doi.org/10.1103/PhysRevD.77.114003}{\emph{Phys. Rev.}
  {\bfseries D77} (2008) 114003},
  [\href{https://arxiv.org/abs/0711.2079}{{\ttfamily 0711.2079}}].

\bibitem{Hoang:2014oea}
A.~H. Hoang, \emph{{The Top Mass: Interpretation and Theoretical
  Uncertainties}},  in \emph{{7th International Workshop on Top Quark Physics
  (TOP2014) Cannes, France, September 28-October 3, 2014}}, 2014,
  \href{https://arxiv.org/abs/1412.3649}{{\ttfamily 1412.3649}},
  \href{http://inspirehep.net/record/1333866/files/arXiv:1412.3649.pdf}{http://inspirehep.net/record/1333866/files/arXiv:1412.3649.pdf}.

\bibitem{Neubert:1993mb}
M.~Neubert, \emph{{Heavy quark symmetry}},
  \href{https://doi.org/10.1016/0370-1573(94)90091-4}{\emph{Phys. Rept.}
  {\bfseries 245} (1994) 259--396},
  [\href{https://arxiv.org/abs/hep-ph/9306320}{{\ttfamily hep-ph/9306320}}].

\bibitem{Manohar:2000dt}
A.~V. Manohar and M.~B. Wise, \emph{{Heavy quark physics}}, {\emph{Camb.
  Monogr. Part. Phys. Nucl. Phys. Cosmol.} {\bfseries 10} (2000) 1--191}.

\bibitem{Hoang:2008yj}
A.~H. Hoang, A.~Jain, I.~Scimemi and I.~W. Stewart, \emph{{Infrared
  Renormalization Group Flow for Heavy Quark Masses}},
  \href{https://doi.org/10.1103/PhysRevLett.101.151602}{\emph{Phys. Rev. Lett.}
  {\bfseries 101} (2008) 151602},
  [\href{https://arxiv.org/abs/0803.4214}{{\ttfamily 0803.4214}}].

\bibitem{Hoang:2017suc}
A.~H. Hoang, A.~Jain, C.~Lepenik, V.~Mateu, M.~Preisser, I.~Scimemi et~al.,
  \emph{{The MSR mass and the $
  \mathcal{O}\left({\Lambda}_{\mathrm{QCD}}\right) $ renormalon sum rule}},
  \href{https://doi.org/10.1007/JHEP04(2018)003}{\emph{JHEP} {\bfseries 04}
  (2018) 003}, [\href{https://arxiv.org/abs/1704.01580}{{\ttfamily
  1704.01580}}].

\bibitem{Nason:2017cxd}
P.~Nason, \emph{{The Top Mass in Hadronic Collisions}},
\newblock 2017.
\newblock \href{https://arxiv.org/abs/1712.02796}{{\ttfamily 1712.02796}}.

\bibitem{Sjostrand:2014zea}
T.~Sj{\"o}strand, S.~Ask, J.~R. Christiansen, R.~Corke, N.~Desai, P.~Ilten
  et~al., \emph{{An Introduction to PYTHIA 8.2}},
  \href{https://doi.org/10.1016/j.cpc.2015.01.024}{\emph{Comput. Phys. Commun.}
  {\bfseries 191} (2015) 159--177},
  [\href{https://arxiv.org/abs/1410.3012}{{\ttfamily 1410.3012}}].

\bibitem{Skands:2014pea}
P.~Skands, S.~Carrazza and J.~Rojo, \emph{{Tuning PYTHIA 8.1: the Monash 2013
  Tune}}, \href{https://doi.org/10.1140/epjc/s10052-014-3024-y}{\emph{Eur.
  Phys. J.} {\bfseries C74} (2014) 3024},
  [\href{https://arxiv.org/abs/1404.5630}{{\ttfamily 1404.5630}}].

\bibitem{Hoang:2017kmk}
A.~H. Hoang, S.~Mantry, A.~Pathak and I.~W. Stewart, \emph{{Extracting a Short
  Distance Top Mass with Light Grooming}},
  \href{https://arxiv.org/abs/1708.02586}{{\ttfamily 1708.02586}}.

\bibitem{Dasgupta:2013ihk}
M.~Dasgupta, A.~Fregoso, S.~Marzani and G.~P. Salam, \emph{{Towards an
  understanding of jet substructure}},
  \href{https://doi.org/10.1007/JHEP09(2013)029}{\emph{JHEP} {\bfseries 09}
  (2013) 029}, [\href{https://arxiv.org/abs/1307.0007}{{\ttfamily 1307.0007}}].

\bibitem{Larkoski:2014wba}
A.~J. Larkoski, S.~Marzani, G.~Soyez and J.~Thaler, \emph{{Soft Drop}},
  \href{https://doi.org/10.1007/JHEP05(2014)146}{\emph{JHEP} {\bfseries 05}
  (2014) 146}, [\href{https://arxiv.org/abs/1402.2657}{{\ttfamily 1402.2657}}].

\bibitem{Kieseler:2015jzh}
J.~Kieseler, K.~Lipka and S.-O. Moch, \emph{{Calibration of the Top-Quark Monte
  Carlo Mass}},
  \href{https://doi.org/10.1103/PhysRevLett.116.162001}{\emph{Phys. Rev. Lett.}
  {\bfseries 116} (2016) 162001},
  [\href{https://arxiv.org/abs/1511.00841}{{\ttfamily 1511.00841}}].

\bibitem{Kim:2017rve}
{\scshape CMS} collaboration, J.~H. Kim, \emph{{Alternative methods for top
  quark mass measurements at the CMS}},
  \href{https://doi.org/10.1051/epjconf/201714108006}{\emph{EPJ Web Conf.}
  {\bfseries 141} (2017) 08006}.

\bibitem{Vos:2016tof}
{\scshape ATLAS, CMS} collaboration, M.~Vos, \emph{{Top-quark mass measurements
  at the LHC: alternative methods}}, {\emph{PoS} {\bfseries TOP2015} (2016)
  035}, [\href{https://arxiv.org/abs/1602.00428}{{\ttfamily 1602.00428}}].

\bibitem{Adomeit:2014yna}
S.~Adomeit, \emph{{Top-quark mass measurements: Alternative techniques (LHC +
  Tevatron)}},  in \emph{{Proceedings, 7th International Workshop on Top Quark
  Physics (TOP2014): Cannes, France, September 28-October 3, 2014}}, 2014,
  \href{https://arxiv.org/abs/1411.7917}{{\ttfamily 1411.7917}},
  \href{http://inspirehep.net/record/1331406/files/arXiv:1411.7917.pdf}{http://inspirehep.net/record/1331406/files/arXiv:1411.7917.pdf}.

\bibitem{Corcella:2017rpt}
G.~Corcella, R.~Franceschini and D.~Kim, \emph{{Fragmentation Uncertainties in
  Hadronic Observables for Top-quark Mass Measurements}},
  \href{https://doi.org/10.1016/j.nuclphysb.2018.02.012}{\emph{Nucl. Phys.}
  {\bfseries B929} (2018) 485--526},
  [\href{https://arxiv.org/abs/1712.05801}{{\ttfamily 1712.05801}}].

\bibitem{Agashe:2016bok}
K.~Agashe, R.~Franceschini, D.~Kim and M.~Schulze, \emph{{Top quark mass
  determination from the energy peaks of b-jets and B-hadrons at NLO QCD}},
  \href{https://doi.org/10.1140/epjc/s10052-016-4494-x}{\emph{Eur. Phys. J.}
  {\bfseries C76} (2016) 636},
  [\href{https://arxiv.org/abs/1603.03445}{{\ttfamily 1603.03445}}].

\bibitem{Biswas:2010sa}
S.~Biswas, K.~Melnikov and M.~Schulze, \emph{{Next-to-leading order QCD effects
  and the top quark mass measurements at the LHC}},
  \href{https://doi.org/10.1007/JHEP08(2010)048}{\emph{JHEP} {\bfseries 08}
  (2010) 048}, [\href{https://arxiv.org/abs/1006.0910}{{\ttfamily 1006.0910}}].

\bibitem{Lester:1999tx}
C.~G. Lester and D.~J. Summers, \emph{{Measuring masses of semiinvisibly
  decaying particles pair produced at hadron colliders}},
  \href{https://doi.org/10.1016/S0370-2693(99)00945-4}{\emph{Phys. Lett.}
  {\bfseries B463} (1999) 99--103},
  [\href{https://arxiv.org/abs/hep-ph/9906349}{{\ttfamily hep-ph/9906349}}].

\bibitem{Matchev:2009ad}
K.~T. Matchev and M.~Park, \emph{{A General method for determining the masses
  of semi-invisibly decaying particles at hadron colliders}},
  \href{https://doi.org/10.1103/PhysRevLett.107.061801}{\emph{Phys. Rev. Lett.}
  {\bfseries 107} (2011) 061801},
  [\href{https://arxiv.org/abs/0910.1584}{{\ttfamily 0910.1584}}].

\bibitem{CMS:2012eya}
{\scshape CMS} collaboration, \emph{{Mass determination in the $t\bar t$ system
  with kinematic endpoints}},  Tech. Rep. CMS-PAS-TOP-11-027, CERN, Geneva,
  2012.

\bibitem{Sirunyan:2017idq}
{\scshape CMS} collaboration, A.~M. Sirunyan et~al., \emph{{Measurement of the
  top quark mass in the dileptonic $t\bar{t}$ decay channel using the mass
  observables $M_{b\ell}$, $M_{T2}$, and $M_{b\ell\nu}$ in pp collisions at
  ${\sqrt{{s}}}$ = 8 TeV}},
  \href{https://doi.org/10.1103/PhysRevD.96.032002}{\emph{Phys. Rev.}
  {\bfseries D96} (2017) 032002},
  [\href{https://arxiv.org/abs/1704.06142}{{\ttfamily 1704.06142}}].

\bibitem{Heinrich:2017bqp}
G.~Heinrich, A.~Maier, R.~Nisius, J.~Schlenk, M.~Schulze, L.~Scyboz et~al.,
  \emph{{NLO and off-shell effects in top quark mass determinations}},
  \href{https://arxiv.org/abs/1709.08615}{{\ttfamily 1709.08615}}.

\bibitem{Gleisberg:2008ta}
T.~Gleisberg, S.~Hoeche, F.~Krauss, M.~Schonherr, S.~Schumann, F.~Siegert
  et~al., \emph{{Event generation with SHERPA 1.1}},
  \href{https://doi.org/10.1088/1126-6708/2009/02/007}{\emph{JHEP} {\bfseries
  02} (2009) 007}, [\href{https://arxiv.org/abs/0811.4622}{{\ttfamily
  0811.4622}}].

\bibitem{Ravasio:2018lzi}
S.~Ferrario~Ravasio, T.~Jezo, P.~Nason and C.~Oleari, \emph{{A Theoretical
  Study of Top-Mass Measurements at the LHC Using NLO+PS Generators of
  Increasing Accuracy}},  \href{https://arxiv.org/abs/1801.03944}{{\ttfamily
  1801.03944}}.

\bibitem{Frixione:2007vw}
S.~Frixione, P.~Nason and C.~Oleari, \emph{{Matching NLO QCD computations with
  Parton Shower simulations: the POWHEG method}},
  \href{https://doi.org/10.1088/1126-6708/2007/11/070}{\emph{JHEP} {\bfseries
  11} (2007) 070}, [\href{https://arxiv.org/abs/0709.2092}{{\ttfamily
  0709.2092}}].

\bibitem{Alioli:2010xd}
S.~Alioli, P.~Nason, C.~Oleari and E.~Re, \emph{{A general framework for
  implementing NLO calculations in shower Monte Carlo programs: the POWHEG
  BOX}}, \href{https://doi.org/10.1007/JHEP06(2010)043}{\emph{JHEP} {\bfseries
  06} (2010) 043}, [\href{https://arxiv.org/abs/1002.2581}{{\ttfamily
  1002.2581}}].

\bibitem{Frixione:2007nw}
S.~Frixione, P.~Nason and G.~Ridolfi, \emph{{A Positive-weight
  next-to-leading-order Monte Carlo for heavy flavour hadroproduction}},
  \href{https://doi.org/10.1088/1126-6708/2007/09/126}{\emph{JHEP} {\bfseries
  09} (2007) 126}, [\href{https://arxiv.org/abs/0707.3088}{{\ttfamily
  0707.3088}}].

\bibitem{Campbell:2014kua}
J.~M. Campbell, R.~K. Ellis, P.~Nason and E.~Re, \emph{{Top-Pair Production and
  Decay at NLO Matched with Parton Showers}},
  \href{https://doi.org/10.1007/JHEP04(2015)114}{\emph{JHEP} {\bfseries 04}
  (2015) 114}, [\href{https://arxiv.org/abs/1412.1828}{{\ttfamily 1412.1828}}].

\bibitem{Jezo:2016ujg}
T.~Jezo, J.~M. Lindert, P.~Nason, C.~Oleari and S.~Pozzorini, \emph{{An NLO+PS
  generator for $t\bar{t}$ and $Wt$ production and decay including non-resonant
  and interference effects}},
  \href{https://doi.org/10.1140/epjc/s10052-016-4538-2}{\emph{Eur. Phys. J.}
  {\bfseries C76} (2016) 691},
  [\href{https://arxiv.org/abs/1607.04538}{{\ttfamily 1607.04538}}].

\bibitem{Bellm:2015jjp}
J.~Bellm et~al., \emph{{Herwig 7.0/Herwig++ 3.0 release note}},
  \href{https://doi.org/10.1140/epjc/s10052-016-4018-8}{\emph{Eur. Phys. J.}
  {\bfseries C76} (2016) 196},
  [\href{https://arxiv.org/abs/1512.01178}{{\ttfamily 1512.01178}}].

\bibitem{Bellm:2017bvx}
J.~Bellm et~al., \emph{{Herwig 7.1 Release Note}},
  \href{https://arxiv.org/abs/1705.06919}{{\ttfamily 1705.06919}}.

\bibitem{Frixione:2014ala}
S.~Frixione and A.~Mitov, \emph{{Determination of the top quark mass from
  leptonic observables}},
  \href{https://doi.org/10.1007/JHEP09(2014)012}{\emph{JHEP} {\bfseries 09}
  (2014) 012}, [\href{https://arxiv.org/abs/1407.2763}{{\ttfamily 1407.2763}}].

\bibitem{Dokshitzer:1991fc}
Y.~L. Dokshitzer, V.~A. Khoze and S.~I. Troian, \emph{{Particle spectra in
  light and heavy quark jets}},
  \href{https://doi.org/10.1088/0954-3899/17/10/003}{\emph{J. Phys.} {\bfseries
  G17} (1991) 1481--1492}.

\bibitem{Dokshitzer:1991fd}
Y.~L. Dokshitzer, V.~A. Khoze and S.~I. Troian, \emph{{On specific QCD
  properties of heavy quark fragmentation ('dead cone')}},
  \href{https://doi.org/10.1088/0954-3899/17/10/023}{\emph{J. Phys.} {\bfseries
  G17} (1991) 1602--1604}.

\bibitem{Maltoni:2016ays}
F.~Maltoni, M.~Selvaggi and J.~Thaler, \emph{{Exposing the dead cone effect
  with jet substructure techniques}},
  \href{https://doi.org/10.1103/PhysRevD.94.054015}{\emph{Phys. Rev.}
  {\bfseries D94} (2016) 054015},
  [\href{https://arxiv.org/abs/1606.03449}{{\ttfamily 1606.03449}}].

\bibitem{Marchesini:1983bm}
G.~Marchesini and B.~R. Webber, \emph{{Simulation of QCD Jets Including Soft
  Gluon Interference}},
  \href{https://doi.org/10.1016/0550-3213(84)90463-2}{\emph{Nucl. Phys.}
  {\bfseries B238} (1984) 1--29}.

\bibitem{Marchesini:1987cf}
G.~Marchesini and B.~R. Webber, \emph{{Monte Carlo Simulation of General Hard
  Processes with Coherent QCD Radiation}},
  \href{https://doi.org/10.1016/0550-3213(88)90089-2}{\emph{Nucl. Phys.}
  {\bfseries B310} (1988) 461--526}.

\bibitem{Catani:1990rr}
S.~Catani, B.~R. Webber and G.~Marchesini, \emph{{QCD coherent branching and
  semiinclusive processes at large x}},
  \href{https://doi.org/10.1016/0550-3213(91)90390-J}{\emph{Nucl. Phys.}
  {\bfseries B349} (1991) 635--654}.

\bibitem{Gieseke:2003rz}
S.~Gieseke, P.~Stephens and B.~Webber, \emph{{New formalism for QCD parton
  showers}}, \href{https://doi.org/10.1088/1126-6708/2003/12/045}{\emph{JHEP}
  {\bfseries 12} (2003) 045},
  [\href{https://arxiv.org/abs/hep-ph/0310083}{{\ttfamily hep-ph/0310083}}].

\bibitem{Krauss:2003cr}
F.~Krauss and G.~Rodrigo, \emph{{Resummed jet rates for e+ e- annihilation into
  massive quarks}},
  \href{https://doi.org/10.1016/j.physletb.2003.09.096}{\emph{Phys. Lett.}
  {\bfseries B576} (2003) 135--142},
  [\href{https://arxiv.org/abs/hep-ph/0303038}{{\ttfamily hep-ph/0303038}}].

\bibitem{Rodrigo:2003ws}
G.~Rodrigo and F.~Krauss, \emph{{Resummed jet rates for heavy quark production
  in $e^+e^-$ annihilation}},
  \href{https://doi.org/10.1140/epjcd/s2004-03-1836-x}{\emph{Eur. Phys. J.}
  {\bfseries C33} (2004) S457--S459},
  [\href{https://arxiv.org/abs/hep-ph/0309325}{{\ttfamily hep-ph/0309325}}].

\bibitem{Catani:1992ua}
S.~Catani, L.~Trentadue, G.~Turnock and B.~R. Webber, \emph{{Resummation of
  large logarithms in e+ e- event shape distributions}},
  \href{https://doi.org/10.1016/0550-3213(93)90271-P}{\emph{Nucl. Phys.}
  {\bfseries B407} (1993) 3--42}.

\bibitem{Abbate:2010xh}
R.~Abbate, M.~Fickinger, A.~H. Hoang, V.~Mateu and I.~W. Stewart, \emph{{Thrust
  at N$^3$LL with Power Corrections and a Precision Global Fit for
  $\alpha_s(M_Z)$}},
  \href{https://doi.org/10.1103/PhysRevD.83.074021}{\emph{Phys. Rev.}
  {\bfseries D83} (2011) 074021},
  [\href{https://arxiv.org/abs/1006.3080}{{\ttfamily 1006.3080}}].

\bibitem{Farhi:1977sg}
E.~Farhi, \emph{{A QCD Test for Jets}},
  \href{https://doi.org/10.1103/PhysRevLett.39.1587}{\emph{Phys. Rev. Lett.}
  {\bfseries 39} (1977) 1587--1588}.

\bibitem{Stewart:2010tn}
I.~W. Stewart, F.~J. Tackmann and W.~J. Waalewijn, \emph{{N-Jettiness: An
  Inclusive Event Shape to Veto Jets}},
  \href{https://doi.org/10.1103/PhysRevLett.105.092002}{\emph{Phys. Rev. Lett.}
  {\bfseries 105} (2010) 092002},
  [\href{https://arxiv.org/abs/1004.2489}{{\ttfamily 1004.2489}}].

\bibitem{Catani:1989ne}
S.~Catani and L.~Trentadue, \emph{{Resummation of the QCD Perturbative Series
  for Hard Processes}},
  \href{https://doi.org/10.1016/0550-3213(89)90273-3}{\emph{Nucl. Phys.}
  {\bfseries B327} (1989) 323--352}.

\bibitem{Korchemsky:1999kt}
G.~P. Korchemsky and G.~F. Sterman, \emph{{Power corrections to event shapes
  and factorization}},
  \href{https://doi.org/10.1016/S0550-3213(99)00308-9}{\emph{Nucl. Phys.}
  {\bfseries B555} (1999) 335--351},
  [\href{https://arxiv.org/abs/hep-ph/9902341}{{\ttfamily hep-ph/9902341}}].

\bibitem{Berger:2003iw}
C.~F. Berger, T.~Kucs and G.~F. Sterman, \emph{{Event shape / energy flow
  correlations}}, \href{https://doi.org/10.1103/PhysRevD.68.014012}{\emph{Phys.
  Rev.} {\bfseries D68} (2003) 014012},
  [\href{https://arxiv.org/abs/hep-ph/0303051}{{\ttfamily hep-ph/0303051}}].

\bibitem{Davison:2008vx}
R.~A. Davison and B.~R. Webber, \emph{{Non-Perturbative Contribution to the
  Thrust Distribution in e+ e- Annihilation}},
  \href{https://doi.org/10.1140/epjc/s10052-008-0836-7}{\emph{Eur. Phys. J.}
  {\bfseries C59} (2009) 13--25},
  [\href{https://arxiv.org/abs/0809.3326}{{\ttfamily 0809.3326}}].

\bibitem{Schwartz:2007ib}
M.~D. Schwartz, \emph{{Resummation and NLO matching of event shapes with
  effective field theory}},
  \href{https://doi.org/10.1103/PhysRevD.77.014026}{\emph{Phys. Rev.}
  {\bfseries D77} (2008) 014026},
  [\href{https://arxiv.org/abs/0709.2709}{{\ttfamily 0709.2709}}].

\bibitem{Becher:2008cf}
T.~Becher and M.~D. Schwartz, \emph{{A precise determination of $\alpha_s$ from
  LEP thrust data using effective field theory}},
  \href{https://doi.org/10.1088/1126-6708/2008/07/034}{\emph{JHEP} {\bfseries
  07} (2008) 034}, [\href{https://arxiv.org/abs/0803.0342}{{\ttfamily
  0803.0342}}].

\bibitem{Almeida:2014uva}
L.~G. Almeida, S.~D. Ellis, C.~Lee, G.~Sterman, I.~Sung and J.~R. Walsh,
  \emph{{Comparing and counting logs in direct and effective methods of QCD
  resummation}}, \href{https://doi.org/10.1007/JHEP04(2014)174}{\emph{JHEP}
  {\bfseries 04} (2014) 174},
  [\href{https://arxiv.org/abs/1401.4460}{{\ttfamily 1401.4460}}].

\bibitem{Manohar:2006nz}
A.~V. Manohar and I.~W. Stewart, \emph{{The Zero-Bin and Mode Factorization in
  Quantum Field Theory}},
  \href{https://doi.org/10.1103/PhysRevD.76.074002}{\emph{Phys. Rev.}
  {\bfseries D76} (2007) 074002},
  [\href{https://arxiv.org/abs/hep-ph/0605001}{{\ttfamily hep-ph/0605001}}].

\bibitem{Hoang:2007vb}
A.~H. Hoang and I.~W. Stewart, \emph{{Designing gapped soft functions for jet
  production}},
  \href{https://doi.org/10.1016/j.physletb.2008.01.040}{\emph{Phys. Lett.}
  {\bfseries B660} (2008) 483--493},
  [\href{https://arxiv.org/abs/0709.3519}{{\ttfamily 0709.3519}}].

\bibitem{Bahr:2008pv}
M.~B{\"a}hr et~al., \emph{{Herwig++ Physics and Manual}},
  \href{https://doi.org/10.1140/epjc/s10052-008-0798-9}{\emph{Eur. Phys. J.}
  {\bfseries C58} (2008) 639--707},
  [\href{https://arxiv.org/abs/0803.0883}{{\ttfamily 0803.0883}}].

\bibitem{Collins:1985xx}
J.~C. Collins and D.~E. Soper, \emph{{The Two Particle Inclusive Cross-section
  in $e^+ e^-$ Annihilation at {PETRA}, {PEP} and {LEP} Energies}},
  \href{https://doi.org/10.1016/0550-3213(87)90035-6}{\emph{Nucl. Phys.}
  {\bfseries B284} (1987) 253--270}.

\bibitem{Platzer:2009jq}
S.~Pl{\"a}tzer and S.~Gieseke, \emph{{Coherent Parton Showers with Local
  Recoils}}, \href{https://doi.org/10.1007/JHEP01(2011)024}{\emph{JHEP}
  {\bfseries 01} (2011) 024},
  [\href{https://arxiv.org/abs/0909.5593}{{\ttfamily 0909.5593}}].

\bibitem{Reichelt:2017hts}
D.~Reichelt, P.~Richardson and A.~Siodmok, \emph{{Improving the Simulation of
  Quark and Gluon Jets with Herwig 7}},
  \href{https://doi.org/10.1140/epjc/s10052-017-5374-8}{\emph{Eur. Phys. J.}
  {\bfseries C77} (2017) 876},
  [\href{https://arxiv.org/abs/1708.01491}{{\ttfamily 1708.01491}}].

\bibitem{Webber:1983if}
B.~R. Webber, \emph{{A QCD Model for Jet Fragmentation Including Soft Gluon
  Interference}},
  \href{https://doi.org/10.1016/0550-3213(84)90333-X}{\emph{Nucl. Phys.}
  {\bfseries B238} (1984) 492--528}.

\bibitem{Contopanagos:1996nh}
H.~Contopanagos, E.~Laenen and G.~F. Sterman, \emph{{Sudakov factorization and
  resummation}},
  \href{https://doi.org/10.1016/S0550-3213(96)00567-6}{\emph{Nucl. Phys.}
  {\bfseries B484} (1997) 303--330},
  [\href{https://arxiv.org/abs/hep-ph/9604313}{{\ttfamily hep-ph/9604313}}].

\bibitem{Beneke:1992ch}
M.~Beneke, \emph{{Large order perturbation theory for a physical quantity}},
  \href{https://doi.org/10.1016/0550-3213(93)90554-3}{\emph{Nucl. Phys.}
  {\bfseries B405} (1993) 424--450}.

\bibitem{Ball:1995ni}
P.~Ball, M.~Beneke and V.~M. Braun, \emph{{Resummation of (beta0 alpha-s)**n
  corrections in QCD: Techniques and applications to the tau hadronic width and
  the heavy quark pole mass}},
  \href{https://doi.org/10.1016/0550-3213(95)00392-6}{\emph{Nucl. Phys.}
  {\bfseries B452} (1995) 563--625},
  [\href{https://arxiv.org/abs/hep-ph/9502300}{{\ttfamily hep-ph/9502300}}].

\bibitem{Beneke:1994rs}
M.~Beneke, \emph{{More on ambiguities in the pole mass}},
  \href{https://doi.org/10.1016/0370-2693(94)01505-7}{\emph{Phys. Lett.}
  {\bfseries B344} (1995) 341--347},
  [\href{https://arxiv.org/abs/hep-ph/9408380}{{\ttfamily hep-ph/9408380}}].

\bibitem{Beneke:1994sw}
M.~Beneke and V.~M. Braun, \emph{{Heavy quark effective theory beyond
  perturbation theory: Renormalons, the pole mass and the residual mass term}},
  \href{https://doi.org/10.1016/0550-3213(94)90314-X}{\emph{Nucl. Phys.}
  {\bfseries B426} (1994) 301--343},
  [\href{https://arxiv.org/abs/hep-ph/9402364}{{\ttfamily hep-ph/9402364}}].

\bibitem{Hoang:2008fs}
A.~H. Hoang and S.~Kluth, \emph{{Hemisphere Soft Function at ${\cal
  O}(\alpha_s^2)$ for Dijet Production in $e^+e^-$ Annihilation}},
  \href{https://arxiv.org/abs/0806.3852}{{\ttfamily 0806.3852}}.

\bibitem{Hoang:2017btd}
A.~H. Hoang, C.~Lepenik and M.~Preisser, \emph{{On the Light Massive Flavor
  Dependence of the Large Order Asymptotic Behavior and the Ambiguity of the
  Pole Mass}}, \href{https://doi.org/10.1007/JHEP09(2017)099}{\emph{JHEP}
  {\bfseries 09} (2017) 099},
  [\href{https://arxiv.org/abs/1706.08526}{{\ttfamily 1706.08526}}].

\bibitem{Hamilton:2006ms}
K.~Hamilton and P.~Richardson, \emph{{A Simulation of QCD radiation in top
  quark decays}},
  \href{https://doi.org/10.1088/1126-6708/2007/02/069}{\emph{JHEP} {\bfseries
  02} (2007) 069}, [\href{https://arxiv.org/abs/hep-ph/0612236}{{\ttfamily
  hep-ph/0612236}}].

\bibitem{Buckley:2010ar}
A.~Buckley, J.~Butterworth, L.~Lonnblad, D.~Grellscheid, H.~Hoeth, J.~Monk
  et~al., \emph{{Rivet user manual}},
  \href{https://doi.org/10.1016/j.cpc.2013.05.021}{\emph{Comput. Phys. Commun.}
  {\bfseries 184} (2013) 2803--2819},
  [\href{https://arxiv.org/abs/1003.0694}{{\ttfamily 1003.0694}}].

\bibitem{Cacciari:2011ma}
M.~Cacciari, G.~P. Salam and G.~Soyez, \emph{{FastJet User Manual}},
  \href{https://doi.org/10.1140/epjc/s10052-012-1896-2}{\emph{Eur. Phys. J.}
  {\bfseries C72} (2012) 1896},
  [\href{https://arxiv.org/abs/1111.6097}{{\ttfamily 1111.6097}}].

\bibitem{Frixione:2002ik}
S.~Frixione and B.~R. Webber, \emph{{Matching NLO QCD computations and parton
  shower simulations}},
  \href{https://doi.org/10.1088/1126-6708/2002/06/029}{\emph{JHEP} {\bfseries
  06} (2002) 029}, [\href{https://arxiv.org/abs/hep-ph/0204244}{{\ttfamily
  hep-ph/0204244}}].

\bibitem{Platzer:2011bc}
S.~Pl{\"a}tzer and S.~Gieseke, \emph{{Dipole Showers and Automated NLO Matching
  in Herwig++}},
  \href{https://doi.org/10.1140/epjc/s10052-012-2187-7}{\emph{Eur. Phys. J.}
  {\bfseries C72} (2012) 2187},
  [\href{https://arxiv.org/abs/1109.6256}{{\ttfamily 1109.6256}}].

\bibitem{Hoeche:2011fd}
S.~Hoeche, F.~Krauss, M.~Schonherr and F.~Siegert, \emph{{A critical appraisal
  of NLO+PS matching methods}},
  \href{https://doi.org/10.1007/JHEP09(2012)049}{\emph{JHEP} {\bfseries 09}
  (2012) 049}, [\href{https://arxiv.org/abs/1111.1220}{{\ttfamily 1111.1220}}].

\bibitem{Nason:2012pr}
P.~Nason and B.~Webber, \emph{{Next-to-Leading-Order Event Generators}},
  \href{https://doi.org/10.1146/annurev-nucl-102711-094928}{\emph{Ann. Rev.
  Nucl. Part. Sci.} {\bfseries 62} (2012) 187--213},
  [\href{https://arxiv.org/abs/1202.1251}{{\ttfamily 1202.1251}}].

\bibitem{Platzer:2012np}
S.~Pl{\"a}tzer and M.~Sj{\"o}dahl, \emph{{Subleading $N_c$ improved Parton
  Showers}}, \href{https://doi.org/10.1007/JHEP07(2012)042}{\emph{JHEP}
  {\bfseries 07} (2012) 042},
  [\href{https://arxiv.org/abs/1201.0260}{{\ttfamily 1201.0260}}].

\bibitem{Collins:1987cp}
J.~C. Collins, \emph{{Spin Correlations in Monte Carlo Event Generators}},
  \href{https://doi.org/10.1016/0550-3213(88)90654-2}{\emph{Nucl. Phys.}
  {\bfseries B304} (1988) 794--804}.

\bibitem{Knowles:1987cu}
I.~G. Knowles, \emph{{Angular Correlations in {QCD}}},
  \href{https://doi.org/10.1016/0550-3213(88)90653-0}{\emph{Nucl. Phys.}
  {\bfseries B304} (1988) 767--793}.

\bibitem{Richardson:2001df}
P.~Richardson, \emph{{Spin correlations in Monte Carlo simulations}},
  \href{https://doi.org/10.1088/1126-6708/2001/11/029}{\emph{JHEP} {\bfseries
  11} (2001) 029}, [\href{https://arxiv.org/abs/hep-ph/0110108}{{\ttfamily
  hep-ph/0110108}}].

\bibitem{Hoche:2010kg}
S.~Hoche, F.~Krauss, M.~Schonherr and F.~Siegert, \emph{{NLO matrix elements
  and truncated showers}},
  \href{https://doi.org/10.1007/JHEP08(2011)123}{\emph{JHEP} {\bfseries 08}
  (2011) 123}, [\href{https://arxiv.org/abs/1009.1127}{{\ttfamily 1009.1127}}].

\bibitem{Beneke:2016cbu}
M.~Beneke, P.~Marquard, P.~Nason and M.~Steinhauser, \emph{{On the ultimate
  uncertainty of the top quark pole mass}},
  \href{https://doi.org/10.1016/j.physletb.2017.10.054}{\emph{Phys. Lett.}
  {\bfseries B775} (2017) 63--70},
  [\href{https://arxiv.org/abs/1605.03609}{{\ttfamily 1605.03609}}].

\bibitem{CMS:2013wfa}
{\scshape CMS} collaboration, \emph{{Projected improvement of the accuracy of
  top-quark mass measurements at the upgraded LHC}},  Tech. Rep.
  CMS-PAS-FTR-13-017, CERN, Geneva, 2013.

\bibitem{CMS:2017gvo}
{\scshape CMS} collaboration, \emph{{ECFA 2016: Prospects for selected standard
  model measurements with the CMS experiment at the High-Luminosity LHC}},
  Tech. Rep. CMS-PAS-FTR-16-006, CERN, Geneva, 2017.

\bibitem{Hoang:1998ng}
A.~H. Hoang, Z.~Ligeti and A.~V. Manohar, \emph{{B decay and the Upsilon
  mass}}, \href{https://doi.org/10.1103/PhysRevLett.82.277}{\emph{Phys. Rev.
  Lett.} {\bfseries 82} (1999) 277--280},
  [\href{https://arxiv.org/abs/hep-ph/9809423}{{\ttfamily hep-ph/9809423}}].

\bibitem{Hoang:1998hm}
A.~H. Hoang, Z.~Ligeti and A.~V. Manohar, \emph{{B decays in the upsilon
  expansion}}, \href{https://doi.org/10.1103/PhysRevD.59.074017}{\emph{Phys.
  Rev.} {\bfseries D59} (1999) 074017},
  [\href{https://arxiv.org/abs/hep-ph/9811239}{{\ttfamily hep-ph/9811239}}].

\bibitem{Hoang:1999ye}
A.~H. Hoang, \emph{{1S and $\overline{\rm MS}$ bottom quark masses from Upsilon
  sum rules}}, \href{https://doi.org/10.1103/PhysRevD.61.034005}{\emph{Phys.
  Rev.} {\bfseries D61} (2000) 034005},
  [\href{https://arxiv.org/abs/hep-ph/9905550}{{\ttfamily hep-ph/9905550}}].

\bibitem{Beneke:1998rk}
M.~Beneke, \emph{{A Quark mass definition adequate for threshold problems}},
  \href{https://doi.org/10.1016/S0370-2693(98)00741-2}{\emph{Phys. Lett.}
  {\bfseries B434} (1998) 115--125},
  [\href{https://arxiv.org/abs/hep-ph/9804241}{{\ttfamily hep-ph/9804241}}].

\bibitem{herwig-web}
\emph{{T}he {H}erwig {E}vent {G}enerator \url{http://herwig.hepforge.org}},
  2018.

\end{thebibliography}\endgroup

\end{document}